\documentclass[twocolappendix,numberedappendix,appendixfloats,iop,apj]{emulateapj}

\usepackage[usenames,dvipsnames]{color}
\usepackage{times, apjfonts}
\usepackage{amsmath, amsthm, amssymb}
\usepackage{graphicx, epsfig}
\usepackage{natbib}
\usepackage{epstopdf} 
\usepackage{CJK} 
\usepackage[colorlinks=true, anchorcolor=blue, linkcolor=blue, urlcolor=Fuchsia, citecolor=NavyBlue, breaklinks=true]{hyperref}
\usepackage[hyphenbreaks]{breakurl}

\shorttitle{COSMIC EVOLUTION OF BLACK HOLES AND SPHEROIDS. V.}
\shortauthors{Park et al.}

\newcommand{\mbh}{$M_{\rm BH}$}
\newcommand{\veldisp}{$\sigma_{*}$}
\newcommand{\Lbul}{$L_{\rm bul}$}
\newcommand{\Lhost}{$L_{\rm host}$}
\newcommand{\Mbul}{$M_{\rm bul}$}
\newcommand{\HST}{{\it HST}}
\newcommand{\msigma}{$M_{\rm BH}-\sigma_{*}$}
\newcommand{\mlbul}{$M_{\rm BH}-L_{\rm bul}$}
\newcommand{\mlhost}{$M_{\rm BH}-L_{\rm host}$}
\newcommand{\mmbul}{$M_{\rm BH}-M_{\rm bul}$}

\newcommand{\msun}{$M_{\odot}$}
\newcommand{\kms}{km~s$^{\rm -1}$}
\newcommand{\ergs}{erg~s$^{\rm -1}$}
\newcommand{\Hb}{H$\beta$}
\newcommand{\iron}{\ion{Fe}{2}}

\begin{document}
\begin{CJK*}{UTF8}{mj}

\title{COSMIC EVOLUTION OF BLACK HOLES AND SPHEROIDS. V. 
THE RELATION BETWEEN BLACK HOLE MASS AND HOST GALAXY LUMINOSITY FOR A SAMPLE OF 79 ACTIVE GALAXIES}
\author{Daeseong Park (박 대 성 )$^{1,2}$}
\author{Jong-Hak Woo (우 종 학 )$^{1,\dag}$}
\author{Vardha N. Bennert$^{3}$}
\author{Tommaso Treu$^{4,6}$}
\author{Matthew W. Auger$^{5}$}
\author{Matthew A. Malkan$^{6}$}

\affil{$^{1}$Astronomy Program, Department of Physics and Astronomy, Seoul National University, Seoul, 151-742, Republic of Korea; \href{mailto:pds2001@astro.snu.ac.kr}{pds2001@astro.snu.ac.kr},\href{mailto:woo@astro.snu.ac.kr}{woo@astro.snu.ac.kr}}
\affil{$^{2}$Department of Physics and Astronomy, University of California, Irvine, CA 92697, USA; \href{mailto:daeseong.park@uci.edu}{daeseong.park@uci.edu}}
\affil{$^{3}$Physics Department, California Polytechnic State University, San Luis Obispo, CA 93407, USA; \href{mailto:vbennert@calpoly.edu}{vbennert@calpoly.edu}}
\affil{$^{4}$Department of Physics, University of California, Santa Barbara, CA 93106, USA; \href{mailto:tt@physics.ucsb.edu}{tt@physics.ucsb.edu}}
\affil{$^{5}$Institute of Astronomy, University of Cambridge, Madingley Road, Cambridge CB3 0HA, UK; \href{mailto:mauger@ast.cam.ac.uk}{mauger@ast.cam.ac.uk}}
\affil{$^{6}$Department of Physics and Astronomy, University of California, Los Angeles, CA 90095, USA; \href{mailto:malkan@astro.ucla.edu}{malkan@astro.ucla.edu}}

\altaffiltext{*}{In this paper, we use the term "bulge" (abbreviated bul) interchangeably to refer to the host galaxy spheroid for
elliptical and lenticular galaxies as well as the bulge component of late-type galaxy.}

\altaffiltext{\dag}{Author to whom any correspondence should be addressed.}

\begin{abstract}
We investigate the cosmic evolution of the black hole (BH) mass -- bulge luminosity relation using 
a sample of 52 active galaxies at $z \sim 0.36$ and $z \sim 0.57$ in the BH mass range of $10^{7.4-9.1}$\msun.
By consistently applying multi-component spectral and structural decomposition to 
high-quality Keck spectra and high-resolution \HST\ images,
BH masses (\mbh) are estimated using the \Hb\ broad emission line combined with the 5100 \AA\ nuclear luminosity, 
and bulge luminosities (\Lbul) are derived from surface photometry.
Comparing the resulting \mlbul\ relation to local active galaxies and taking into account selection effects,
we find evolution of the form $M_{\rm BH} / L_{\rm bul} \propto (1+z)^{\gamma}$ 
with $\gamma=1.8\pm0.7$, consistent with BH growth preceding that of the host galaxies.
Including an additional sample of 27 active galaxies with $0.5<z<1.9$ taken from the literature and measured in a consistent way, 
we obtain $\gamma=0.9\pm0.7$ for the \mlbul\ relation and $\gamma=0.4\pm0.5$ for the \mbh--total host galaxy luminosity (\Lhost) relation.
The results strengthen the findings from our previous studies and provide additional evidence for host-galaxy bulge growth 
being dominated by disk-to-bulge transformation via minor mergers and/or disk instabilities.
\end{abstract}
\keywords{black hole physics –-- galaxies: active –-- galaxies: nuclei –-- galaxies: evolution}

\section{INTRODUCTION} \label{sec:intro}
The co-evolution of supermassive black holes (BHs) and their host galaxies,
suggested to explain the tight correlations between BH mass (\mbh)
and host-galaxy properties, such as the \mbh-stellar velocity dispersion (\veldisp),
\mbh-bulge luminosity (\Lbul) and \mbh-bulge mass (\Mbul) relations
discovered in the local Universe 
\citep[e.g.,][]{Magorrian+98,Ferrarese&Merritt00,Gebhardt+00,Marconi&Hunt03,Haring&Rix04} (see also recent studies by \citealt{Gultekin+09,Graham+11,Beifiori+12,McConnell&Ma13,Graham&Scott13,Lasker+14})
can be considered a key element in our understanding of galaxy formation and evolution \citep[see][]{Ferrarese&Ford05,Kormendy&Ho13}.\altaffilmark{*}
In theoretical models, AGN feedback has been considered as a promising physical driver for these correlations
\citep[e.g.,][]{Kauffmann&Haehnelt00,Volonteri+03,DiMatteo05,Croton+06,Hopkins+09,Dubois+13}.
Another possibility is statistical convergence from hierarchical merging that 
reproduces the observed correlations without the need of a physical coupling \citep[e.g.,][]{Peng07,Hirschmann+10,Jahnke&Maccio11}.
Recently, \citet{Angles-Alcazar+13a,Angles-Alcazar+13b} have shown that the scaling relations 
can also be achieved in the galaxy-scale torque-limited BH accretion model as an alternative to self-regulated BH growth models driven by AGN feedback.

However, given the assumptions and approximations involved in the theoretical models which lead to degeneracies of the underlying parameters,
the origin of the BH mass-host galaxy coupling is still an open question.
Observations, on the other hand, can provide direct constraints 
on how BHs and galaxies co-evolve by probing the scaling relations over cosmic time.
Such an empirical evidence is essential to determine the underlying fundamental physical processes at work
and to guide the models of galaxy formation and evolution.

To measure BH masses in the distant Universe, observational studies have to rely on 
galaxies with actively accreting BHs (also known as Active Galactic Nuclei = AGNs)
and in particular broad-line (Type I) AGNs to apply
the virial method.\footnote{See recent reviews by \citet{Shen13} and \citet{Peterson13} for BH mass measurements in active galaxies.} 
The majority of these studies have found an evolution in which the BH growth precedes the growth of the host-galaxy bulge
\citep[e.g.,][]{Treu+04,Treu+07,McLure+06,Shields+06,Peng+06,Woo+06,Woo+08,Salviander+07,Jahnke+09,Decarli+10,Merloni+10,Bennert+10,Bennert+11b,Cisternas+11,Hiner+12,Canalizo+12,Bongiorno+14}.
However, some studies are consistent with no evolution  \citep[e.g.,][]{Shields+03,ShenJ+08,Schramm&Silverman13,Salviander&Shields13,Salviander+14}, 
and others even report an opposite trend, i.e., under-massive BHs given their host galaxies \citep[e.g.,][]{Alexander+08,Shapiro+09,Urrutia+12,Busch+14}.

Despite the great amount of effort put towards determining the evolution of the BH mass scaling relations, 
uncertainties remain, largely due to the inherent uncertainties in BH mass estimates using the virial method
\citep[e.g.,][]{Woo+10,Park+12a,Park+12b} together with measurement systematics in host-galaxy properties \citep[e.g.,][]{Woo+06,Kim+08a,Kim+08b},
and small sample sizes and limited dynamic ranges.
Making use of high quality data of a large sample covering a wide dynamic range
and taking into account systematic uncertainties 
as well as observational biases
(e.g., \citealt{Lauer+07,Shen&Kelly10,Schulze&Wisotzki11}, see also \citealt{Lamastra+10})
is essential to make progress in understanding the cosmic evolution of the BH mass scaling relations.
Following the footsteps of our previous work, the current paper represents another step towards this goal.

The evolution of the BH mass scaling rations has been the main focus of our team effort.
While the 
$M_{\rm BH}$ - $\sigma_{*}$ relation at lookback time of 4-6 Gyr has been probed based on the
high quality Keck spectra \citep{Treu+04,Woo+06,Woo+08}, 
our group made the first attempt in \citet{Treu+07} to study the evolution of the \mlbul\ relation
using a carefully selected sample of 17 active galaxies at $z\sim0.36$, 
determining both BH masses and host-galaxy properties 
by combining high quality Keck spectra and high-resolution Hubble Space Telescope (\HST) Advanced Camera for Surveys (ACS)
images. The results revealed a significant offset of the high-redshift sample 
from the local \mlbul\ relation corresponding to an evolution of the form $M_{\rm BH}/L_{\rm bul} \propto (1+z)^{1.5\pm1.0}$,
with selection effects being negligible. 

\citet{Bennert+10} went a step further by including 23 new galaxies
(17 at $z\sim0.36$, six at $z\sim0.57$)
imaged with the \HST\ Near Infrared Camera and Multi-Object Spectrometer (NICMOS).
Thus, the total number of objects in the sample was 40 ($=17+23$).
Furthermore, a local comparison sample of reverberation-mapped (RM) active galaxies 
measured in a consistent manner to minimize biases was used as a local baseline.
An evolutionary trend of the form $M_{\rm BH}/L_{\rm bul} \propto (1+z)^{1.4\pm0.2}$
was derived, taking into account selection effects via a Monte Carlo approach.
In contrast, the \mlhost\ relation showed apparently no evolution (at least out to a redshift of $\sim$1), suggestive
of dominant bulge growth through secular evolution by a re-distribution of disk stars.

Here, we continue these efforts by adding 12 new galaxies (three at $z\sim0.36$, nine at $z\sim0.57$) based on \HST\ 
Wide Field Camera 3 (WFC3) images, and finalize the result on the evolution of the \mlbul\ relation
by updating all \mbh\ and \Lbul\ measurements.
To minimize possible measurement systematics, we perform a consistent analysis 
for the entire sample ($40+12=52$ objects total) to obtain BH masses and bulge luminosities.
In addition, in contrast to our previous analysis \citep{Woo+06,Woo+08},
we improved the spectral decomposition method by taking into account host-galaxy starlight 
and broad iron emission contribution for a more accurate emission-line width measurement
\citep[see also][]{Park+12b}.
Finally, the photometric decomposition now takes advantage of a Markov Chain Monte Carlo 
(MCMC) sampler for better optimization in the large parameter space,
simultaneously allowing for linear combinations of different point-spread function (PSF) 
models to account for a possible PSF mismatch.
Including a sample of 27 objects taken from the literature
and analyzed in a consistent way, our final sample consists of 79 active galaxies for which
we derive the evolution of the \mlbul\ relation,
taking into account selection effects with a revised Monte Carlo technique.

The paper is organized as follows. 
Sample selection, observations, and data reduction are described in Section~\ref{sec:sample}.
Section~\ref{sec:meas} summarizes the analysis of the Keck spectra for an estimation of BH mass
and surface photometry of \HST\ images for bulge and host-galaxy luminosity measurements.
Section~\ref{sec:localsam} describes the adopted local comparison sample.
In Section~\ref{sec:results}, we present our main results, namely constraints on the 
redshift evolution of the \mlbul\ relation, including selection effects and estimates for possible BH mass growth by accretion.
We summarize our work and discuss its implications in Section~\ref{sec:conclusion}.
The updated measurements for the previous sample of 40 galaxies are given in Appendix~\ref{app:reanalysis}.
Appendix~\ref{app:compare_L5100} compares AGN continuum luminosities measured from spectra and images.

Throughout this paper, the following cosmological parameters were adopted:
$H_0 = 70$~km~s$^{-1}$~Mpc$^{-1}$, $\Omega_{\rm m} = 0.30$, and $\Omega_{\Lambda} = 0.70$.
Magnitudes are given in the AB system.

\section{SAMPLE SELECTION, OBSERVATIONS, AND DATA REDUCTION} \label{sec:sample}
We here summarize sample selection, observations, and data reduction for the full sample of 52 objects.

\subsection{Sample Selection} \label{sec:sampleselection}
To simultaneously determine BH masses (\mbh) from broad emission-line width and continuum luminosity, 
stellar velocity dispersions (\veldisp) from absorption lines, and host-galaxy bulge luminosities (\Lbul), 
high signal-to-noise (S/N) spectra and high-resolution images of objects with comparable 
nuclear and stellar light fractions are essential.
For that purpose, a sample of moderate-luminosity broad-line AGNs 
was carefully selected from the Sloan Digital Sky Survey (SDSS) database with specific redshift windows of $0.35<z<0.37$ (named to S object)
and $0.56<z<0.58$ (named to W object) to minimize the uncertainties from strong sky features.
The following selection criteria were applied:
(i) \Hb\ equivalent width and Gaussian width greater than 5\AA\ in the rest-frame,
(ii) spatially resolved in the SDSS images,
and (iii) $g' - r' > 0.1$ and $r' - i' > 0.3$ for a non-negligible stellar light fraction.
Object showing strong \iron\ nuclear emission were eliminated from the sample after visual inspection of the SDSS spectra.
In addition, supplementary objects at $0.35<z<0.37$ (named to SS object) were selected to extend the BH mass dynamic range to low-mass range
with the additional selection criterion $M_{\rm BH} \lesssim 10^8M_{\odot}$ using the measurements from the SDSS spectra and the BH mass calibration by \citet{McGill+08}.

Our final sample contains a total of 52 moderate-luminosity 
($\lambda L_{5100} \sim 10^{44}$\ergs) AGNs at intermediate-redshifts (37 at $z \sim 0.36$ and 15 at $z \sim 0.57$).
Out of those, 40 objects were already analyzed and presented in the series of our previous papers 
\citep{Treu+04,Woo+06,Treu+07,Woo+08,Bennert+10}.
We here analyze the new 12 AGNs (three at $z\sim0.36$, nine at $z\sim0.57$)
observed with \HST\ WFC3 as well as re-analyze those 40 objects in a consistent manner.
Table~\ref{tab:objlist} lists all 52 objects.

\subsection{Observations and Data Reduction}\label{sec:obs_reduc}

We obtained high quality spectra for the entire sample using the Low Resolution Imaging Spectrometer (LRIS) at the Keck I telescope. The spectroscopic observations and data reductions 
were described by \citet{Woo+06, Woo+08}, and here we briefly summarize the procedure.
We used two spectroscopic setups, namely, the 900 lines mm$^{-1}$ gratings with a Gaussian velocity resolution of $\sim$55 km s$^{-1}$ and the 831 lines mm$^{-1}$ gratings with a Gaussian velocity resolution of $\sim$58 km s$^{-1}$, respectively for objects at z$\sim$0.36 and z$\sim$0.57.
Total exposure time ranges from 600 s to 4.5 hr for each object. 
After performing the standard spectroscopic reduction procedures using 
a series of \texttt{IRAF} scripts, one-dimensional spectra were extracted with 
a window of $4-5$ pixels ($\sim 1\arcsec$).  
To minimize the uncertainties of long-slit spectrophotometry due to slit losses and seeing effects, we performed a re-calibration of 
the flux scale based on the corresponding SDSS DR7 spectra.
We then applied a Galactic extinction correction to the spectra using the $E(B-V)$ values from \citet{Schlafly&Finkbeiner11} listed in
the NASA/IPAC Extragalactic Database (NED\footnote{\url{http://ned.ipac.caltech.edu/}}) and the reddening curve of \citet{Fitzpatrick99}.
The final reduced spectra are presented in Figure~\ref{fig:specfit12} \& \ref{fig:specfit40_1}.
The average S/N at rest-frame 5100\AA\ of the spectra is $\rm S/N \approx 61$ pixel$^{-1}$ (see Table~\ref{tab:specmeas}).

The \HST\ imaging data for the three (nine) objects at $z = 0.36$ ($z = 0.57$) 
were obtained as part of GO-11166, PI: Woo (GO-11208; PI: Treu).
All 12 objects were observed with WFC3 aboard \HST\ in the F110W filter (wide $YJ$ band) for a total 
exposure time of 2397 sec per object.
Four separate exposures for each target were dither-combined using \texttt{MultiDrizzle} within the \texttt{PyRAF} environment.
A final pixel scale of $0.09''$ and a pixfrac of 0.9 were adopted for the \texttt{MultiDrizzle} task.
The \HST\ imaging observations and data reductions for the previous 40 objects were presented in \citet{Treu+07} and \citet{Bennert+10}. 
The final drizzled (i.e., distortion corrected, cosmic rays and defects removed, sky background subtracted) images for 12 objects (40 objects)
are shown in the first column of Figure~\ref{fig:imgfit1dSBP_WFC3_1} (Figure~\ref{fig:imgfit1dSBP_pre40_1}).

\section{DERIVED QUANTITIES} \label{sec:meas}
To investigate the evolution of the BH mass scaling relations over cosmic time, both the BH mass and 
host-galaxy properties (here, \mbh\ and \Lbul) as a function of redshift are required. 
In this section, we present estimates of \mbh\ from a combination of spectral and imaging analysis, and \Lbul\ measurements from high-resolution images.

\subsection{Black Hole Mass} \label{sec:MBH}
To estimate BH masses, we applied the multi-component spectral decomposition technique,
which was based on our previous work \citet{Woo+06}, and significantly improved 
by \citet{Park+12b}, including host galaxy stellar population models. 
The spectra were first converted to rest-frame wavelengths using redshifts from \citet{Hewett&Wild10} (Table~\ref{tab:objlist}).
The observed continuum was then modeled by a combination of a single power-law, an \iron\ template, and a host-galaxy template, respectively, for the featureless AGN continuum, the AGN \iron\ emission blends, and the host-galaxy starlight in the regions of 4430-4770\AA\ and 5080-5450\AA~(slightly adjusted for each spectrum to avoid including wings of 
adjacent broad emission lines and some absorption features).
Weak AGN narrow emission lines (e.g., \ion{He}{1} $\lambda4471$, [\ion{Fe}{7}] $\lambda5160$, [\ion{N}{1}] $\lambda5201$, 
[\ion{Ca}{5}] $\lambda5310$) and the broad \ion{He}{2} $\lambda4686$~line were masked out
during the fitting process.

The \iron\ template was adopted from the I Zw 1 \iron\ template of \citet{Boroson&Green92}.
The stellar template is composed of seven stellar spectra of G and K giants with
various temperatures from the Indo-US spectral library\footnote{\url{http://www.noao.edu/cflib/}} \citep{Valdes+04}, 
which have been widely used for stellar-velocity
dispersion measurements on Keck spectra in many studies \citep[e.g.,][]{Wolf&Sheinis08,Suyu+10,Bennert+11a,FernandezLorenzo+11,Harris+12,Suyu+13}.
These high-resolution stellar template spectra ($\sim34$ \kms; \citealt{Beifiori+11}) were degraded to match the Keck spectral resolution.
Note that our template for the host-galaxy starlight is different from that of \citet[][a single synthetic template with solar
metallicity and 11 Gyr old from \citealt{Bruzual&Charlot03}]{Park+12b},
since our spectral fitting range is dominated by features of late-type stellar spectra such as 
Mg \textit{b} triplet ($\sim5175$~\AA) and Fe (5270~\AA) absorption lines.
Moreover, using a combination of stellar templates resulted in 
smaller $\chi^2$ values and residuals compared to a single synthetic galaxy template.

The best-fit continuum models were determined by $\chi^2$ minimization
using the nonlinear Levenberg-Marquardt least-squares fitting routine \texttt{mpfit} \citep{Markwardt09} in IDL to optimize the following parameters:
the normalization and slope of the power-law model and 
the velocity shifts and widths of the Gaussian broadening kernels for the convolution of the \iron\ and host-galaxy templates.
The weights for a linear combination of the \iron\ and stellar templates were internally optimized using a bounded-variable least-squares solver
(\texttt{bvls}\footnote{Implemented in IDL by Michele Cappellari and available at \url{http://www-astro.physics.ox.ac.uk/~mxc/software/}.}) 
with the constraint of non-negative values during the fitting.
We measured the AGN continuum luminosity at 5100\AA\ from the power-law model
for comparing with the AGN continuum luminosity measured from the \HST\ imaging
(see Appendix~\ref{app:compare_L5100} for details).

\begin{figure*}
\centering
    \includegraphics[width=\textwidth]{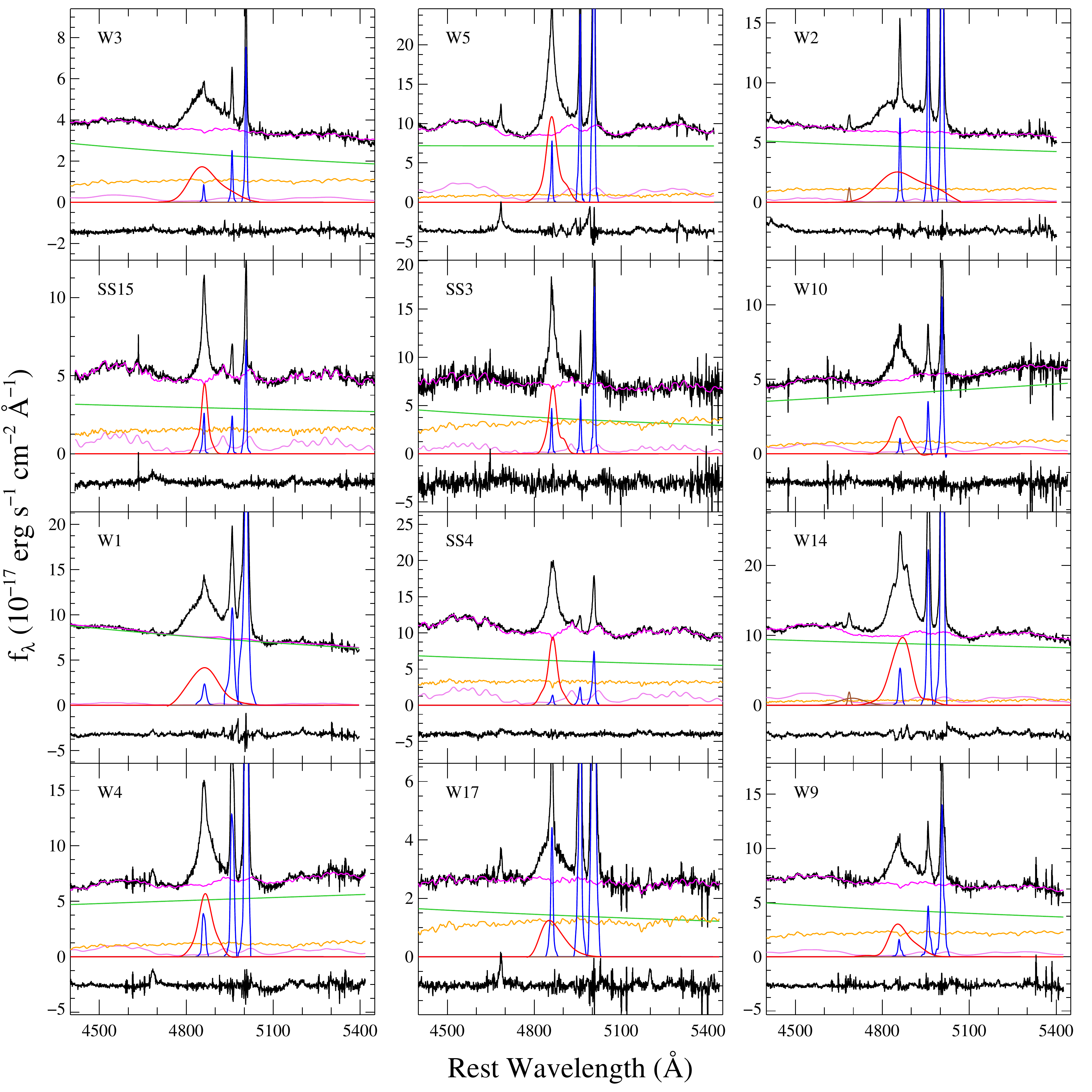}
    \caption{
    Multi-component spectral decomposition for 12 objects.
    The observed spectra are shown along with the best-fit models.
    In each panel, observed spectra
    (black) and the continuum$+$\ion{Fe}{2}$+$stellar best-fit model
    (magenta) are shown in the upper part, and the best-fit power-law
    continuum (green), stellar template (yellow), and \ion{Fe}{2}
    template (violet) models are presented in the middle part.  Three
    narrow lines [\Hb, [\ion{O}{3}] $\lambda\lambda 4959,5007$
    (blue)], broad \Hb\ (red), and the broad and narrow \ion{He}{2}
    $\lambda4686$ components (brown; only included if blended with H$\beta$) are presented in the bottom part.
    The residuals (black), representing the difference between the
    observed spectra and the sum of all model components, are
    arbitrarily shifted downward for clarity.
    \label{fig:specfit12}}
\end{figure*}

After subtracting the best-fit continuum model, the \Hb\ emission line region complex was modeled with a combination of 
a sixth-order Gauss-Hermite series for the \Hb\ broad component, a tenth-order Gauss-Hermite series with different flux scaling ratios 
for the \Hb\ narrow component and [\ion{O}{3}] $\lambda\lambda4959, 5007$ narrow lines, 
and two Gaussian functions for the \ion{He}{2} $\lambda4686$ line whenever it blends with the \Hb\ profile.
Figure~\ref{fig:specfit12} shows the observed spectra with the best-fit models for our sample of 12 objects 
(see Figure~\ref{fig:specfit40_1} for the previous 40 objects).
We measured line widths ($\Delta V$), 
Full Width at Half Maximum
(FWHM) and line dispersion ($\sigma$), for the \Hb\ broad emission line 
from the best-fit profile of the sixth-order Gauss-Hermite series.
The measured line widths were finally corrected for instrumental resolution.

Using the method described above we performed the multi-component spectral
decomposition for all 52 objects in our sample (Table~\ref{tab:specmeas}).
We have thus updated spectral measurements for the samples presented in our
previous works \citep[][see Appendix~\ref{app:reanalysis} for a comparison between previous and updated measurements]{Woo+06,Treu+07,Woo+08,Bennert+10}. 

For the \mbh~estimation, we use the following formalism, derived by combining the recent calibrations for 
the size-luminosity ($R-L$) relationship \citep[$R_{\rm BLR} \propto L^{0.519}$,][]{Bentz+09a} and 
the virial factor \citep[$\log f = 0.71$,][]{Park+12a,Woo+13} from the virial equation ($M_{\rm BH} = f R_{\rm BLR} {\varDelta V}^2/G$
where $G$ is the gravitational constant):
\begin{eqnarray}\label{eq:MBH}
\log \left(\frac{M_{\rm BH} }{M_\odot}\right)
&~=~& 7.536 
+ 0.519 ~ \log\left(\frac{\lambda L_{5100}}{10^{44}~\rm erg~s^{-1}}\right) \nonumber\\
&~+~& 2 ~ \log \left(\frac{\sigma_{\textrm{\Hb}}}{1000~ \rm km~s^{-1}}\right) ,
\end{eqnarray}
where the overall uncertainty of single-epoch (SE) BH masses is assumed to be 0.4 dex, 
estimated by summing in quadrature each source of uncertainties, i.e., 
0.31 dex scatter of the virial factor \citep{Woo+10}, 
0.2 dex additional variation of the virial factor based on the direction of regression in its calibration \citep{Park+12a}, 
0.05 dex scatter due to AGN variability \citep{Park+12b}, 
and 0.15 dex scatter of the size-luminosity relation \citep{Bentz+09a}.
Although the $R-L$ relation has recently been updated 
with nine new low-mass RM AGNs by \citet{Bentz+13}, we use the calibration of \citet{Bentz+09a} 
for consistency with the local RM AGN sample adopted from 
\citet[][re-analysed in \citealt{Bennert+10}]{Bentz+09b}. 
The results do not change within the uncertainties even if we adopt the latest $R-L$ calibration.
Note that we use the AGN continuum luminosity measured from \HST\ images, 
as described in the following section, for the final \mbh\ estimates given in Table \ref{tab:final_quan} 
(see Appendix~\ref{app:compare_L5100} for a comparison between luminosity estimates from spectra and images).

\subsection{Bulge Luminosity} \label{sec:Lbul}
To determine AGN and bulge luminosities of the host galaxies,
we performed two-dimensional surface photometry on \HST\ imaging data for the entire
sample including the 12 new objects, using a modified version of the image fitting code
``Surface Photometry and Structural Modeling of Imaging Data''
\citep[SPASMOID][]{Bennert+11a,Bennert+11b}
written by Matthew W. Auger.
The code allows for a linear combinations of different PSFs to model the AGN,
accounting for any potential PSF mismatch,
which is particularly important for the \HST\ image analysis of host galaxies with a central bright point source \citep{Kim+08b}.
To efficiently explore the multi-parameter space, the code adopts an adaptive simulated annealing algorithm with an MCMC
sampler in the \texttt{pymc}\footnote{\url{https://github.com/pymc-devs/pymc}} framework,
which is superior to a local $\chi^2$ minimization method due to less sensitivity to initial guesses
and less likely to get stuck in local minima and thus achieving better convergence on a global minimum over the posterior distribution, at the cost of longer execution time.

In this section we focus on the analysis of the new 12 objects.
We created a library of 16 PSFs from nearby bright, isolated, unsaturated stars carefully selected over the science fields, 
normalized and shifted relative to each other using spline interpolation to obtain centroid images.
Empirical stellar PSFs are generally considered better than synthetic TinyTim 
PSFs given that they were observed simultaneously with the science 
target and reduced and analyzed in the same way \citep{Kim+08b,Canalizo+12}.
The central point source (i.e., AGN) was then modeled as a scaled linear combination of these different PSFs. 
On average, a combination of four PSF images was chosen for the AGN.
If a single arbitrarily chosen PSF model from the library is adopted for each object, the derived AGN (bulge) luminosity can be incorrectly shifted by up to $\sim0.2$ ($\sim0.3$) mag compared to that of the multiple PSF model.
If the single largest amplitude PSF model, taken from the selected PSF combinations of the multiple PSF fits, is adopted,
there is on average $\sim0.06$ ($\sim0.09$) mag scatter for the AGN (bulge) luminosity estimates.

The host galaxy was then fitted with a \citet{deVau48} profile to model the bulge component.
After carefully examining the original and residual images
\citep[following a similar strategy adopted by][]{Treu+07, Kim+08a, Bennert+10},
an exponential disk profile was added if deemed necessary (i.e., if an extended structure was clearly visible in the original and residual images and the resulting parameters were physically acceptable when fitted with the additional disk component).
Five out of 12 objects were modeled with an additional disk component.
All model components for the host galaxy are concentric, 
but an offset between the AGN and host galaxy centroid is allowed. 
The minimum radius of the \citet{deVau48} profile was set to be 2.5 pixels (i.e., the minimum resolvable size given the PSFs).
The normalization of each profile (i.e., magnitude of each model component) 
is optimized by fitting a linear combination of all models 
given the structural parameters (i.e., centroid, effective radius, axis ratio, and position angle) 
to data with a non-negative least squares solver \citep[\texttt{nnls};][]{Lawson&Hanson87}.
Note that all model components were fitted simultaneously.

Out of the 12 objects, four bulge component fits (i.e., W3, SS3, W17, and W9) 
resulted in small effective radii, approaching the minimum size.
Thus, we assign an upper limit to the bulge luminosities of these objects.
To estimate the bulge luminosity from the upper limit, 
we applied the same method described in \citet{Bennert+10}.
In brief, by taking advantage of the prior knowledge of the bulge-to-total luminosity ratios, 
measured by \citet{Benson+07} for a sample of 8839 SDSS galaxies, 
we derived the posterior distribution by combining the prior and likelihood for the B/T ratios as shown in Figure~\ref{fig:BTprior}.
A non-zero step function up to the measured upper limit B/T was adopted for the likelihood function.
The prior was determined by using the B/T distribution of galaxies 
from \citet{Benson+07}
whose total galaxy magnitudes are within $\pm$0.5 mag 
of the total host galaxy magnitude of the sample here.
(Note that even if the bulge magnitudes are upper limits, the total host-galaxy magnitudes
are robust.)
For each object, the mean value from the B/T posterior distribution was adopted 
to calculate the final bulge luminosity from the total host galaxy luminosity.
Note that the 14 upper limit objects in our previous work \citep{Bennert+10} were 
also consistently re-analyzed.

\begin{figure*}
\centering
    \includegraphics[width=0.75\textwidth]{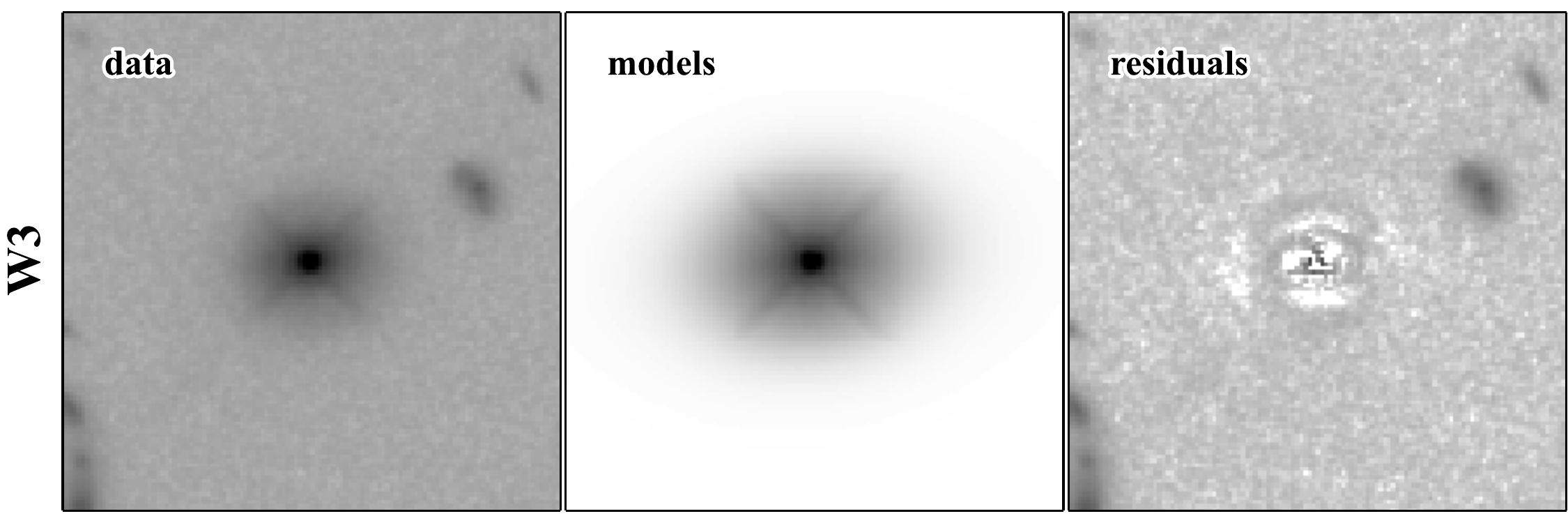}\includegraphics[width=0.25\textwidth]{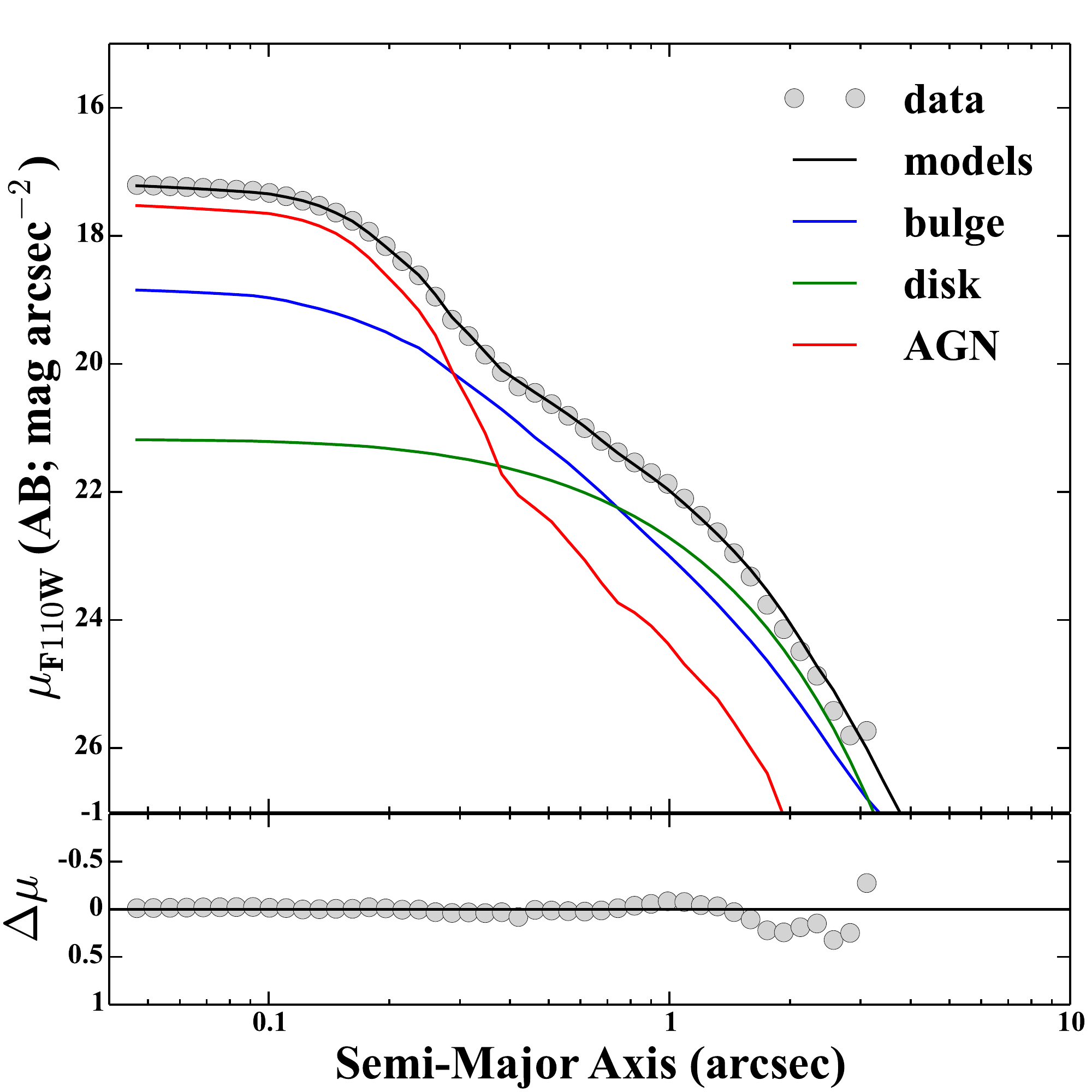}\\
    \includegraphics[width=0.75\textwidth]{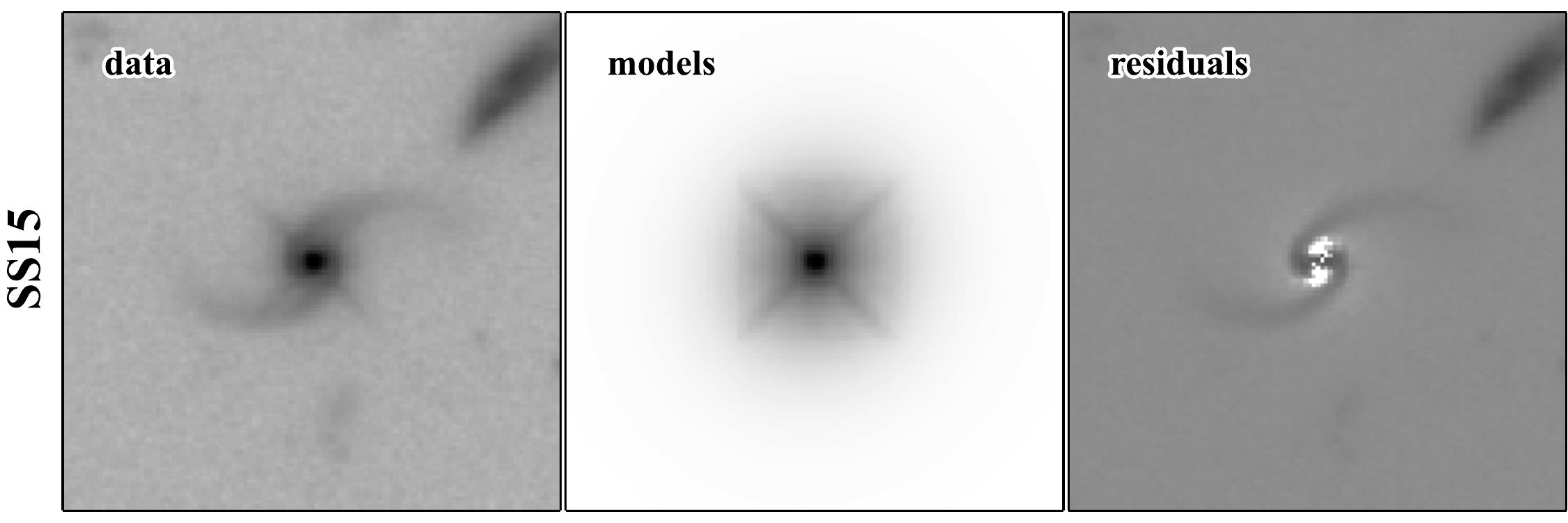}\includegraphics[width=0.25\textwidth]{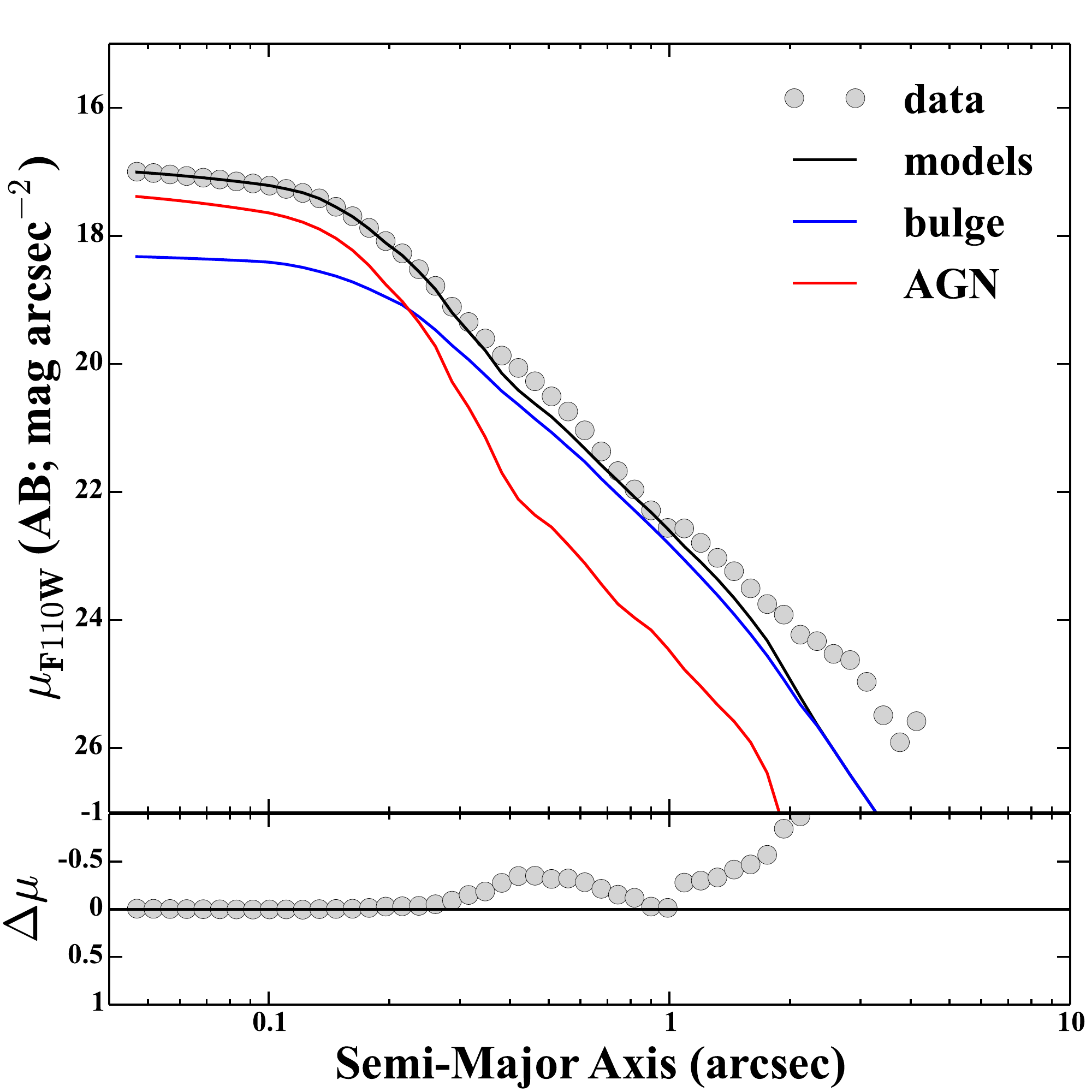}\\
    \includegraphics[width=0.75\textwidth]{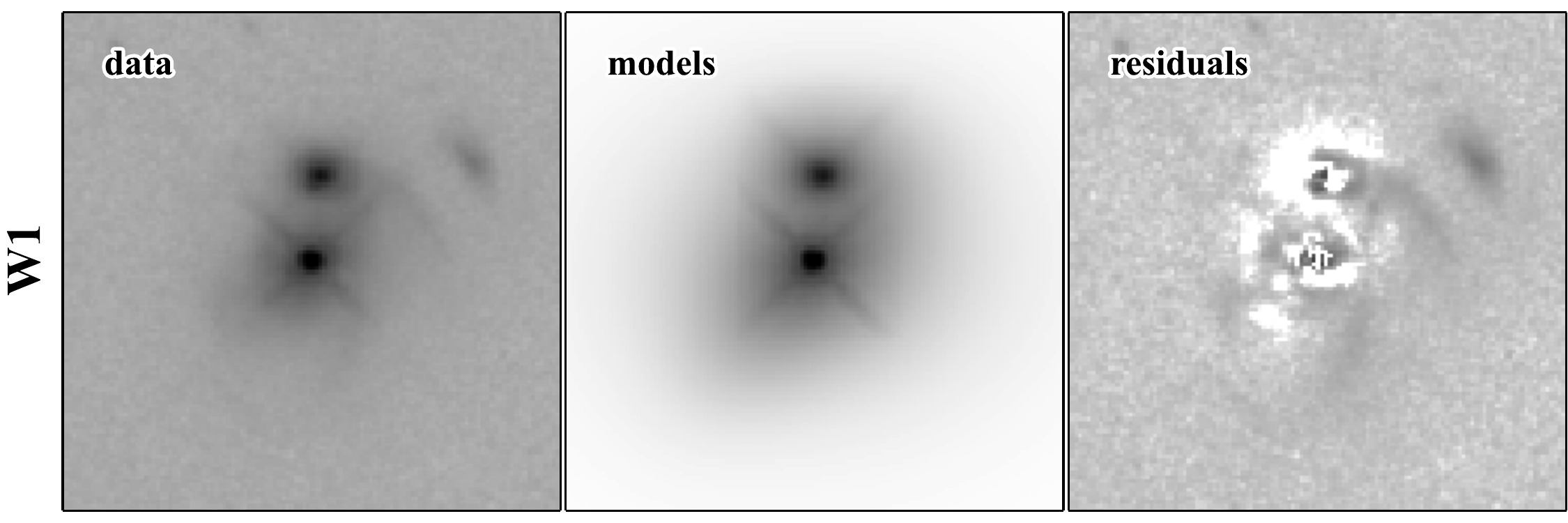}\includegraphics[width=0.25\textwidth]{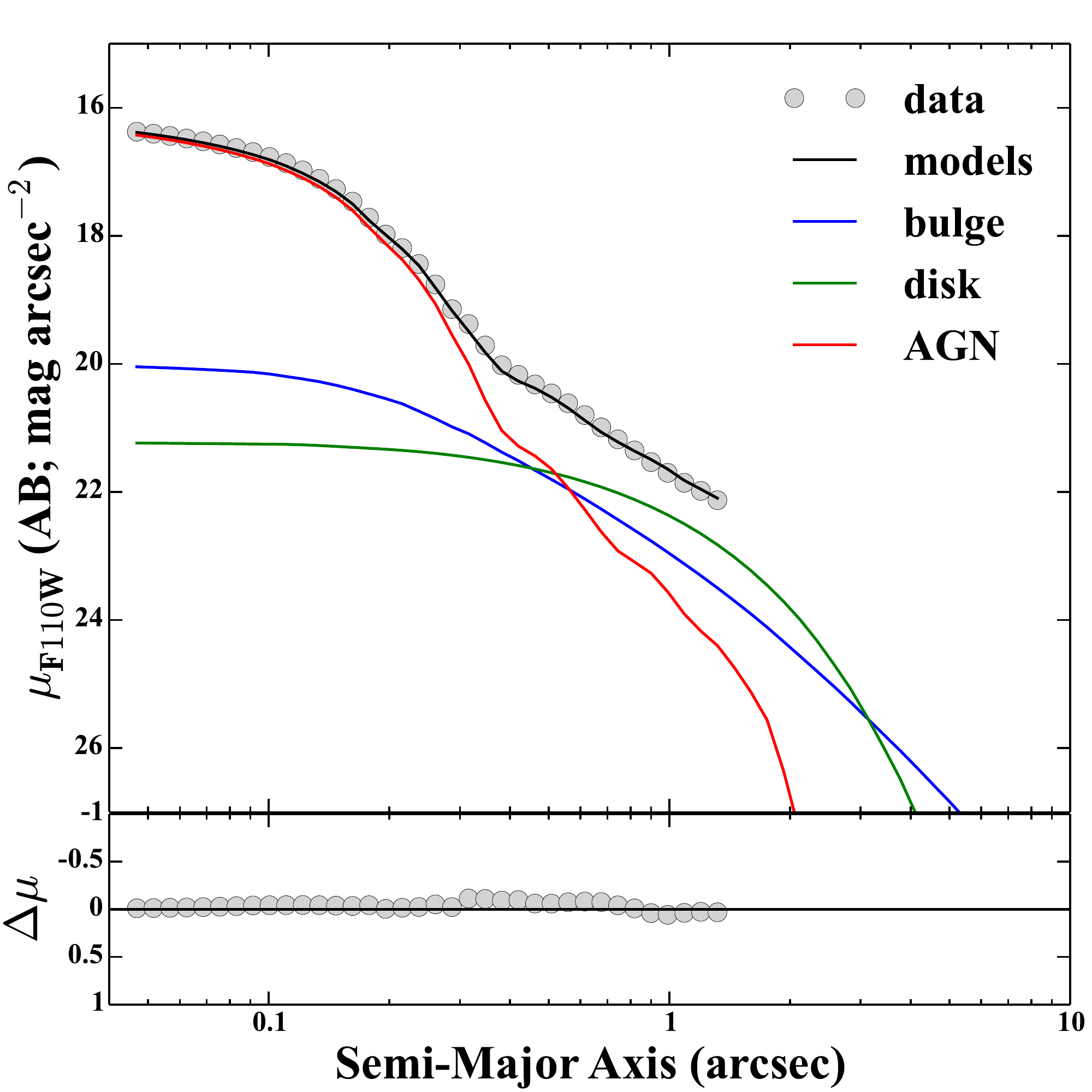}\\
    \includegraphics[width=0.75\textwidth]{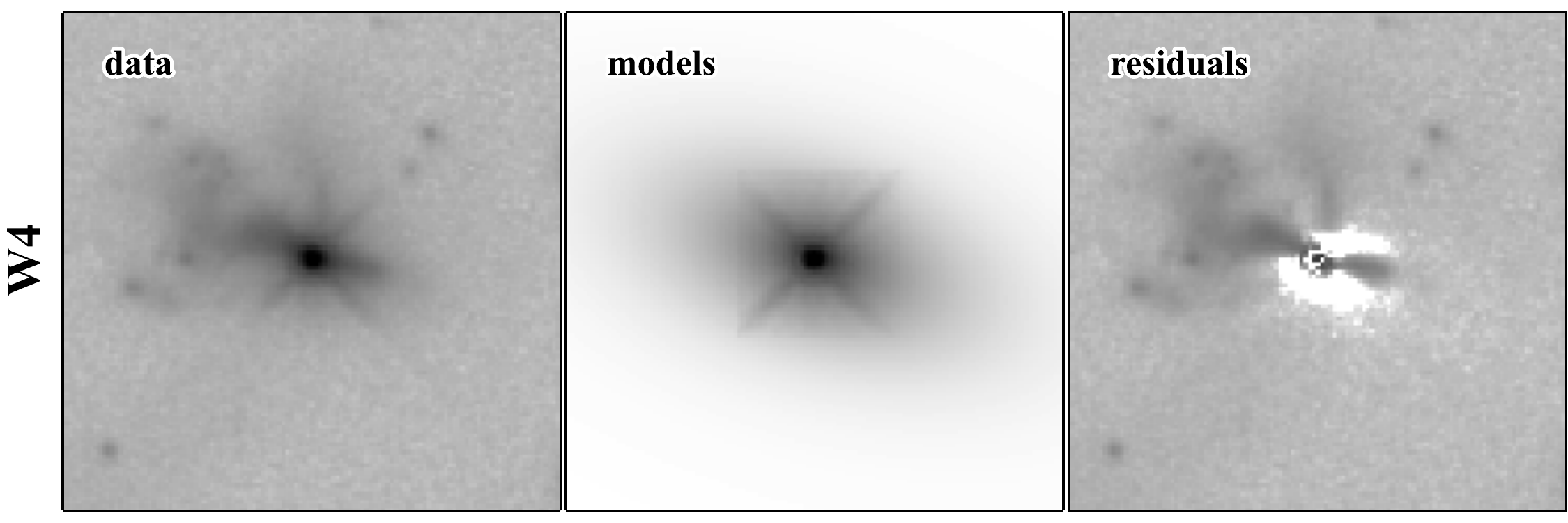}\includegraphics[width=0.25\textwidth]{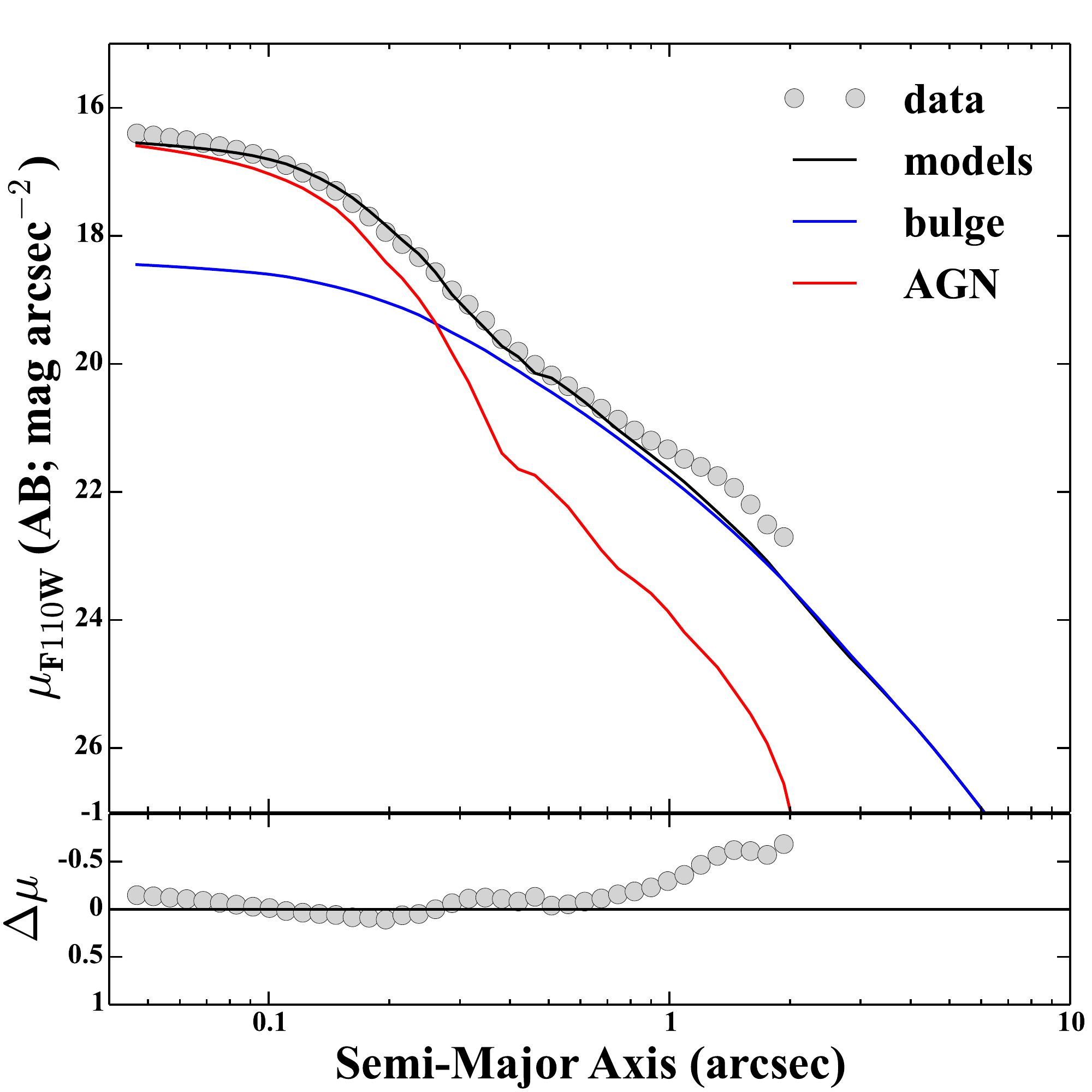}
        \caption{
	    \HST\ WFC3 F110W images for 12 objects. In each row, observed data (first column), best-fit models (second column), and residuals (third column)
	    are presented with the object name. All images are $10.8\arcsec\times10.8\arcsec$ in size and
 displayed with an inverted asinh stretch.
	    The fourth column shows the corresponding one-dimensional surface brightness profiles. In each top panel, the profiles measured from the data (open circles), the best-fit model (black solid line), and the sub-components of the model for bulge (blue solid line), disk (green solid line), AGN (red solid line) are shown. Residuals (gray circles), the difference of the profiles between the data and the best-fit model, are presented in each bottom panel. Note that the one-dimensional surface brightness profiles are shown for illustration purposes only, the actual fitting made use of the full two-dimensional images.
        \label{fig:imgfit1dSBP_WFC3_1}}
\end{figure*}

For one target (W1), a nearby object was fitted simultaneously since its light profile overlaps with that of the science target.
In all other cases, surrounding objects were masked-out during the fitting process.
In Figure~\ref{fig:imgfit1dSBP_WFC3_1}, we show the images, best-fit models, and residuals for the 12 objects.
For illustration purposes only, 
one-dimensional surface brightness profiles obtained with the IRAF \texttt{ellipse} task are shown in Figure~\ref{fig:imgfit1dSBP_WFC3_1}.
The 40 objects presented in the previous papers of the series
were consistently re-measured using the same method (see Appendix~\ref{app:reanalysis}).

The apparent AB magnitudes were determined by converting counts to magnitude using
equation 11 in \citet{Sirianni+05}, i.e., 
$\rm ABmag = -2.5 \log ( counts [e^{-1} s^{-1}]) + zero-point,$
with zero-point = 26.8223 mag for WFC3/F110W.
To obtain rest-frame $V$-band luminosities of the host-galaxy bulges, 
we first corrected for Galactic extinction 
using $E(B-V)$ values from \citet{Schlafly&Finkbeiner11} listed in NED and assuming $A_{\rm F110W} = 0.902\ E(B-V)$ \citep{Schlegel+98}.
The extinction-corrected F110W AB magnitudes were then transformed to rest-frame $V$-band by applying $K$-correction with an early-type galaxy
template spectrum\footnote{This empirical observed SED templates are available at \url{http://webast.ast.obs-mip.fr/hyperz/}.} of \citet{Coleman+80} extended to UV and IR regions using the spectral evolutionary models of \cite{Bruzual&Charlot93}.
We estimate an uncertainty of the template choice as $<0.06$ mag (i.e., 0.02 dex in luminosity) 
using the scatter from 14 single stellar population templates with ages ranging from 2 to 8.5 Gyr.
The $V$-band luminosities are given by $\log L_{V}/L_{V,\odot} = 0.4(M_{V,\odot} - M_{V})$ where $M_{V,\odot}=4.83$.
We adopt a conservative total uncertainty of 0.2 dex ($\sim 0.5$ mag) for the bulge luminosity estimates as 
discussed in \citet{Treu+07} and \citet{Bennert+10}.
Note that the F110W band corresponds to rest-frame $R$ and $I$ 
bands for the redshift range covered by our sample,
allowing for a robust decomposition between the bulge
and the blue AGN light that would dominate shorter bandpasses 
while also minimizing dust attenuation.
The scatter of red colors of bulges (i.e., $V-R$ and $V-I$) are known to be small.
For a more direct comparison with local samples,
we correct for passive luminosity evolution due to the aging of the stellar populations,
by applying the following equation as previously adopted in \citet{Treu+07} and \citet{Bennert+10}:
\begin{equation}
\log L_{V,0} = \log L_{V} - (0.62\pm0.09) \times z.
\end{equation}

To derive the AGN 5100\AA\ continuum luminosity ($\lambda L_{5100}^{\rm image}$) from the \HST\ image analysis, 
we transformed the extinction-corrected PSF F110W AB magnitude to rest-frame 
5100\AA\ by assuming a single power-law SED ($f_{\nu} \propto \nu^{-0.5}$) as adopted by \citet{Bentz+06} and \citet{Bennert+10, Bennert+11a}.
The slope of the power-law continuum is the same  as the
median value of the power-law continuum slopes measured from our 52 spectra,
although the slopes are based on a limited wavelength range ($\sim4400-5500$\AA), and show a large scatter.
However, by varying the adopted slope between $-0.2$ and $-1$,
the reported range in the literature (see \citealt{Bennert+11a} and references therein),
we estimate that the uncertainty in the derived luminosity due to the choice of a fixed slope of $-0.5$
is $\pm0.05$ dex on average, 
thus negligible compared to the adopted total uncertainty for \mbh\ (i.e., 0.4 dex).
Note that $\lambda L_{5100}^{\rm image}$ is preferred over $\lambda L_{5100}^{\rm spec}$ 
since it is not affected by the uncertainties from slit losses, seeing effects, and the difficulty of absolute spectrophotometric calibration in spectral measurements
(see Figure~\ref{fig:compare_L5100} and Appendix~\ref{app:compare_L5100} for comparison between $\lambda L_{5100}^{\lowercase{\rm image}}$ and $\lambda L_{5100}^{\lowercase{\rm spec}}$).

The measured quantities from the \HST\ image analysis for the full sample are listed in Table~\ref{tab:imgmeas}.
Table~\ref{tab:final_quan} provides the final quantities of BH mass,
as derived from equation~(\ref{eq:MBH}) using $\sigma_{\rm H\beta}$ and $\lambda L_{5100}^{\rm image}$,
and host-galaxy properties.
The bulge luminosities with and without correction for passive evolution are given.

\section{LOCAL COMPARISON SAMPLES}\label{sec:localsam}
Adopting a robust local baseline is crucial for an accurate characterization of the evolution of the scaling relation.
We could adopt the local baseline relation either from local active galaxies \citep{Bennert+10} or from local quiescent galaxies \citep{McConnell&Ma13}.

The local active galaxy sample consists of RM AGNs
for which both reliable BH masses and host-galaxy properties from 
\HST\ images are available.
We take the RM AGN properties from Table 3 in \citet{Bennert+10}
who re-analyzed the host galaxies presented in \citet{Bentz+09b}
in a manner comparable to the analysis of the higher $z$ samples.
This choice is made in order to reduce systematic uncertainties involved in bulge luminosity measurements.
The dynamic ranges of \mbh\ and \Lbul\ for our intermediate-$z$ 
sample are comparable and well covered by those of the local RM AGNs.

A direct comparison of our intermediate-$z$ active galaxies,
selected based on BH property
 (e.g., nuclear luminosity and broad emission line, hence \mbh),
 to the local quiescent galaxies, selected by galaxy property (e.g., galaxy luminosity) ,
is not straightforward, since the samples are subject to different selection functions \citep{Lauer+07},
which could introduce a substantial effect on the evolutionary signal, if not properly taken into account.
In addition, the recent sample of local quiescent galaxies compiled in \citet{McConnell&Ma13} 
suffers from a lack of low-mass objects (i.e., $M_{\rm BH} \lesssim 10^8M_{\odot}$) and is limited to early-type galaxies in the \mlbul\ plane.
A direct comparison of the \mlbul\ relation between active and quiescent galaxies is further complicated 
by the normalization of the BH mass scale (i.e., the virial factor) for active galaxies,
which forces the local RM AGNs into agreement with the \msigma\ relation of local quiescent galaxies 
\citep[e.g.,][]{Onken+04,Woo+10,Graham+11,Park+12a,Woo+13,Grier+13} instead of the the \mlbul\ relation,
because of the smaller intrinsic scatter of the former.

We thus consider the local RM AGN sample as the better suited comparison sample
and use it as the fiducial local baseline.
Note that we consistently apply the same virial factor for both samples of local and distant active galaxies,  
assuming that the virial factor does not change with redshift.

\section{RESULTS} \label{sec:results}

\subsection{$M_{\rm BH}-L_{\rm bul}$ Relation}\label{sec:MLrelation}
Figure~\ref{fig:M-Lbul_reestimated_scatter} 
shows the resulting BH mass--bulge luminosity relation for a total of 
52 intermediate-$z$ objects as well as the local comparison sample.
Figure~\ref{fig:M0-Lbul0_offset} shows the offset from the fiducial local relation as a function of redshift.
As a comparison, we show the local RM AGNs 
with black squares and intrinsic dispersion (i.e., $0.21$ dex) of the local baseline as a gray shaded region.
Overall, BHs are overly massive compared to the expectation from the local relation. 
When modeling the redshift evolution of the offset as $\Delta \log M_{\rm BH} = \gamma \log (1+z)$,
without taking into account selection effects, 
we find $\gamma = +1.3\pm0.4$ with an intrinsic scatter of $0.2\pm0.1$ dex using
the \texttt{FITEXY} estimator implemented in \citet{Park+12a}.

\subsection{Host-Galaxy Morphology} \label{sec:SamDep}
When classifying the host galaxies as ellipticals (fitted by a \citealt{deVau48} profile only),
spirals (fitted by a \citealt{deVau48} + exponential profile) or merging/interacting,
our sample consists of comparable numbers of each type 
(i.e., 18 for ellipticals, 18 for spirals, and 16 for merging/interacting galaxies). 
To probe whether the observed offset in BH mass depends on a specific morphological type of our sample,
we show the offset as a function of this simple morphological classification in Figure~\ref{fig:M-Lbul_offset_type}.
No clear dependency on morphological type is observed.
The objects containing a bar component (i.e., 7 out of 52)
seem to have a marginally larger offset in BH mass than average.
However, the sample size is too small, especially when split into sub-samples,
for a conclusive result.

\subsection{Redshift Evolution Including Selection Effects}
\label{redshiftevolution}
Improper accounting for the selection function can introduce a bias in the inferred evolution of the scaling relations
\citep[e.g.,][]{Treu+07, Lauer+07}.
Our sample of intermediate-$z$ AGN host galaxies is selected based on nuclear 
(AGN) luminosity and width of the \Hb\ broad emission line (i.e., BH mass).
Given the steeply declining bulge luminosity function and the intrinsic dispersion of the \mlbul\ relation,
this will favor selecting galaxies with under-luminous bulges at a given BH mass, similar to the well-known Malmquist bias.
The distribution of BH masses (i.e., lower and upper limits) 
of our sample relative to the entire mass distribution of the supermassive 
BH population is also an important factor to take into account.
Note that our samples at 
$z=0.36$ and $z=0.57$ have different selection criteria on BH mass (see Section \ref{sec:sampleselection}).
The SS* objects (16 at $z \sim 0.36$; the blue plus signs in Fig.~\ref{fig:M-Lbul_reestimated_scatter}) were selected with an additional constraint of $M_{\rm BH} \lesssim 10^8M_{\odot}$ 
to extend the dynamic range to lower masses compared to the initial sample 
(S* and W* objects; 21 at $z \sim 0.36$ and 15 at $z \sim 0.57$).
High mass objects which could introduce an offset above the \mlbul\ relation
were thus purposefully selected against for this particular sub-sample.

To constrain evolution and intrinsic scatter taking into account the effects mentioned above,
we adopt the Monte Carlo simulation method introduced by \citet{Treu+07} and \citet{Bennert+10}
with a slight modification as described below. 
First, we generate samples of the joint distribution of 
BH mass and bulge luminosity from a combination of the local active BH mass function from \citet[][the modified Schechter function fit in their Table 3]{Schulze&Wisotzki10} and the local \mlbul\ relation from \citet[][the linear fit in their Table 4]{Bennert+10}.
Since we are using an active galaxy sample, it is also important to take into account for the active fraction bias as suggested by \citet{Schulze&Wisotzki11}.
This is easily done, however, assuming that the active fraction is not a strong function of redshift over the range covered here.
It is sufficient to start from the BH mass function of active galaxies to generate simulated samples.
This allows us to directly compare the local simulated active galaxies to the high-$z$ observed active 
galaxies, avoiding the currently uncertain prediction of the active fraction
(in other words, we assume that the mass-dependent effect of the active faction cancels out between local and higher-$z$ samples).

Next, simulated samples with Gaussian random noise added on both axes
are constructed as a function of the two free parameters $\gamma$ and $\sigma_{\rm int}$.
We then consider the observational selection on $\log M_{\rm BH}$, which are simply modeled by lower and upper limits of [7.3, 8.2] for SS* objects (16 out of total 52) and [7.7, 9.1] for S* and W* objects (36 out of total 52), respectively, from the observed distributions of $\log M_{\rm BH}$.
Note that adopting such a simple threshold is a practical approach,
given the difficulty of deriving a more precise selection function by 
including all the details involved in the observation and sampling processes.
The likelihood of the observed BH mass for the given bulge luminosity for each object is calculated 
from the probability distribution of the BH masses of the simulated sample at the given $\gamma$ and $\sigma_{\rm int}$ 
with corresponding bulge luminosity within the measurement uncertainty.
By adopting un-informative uniform priors, 
we evaluate the posterior distribution function
and take the best-fit values at the maximum of the one-dimensional marginalized probability distribution with 1$\sigma$ uncertainties.

Figure~\ref{fig:MC_seleff} shows the results of the Monte Carlo simulations in the two-dimensional plane spanned by $\gamma$ and $\sigma_{\rm int}$.
For a uniform prior of $\sigma_{\rm int}$, the parameters are not well constrained since 
the dynamic range in redshifts of our sample is insufficient to determine $\gamma$ and $\sigma_{\rm int}$ simultaneously.
If we adopt the log-normal prior from \citet[][$\sigma_{\rm int}=0.21\pm0.08$]{Bennert+10}
under the assumption that the intrinsic scatter has a similar magnitude as that of the local sample,
the slope is found to be $\gamma = +1.8\pm0.7$ with $\sigma_{\rm int}=0.3\pm0.1$.
The obtained slope is rather steeper than that derived without taking into account selection effects in Sec.~\ref{sec:MLrelation}. This increase of the slope mainly results from proper accounting for the selection function of the SS* objects, which consequently leads to a positive offset on the result.
We obtain consistent estimates for the slope, $\gamma = +1.8\pm0.9$ and $\gamma = +2.0\pm1.1$, if we adopt the log-normal priors for $\sigma_{\rm int}$ from \citet[][$\sigma_{\rm int}=0.38\pm0.09$]{Gultekin+09} and  \citet[][$\sigma_{\rm int}=0.52\pm0.06$]{McConnell&Ma13}, respectively.
We also obtain a consistent estimate for the slope, $\gamma = +1.7\pm0.6$, if we broaden the mass interval of the selection function by as much as $0.4$ dex (i.e., the adopted uncertainty of SE BH masses).
This trend can also be expressed as $M_{\rm BH}/L_{\rm bul} \propto (1+z)^{1.8\pm0.7}$,
consistent with our previous results, and with that 
BH growth precedes bulge assembly (\citealt{Woo+06,Woo+08,Treu+07,Bennert+10,Bennert+11b}; see also, \citealt{Canalizo+12}).
If our intermediate-$z$ galaxies are to fall on the local relation as evolutionary end-point,
their bulge luminosities have to increase by $0.24$ dex (i.e., $\sim70$\%) 
and $0.35$ dex (i.e., more than a factor of two) by today from $z=0.36$ ($\sim$ 4 Gyr) and $z=0.57$ ($\sim$ 6 Gyr), respectively.
This requires formation of new stars or injection of young and old stars into the bulge component without a significant BH growth.

To increase the redshift range studied,
we include two literature samples from \citet[][a sample of 11 X-ray selected AGNs in $1<z<1.9$]{Bennert+11b} and \citet[][a sample of 18 X-ray selected AGNs in $0.5<z<1.1$]{Schramm&Silverman13} with a similar approach to our work,
thus minimizing possible measurement systematics.
(Note that we use the measurements provided by \citet{Bennert+11b} for two overlapping objects between the samples.)
Taking advantage of this increased sample size of a total of 79 objects and extended redshift distribution of $0.5<z<1.9$, 
the evolutionary slope, $\gamma$, can be constrained without the need for informative priors for the intrinsic scatter.
Note that these samples have different selection functions compared to our mass-selected sample 
since they were selected from X-ray flux limited surveys.
Given the difficulty of deriving exact selection functions, 
we practically apply mass selections on $\log M_{\rm BH}$ in the same manner of our sample, i.e., 
with mass limits of [7.8, 9.3] for the sample of \citet{Bennert+11b} and [7.1, 9.3] for that of \citet{Schramm&Silverman13}.
Figure~\ref{fig:offset_bul_host_addsample} shows the offset in BH mass
for all 79 active galaxies for both the bulge luminosity and host-galaxy luminosity.
For the bulge luminosity, the resulting evolution
($M_{\rm BH}/L_{\rm bul} \propto (1+z)^{0.9\pm0.7}$ with $\sigma_{\rm int}=0.6\pm0.2$) 
is consistent with the results obtained above within the uncertainties.
However, for the host-galaxy luminosity we find 
a milder evolution that can even be considered zero evolution, given the uncertainties 
($M_{\rm BH}/L_{\rm host} \propto (1+z)^{0.4\pm0.5}$ with $\sigma_{\rm int}=0.4\pm0.2$).
If we include only the sample from \citet{Bennert+11b}, 
which is based on an almost identical analysis, the slope is found to be
$(1+z)^{1.2\pm0.9}$ ($(1+z)^{0.7\pm0.7}$) for the bulge (host-galaxy) luminosity.
These results are in broad agreement with those of previous studies \citep[e.g.,][]{Jahnke+09,Merloni+10,Bennert+11b,Cisternas+11,Schramm&Silverman13} and provide
further evidence in support of a scenario in which secular processes, which lead to galaxy-structure evolution by 
a re-distribution of stars from disk to bulge, play the dominant role in bulge growth mechanism \citep[e.g.,][]{Croton06,Parry+09}.

\subsection{\mbh\ Growth By Accretion}
For a direct comparison with the local sample, 
we need to account the possible additional BH growth through accretion
since $z=0.36$ and $z=0.57$, respectively.
Although it is uncertain to estimate the BH mass growth rate and lifetime 
for individual AGNs, we adopt a common approach in the following manner.

First, we estimate the bolometric luminosities of the AGNs
as $L_{\rm bol} = 9.26 \times \lambda L_{5100}^{\rm image}$
(see \citealt{Shen+08} and references therein).
The resulting Eddington ratios of our sample
range from $0.01$ to $0.24$, with an average of $\sim$$0.08$. 
Then, the BH mass growth rate is estimated as
\begin{equation}
\dot{M}_{\rm BH} = \dot{M}_{\rm infall} (1-\epsilon)=\frac{L_{\rm bol} (1-\epsilon)}{\epsilon c^2} ,
\end{equation}
where $L_{\rm bol}=\epsilon \dot{M}_{\rm infall} c^2$ is the bolometric luminosity and $\epsilon$ is the radiative efficiency 
(i.e., fraction of accreted mass converted into radiation).
By assuming the standard average radiative efficiency of $10\%$ (\citealt{Yu&Tremaine02}; but see also \citealt{Wang+09, Davis&Laor11,Li+12}),
the growth rate for the sample of our 52 objects is in the range of $0.05-0.7$ \msun/year with an average of $0.2$ \msun/year.

Finally, we estimate AGN lifetimes; estimates for the typical AGN lifetime 
found in the literature range from $\sim 1$ Myr to $\sim 1$ Gyr \citep[e.g.,][]{Martini&Weinberg01,Yu&Tremaine02,Marconi+04, Martini04,Porciani+04, Shankar+04,Yu&Lu04,Hopkins+05, Shen+07,Wang+08,Croton09,Gilli+09,Hopkins&Hernquist09,Cao10,Kelly+10,Furlanetto&Lidz11,Richardson+13}.
However, AGN lifetime is likely a function of luminosity and/or mass, and not a single value for the entire population,
given the diverse physical properties of the AGN population.
The AGN lifetime can be estimated as $t_{\rm AGN} \equiv \delta \times t_{\rm H}(z)$ where $\delta$ is the duty cycle and $t_{\rm H}(z)$ is the Hubble time at the given redshift. 
We here adopt the semi-analytic prediction for the duty cycle as a function of BH mass and redshift, $\delta=\delta(M_{\rm BH},z)$,
given in Table 4 of \citet[][see also their Figure 7]{Shankar+09b}.
This reflects AGN downsizing: a higher mass and higher activity population has a shorter lifetime, 
thus completing its BH mass growth by accretion at an earlier epoch (i.e., anti-hierarchical BH growth).
The estimated lifetimes for our sample range from $3$ Myr to $65$ Myr with an average of $24$ Myr.

These lifetime estimates along with the growth rates 
lead to BH mass growth by on average $0.02$ dex for our sample with a maximum of $0.08$ dex.
If we consistently estimate the BH mass growth for the sample of local RM AGNs, 
the average mass growth will also be $\sim 0.02$ dex. 
This insignificant BH mass growth implies that the previously inferred evolution 
(Section~\ref{redshiftevolution}) is dependent on bulge growth only.

\section{DISCUSSION AND CONCLUSIONS}\label{sec:conclusion}
We study the cosmic evolution of the BH mass--bulge luminosity relation by 
performing a uniform and consistent analysis of high-quality Keck spectra and high-resolution \HST\ images
for a sample of 52 active galaxies at $z \sim 0.36$ and $z \sim 0.57$, corresponding to look-back times of 4-6 Gyrs.
Using Monte Carlo simulations to take into account selection effects, 
we find an evolutionary trend of the form $M_{\rm BH} / L_{\rm bul} \propto (1+z)^{\gamma}$ with $\gamma=1.8\pm0.7$.
By combining our sample with a literature sample of 27 AGNs at $0.5<z<1.9$ (taken from \citealt{Bennert+11a} and \citealt{Schramm&Silverman13}), we find a weaker, but consistent within the uncertainties, evolution of $\gamma=0.9\pm0.7$.

The overall evolutionary trend we find is consistent with those reported by 
\citet[][$\gamma=1.5\pm1.0$]{Treu+07} and \citet[][$\gamma=1.4\pm0.2$]{Bennert+10} based on the \mlbul\ relation and 
\citet[][$\gamma=2.07\pm0.76$]{McLure+06}, \citet[][$\gamma=1.2$]{Jahnke+09}, \citet[][$\gamma=1.4$]{Decarli+10}, \citet[][$\gamma=1.15\pm0.34$]{Cisternas+11}, \citet[][$\gamma=1.96\pm0.55$]{Bennert+11b} based on the \mmbul\ relation and
\citet[][$\gamma=1.66\pm0.43$]{Woo+06}, \citet[][$\gamma=3.1\pm1.5$]{Woo+08} based on the \msigma\ relation.
From a theoretical approach using a self-regulated BH growth model
\citet{Wyithe&Loeb03} also expect $M_{\rm BH} / M_{\rm bul} \propto (1+z)^{3/2}$.
\citet{Merloni+04} present a weaker evolution of $M_{\rm BH} / M_{\rm bul} \propto (1+z)^{1/2}$ 
based on empirical models for the joint evolution of the stellar and BH mass densities.
Using global constrains on the BH mass density evolution from the galaxy distribution functions and the AGN luminosity function,
\citet{Shankar+09a} and \citet{Zhang+12} find a mild evolution of $\gamma = 0.33$ and $\gamma = 0.64\pm0.28$, respectively.
Recently, \citet{Shankar+13} predicted evolution for both the \msigma\ and \mmbul\ relations based on the Munich semi-analytic model of galaxy formation and evolution. 

Our results indicate that BHs in the distant Universe tend to reside
in smaller bulges than today.
Interpreted in the framework of co-evolution of BHs and their host galaxies
and assuming that the local relation is the final product,
BHs grow first and their host galaxies need to catch up.
Thus, a substantial bulge growth is expected between the observed intermediate-$z$ epochs and today.
Out of our sample of 52 active galaxies, $\sim 30$\% show signs of (major) mergers/interactions --
a promising way to grow the bulge.
\citet{Croton06} suggested that a merger with a disk-dominated system 
containing no BH can explain substantial growth of bulge luminosity by transferring stars in a disk to a bulge.
However, this would only work for a fraction of our sample.
Recently, secular evolution driven by disk instabilities and/or minor merging 
has also been suggested for the bulge growth mechanism by redistributing mass into 
the bulge component without a significant growth of BH
\cite[e.g.,][]{Parry+09,Jahnke+09,Cisternas+11,Bennert+10,Bennert+11b,Schramm&Silverman13}.

Selection effects can mimic an evolutionary trend (\citealt{Lauer+07,Shen&Kelly10,Schulze&Wisotzki11}, see also \citealt{Merloni+10, Volonteri&Stark11,Portinari+12,Salviander&Shields13,Schulze&Wisotzki14}).
Thus, we here consider three kinds of selection effects in the analysis.
(i) Performing Monte Carlo simulations, we take into account the potential bias 
that might arise when selecting a broad-line AGN sample based on their luminosities (i.e., BH masses)
\citep[][]{Treu+07,Lauer+07}.
Given the presence of intrinsic scatter of the scaling relations,
particularly in the high-luminosity regime where the galaxy (and bulge) luminosity function is steeply decreasing,
this can lead to a preferential selection of higher mass BHs. 

(ii) In the same simulations, we also take into account the selection effect introduced
by the large uncertainties on BH mass measured from the SE method
(\citealt{Shen&Kelly10}; but see also \citealt{Schulze&Wisotzki11}).
It is more likely to detect massive BHs at a given bulge luminosity 
since the true lower mass BHs have a higher chance of being scatted into the higher SE mass bin 
through the SE mass estimates with large uncertainty than the intrinsically higher-mass BHs, under the steeply declining BH mass function.
Thus, this will lead to a positive bias. 
On the contrary, a negative bias may be expected from the uncertainty of the bulge luminosity -- 
given the steeply declining galaxy luminosity function, for a given BH mass,
there will be a higher chance of scattering  effectively less luminous galaxies into the brighter luminosity bins.

(iii) Lastly, we consider the active fraction selection function suggested by \citet{Schulze&Wisotzki11}
that can cause a negative offset in a sample of AGNs by preferentially observing less massive BHs for a given bulge luminosity
in the presence of intrinsic scatter of the scaling relation,
since the active fraction (i.e., the probability of BHs to be observed as active galaxies) decreases as a function of mass.
Since the details of mass and redshift dependence of the active fraction is not well-known,
we by-pass this bias by performing Monte Carlo simulations based on active BH mass function,
assuming that the active fraction is independent of redshift for the redshift range covered by our sample.

Aside from these selection effects, there are other limitations that need to be addressed
for a better estimation of the evolution of the scaling relations.
First, BH mass measurements for distant active galaxies have to rely on the empirically calibrated SE method
which is subject to relatively large random and systematic uncertainties
(see a review by \citealt{Shen13} and references therein).
The largest systematic uncertainty stems from the virial factor 
that depends on the unknown kinematics and geometry of the BLR
and is currently adopted from an empirically-calibrated average virial factor for the entire BH population \citep[see, e.g.,][]{Woo+10,Park+12b,Woo+13}.
A direct assessment of the virial factor for each active galaxy will greatly reduce the uncertainties in \mbh\ measurements 
\citep[see, e.g.,][]{Pancoast+11, Pancoast+12,Pancoast+13, Brewer+11,Li+13}.

Second, the results from our own image decomposition might be systematically different to those from  
other published studies \citep[e.g., using GALFIT;][]{Peng+02, Peng+10};
however, a thorough comparison is beyond the scope of this work.

Third, the sample of local RM AGNs is small and covers a small dynamic range.
The extension of this sample and a more complete establishment of the local scaling 
relation will ultimately shed light on the accurate characterization of the BH-galaxy co-evolution.
Although the BH mass range covered in our sample and the local RM AGNs are almost the same,
we need to extend our sample to higher and lower \Lbul\ regimes for a more direct comparison to the local RM AGNs.
Extending the sample toward the low-mass regime ($M_{\rm BH}\lesssim10^{7.5}$\msun) where the magnitude of selection biases is expected to be smaller is essential.

Properly taking into account the selection effects, we have derived the overall positive evolutionary trend, although the result is subject to the adopted prior for the intrinsic scatter because we cannot constrain the slope and intrinsic scatter simultaneously due to the insufficient dynamic range of our sample. At this point, it is difficult to distinguish between a mean evolution of the scaling relations (normalization) and an evolution of their intrinsic scatter \citep[see also][]{Merloni+10} with our sample ; larger data sets of uniformly selected and consistently measured samples are necessary.

\acknowledgments
This work has been supported by the National Research Foundation of Korea (NRF) grant funded by the Korea government (No.  2012-R1A2A2A01006087). D.P. thanks Hyung Mok Lee, Aaron J. Barth, and Daniel J. Carson for helpful comments and Hee Il Kim for practical 
help to use computational resources, \texttt{Zenith} and \texttt{Gmunu} linux clusters at SNU. VNB acknowledges assistance from a National Science Foundation (NSF) Research at Undergraduate Institutions (RUI) grant AST-1312296.
Note that findings and conclusions do not necessarily represent views of the NSF.
This work is based on data obtained with the {\it Hubble Space Telescope} and the 10m W. M. Keck Telescope.
We acknowledge financial support from NASA through \HST\ proposals GO-10216, GO-11166, GO-11208, and GO-11341.
We thank the anonymous referee for useful comments and suggestions that have improved the paper.

\begin{figure*}
\centering
    \includegraphics[width=0.75\textwidth]{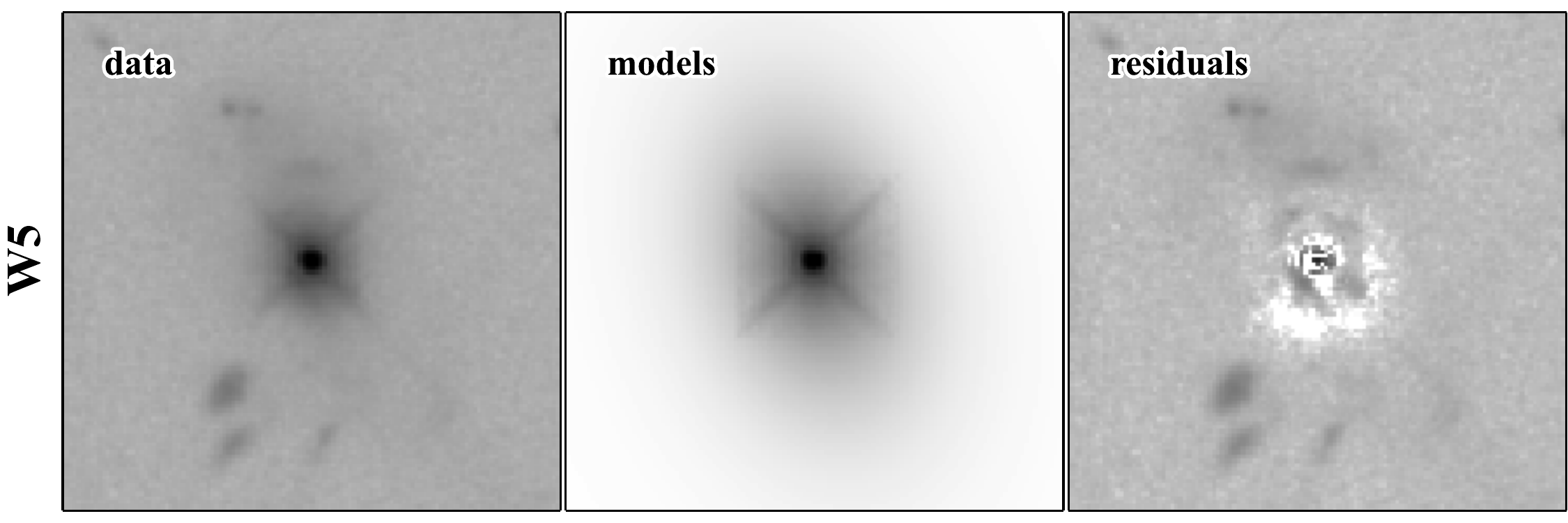}\includegraphics[width=0.25\textwidth]{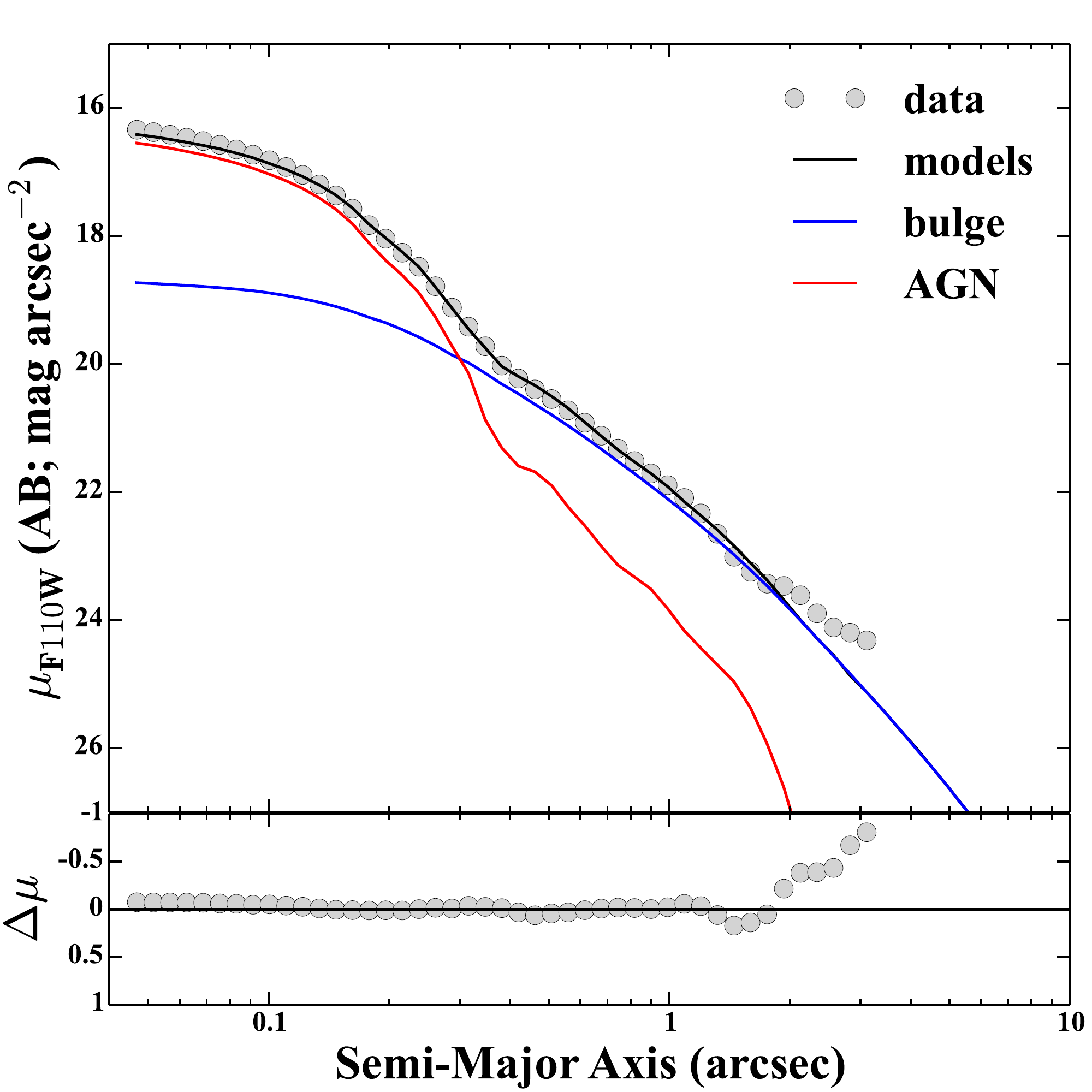}\\
    \includegraphics[width=0.75\textwidth]{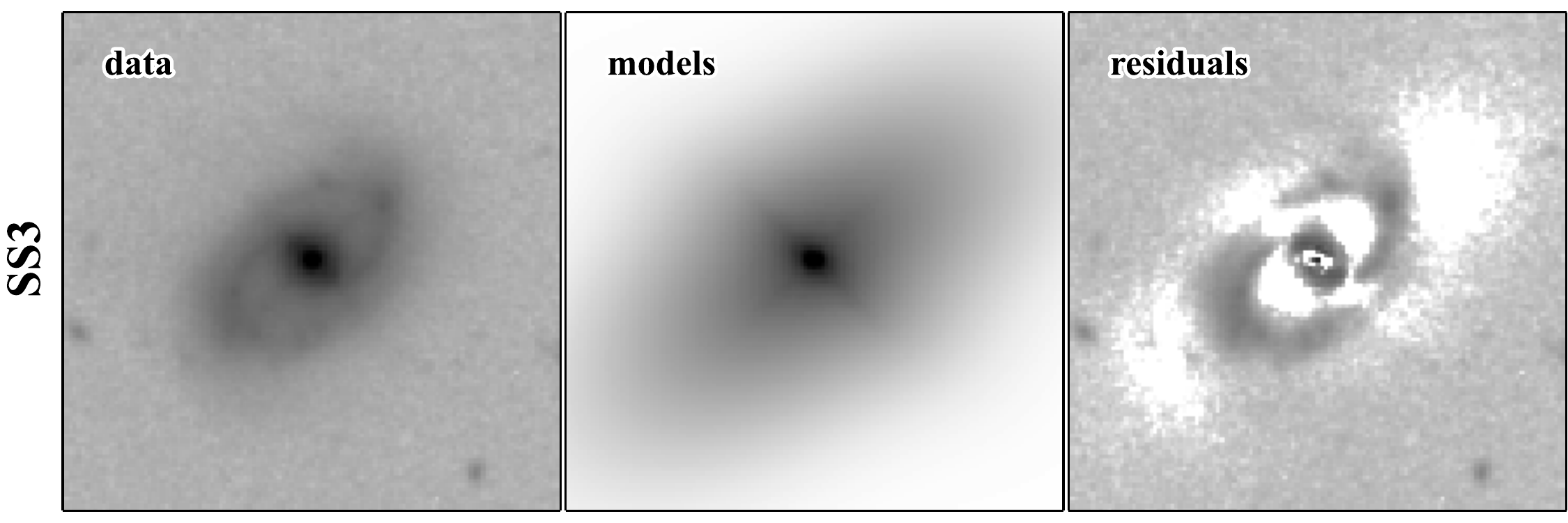}\includegraphics[width=0.25\textwidth]{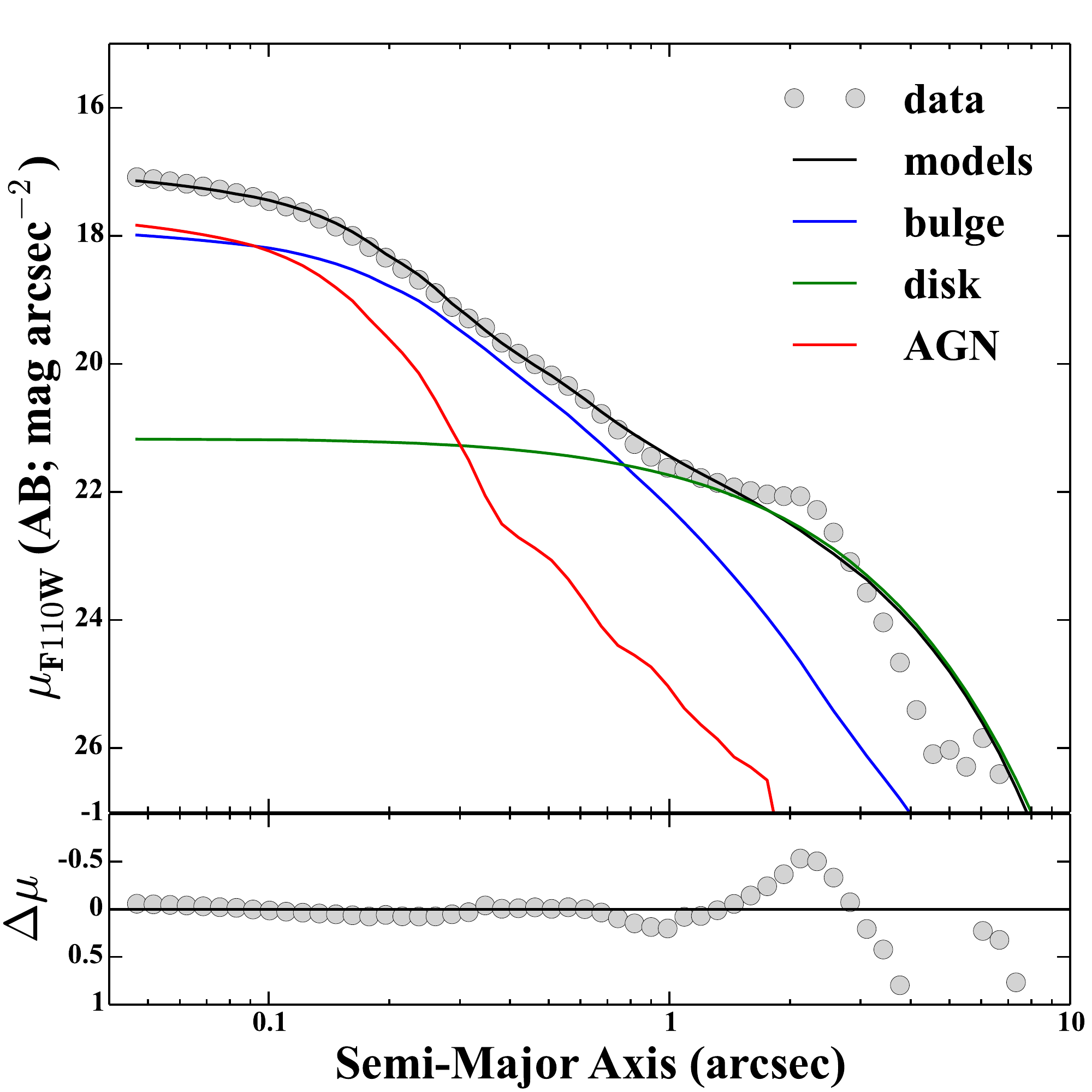}\\
    \includegraphics[width=0.75\textwidth]{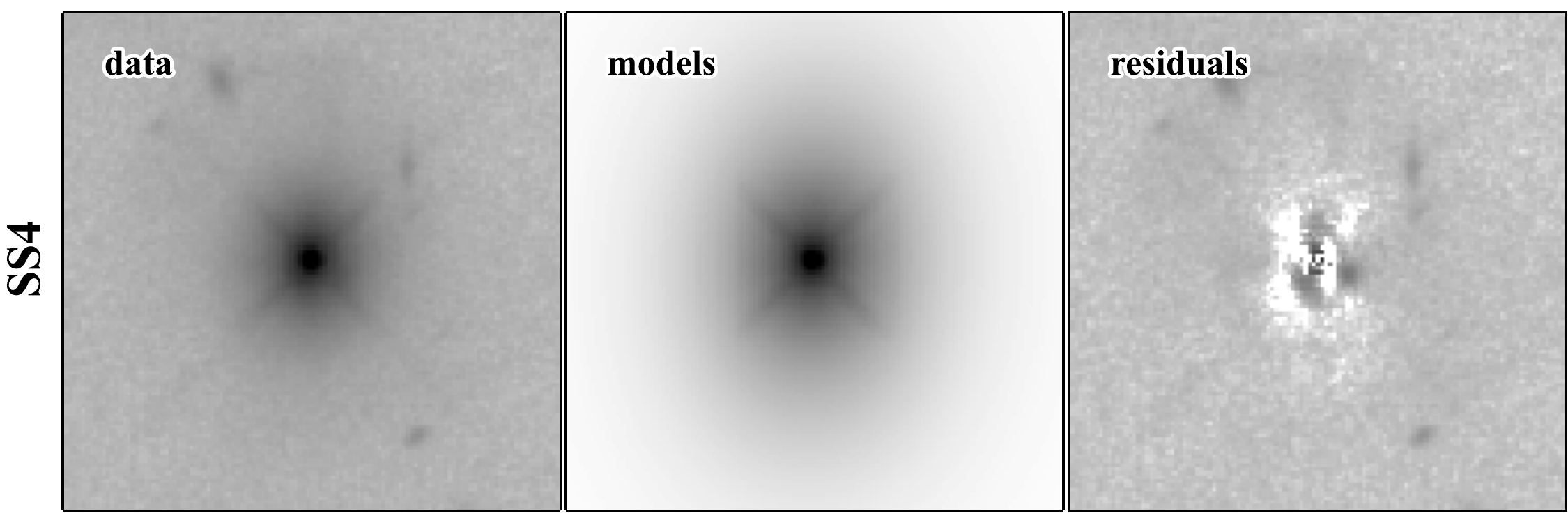}\includegraphics[width=0.25\textwidth]{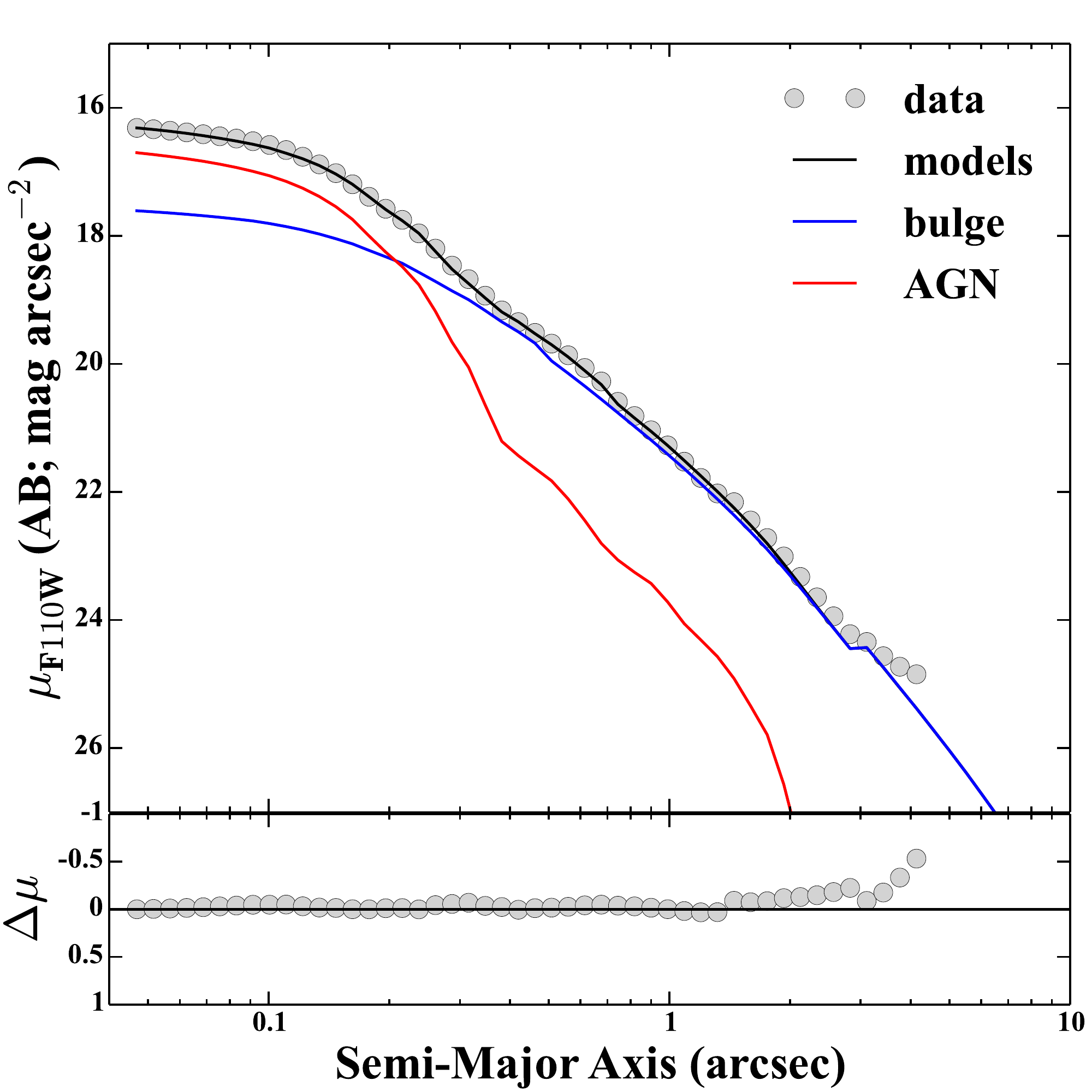}\\
    \includegraphics[width=0.75\textwidth]{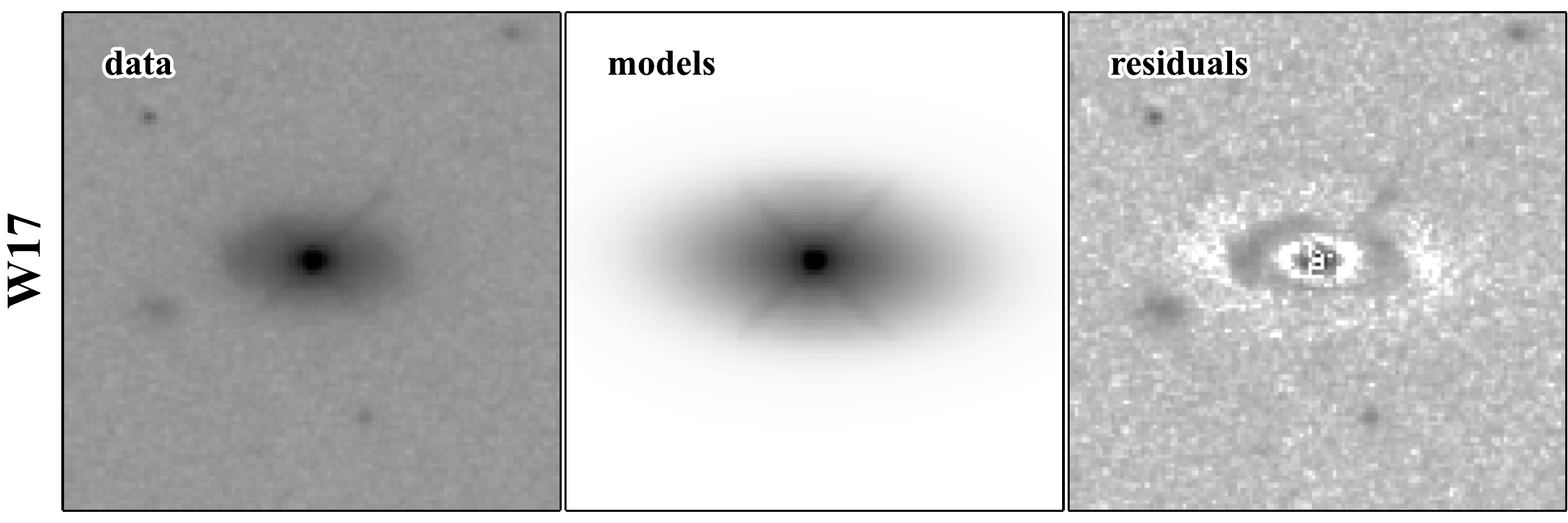}\includegraphics[width=0.25\textwidth]{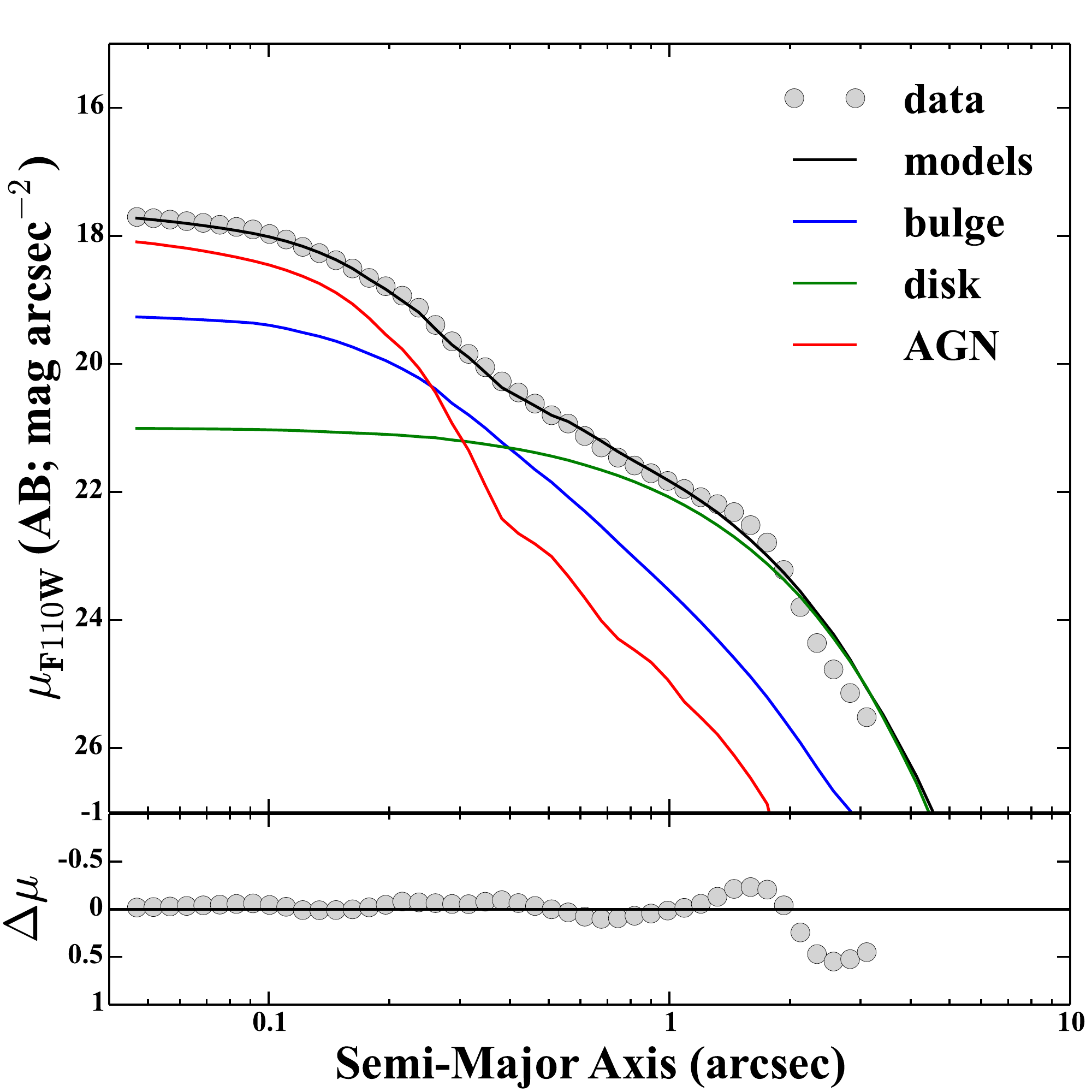}
    \figurenum{2}
        \caption{
        \it Continued.
        \label{fig:imgfit1dSBP_WFC3_2}}
\end{figure*}

\begin{figure*}
\centering
    \includegraphics[width=0.75\textwidth]{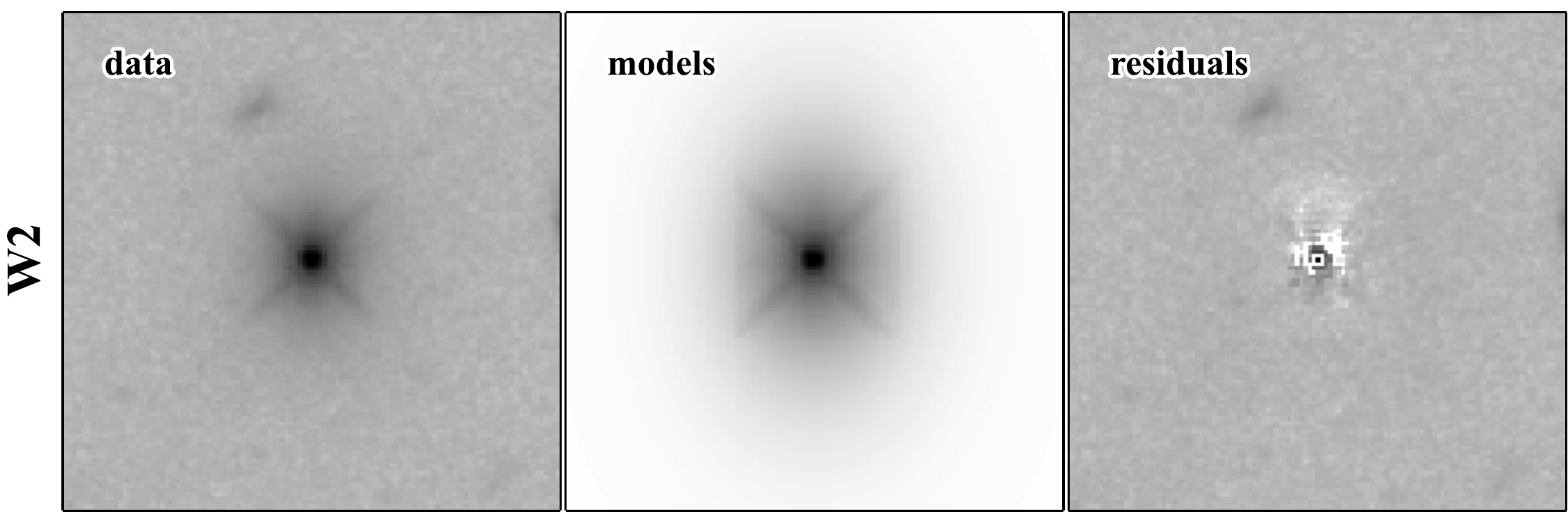}\includegraphics[width=0.25\textwidth]{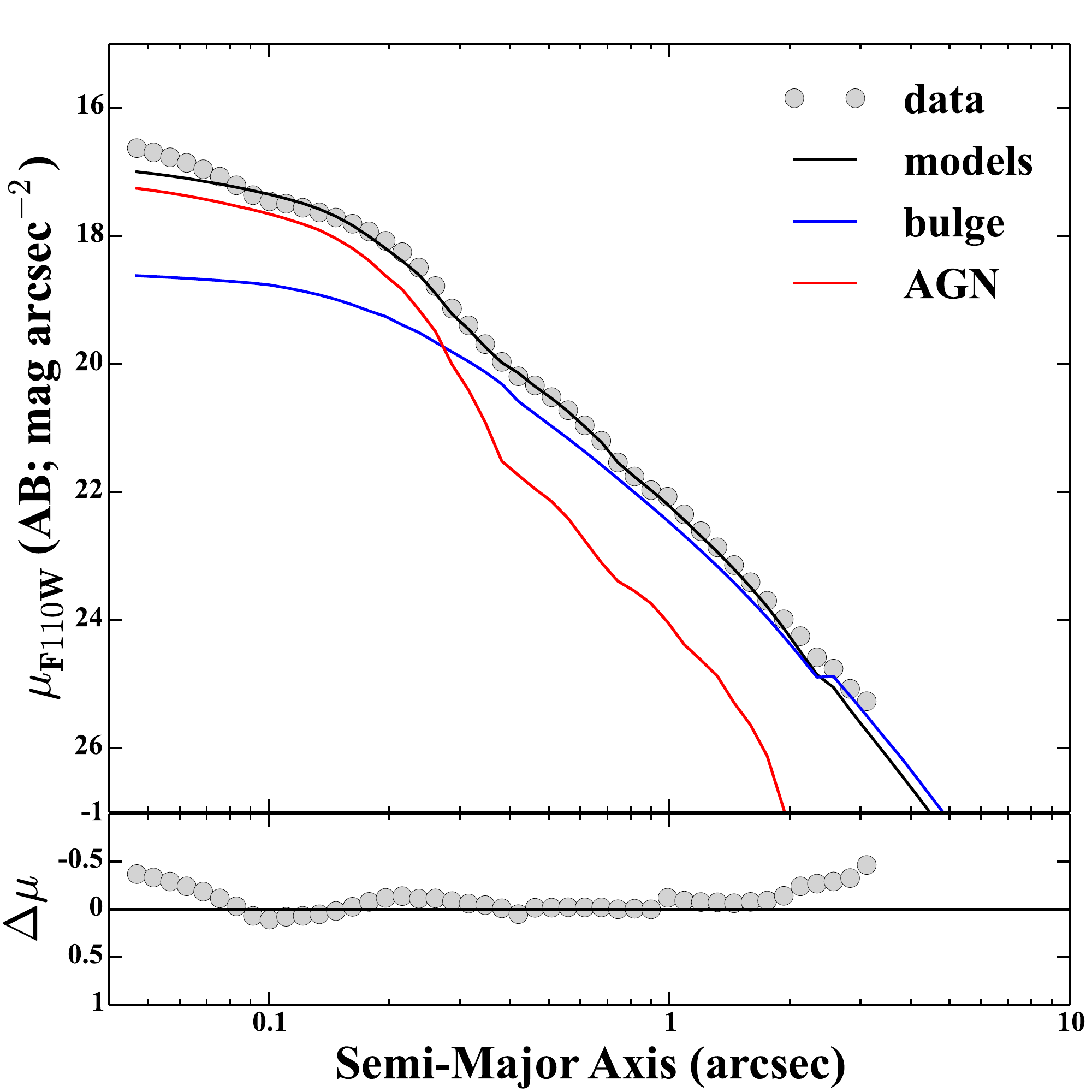}\\
    \includegraphics[width=0.75\textwidth]{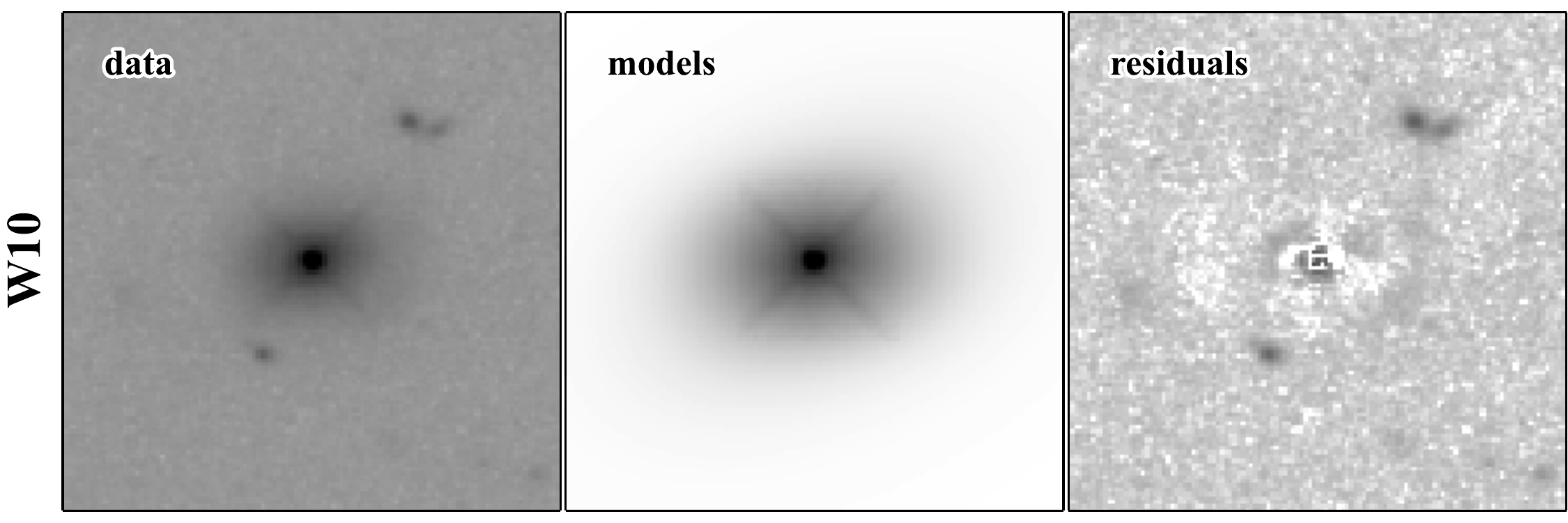}\includegraphics[width=0.25\textwidth]{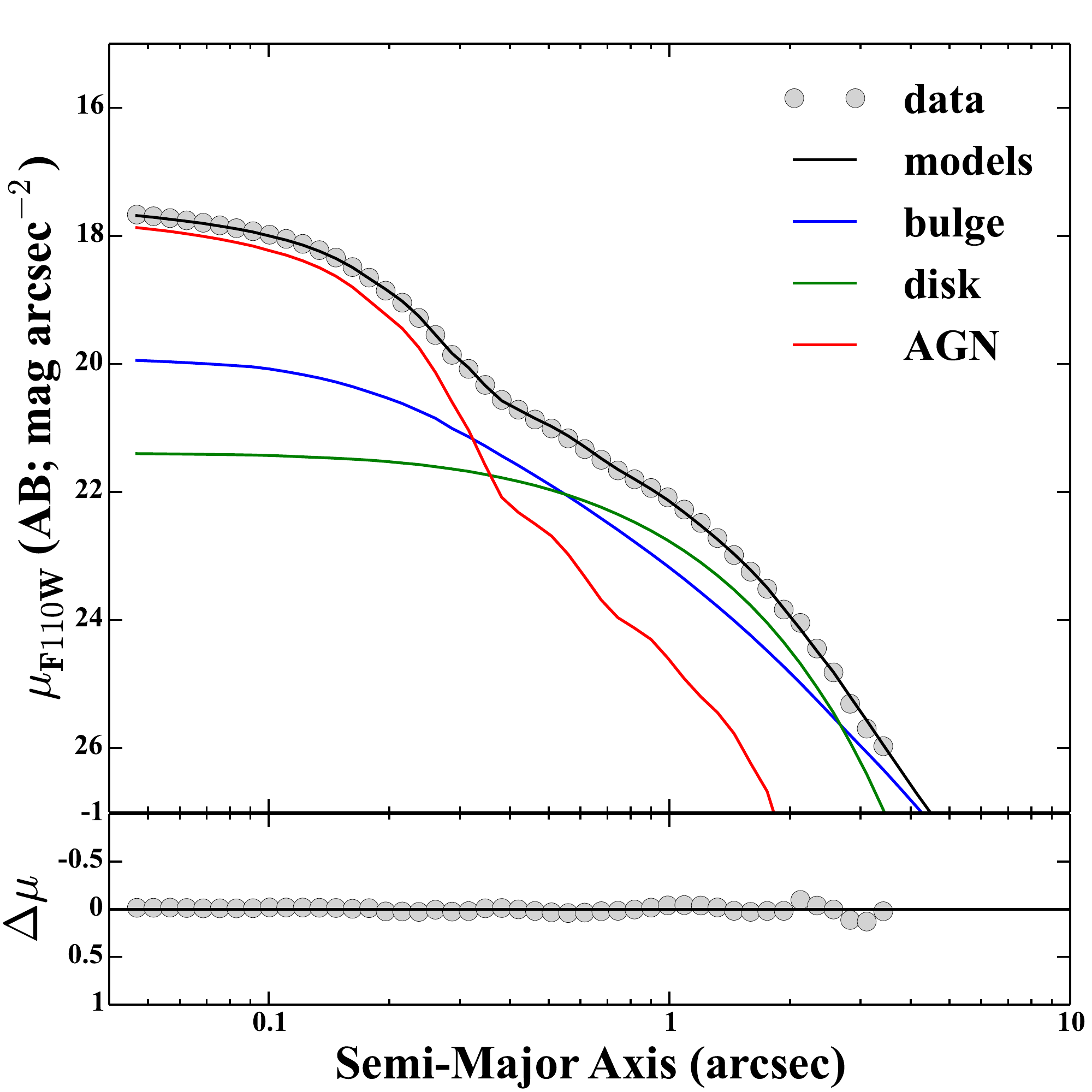}\\
    \includegraphics[width=0.75\textwidth]{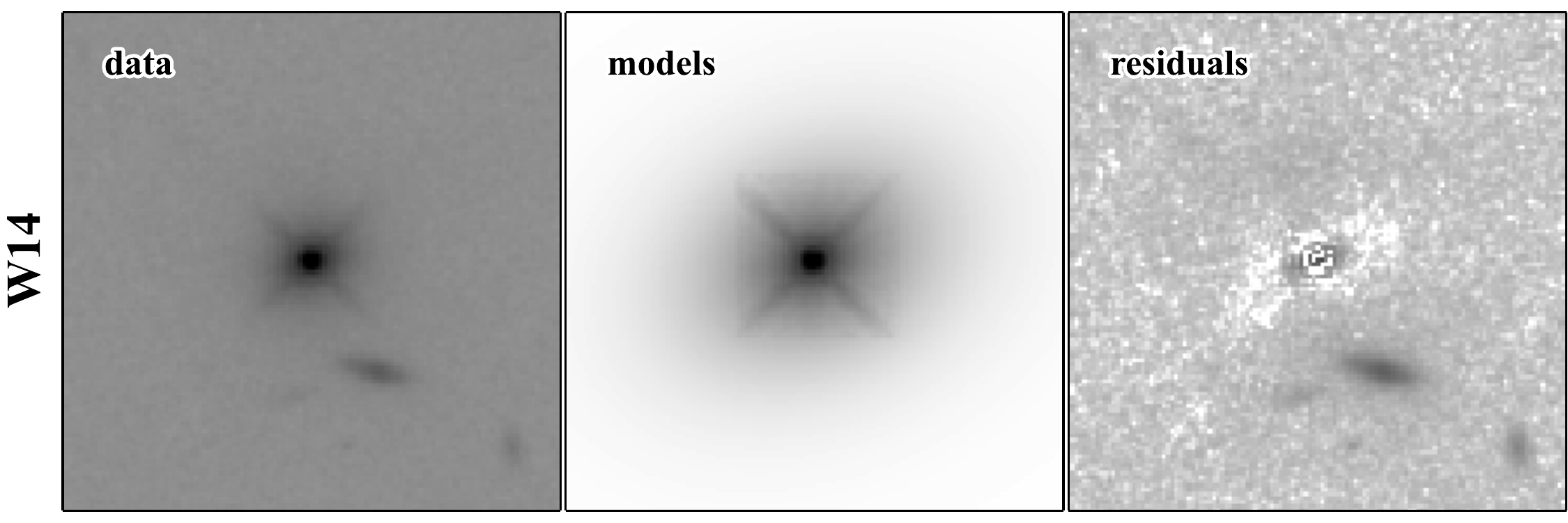}\includegraphics[width=0.25\textwidth]{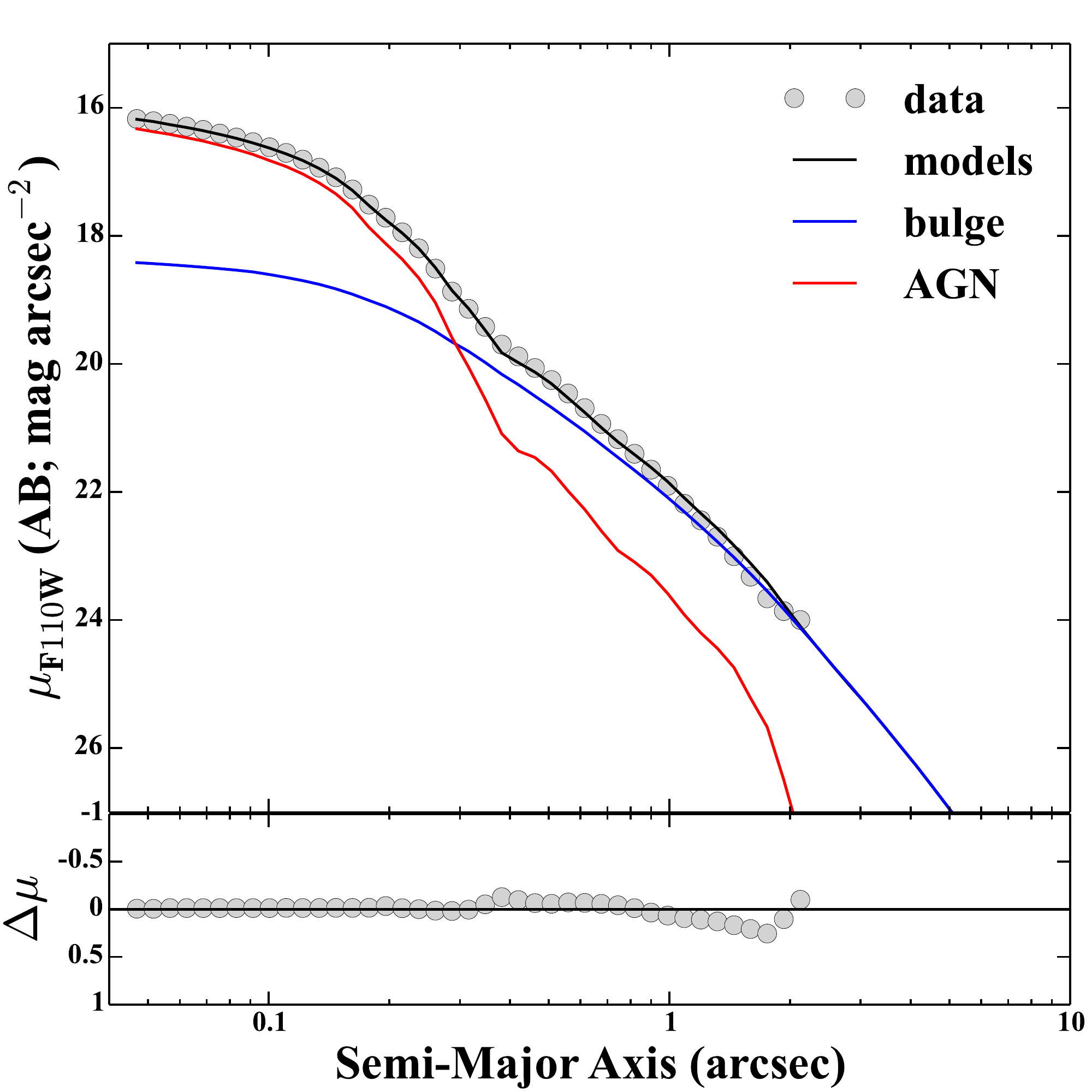}\\
    \includegraphics[width=0.75\textwidth]{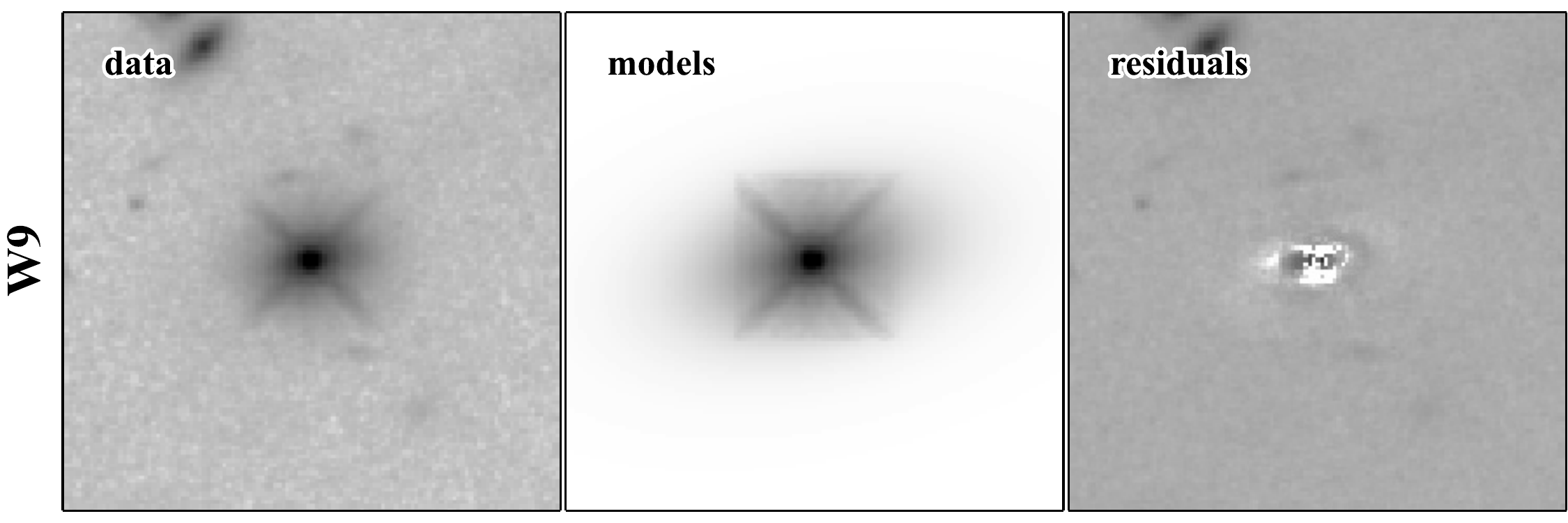}\includegraphics[width=0.25\textwidth]{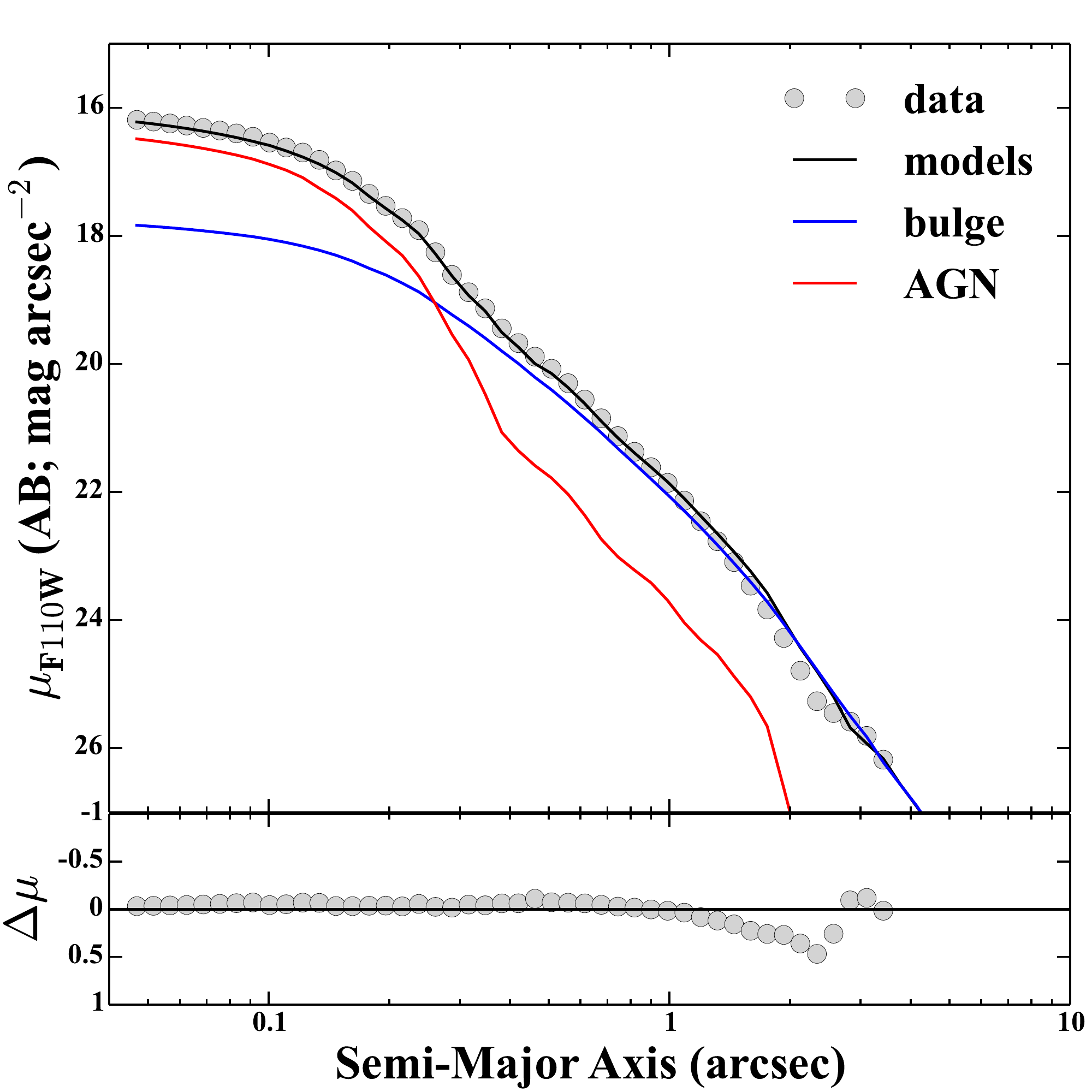}
    \figurenum{2}
        \caption{
        \it Continued.
        \label{fig:imgfit1dSBP_WFC3_3}}
\end{figure*}

\begin{figure*}
\centering
    \includegraphics[width=\textwidth]{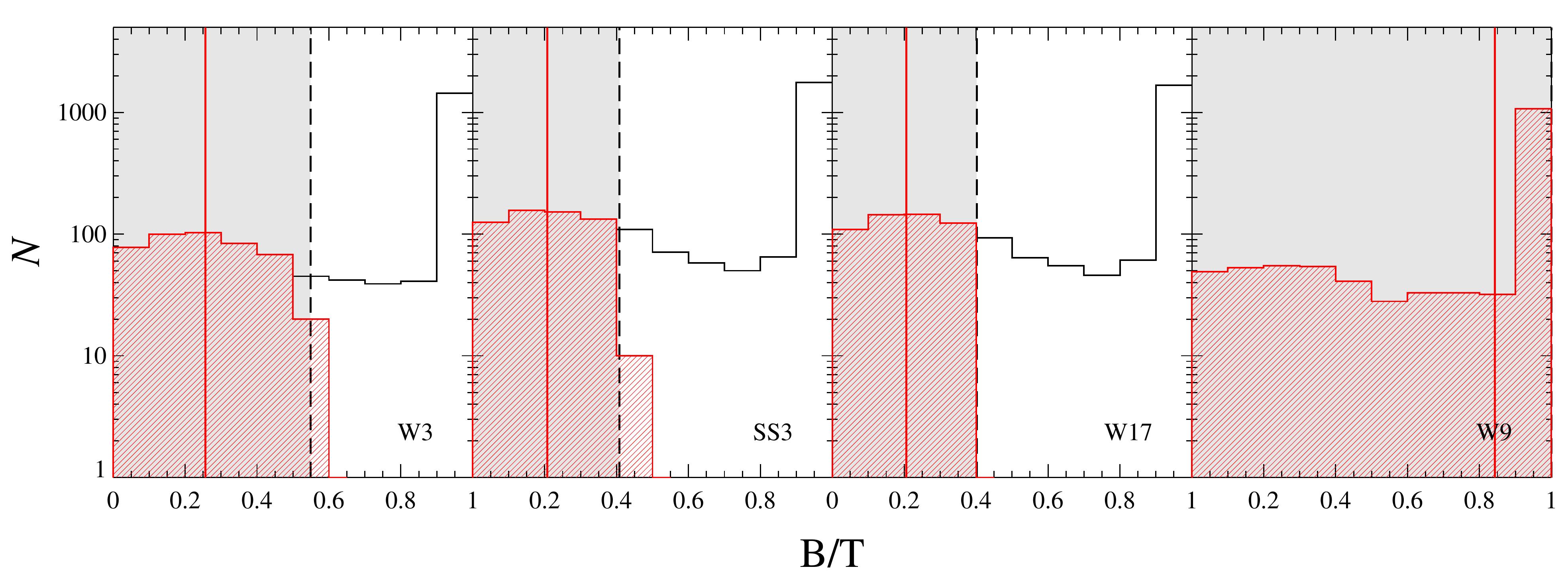}
    \caption{
    Bulge-to-total (B/T) luminosity ratio distributions using 
informative priors from \citet{Benson+07}  to estimate \Lbul~for those four objects with upper limits.
    The black histograms indicate B/T prior distributions from SDSS galaxies that have total magnitudes within $\pm0.5$ mag to those of our active galaxy sample.
    The vertical black dashed line shows an upper limit value for the B/T measured from our surface photometry and 
    the B/T likelihood function as the form of a step function is displayed as a grey shade.
    The posterior distribution for the B/T ratios, derived by combining the prior (black histogram) and likelihood (gray shade), is plotted as a red hashed histogram with its mean value (vertical red solid line) in each panel.
    \label{fig:BTprior}}
\end{figure*}

\begin{figure*}
\centering
    \includegraphics[width=0.6\textwidth]{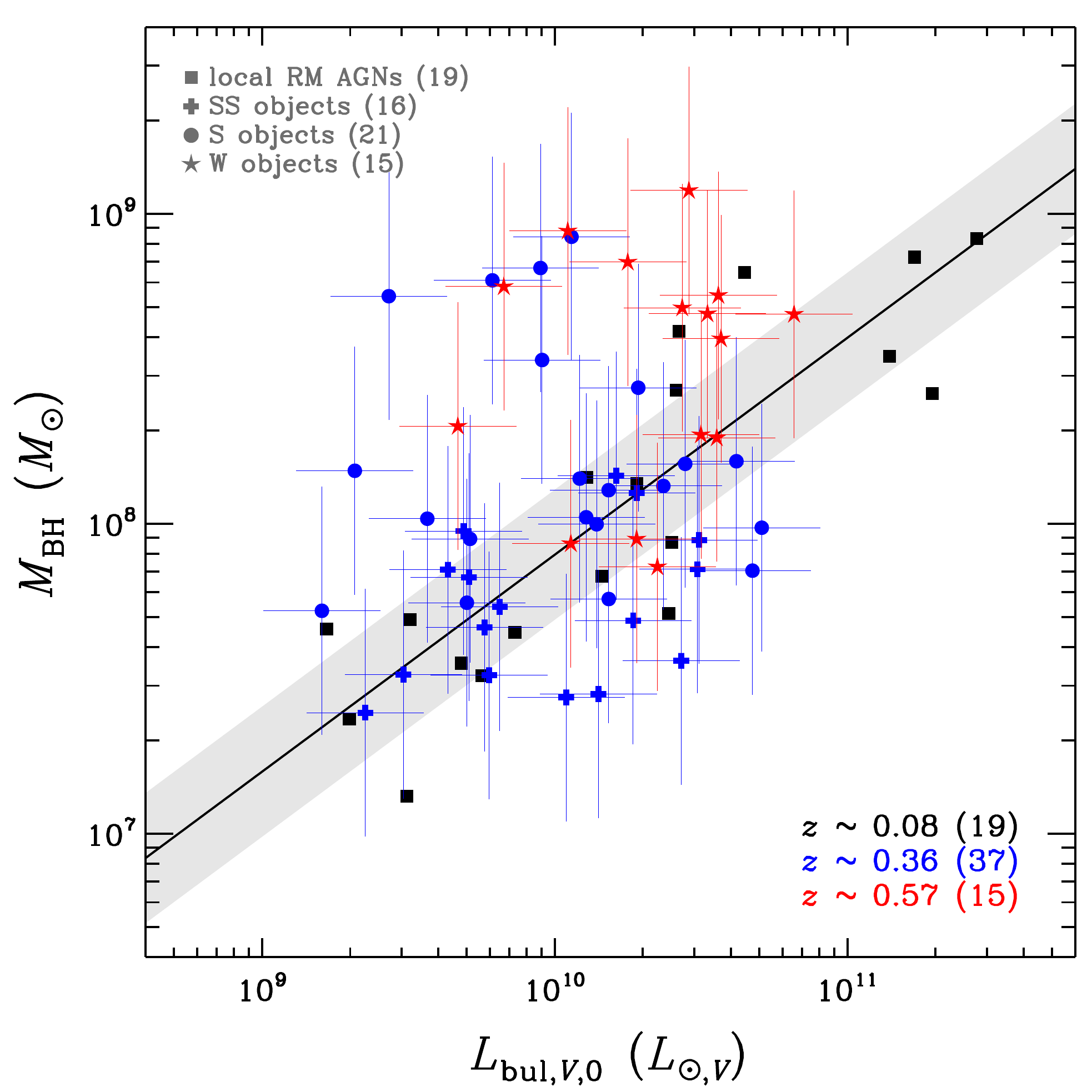}
    \caption{
    \mlbul\ relation. Colored symbols indicate our intermediate-$z$ sample 
(plus signs: SS objects; circles: S objects; stars: W objects; see Sec.~\ref{sec:sampleselection} for the details of the sample). 
   Corresponding redshifts of the samples are expressed by different colors (black: local ($\bar{z} \sim 0.08$); blue: $z = 0.36$; red: $z=0.57$). The black filled squares are the local RM AGNs taken from \citet{Bennert+10} with the best-fit relation (black solid line) and its intrinsic scatter (0.21 dex; gray shaded region).
     \label{fig:M-Lbul_reestimated_scatter}}
\end{figure*}

\begin{figure*}
\centering
    \includegraphics[width=0.7\textwidth]{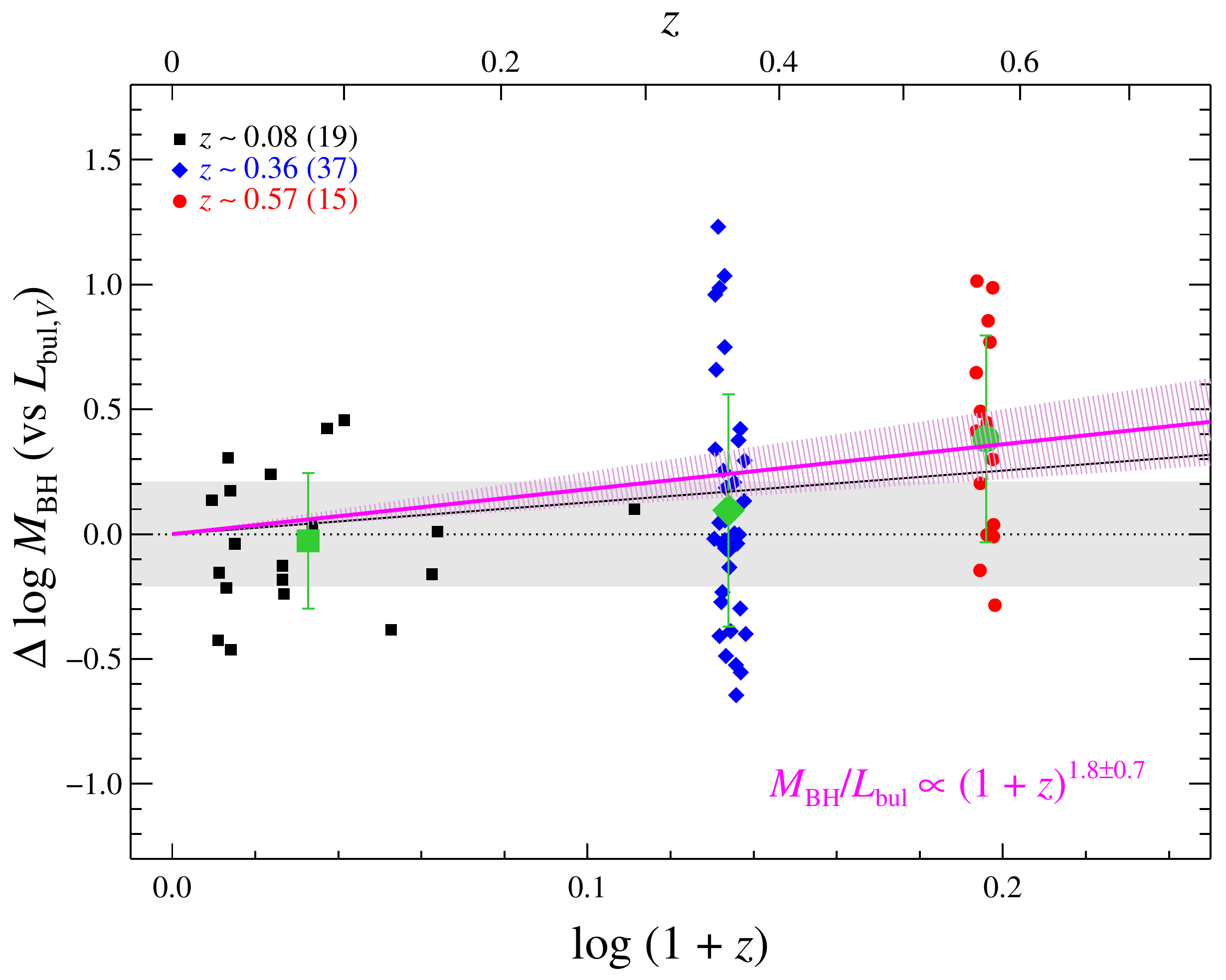}
    \caption{
    Redshift evolution of the offset in $\log M_{\rm BH}$ for a given \Lbul\ with respect to local baseline \mlbul\ relation (black dotted line with gray shaded region showing the intrinsic scatter).
    Colored symbols indicate local RM AGNs (black squares at $\bar{z} \sim 0.08$) and our intermediate-$z$ sample at $z = 0.36$ (blue diamonds) and at $z = 0.57$ (red circles).
    The mean and root-mean-square (rms) scatter of offsets for each sample are shown as green big symbols with error bars.
    The black solid line represents the best-fit trend for all intermediate-$z$ objects in the functional form of $\Delta \log M_{\rm BH} = \gamma \log (1+z)$ without taking into account for selection effects.
    The magenta solid line with the hatched $1\sigma$ confidence range shows the result when taking into account selection effects.     The corresponding best-fit value for the evolution slope is given in the lower right corner.
    \label{fig:M0-Lbul0_offset}}
\end{figure*}

\begin{figure*}
\centering
    \includegraphics[width=0.6\textwidth]{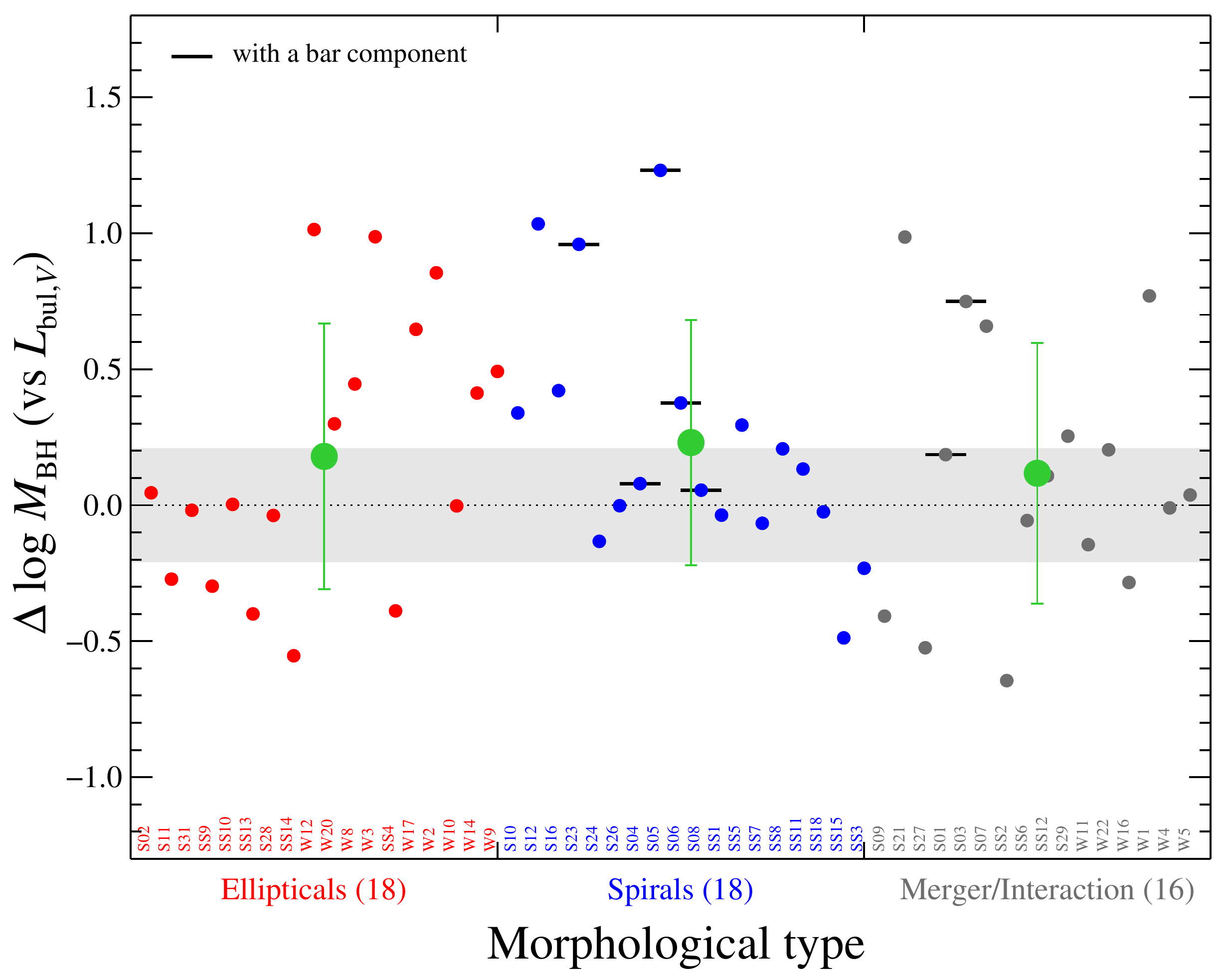}
    \caption{
    The measured offset in $\log M_{\rm BH}$ for a given \Lbul\ with respect to local baseline \mlbul\ relation (black dotted line with gray shaded region for the intrinsic scatter) with simple morphological type classification based on the visual inspection of \HST\ images.
	Objects containing a bar component are indicated with a black horizontal bar. 
    \label{fig:M-Lbul_offset_type}}
\end{figure*}

\begin{figure*}
\centering
    \includegraphics[width=0.45\textwidth]{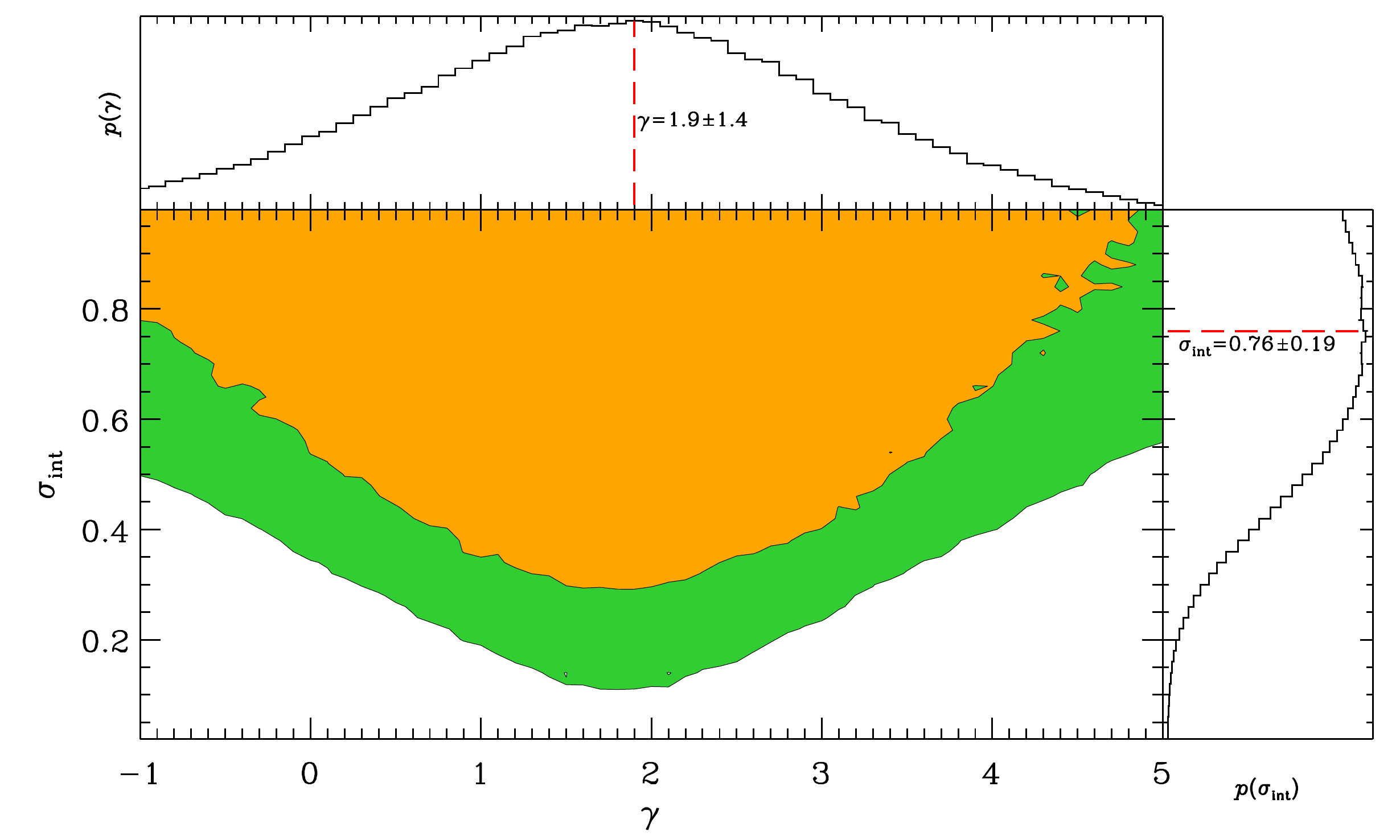}\\
    \includegraphics[width=0.45\textwidth]{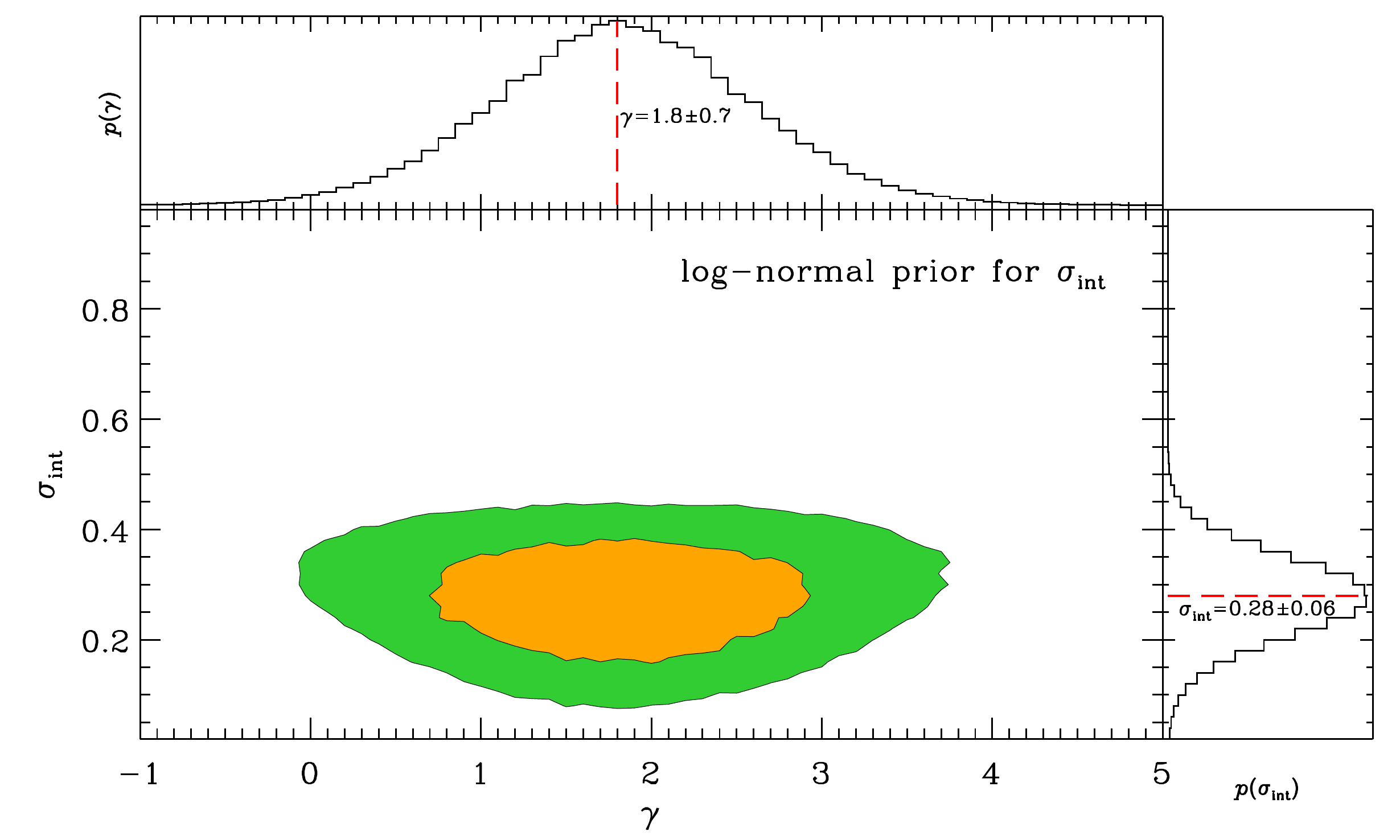}
    \caption{
    Monte Carlo simulation results constraining the evolution of
$\Delta \log M_{\rm BH} = \gamma \log (1+z)$ with intrinsic scatter $\sigma_{\rm int}$,
by taking into account selection effects.
    Upper panel: Evolutionary trend assuming uniform priors; neither slope nor scatter are well constrained.
    Bottom panel: the same as in the upper panel, but assuming a log-normal prior for $\sigma_{\rm int}$
\citep[][$\sigma_{\rm int}=0.21\pm0.08$]{Bennert+10}.
    The 2D posterior distributions of $\gamma$ and $\sigma_{\rm int}$ are plotted with a yellow (green) filled contour 
    corresponding to 1$\sigma$ (2$\sigma$) confidence level.
    The marginalized 1D distributions for each parameter are shown in the top and right sides in each panel
	with the adopted best-fit values (red dashed lines) and 1$\sigma$ uncertainties.
    \label{fig:MC_seleff}}
\end{figure*}

\begin{figure*}
\centering
    \includegraphics[width=0.45\textwidth]{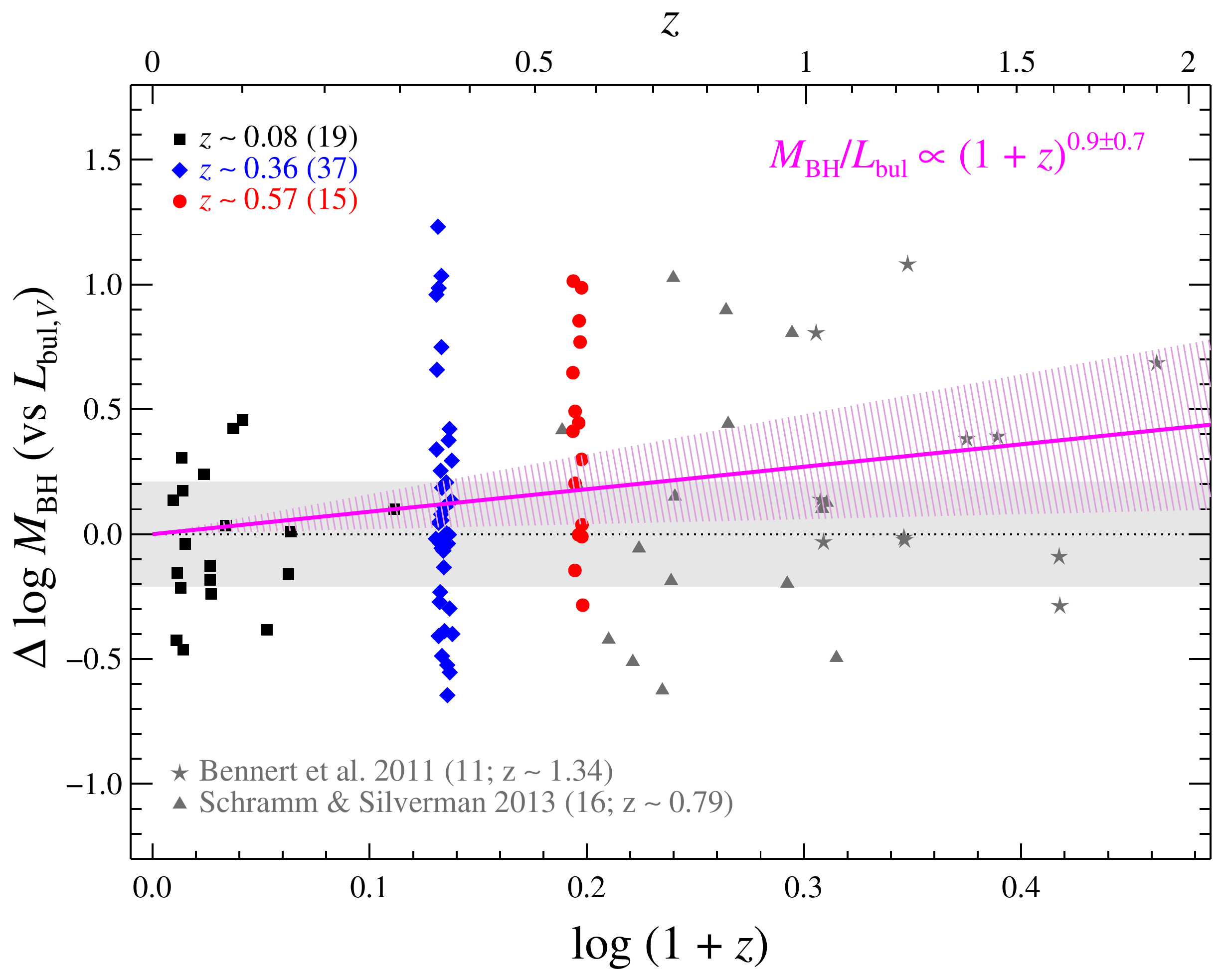}
    \includegraphics[width=0.45\textwidth]{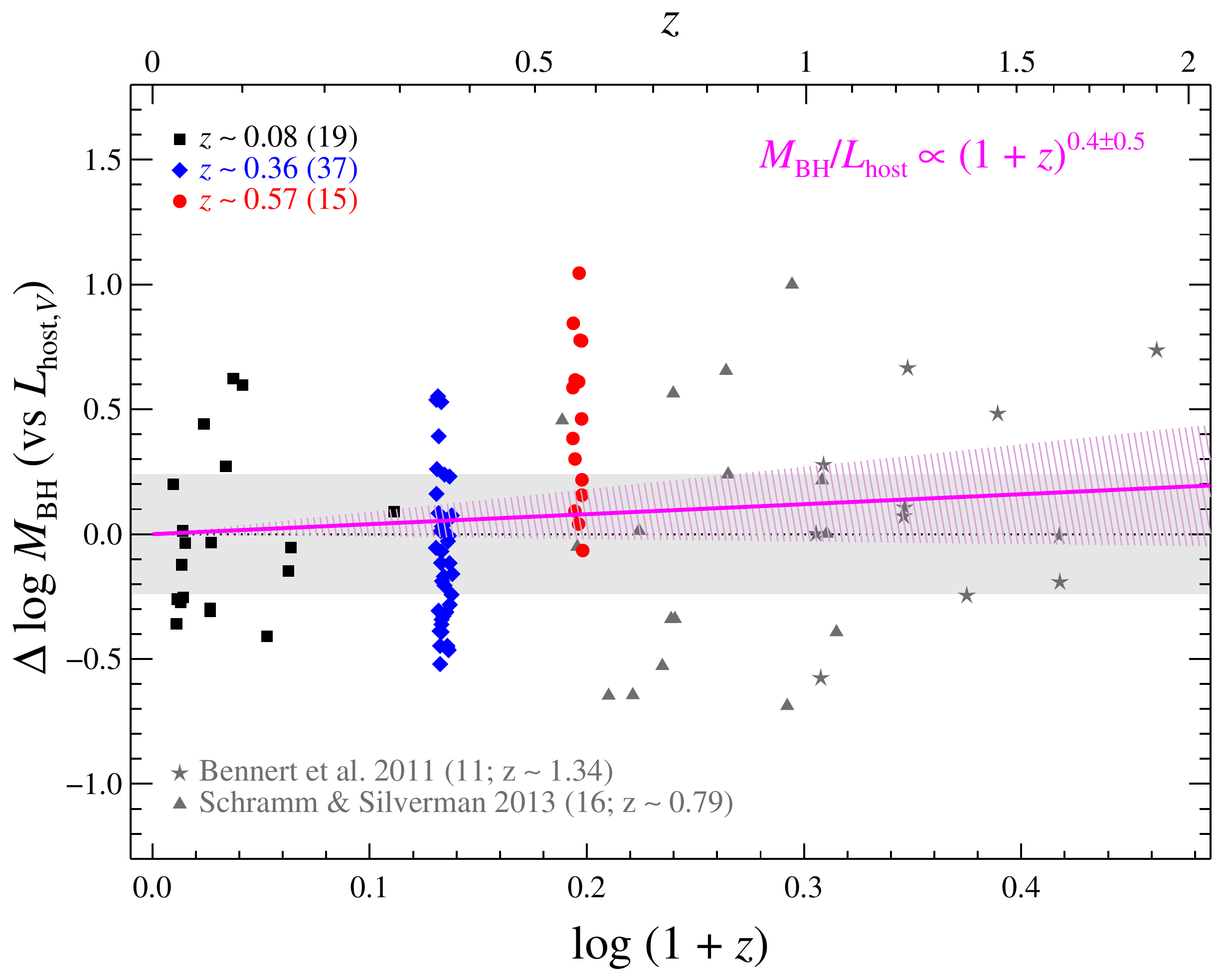}
    \caption{
    Same as Figure~\ref{fig:M0-Lbul0_offset}, but with the additional samples from 
\citet{Bennert+11b} and \citet{Schramm&Silverman13}.
    Left (right) panel shows the evolution of the mass offset for a given \Lbul\ (\Lhost) with respect to the local baseline \mlbul\ (\mlhost) relation. 
    The best-fit evolution slope ($\gamma$) estimated from the Monte Carlo simulation incorporating selection effects is given at each upper right corner and over-plotted as a magenta solid line with a hatched $1\sigma$ confidence range. 
    \label{fig:offset_bul_host_addsample}}
\end{figure*}

\begin{deluxetable*}{llccc}
\tablecolumns{5}
\tablewidth{0pt}
\tablecaption{Sample} 
\tablehead{ 
\colhead{Object} &
\colhead{SDSS name} &
\colhead{$z$} & 
\colhead{$D_{\rm L}$} &
\colhead{$E(B-V)$} \\ 
\colhead{} &
\colhead{} &
\colhead{} &
\colhead{(Mpc)} &
\colhead{(mag)} \\
\colhead{(1)} &
\colhead{(2)} &
\colhead{(3)} &
\colhead{(4)} &
\colhead{(5)} 
} 
\startdata
\multicolumn{5}{c}{Sample presented in \citet{Treu+07}}\\
\\
S09    &  $\rm SDSS-J005916.10+153816.0$  &  0.354488  &  1884.8  &  0.089  \\
S10    &  $\rm SDSS-J010112.06-094500.7$  &  0.351342  &  1865.3  &  0.030  \\
S12    &  $\rm SDSS-J021340.59+134756.0$  &  0.358309  &  1908.6  &  0.104  \\
S21    &  $\rm SDSS-J110556.18+031243.1$  &  0.354551  &  1885.2  &  0.048  \\
S16    &  $\rm SDSS-J111937.58+005620.3$  &  0.370213  &  1983.1  &  0.033  \\
S23    &  $\rm SDSS-J140016.65-010822.1$  &  0.351314  &  1865.1  &  0.039  \\
S24    &  $\rm SDSS-J140034.70+004733.3$  &  0.361910  &  1931.1  &  0.032  \\
S26    &  $\rm SDSS-J152922.24+592854.5$  &  0.369242  &  1977.0  &  0.014  \\
S27    &  $\rm SDSS-J153651.27+541442.6$  &  0.366873  &  1962.1  &  0.020  \\
S01    &  $\rm SDSS-J153916.24+032322.0$  &  0.359351  &  1915.1  &  0.058  \\
S02    &  $\rm SDSS-J161111.66+513131.1$  &  0.354384  &  1884.2  &  0.021  \\
S03    &  $\rm SDSS-J173203.08+611751.8$  &  0.358429  &  1909.3  &  0.040  \\
S04    &  $\rm SDSS-J210211.50-064645.0$  &  0.357906  &  1906.1  &  0.076  \\
S05    &  $\rm SDSS-J210451.83-071209.4$  &  0.353505  &  1878.7  &  0.086  \\
S06    &  $\rm SDSS-J212034.18-064122.2$  &  0.368817  &  1974.3  &  0.186  \\
S07    &  $\rm SDSS-J230946.07+000048.9$  &  0.351999  &  1869.3  &  0.041  \\
S08    &  $\rm SDSS-J235953.44-093655.6$  &  0.358619  &  1910.5  &  0.030  \\
\hline
\multicolumn{5}{c}{Sample presented in \citet{Bennert+10}}\\
\\
S11    &  $\rm SDSS-J010715.97-083429.4$  &  0.355877  &  1893.4  &  0.049  \\
SS1    &  $\rm SDSS-J080427.99+522306.2$  &  0.356555  &  1897.7  &  0.043  \\
SS2    &  $\rm SDSS-J093455.60+051409.1$  &  0.367083  &  1963.4  &  0.033  \\
SS5    &  $\rm SDSS-J100706.26+084228.4$  &  0.373450  &  2003.5  &  0.029  \\
S31    &  $\rm SDSS-J101527.26+625911.5$  &  0.350568  &  1860.5  &  0.006  \\
SS6    &  $\rm SDSS-J102103.58+304755.9$  &  0.358781  &  1911.5  &  0.025  \\
SS7    &  $\rm SDSS-J104331.50-010732.8$  &  0.361284  &  1927.1  &  0.046  \\
SS8    &  $\rm SDSS-J104610.60+035031.2$  &  0.365515  &  1953.6  &  0.039  \\
SS9    &  $\rm SDSS-J125838.71+455515.5$  &  0.370188  &  1982.9  &  0.012  \\
SS10   &  $\rm SDSS-J133414.84+114221.5$  &  0.365808  &  1955.5  &  0.023  \\
SS11   &  $\rm SDSS-J135226.90+392426.8$  &  0.373111  &  2001.3  &  0.016  \\
SS12   &  $\rm SDSS-J150116.82+533102.1$  &  0.362919  &  1937.4  &  0.013  \\
SS13   &  $\rm SDSS-J150541.79+493520.0$  &  0.374316  &  2008.9  &  0.013  \\
S28    &  $\rm SDSS-J161156.29+451610.9$  &  0.367841  &  1968.2  &  0.011  \\
SS14   &  $\rm SDSS-J211531.68-072627.5$  &  0.370558  &  1985.3  &  0.117  \\
S29    &  $\rm SDSS-J215841.92-011500.3$  &  0.357366  &  1902.7  &  0.083  \\
SS18   &  $\rm SDSS-J234050.52+010635.5$  &  0.358543  &  1910.0  &  0.029  \\
W11    &  $\rm SDSS-J015516.18-094556.0$  &  0.565000  &  3282.3  &  0.019  \\
W22    &  $\rm SDSS-J034229.70-052319.4$  &  0.565167  &  3283.5  &  0.042  \\
W12    &  $\rm SDSS-J143955.10+355305.3$  &  0.562309  &  3263.4  &  0.010  \\
W20    &  $\rm SDSS-J150014.81+322940.4$  &  0.576130  &  3360.7  &  0.014  \\
W16    &  $\rm SDSS-J152654.93-003243.3$  &  0.578015  &  3374.0  &  0.106  \\
W8     &  $\rm SDSS-J163252.42+263749.1$  &  0.571209  &  3326.0  &  0.043  \\
\hline
\multicolumn{5}{c}{Sample presented here}\\
\\
W3     &  $\rm SDSS-J002005.69-005016.3$  &  0.576049  &  3360.1  &  0.024  \\
SS15   &  $\rm SDSS-J014412.77-000610.5$  &  0.359329  &  1914.9  &  0.024  \\
W1     &  $\rm SDSS-J083654.98+075712.4$  &  0.573637  &  3343.1  &  0.026  \\
W4     &  $\rm SDSS-J093210.96+433813.1$  &  0.576601  &  3364.0  &  0.018  \\
W5     &  $\rm SDSS-J094852.73+363120.5$  &  0.576728  &  3364.9  &  0.012  \\
SS3    &  $\rm SDSS-J095553.14+633742.8$  &  0.356623  &  1898.1  &  0.028  \\
SS4    &  $\rm SDSS-J095850.15+400342.3$  &  0.362909  &  1937.3  &  0.011  \\
W17    &  $\rm SDSS-J100728.38+392651.8$  &  0.561690  &  3259.0  &  0.012  \\
W2     &  $\rm SDSS-J110641.86+614146.5$  &  0.572026  &  3331.7  &  0.008  \\
W10    &  $\rm SDSS-J111415.83-005920.4$  &  0.571076  &  3325.0  &  0.035  \\
W14    &  $\rm SDSS-J125631.89-023130.6$  &  0.561702  &  3259.1  &  0.019  \\
W9     &  $\rm SDSS-J155227.81+562236.4$  &  0.565356  &  3284.8  &  0.010  
\enddata
\label{tab:objlist}
\tablecomments{
Column 1: Object ID.
Column 2: SDSS name.
Column 3: Redshifts as listed in NED from improved redshifts by \citet{Hewett&Wild10}.
Column 4: Luminosity distance.
Column 5: $E(B-V)$ as listed in NED from the 
\citet{Schlafly&Finkbeiner11} recalibration of the 
\citet{Schlegel+98} infrared-based dust map.
}
\end{deluxetable*}

\begin{deluxetable*}{lccccc}
\tablecolumns{6}
\tablewidth{0pt}
\tablecaption{Results from Keck spectroscopic analysis}
\tablehead{ 
\colhead{Object} &
\colhead{S/N} &
\colhead{FWHM$_{\rm H\beta}$} & 
\colhead{$\sigma_{\rm H\beta}$} &
\colhead{$\lambda L_{5100}^{\rm spec}$} &
\colhead{$\log M_{\rm BH}^{\rm spec}$}  \\
\colhead{ } &
\colhead{(pix$^{-1}$)} &
\colhead{(\kms)} &
\colhead{(\kms)} &
\colhead{($10^{44}$\ergs)} &
\colhead{($M_{\odot}$)} \\
\colhead{(1)} &
\colhead{(2)} &
\colhead{(3)} &
\colhead{(4)} &
\colhead{(5)} &
\colhead{(6)}
} 
\startdata
S09    &     39  &    2655  &    1748  &   1.76  &   8.15  \\
S10    &     96  &    4850  &    2597  &   2.77  &   8.59  \\
S12    &     40  &    8800  &    4256  &   1.82  &   8.93  \\
S21    &     75  &    8296  &    3897  &   5.33  &   9.09  \\
S16    &      6  &    3749  &    1867  &   0.69  &   8.00  \\
S23    &    108  &    9629  &    4251  &   1.78  &   8.92  \\
S24    &    100  &    7061  &    2635  &   1.49  &   8.47  \\
S26    &     50  &    5386  &    1914  &   0.83  &   8.06  \\
S27    &     42  &    2508  &    1409  &   1.26  &   7.89  \\
S01    &     69  &    4662  &    2194  &   1.37  &   8.29  \\
S02    &     44  &    4841  &    2274  &   1.25  &   8.30  \\
S03    &     88  &    3018  &    1716  &   2.11  &   8.17  \\
S04    &     46  &    2821  &    1749  &   1.19  &   8.06  \\
S05    &    119  &    4908  &    3333  &   2.23  &   8.76  \\
S06    &     31  &    4527  &    1413  &   1.10  &   7.86  \\
S07    &    108  &    4635  &    2547  &   1.81  &   8.48  \\
S08    &     54  &    2909  &    1217  &   1.59  &   7.81  \\
S11    &    114  &    2595  &    1354  &   1.57  &   7.90  \\
SS1    &     26  &    2620  &    1501  &   1.04  &   7.90  \\
SS2    &     32  &    2815  &    1316  &   0.83  &   7.73  \\
SS5    &     46  &    2790  &    1612  &   1.40  &   8.03  \\
S31    &     79  &    4012  &    2117  &   0.93  &   8.17  \\
SS6    &     48  &    1947  &    1031  &   0.69  &   7.48  \\
SS7    &     54  &    2959  &    1371  &   0.98  &   7.81  \\
SS8    &     82  &    2733  &    1532  &   1.54  &   8.00  \\
SS9    &     70  &    2787  &    1569  &   1.25  &   7.98  \\
SS10   &     84  &    2232  &    1431  &   4.09  &   8.16  \\
SS11   &     49  &    3505  &    1466  &   2.07  &   8.03  \\
SS12   &    116  &    2101  &    1371  &   4.34  &   8.14  \\
SS13   &    108  &    2169  &    1143  &   1.49  &   7.74  \\
S28    &     73  &    4600  &    2532  &   0.97  &   8.33  \\
SS14   &     51  &    2143  &    1212  &   0.65  &   7.60  \\
S29    &     54  &    3533  &    1847  &   1.20  &   8.11  \\
SS18   &     63  &    1631  &    1029  &   1.90  &   7.71  \\
W11    &     18  &    3812  &    2026  &   0.78  &   8.09  \\
W22    &     81  &    5835  &    2654  &   4.65  &   8.73  \\
W12    &     63  &    7698  &    3859  &   3.62  &   9.00  \\
W20    &     26  &   10861  &    3806  &   1.33  &   8.76  \\
W16    &     37  &    2392  &    1564  &   1.05  &   7.94  \\
W8     &     57  &    7340  &    2977  &   4.17  &   8.81  \\
W3     &     59  &    7461  &    3508  &   1.47  &   8.71  \\
SS15   &     46  &    1604  &    1000  &   0.64  &   7.43  \\
W1     &     80  &    7378  &    3152  &   4.71  &   8.88  \\
W4     &     51  &    3490  &    1728  &   3.68  &   8.30  \\
W5     &     72  &    2722  &    1738  &   4.94  &   8.38  \\
SS3\tablenotemark{a}    &     13  &    1953  &    1252  &   0.74  &   7.66  \\
SS4    &     64  &    2213  &    1378  &   1.35  &   7.88  \\
W17    &     24  &    5556  &    2483  &   0.86  &   8.29  \\
W2     &     66  &   12647  &    4811  &   3.03  &   9.15  \\
W10    &     31  &    3636  &    1477  &   2.92  &   8.12  \\
W14    &     76  &    5001  &    2616  &   5.56  &   8.76  \\
W9     &     62  &    5273  &    2747  &   2.64  &   8.63  
\enddata
\label{tab:specmeas}
\tablecomments{
Column 1: Object ID.
Column 2: S/N averaged at rest wavelength range of 5080--5120 \AA.
Column 3: FWHM of \Hb~broad emission line.
Column 4: Line dispersion of \Hb~broad emission line.
Column 5: Continuum luminosities at 5100 \AA\ as measured from spectra.
Column 6: BH mass derived from Eq~(\ref{eq:MBH}) using $\sigma_{\rm H\beta}$ and $\lambda L_{5100}^{\rm spec}$ measurements. Note that all spectroscopic properties and BH mass estimates
are updated from \citet{Woo+06,Woo+08}. 
}
\tablenotetext{a}{For this object, the results are based on the SDSS DR7 spectrum because no Keck spectrum is available.}
\end{deluxetable*}

\begin{deluxetable*}{lccccccccccccc}
\tablecolumns{13}
\tablewidth{0pt}
\tablecaption{Results from \HST\ image analysis}
\tablehead{ 
\colhead{Object} &
\colhead{Instrument/Filter} &
\colhead{N$_{\rm comp.}$} &
\colhead{Total} &
\colhead{PSF} &
\colhead{Host} &
\colhead{Bulge} &
\colhead{$r_{\rm eff, bul}$} &
\colhead{$r_{\rm eff, bul}$} &
\colhead{$f_{\rm AGN}$} &
\colhead{$\lambda L_{5100}^{\rm image}$} &
\colhead{$\log L_{{\rm host},V}$} &
\colhead{$\log L_{{\rm bul},V}$}  \\
\colhead{ } &
\colhead{ } &
\colhead{ } &
\colhead{(mag)} &
\colhead{(mag)} &
\colhead{(mag)} &
\colhead{(mag)} &
\colhead{($\arcsec$)} &
\colhead{(kpc)} &
\colhead{ } &
\colhead{($10^{44}$\ergs)} &
\colhead{($L_{\odot,V}$)} &
\colhead{($L_{\odot,V}$)} \\
\colhead{(1)} &
\colhead{(2)} &
\colhead{(3)} &
\colhead{(4)} &
\colhead{(5)} &
\colhead{(6)} &
\colhead{(7)} &
\colhead{(8)} &
\colhead{(9)} &
\colhead{(10)} &
\colhead{(11)} &
\colhead{(12)} &
\colhead{(13)} 
} 
\startdata
S09    &  ACS/F775W     &  3  &  18.10  &  19.67  &  18.39  &  18.46  &  2.60  &  12.97  &  0.23  &  0.86  &  10.95  &  10.93                   \\
S10    &  ACS/F775W     &  3  &  17.96  &  19.13  &  18.41  &  19.49  &  0.11  &   0.52  &  0.34  &  1.38  &  10.93  &  10.50                   \\
S12    &  ACS/F775W     &  3  &  18.17  &  19.58  &  18.52  &  20.79  &  0.17  &   0.85  &  0.27  &  0.96  &  10.92  &  10.01                   \\
S21    &  ACS/F775W     &  3  &  17.32  &  18.51  &  17.77  &  19.07  &  0.10  &   0.50  &  0.34  &  2.52  &  11.20  &  10.68\tablenotemark{a}  \\
S16    &  ACS/F775W     &  3  &  19.11  &  19.91  &  19.82  &  21.42  &  0.41  &   2.12  &  0.48  &  0.76  &  10.43  &   9.79                   \\
S23    &  ACS/F775W     &  4  &  18.01  &  19.33  &  18.39  &  20.33  &  0.24  &   1.18  &  0.30  &  1.15  &  10.94  &  10.17                   \\
S24    &  ACS/F775W     &  3  &  18.10  &  20.41  &  18.24  &  18.72  &  1.83  &   9.24  &  0.12  &  0.46  &  11.04  &  10.85                   \\
S26    &  ACS/F775W     &  3  &  18.84  &  20.10  &  19.24  &  19.97  &  0.24  &   1.25  &  0.31  &  0.64  &  10.66  &  10.37                   \\
S27    &  ACS/F775W     &  3  &  18.11  &  19.52  &  18.45  &  18.62  &  4.71  &  23.98  &  0.27  &  1.07  &  10.97  &  10.90                   \\
S01    &  ACS/F775W     &  4  &  18.53  &  19.89  &  18.89  &  20.05  &  0.97  &   4.87  &  0.28  &  0.72  &  10.77  &  10.31                   \\
S02    &  ACS/F775W     &  3  &  19.03  &  20.61  &  19.32  &  19.96  &  0.45  &   2.25  &  0.23  &  0.36  &  10.58  &  10.33                   \\
S03    &  ACS/F775W     &  4  &  17.89  &  18.74  &  18.56  &  21.36  &  0.10  &   0.51  &  0.46  &  2.08  &  10.90  &   9.78\tablenotemark{a}  \\
S04    &  ACS/F775W     &  4  &  18.07  &  19.11  &  18.60  &  19.79  &  0.41  &   2.03  &  0.38  &  1.47  &  10.88  &  10.41                   \\
S05    &  ACS/F775W     &  4  &  17.97  &  18.77  &  18.68  &  21.06  &  0.10  &   0.50  &  0.48  &  1.96  &  10.84  &   9.88\tablenotemark{a}  \\
S06    &  ACS/F775W     &  4  &  18.51  &  20.17  &  18.78  &  21.69  &  0.10  &   0.53  &  0.22  &  0.59  &  10.85  &   9.68\tablenotemark{a}  \\
S07    &  ACS/F775W     &  3  &  17.78  &  18.62  &  18.44  &  20.32  &  0.24  &   1.20  &  0.46  &  2.22  &  10.92  &  10.18                   \\
S08    &  ACS/F775W     &  4  &  18.31  &  19.35  &  18.83  &  21.00  &  0.17  &   0.88  &  0.38  &  1.18  &  10.79  &   9.92                   \\
S11    &  NICMOS/F110W  &  3  &  17.86  &  19.54  &  18.11  &  19.10  &  0.10  &   0.51  &  0.21  &  0.83  &  10.80  &  10.40                   \\
SS1    &  NICMOS/F110W  &  3  &  17.88  &  20.10  &  18.03  &  19.37  &  0.10  &   0.48  &  0.13  &  0.50  &  10.84  &  10.30\tablenotemark{a}  \\
SS2    &  NICMOS/F110W  &  2  &  18.37  &  20.46  &  18.55  &  18.55  &  0.38  &   1.96  &  0.15  &  0.38  &  10.66  &  10.66                   \\
SS5    &  NICMOS/F110W  &  3  &  18.31  &  19.33  &  18.85  &  19.65  &  0.10  &   0.50  &  0.39  &  1.12  &  10.56  &  10.24\tablenotemark{a}  \\
S31    &  NICMOS/F110W  &  3  &  17.56  &  19.27  &  17.81  &  18.41  &  1.25  &   6.17  &  0.21  &  1.02  &  10.90  &  10.66                   \\
SS6    &  NICMOS/F110W  &  3  &  18.84  &  20.19  &  19.20  &  20.38  &  0.10  &   0.48  &  0.29  &  0.46  &  10.37  &   9.90\tablenotemark{a}  \\
SS7    &  NICMOS/F110W  &  3  &  18.30  &  20.07  &  18.54  &  19.41  &  0.10  &   0.49  &  0.20  &  0.53  &  10.64  &  10.30\tablenotemark{a}  \\
SS8    &  NICMOS/F110W  &  3  &  17.89  &  19.67  &  18.12  &  19.95  &  0.10  &   0.49  &  0.19  &  0.78  &  10.82  &  10.09\tablenotemark{a}  \\
SS9    &  NICMOS/F110W  &  2  &  18.02  &  19.33  &  18.41  &  18.41  &  0.31  &   1.57  &  0.30  &  1.09  &  10.72  &  10.72                   \\
SS10   &  NICMOS/F110W  &  3  &  17.55  &  18.19  &  18.42  &  18.92  &  0.10  &   0.51  &  0.55  &  3.05  &  10.71  &  10.51                   \\
SS11   &  NICMOS/F110W  &  3  &  18.11  &  19.65  &  18.41  &  19.75  &  0.10  &   0.49  &  0.24  &  0.83  &  10.73  &  10.19\tablenotemark{a}  \\
SS12   &  NICMOS/F110W  &  2  &  17.37  &  17.73  &  18.75  &  18.75  &  0.10  &   0.48  &  0.72  &  4.61  &  10.56  &  10.56\tablenotemark{a}  \\
SS13   &  NICMOS/F110W  &  2  &  18.38  &  19.29  &  19.00  &  19.00  &  0.23  &   1.16  &  0.43  &  1.17  &  10.50  &  10.50                   \\
S28    &  NICMOS/F110W  &  3  &  18.05  &  20.48  &  18.17  &  18.70  &  0.32  &   1.64  &  0.11  &  0.37  &  10.81  &  10.60                   \\
SS14   &  NICMOS/F110W  &  2  &  19.00  &  20.65  &  19.27  &  19.27  &  0.29  &   1.49  &  0.22  &  0.33  &  10.38  &  10.38                   \\
S29    &  NICMOS/F110W  &  3  &  18.36  &  19.92  &  18.66  &  19.50  &  0.10  &   0.48  &  0.24  &  0.59  &  10.59  &  10.25\tablenotemark{a}  \\
SS18   &  NICMOS/F110W  &  3  &  18.39  &  19.58  &  18.83  &  20.13  &  0.10  &   0.48  &  0.33  &  0.81  &  10.52  &  10.00\tablenotemark{a}  \\
W11    &  NICMOS/F110W  &  2  &  19.62  &  21.41  &  19.85  &  19.85  &  0.28  &   1.83  &  0.19  &  0.41  &  10.63  &  10.63                   \\
W22    &  NICMOS/F110W  &  2  &  17.99  &  19.05  &  18.50  &  18.50  &  1.15  &   7.46  &  0.38  &  3.65  &  11.17  &  11.17                   \\
W12    &  NICMOS/F110W  &  3  &  18.51  &  19.31  &  19.21  &  19.59  &  0.10  &   0.62  &  0.48  &  2.84  &  10.88  &  10.73\tablenotemark{a}  \\
W20    &  NICMOS/F110W  &  2  &  18.98  &  20.99  &  19.17  &  19.17  &  0.49  &   3.20  &  0.16  &  0.64  &  10.93  &  10.93                   \\
W16    &  NICMOS/F110W  &  2  &  19.38  &  20.82  &  19.72  &  19.72  &  0.17  &   1.12  &  0.27  &  0.75  &  10.71  &  10.71                   \\
W8     &  NICMOS/F110W  &  2  &  18.46  &  19.26  &  19.17  &  19.17  &  0.23  &   1.49  &  0.48  &  3.07  &  10.91  &  10.91                   \\
W3     &  WFC3/F110W    &  3  &  18.84  &  19.80  &  19.43  &  20.08  &  0.23  &   1.48  &  0.41  &  1.85  &  10.78  &  10.52\tablenotemark{a}  \\
SS15   &  WFC3/F110W    &  2  &  18.81  &  19.79  &  19.38  &  19.38  &  0.26  &   1.32  &  0.41  &  0.65  &  10.26  &  10.26                   \\
W1     &  WFC3/F110W    &  3  &  18.33  &  18.96  &  19.22  &  19.84  &  1.17  &   7.67  &  0.56  &  3.97  &  10.85  &  10.61                   \\
W4     &  WFC3/F110W    &  2  &  18.38  &  19.19  &  19.09  &  19.09  &  0.48  &   3.15  &  0.48  &  3.26  &  10.91  &  10.91                   \\
W5     &  WFC3/F110W    &  2  &  18.44  &  19.17  &  19.22  &  19.22  &  0.60  &   3.91  &  0.51  &  3.32  &  10.86  &  10.86                   \\
SS3    &  WFC3/F110W    &  3  &  18.16  &  20.37  &  18.31  &  19.28  &  0.23  &   1.13  &  0.13  &  0.38  &  10.68  &  10.29\tablenotemark{a}  \\
SS4    &  WFC3/F110W    &  2  &  17.88  &  19.17  &  18.28  &  18.28  &  0.50  &   2.53  &  0.31  &  1.18  &  10.71  &  10.71                   \\
W17    &  WFC3/F110W    &  3  &  19.15  &  20.46  &  19.53  &  20.52  &  0.23  &   1.46  &  0.30  &  0.95  &  10.70  &  10.31\tablenotemark{a}  \\
W2     &  WFC3/F110W    &  2  &  18.70  &  19.61  &  19.31  &  19.31  &  0.47  &   3.04  &  0.43  &  2.17  &  10.81  &  10.81                   \\
W10    &  WFC3/F110W    &  3  &  19.12  &  20.15  &  19.65  &  20.31  &  0.68  &   4.42  &  0.39  &  1.31  &  10.67  &  10.41                   \\
W14    &  WFC3/F110W    &  2  &  18.27  &  18.93  &  19.12  &  19.12  &  0.49  &   3.19  &  0.54  &  3.89  &  10.87  &  10.87                   \\
W9     &  WFC3/F110W    &  2  &  18.36  &  19.06  &  19.16  &  19.16  &  0.23  &   1.46  &  0.52  &  3.50  &  10.86  &  10.86\tablenotemark{a} 
\enddata
\label{tab:imgmeas}
\tablecomments{
Column 1: Object ID.
Column 2: \HST\ instrument and filter.
Column 3: Number of model components fitted (2=PSF+Bulge; 3=PSF+Bulge+Disk; 4=PSF+Bulge+Disk+Bar).
Column 4: Total extinction-corrected AB magnitude (Total=PSF+Bulge+(Disk)+(Bar)).
Column 5: AGN extinction-corrected AB magnitude (from PSF).
Column 6: Host-galaxy extinction-corrected AB magnitude (Host=Bulge+(Disk)+(Bar)).
Column 7: Bulge extinction-corrected AB magnitude.
Column 8: Bulge effective radius in arcsec.
Column 9: Bulge effective radius in kpc.
Column 10: AGN-to-total light fraction.
Column 11: AGN continuum luminosities at rest-frame 5100 \AA\ in $10^{44}$ \ergs\ measured from images.
Column 12: Host-galaxy luminosity in rest-frame $V$ (solar units), not corrected for evolution.
Column 13: Bulge luminosity in rest-frame $V$ (solar units), not corrected for evolution.
}
\tablenotetext{a}{This bulge luminosity is an upper limit value.}
\end{deluxetable*}

\begin{deluxetable*}{lccc}
\tablecolumns{4}
\tablewidth{0pt}
\tablecaption{Resulting \mbh\ and \Lbul}
\tablehead{ 
\colhead{Object} &
\colhead{$\log M_{\rm BH}$} &
\colhead{$\log L_{{\rm bul},V}$} &
\colhead{$\log L_{{\rm bul},V,0}$} \\
\colhead{} &
\colhead{($M_{\odot}$)} &
\colhead{($L_{\odot,V}$)} &
\colhead{($L_{\odot,V}$)} \\
\colhead{(1)} &
\colhead{(2)} &
\colhead{(3)} &
\colhead{(4)}
} 
\startdata
S09    &   7.99  &  10.93  &  10.71  \\
S10    &   8.44  &  10.50  &  10.29  \\
S12    &   8.78  &  10.01  &   9.79  \\
S21    &   8.93  &  10.28  &  10.06  \\
S16    &   8.02  &   9.79  &   9.56  \\
S23    &   8.82  &  10.17  &   9.95  \\
S24    &   8.20  &  10.85  &  10.62  \\
S26    &   8.00  &  10.37  &  10.14  \\
S27    &   7.85  &  10.90  &  10.68  \\
S01    &   8.15  &  10.31  &  10.09  \\
S02    &   8.02  &  10.33  &  10.11  \\
S03    &   8.17  &   9.54  &   9.32  \\
S04    &   8.11  &  10.41  &  10.18  \\
S05    &   8.73  &   9.65  &   9.43  \\
S06    &   7.72  &   9.43  &   9.20  \\
S07    &   8.53  &  10.18  &   9.96  \\
S08    &   7.74  &   9.92  &   9.70  \\
S11    &   7.76  &  10.40  &  10.18  \\
SS1    &   7.73  &  10.03  &   9.81  \\
SS2    &   7.56  &  10.66  &  10.43  \\
SS5    &   7.98  &   9.92  &   9.69  \\
S31    &   8.19  &  10.66  &  10.45  \\
SS6    &   7.39  &   9.57  &   9.35  \\
SS7    &   7.67  &   9.98  &   9.76  \\
SS8    &   7.85  &   9.86  &   9.63  \\
SS9    &   7.95  &  10.72  &  10.49  \\
SS10   &   8.10  &  10.51  &  10.28  \\
SS11   &   7.83  &   9.94  &   9.71  \\
SS12   &   8.15  &  10.44  &  10.21  \\
SS13   &   7.69  &  10.50  &  10.27  \\
S28    &   8.12  &  10.60  &  10.37  \\
SS14   &   7.45  &  10.38  &  10.15  \\
S29    &   7.95  &   9.93  &   9.71  \\
SS18   &   7.51  &   9.71  &   9.48  \\
W11    &   7.95  &  10.63  &  10.28  \\
W22    &   8.68  &  11.17  &  10.82  \\
W12    &   8.94  &  10.39  &  10.04  \\
W20    &   8.60  &  10.93  &  10.57  \\
W16    &   7.86  &  10.71  &  10.35  \\
W8     &   8.74  &  10.91  &  10.56  \\
W3     &   8.76  &  10.18  &   9.83  \\
SS15   &   7.44  &  10.26  &  10.04  \\
W1     &   8.84  &  10.61  &  10.25  \\
W4     &   8.28  &  10.91  &  10.55  \\
W5     &   8.29  &  10.86  &  10.50  \\
SS3    &   7.51  &  10.00  &   9.78  \\
SS4    &   7.85  &  10.71  &  10.49  \\
W17    &   8.31  &  10.02  &   9.67  \\
W2     &   9.07  &  10.81  &  10.46  \\
W10    &   7.94  &  10.41  &  10.05  \\
W14    &   8.68  &  10.87  &  10.52  \\
W9     &   8.70  &  10.79  &  10.44  
\enddata
\label{tab:final_quan}
\tablecomments{
Column 1: Object ID.
Column 2: BH mass derived from Eq.~(\ref{eq:MBH}) using $\sigma_{\rm H\beta}$ and $\lambda L_{5100}^{\rm image}$ (in solar units).
Column 3: Bulge luminosity in rest-frame $V$ (in solar units). For 18 objects with upper limits, the bulge luminosity was
derived using informative priors (see Section~\ref{sec:Lbul} for details).
Column 4: Final bulge luminosity corrected for evolution by aging of the stellar population.}
\end{deluxetable*}

\clearpage
\begin{appendix}

\section{Updated Measurements of Previous Sample}
\label{app:reanalysis}
We performed a consistent spectral and image analysis for 40 objects 
presented by \citet{Treu+07} and \citet{Bennert+10},
using the same methods described in the main text 
to minimize measurement systematics (see Figure~\ref{fig:specfit40_1} and \ref{fig:imgfit1dSBP_pre40_1}).

We compare the previous and new measurements for BH masses and bulge luminosities in Figure~\ref{fig:compare_pre_new}.
On average we obtained consistent measurements with previous results (i.e., close to zero offsets).
However, there is a considerable scatter ($\sim0.18$ dex for \mbh\ and $\sim0.22$ dex for \Lbul),
indicating the necessity of a homogeneous and careful analysis.
We consider the results presented here more robust, given several improvements in the analysis.
For one, the multi-component spectral decomposition applied here
takes into account host-galaxy starlight contribution as well as iron emission blends 
for a better isolation of the broad H$\beta$ emission line, resulting in a more accurate measurement of BH mass.
The difference between the previous and new line width ($\sigma_{\rm line}$) is $\sim0.08$ dex scatter.
Second, the current multi-component image decomposition has advantages over the previous approach.
It not only achieves a better optimization by probing the true global minimum over parameter spaces, 
but the PSF model consisting of a linear combination of several field stars
minimizes any PSF mismatch and arguably provides more accurate structural decomposition results.
Moreover, in contrast to the previous approach, our model allows off-centered AGN and galaxy components for a given object.

\begin{figure*}
\centering
    \includegraphics[width=\textwidth]{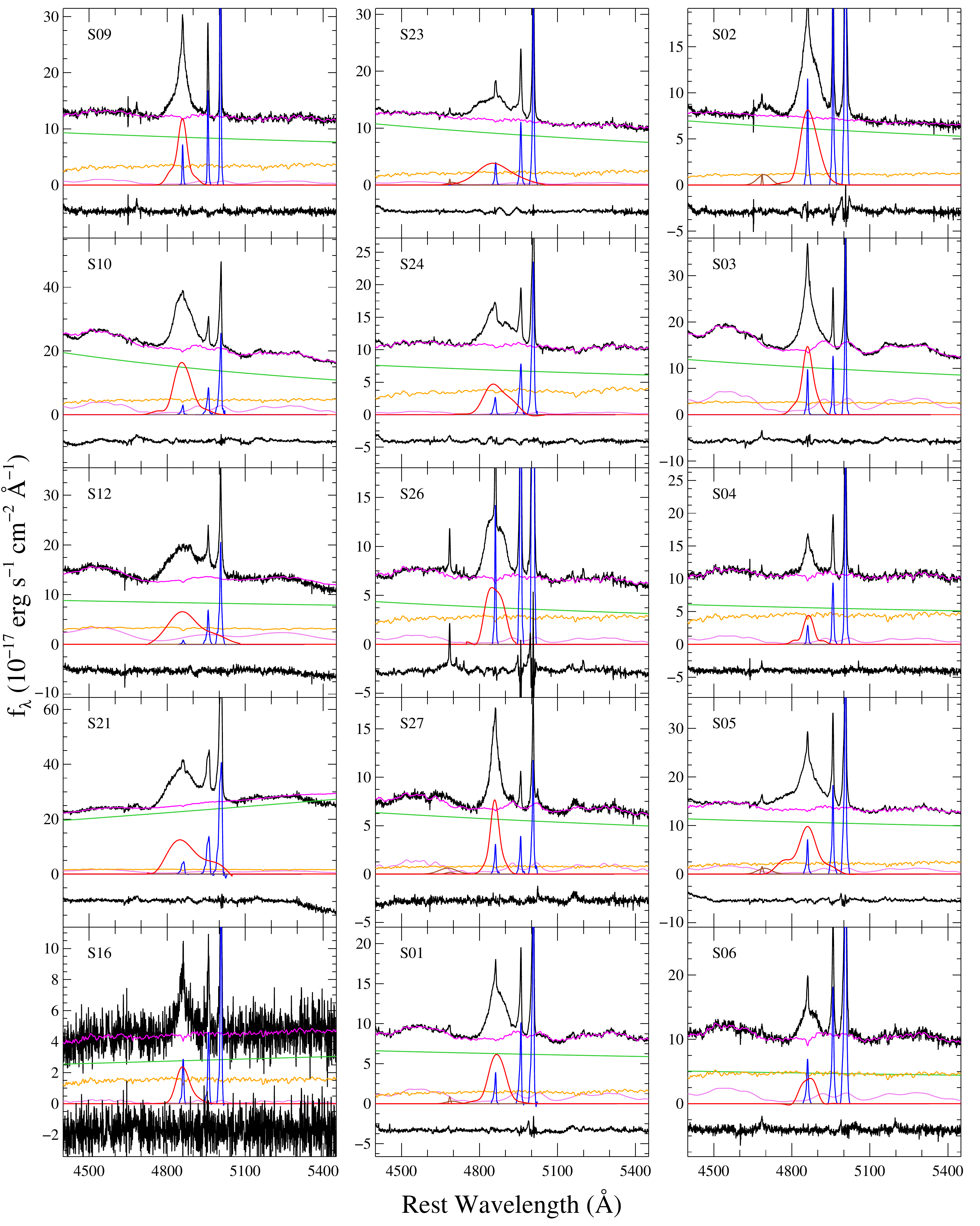}
    \caption{
    Same as Figure~\ref{fig:specfit12}, but for the previous sample of 40 objects.
    \label{fig:specfit40_1}}
\end{figure*}

\begin{figure*}
\centering
    \includegraphics[width=\textwidth]{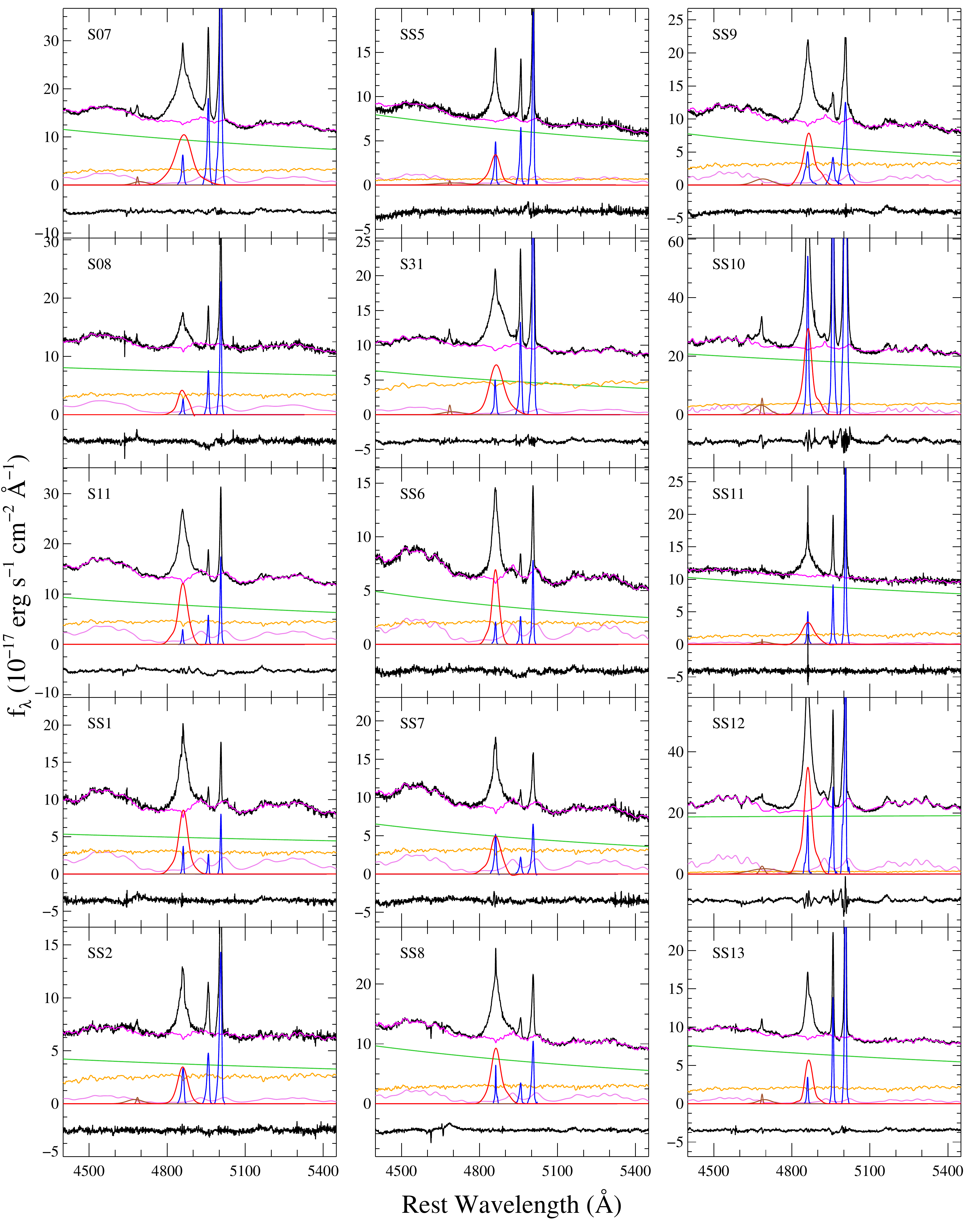}
    \figurenum{A\-1}
    \caption{
    \it Continued.
    \label{fig:specfit40_2}}
\end{figure*}

\begin{figure*}
\centering
    \includegraphics[width=\textwidth]{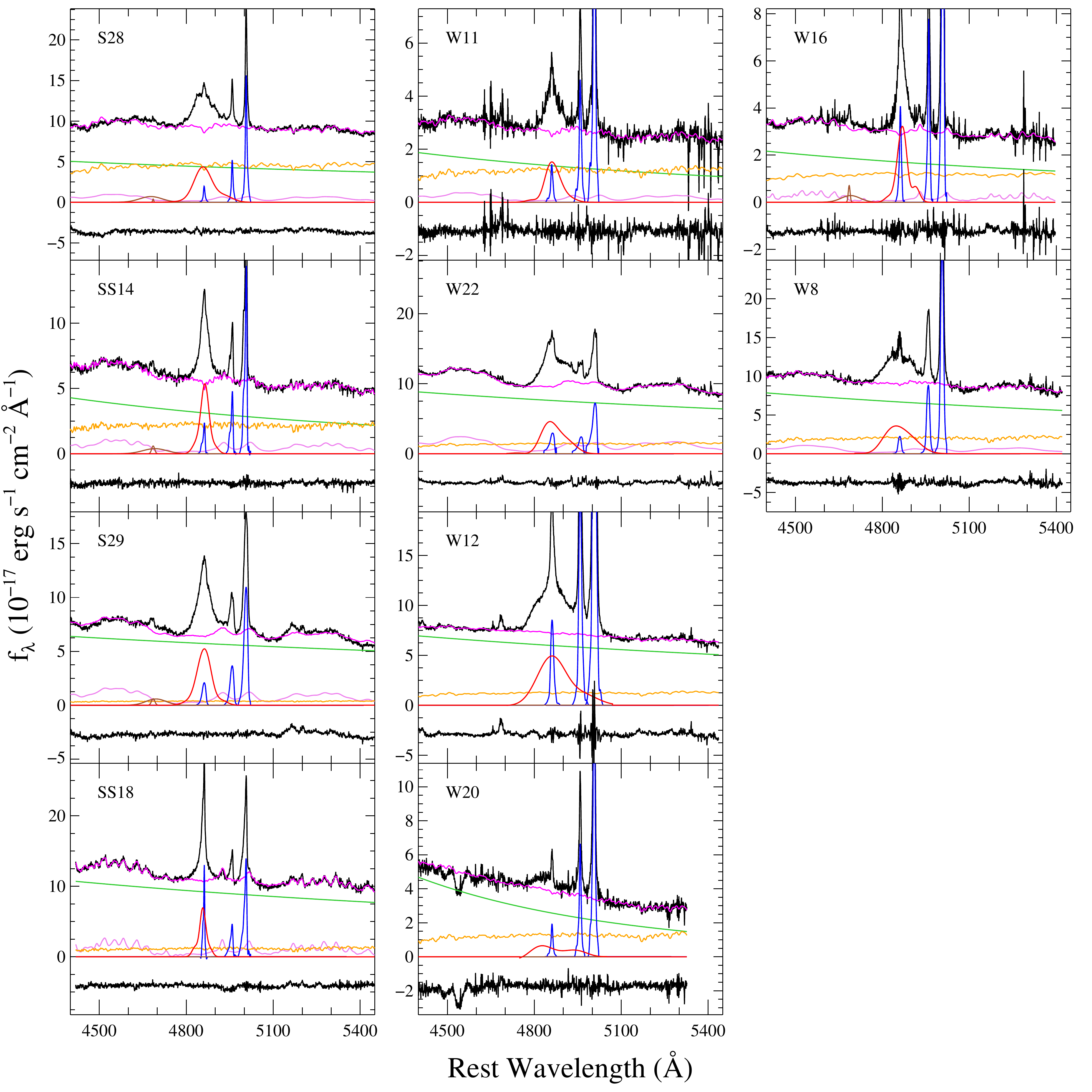}
    \figurenum{A\-1}
    \caption{
	\it Continued.
    \label{fig:specfit40_3}}
\end{figure*}

\begin{figure*}
\centering
    \includegraphics[width=0.75\textwidth]{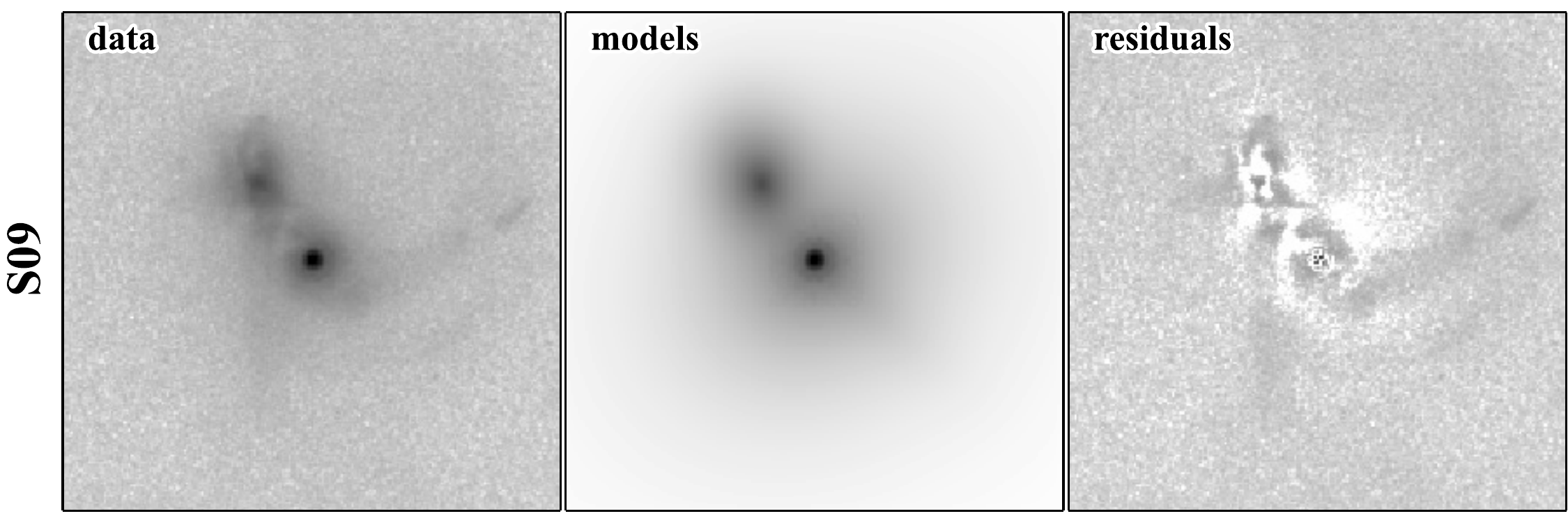}\includegraphics[width=0.25\textwidth]{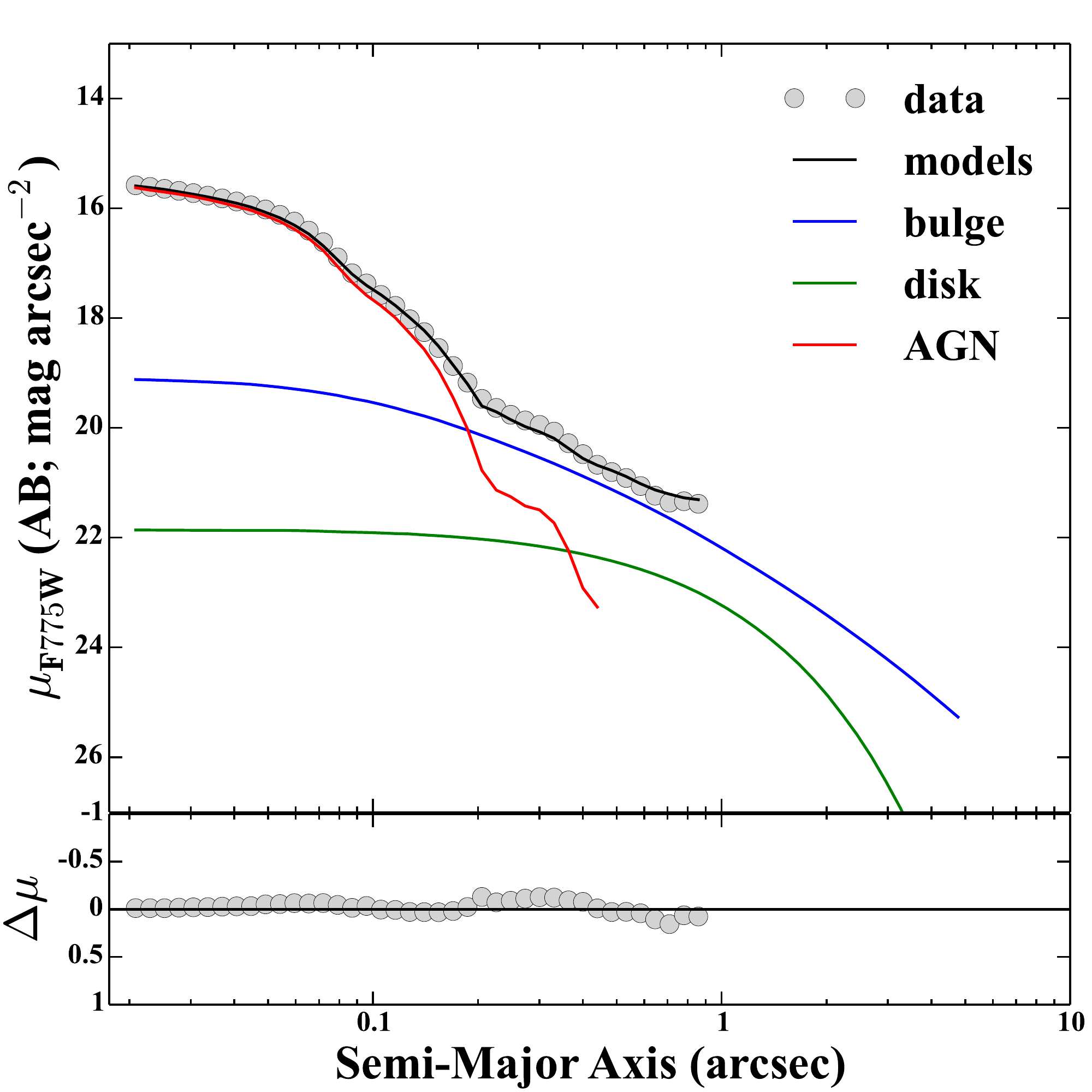}\\
    \includegraphics[width=0.75\textwidth]{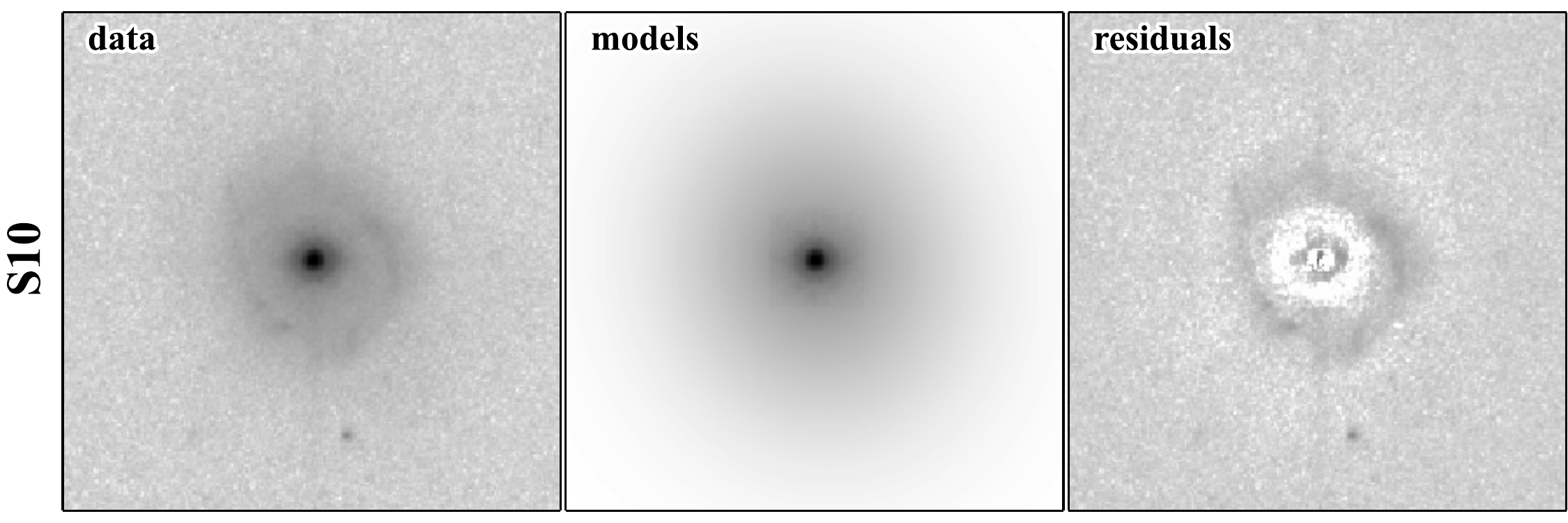}\includegraphics[width=0.25\textwidth]{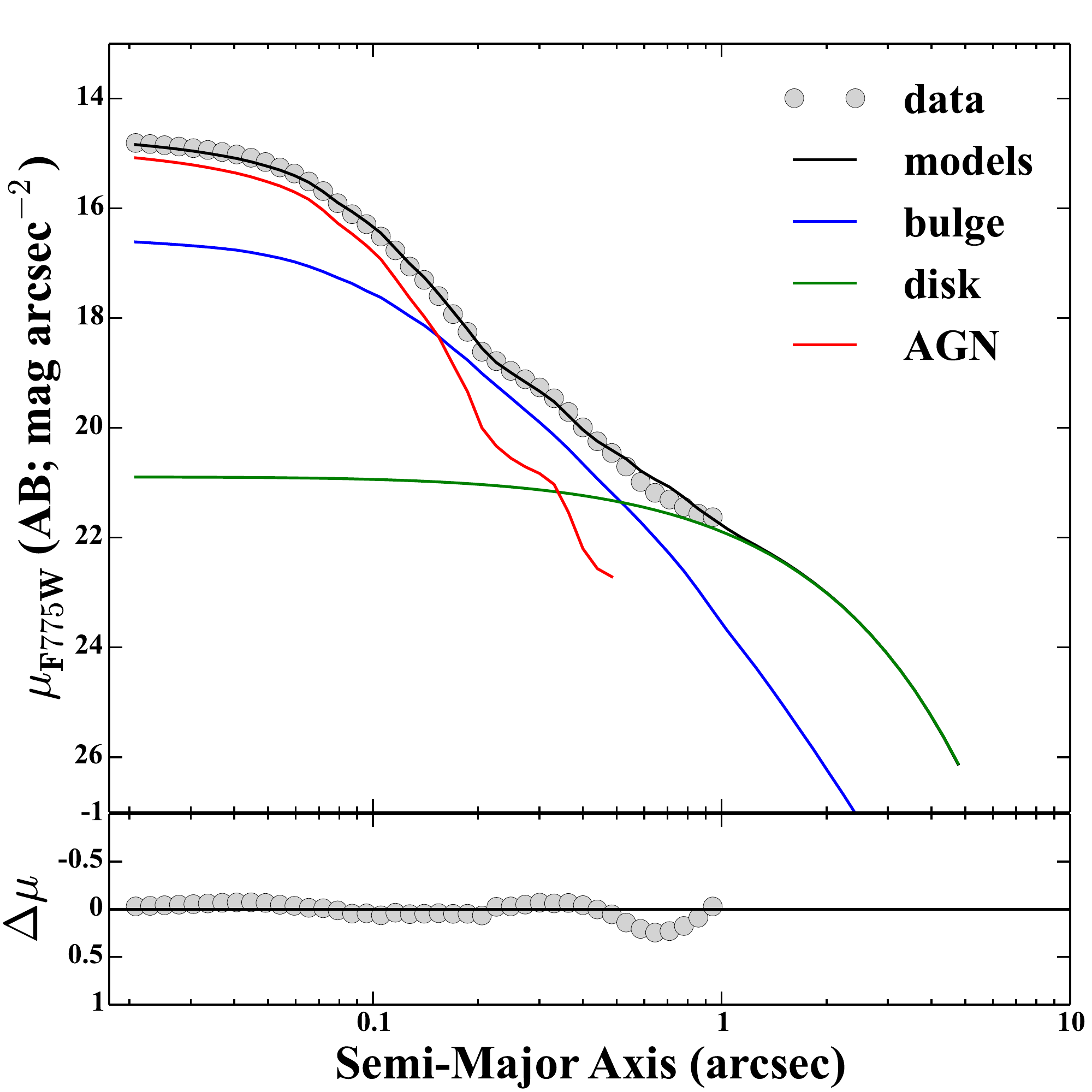}\\
    \includegraphics[width=0.75\textwidth]{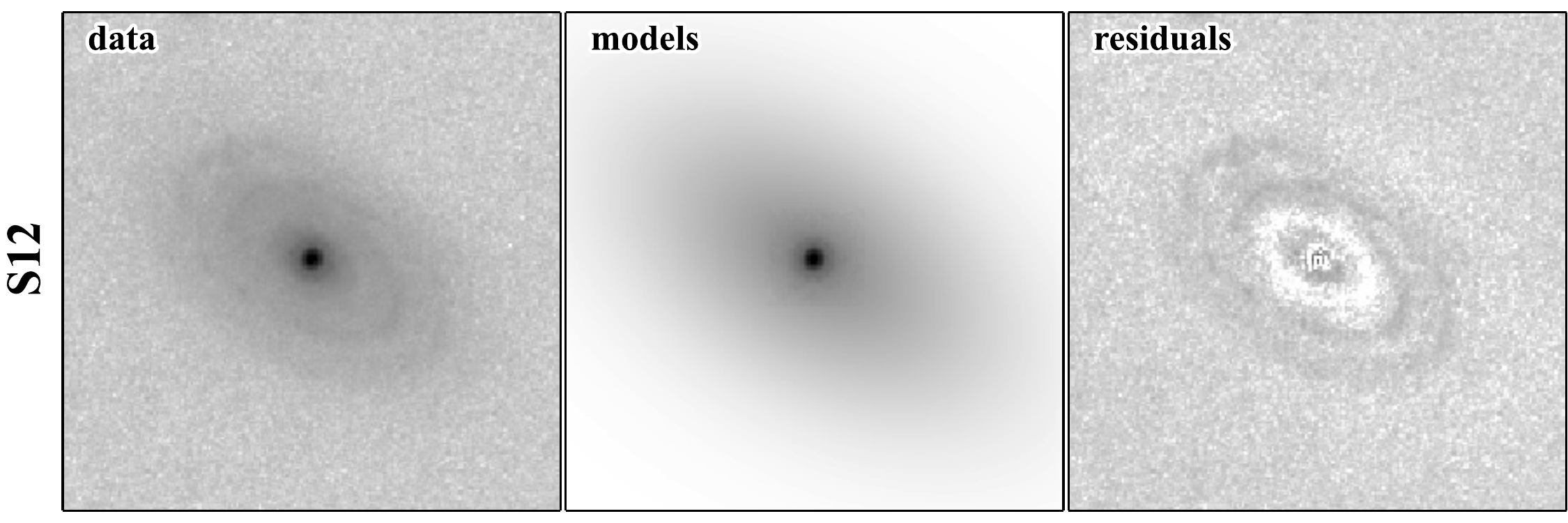}\includegraphics[width=0.25\textwidth]{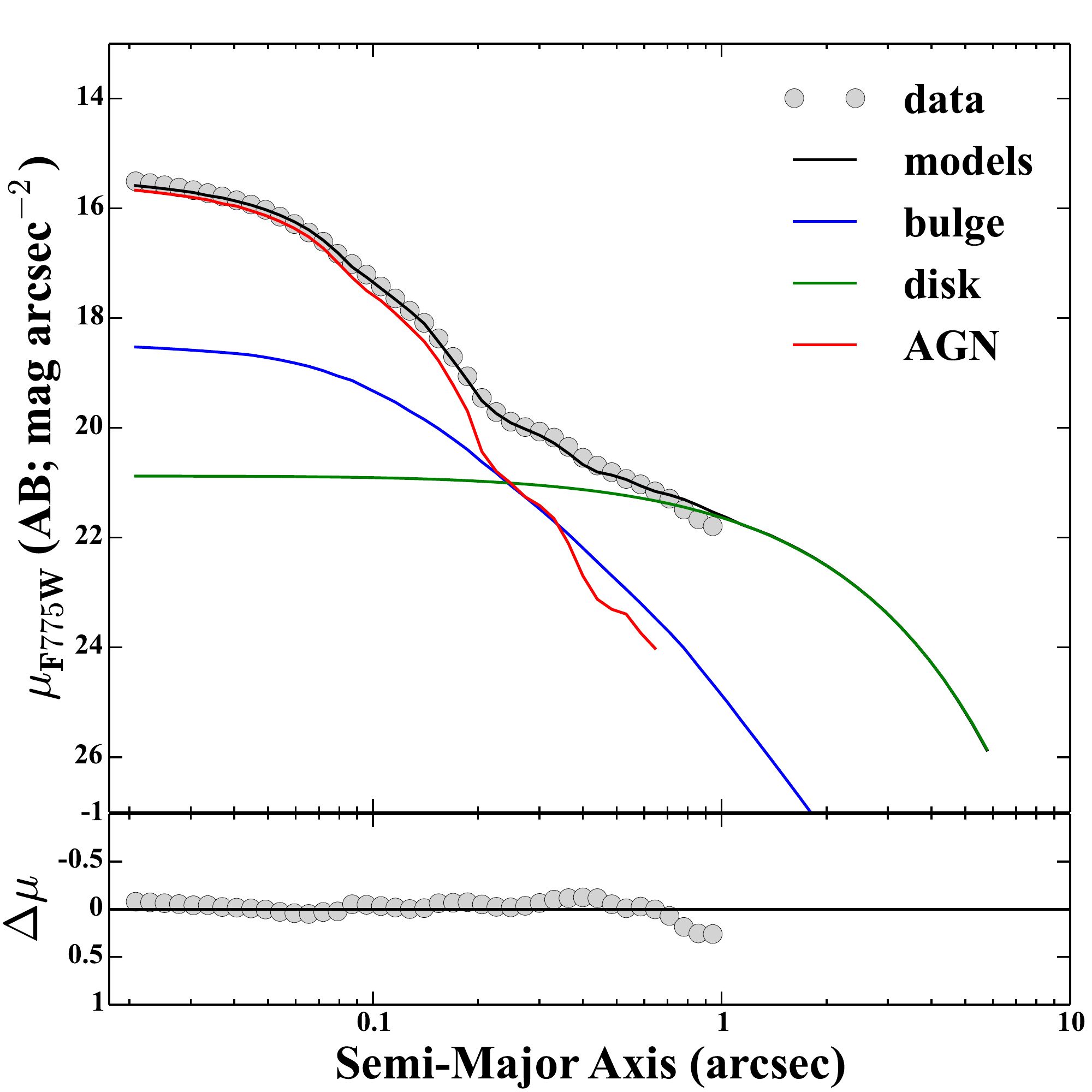}\\
    \includegraphics[width=0.75\textwidth]{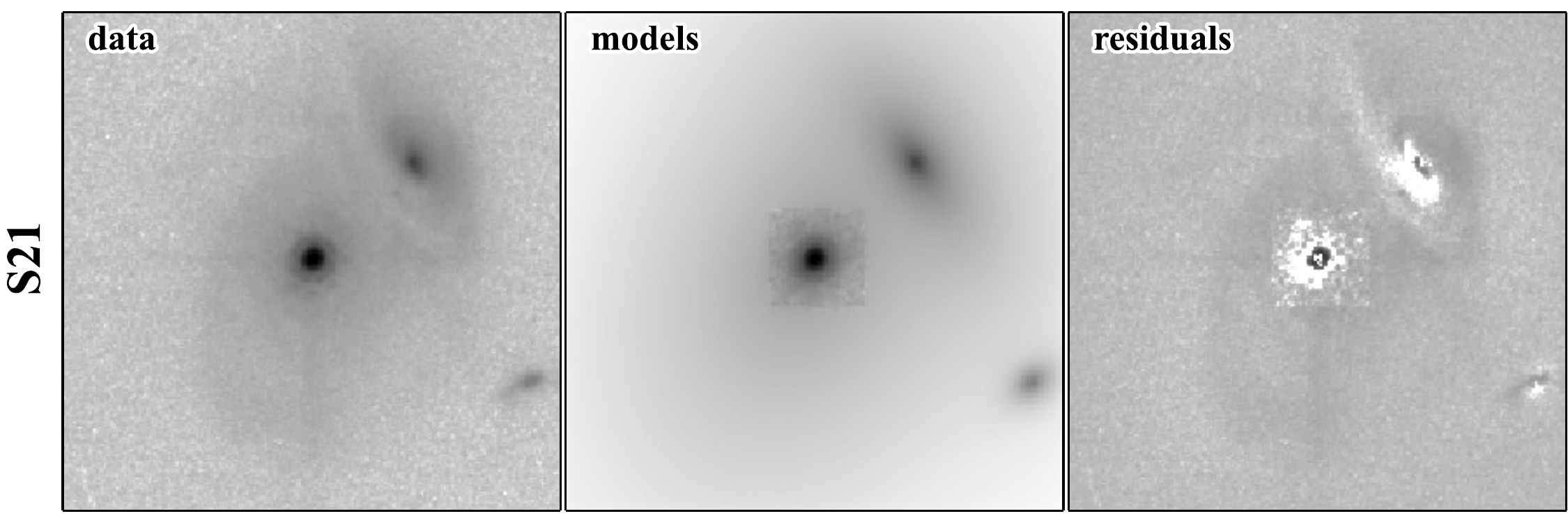}\includegraphics[width=0.25\textwidth]{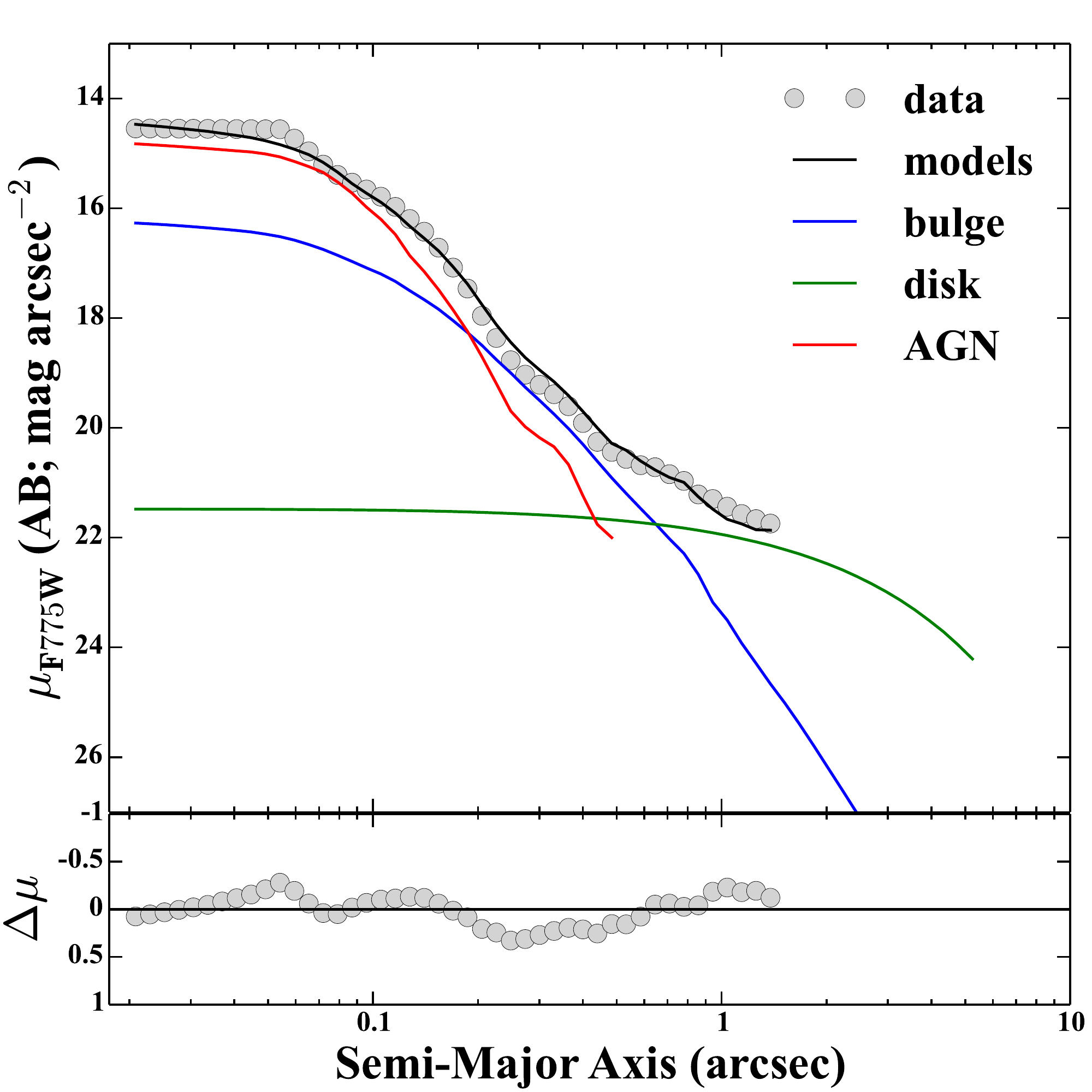}\\
    \includegraphics[width=0.75\textwidth]{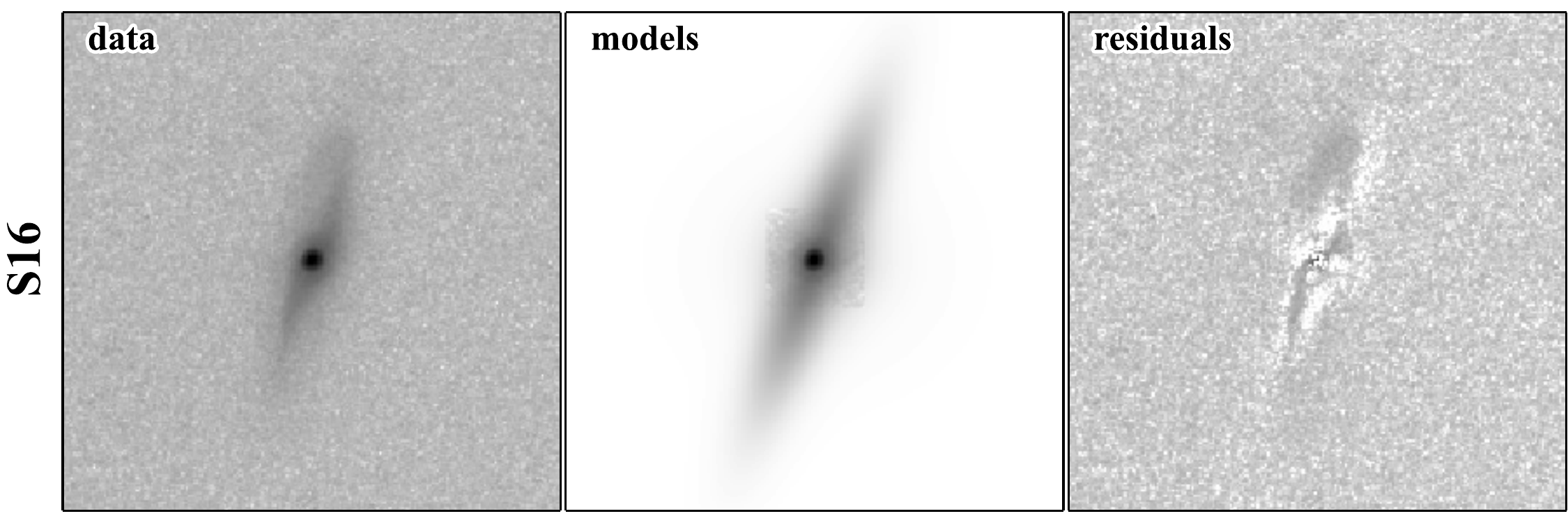}\includegraphics[width=0.25\textwidth]{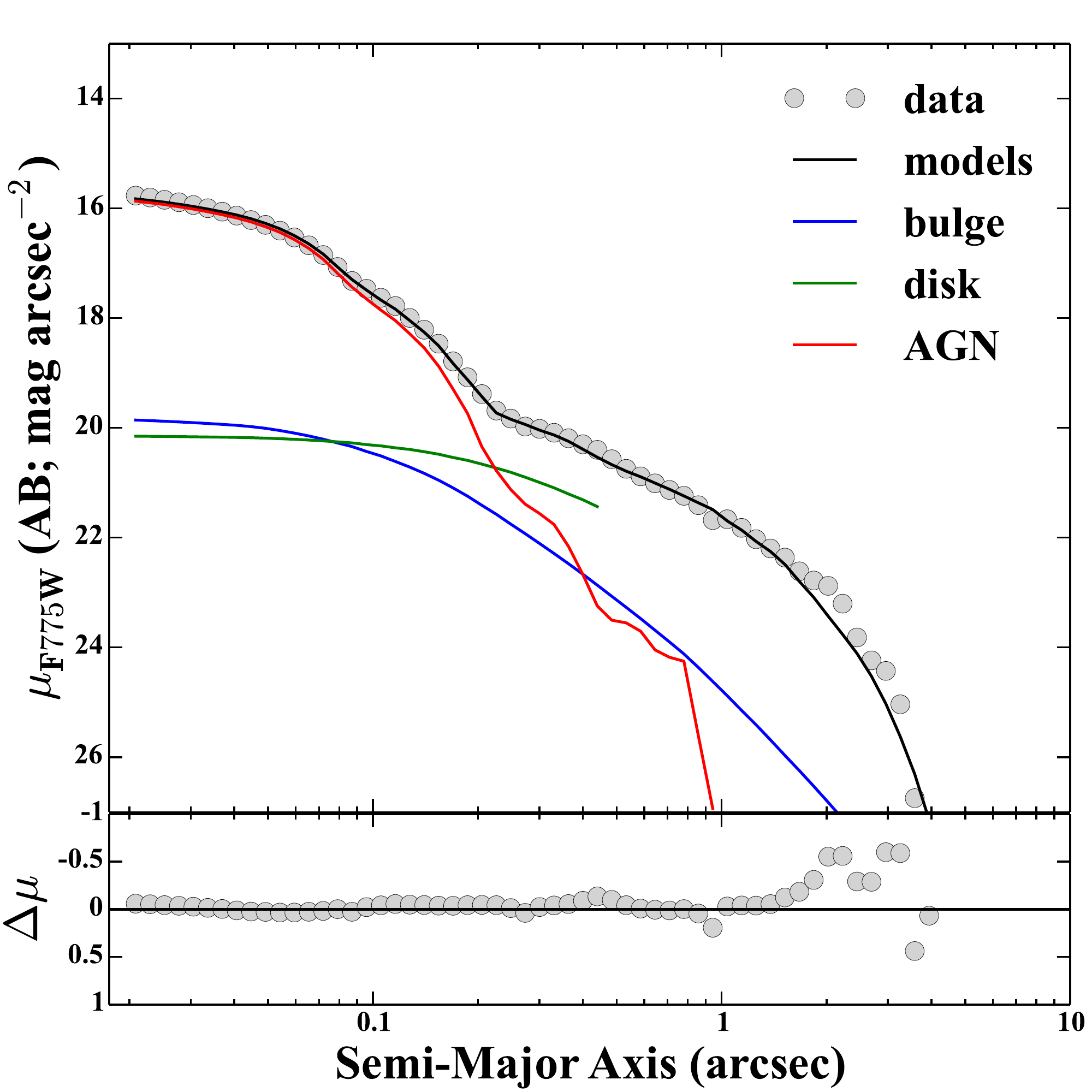}
        \caption{
   		 Same as Figure~\ref{fig:imgfit1dSBP_WFC3_1}, but for the previous sample of 40 objects with displayed image sizes of $8\arcsec\times8\arcsec$ (\HST\ ACS images; first 17 objects) and $7.6\arcsec\times7.6\arcsec$ (\HST\ NICMOS images; next 23 objects).   
        \label{fig:imgfit1dSBP_pre40_1}}
\end{figure*}

\begin{figure*}
\centering
    \includegraphics[width=0.75\textwidth]{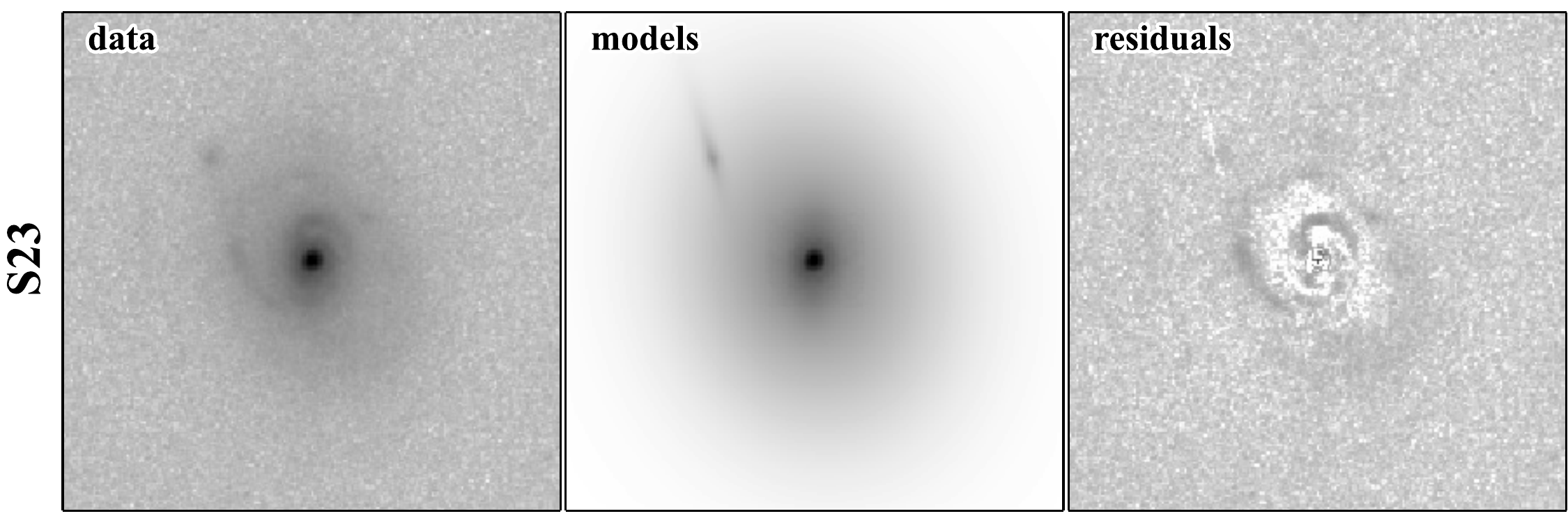}\includegraphics[width=0.25\textwidth]{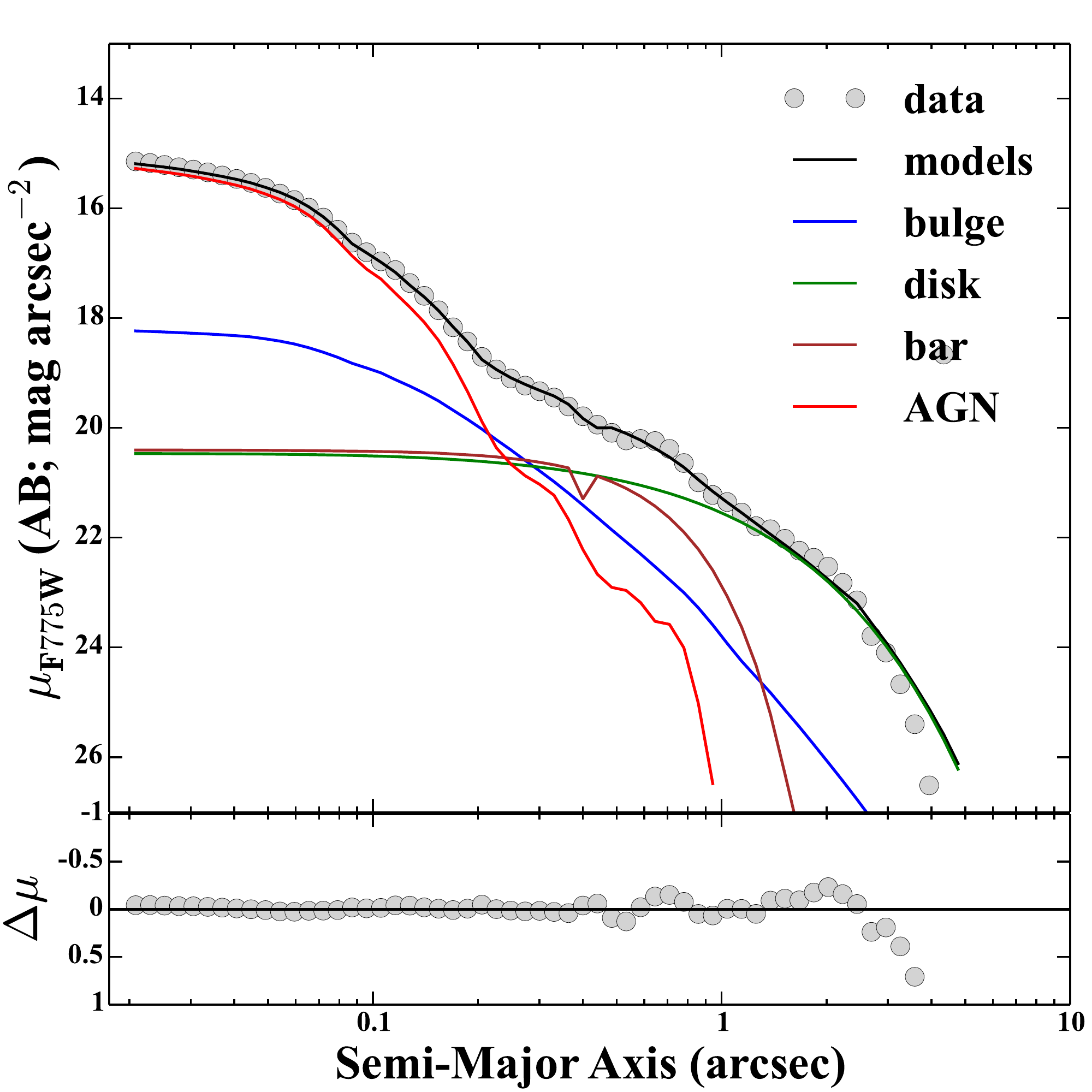}\\
    \includegraphics[width=0.75\textwidth]{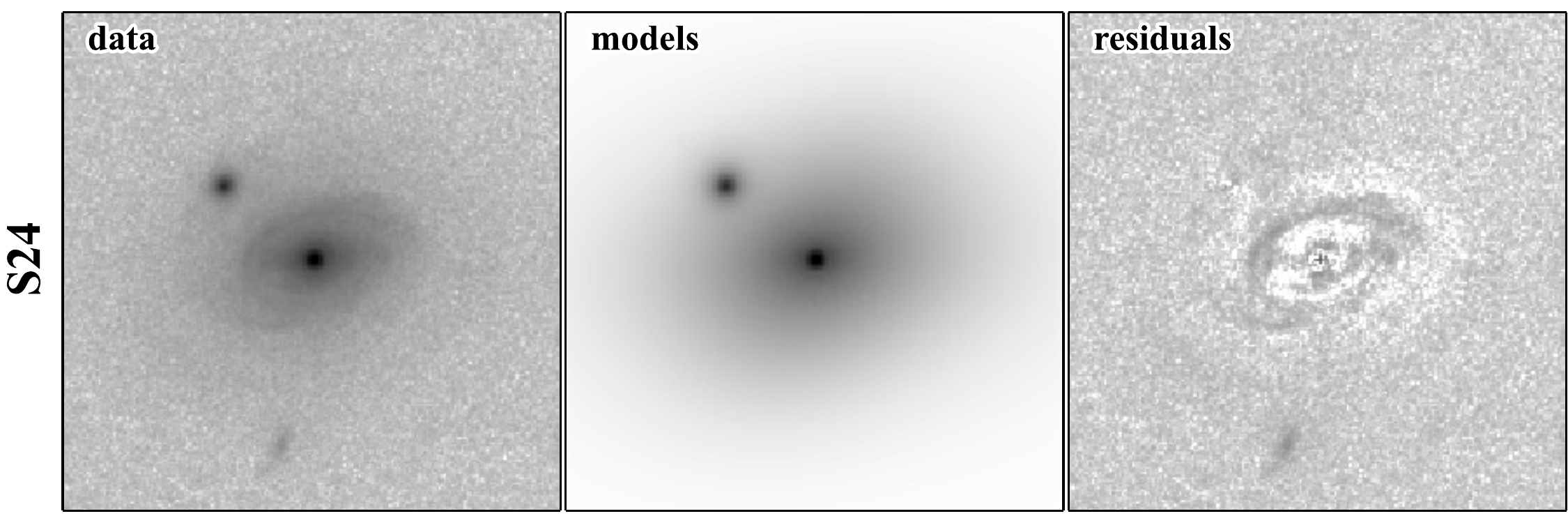}\includegraphics[width=0.25\textwidth]{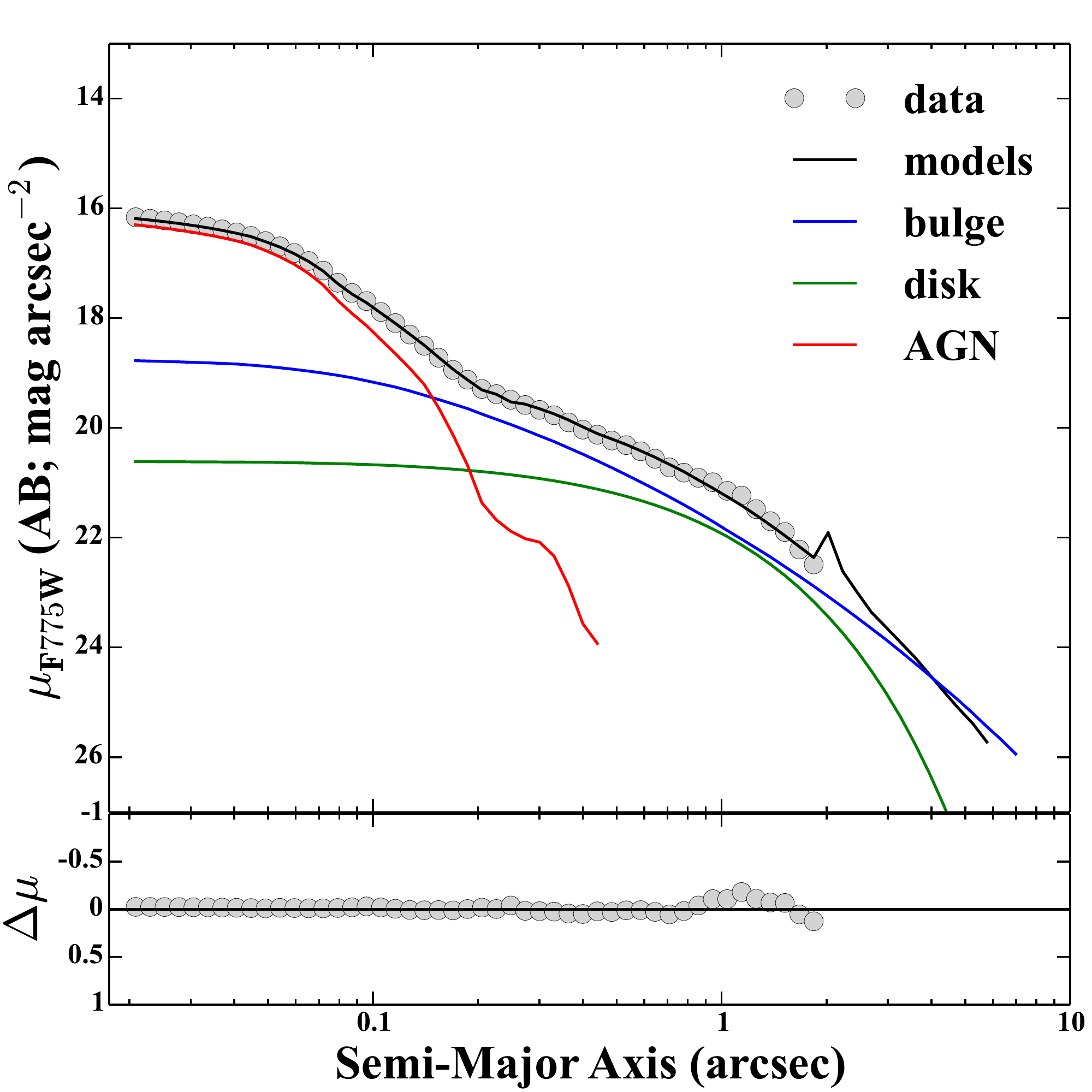}\\
    \includegraphics[width=0.75\textwidth]{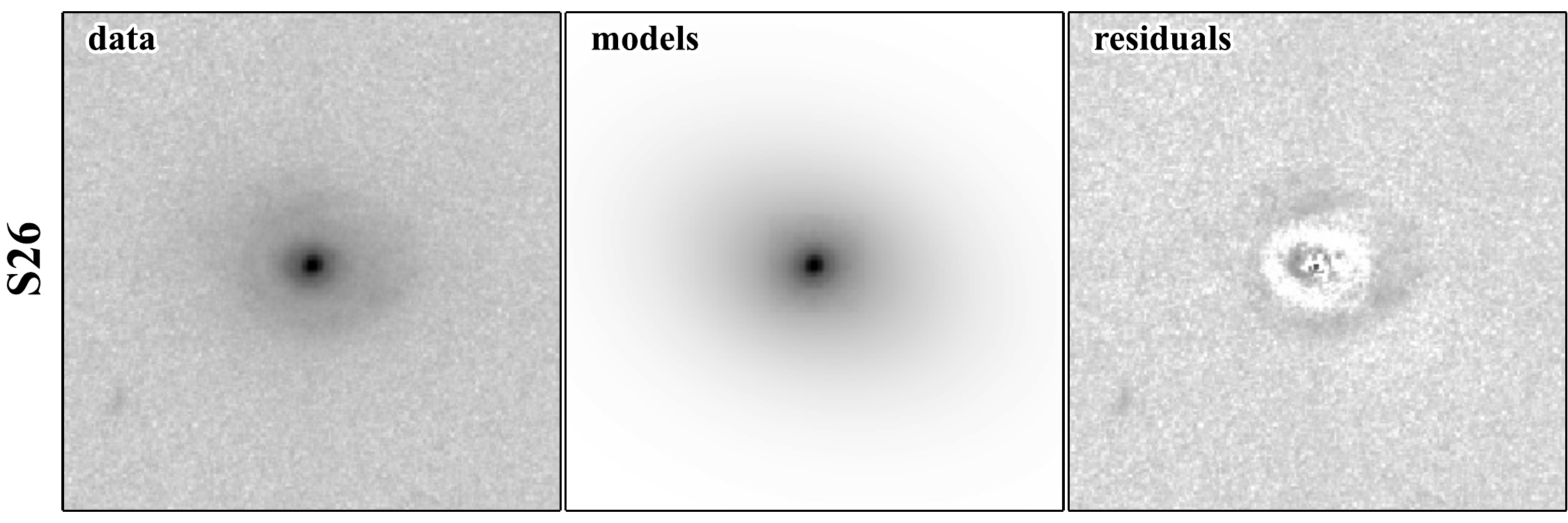}\includegraphics[width=0.25\textwidth]{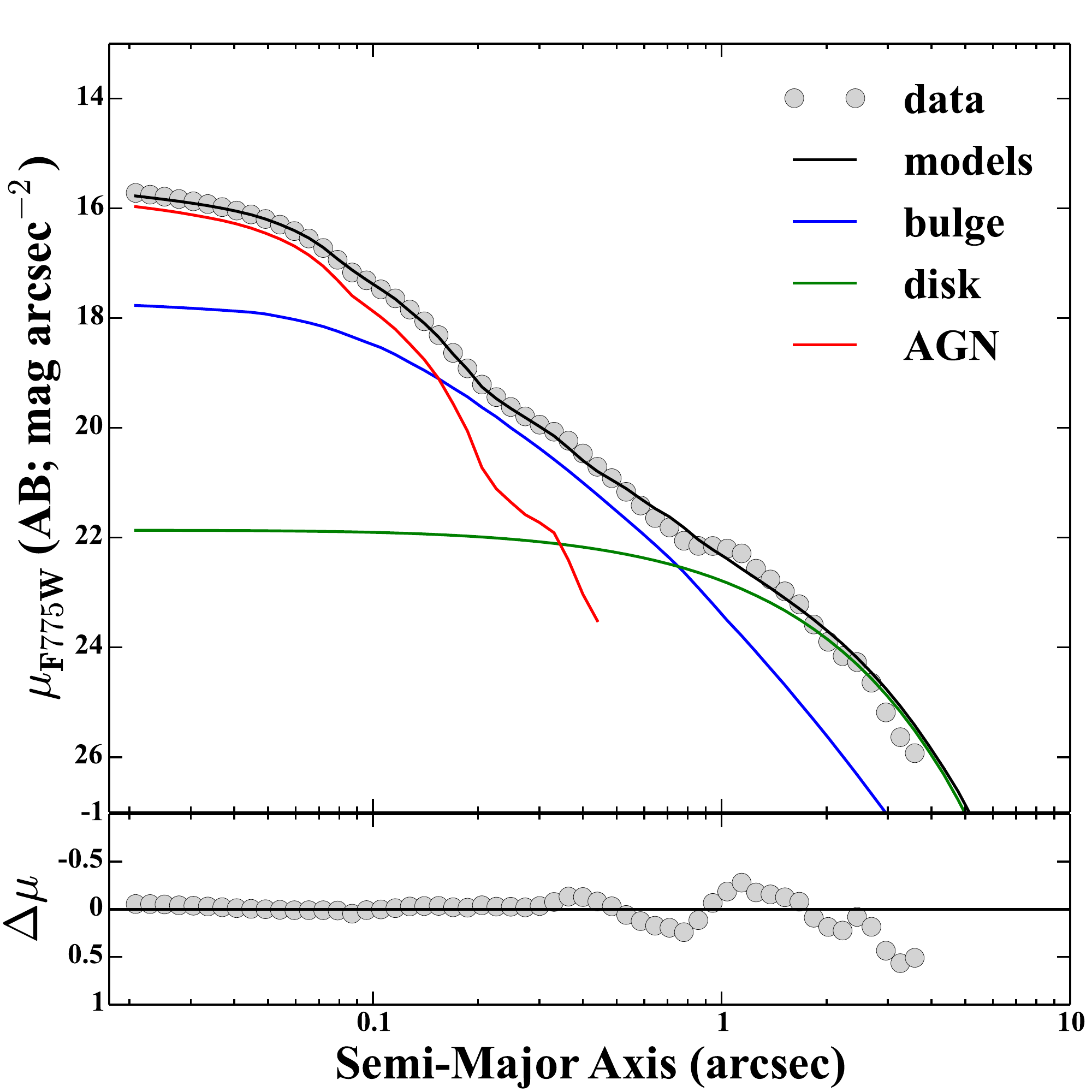}\\
    \includegraphics[width=0.75\textwidth]{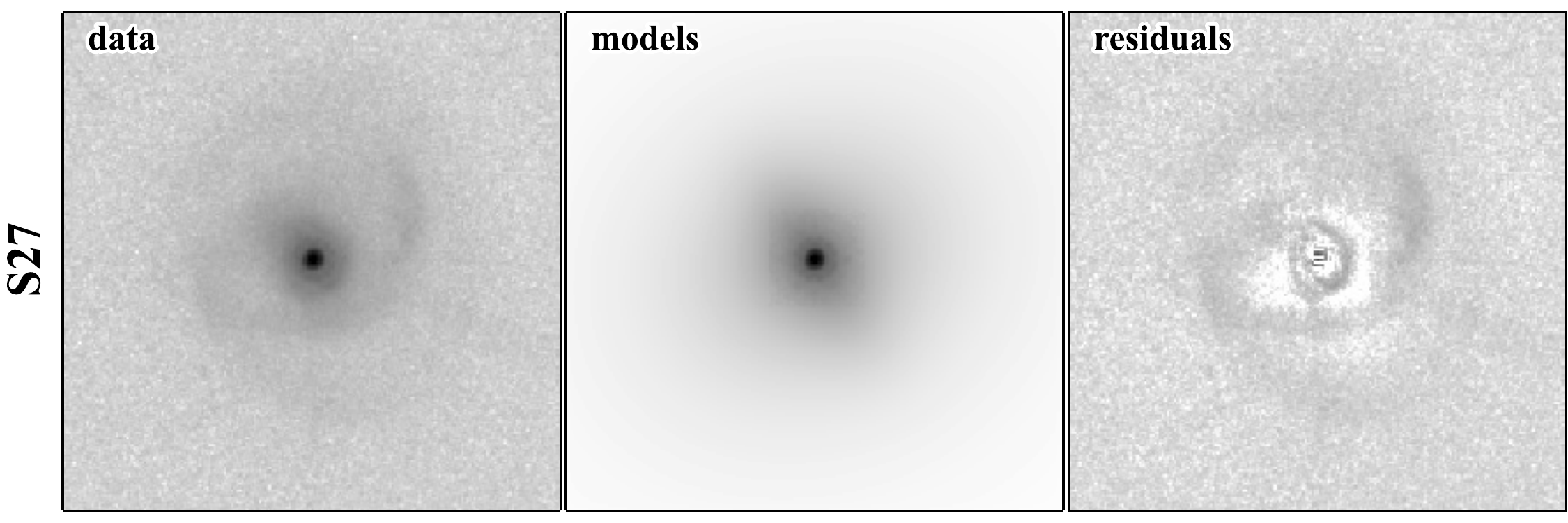}\includegraphics[width=0.25\textwidth]{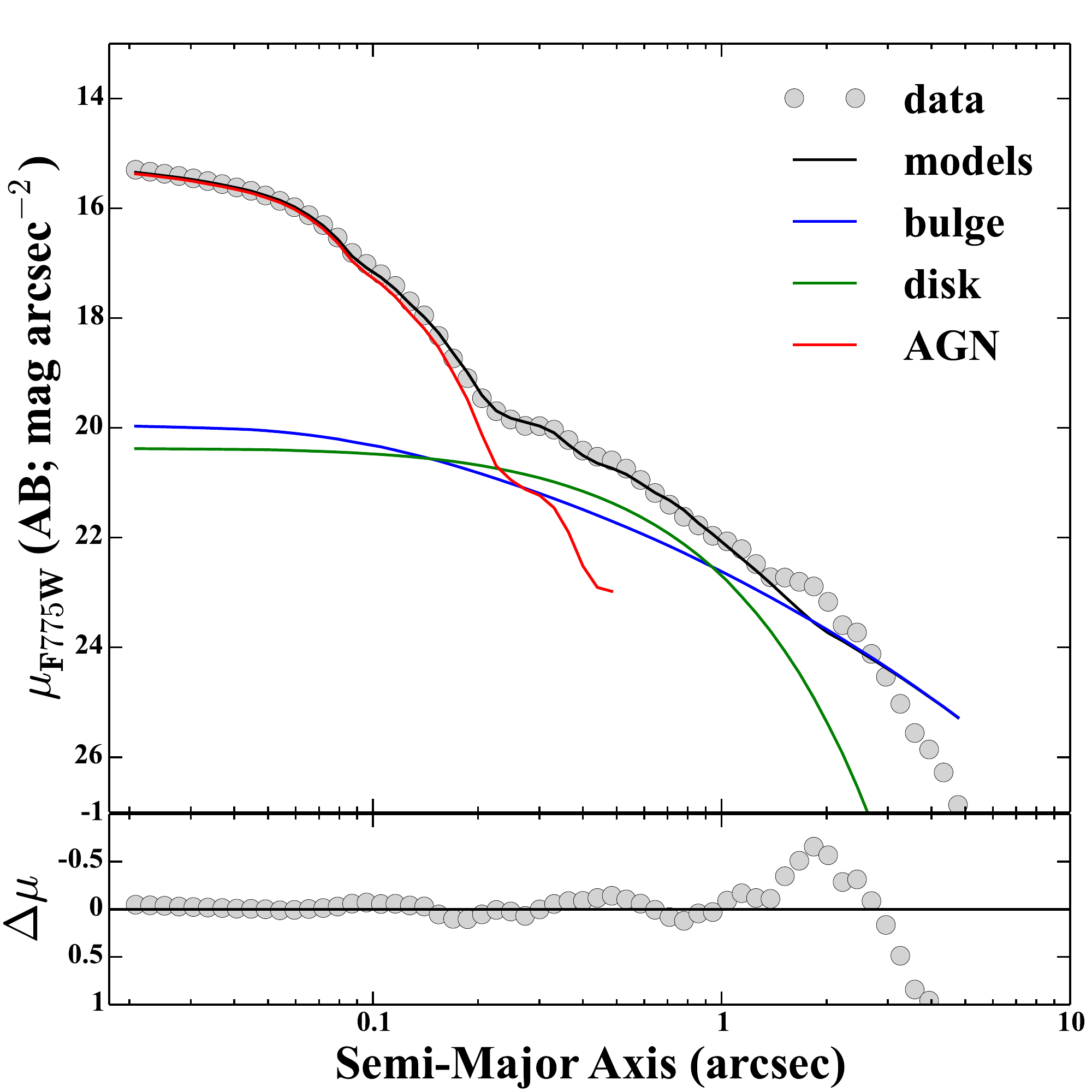}\\
    \includegraphics[width=0.75\textwidth]{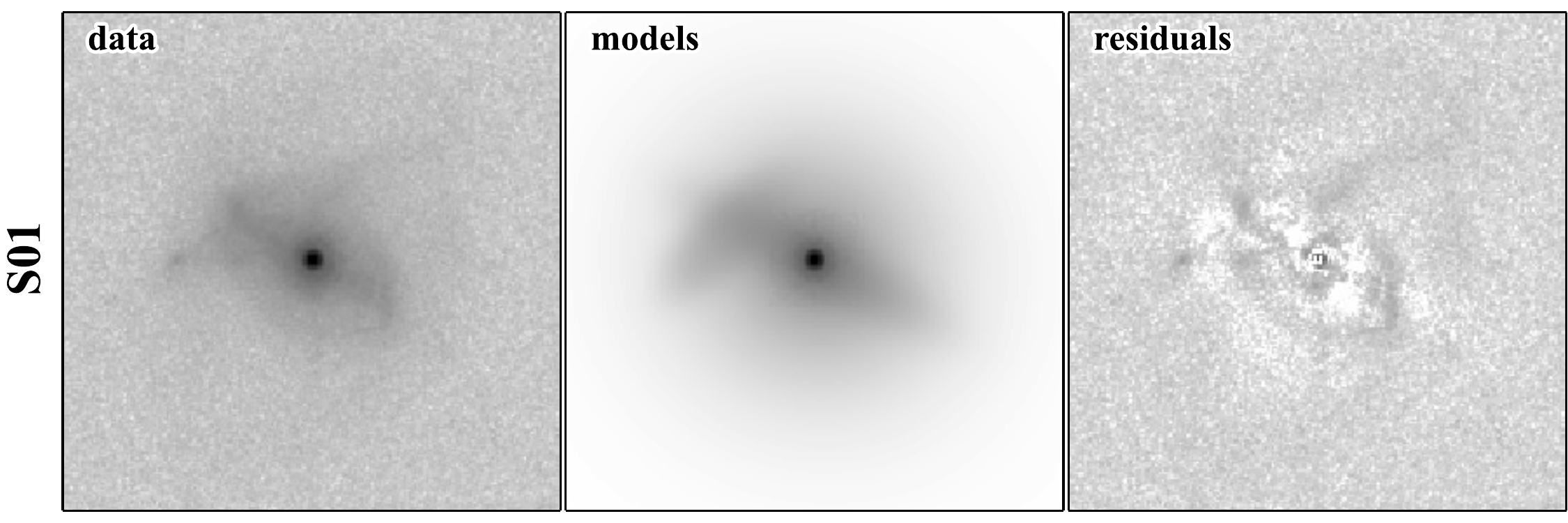}\includegraphics[width=0.25\textwidth]{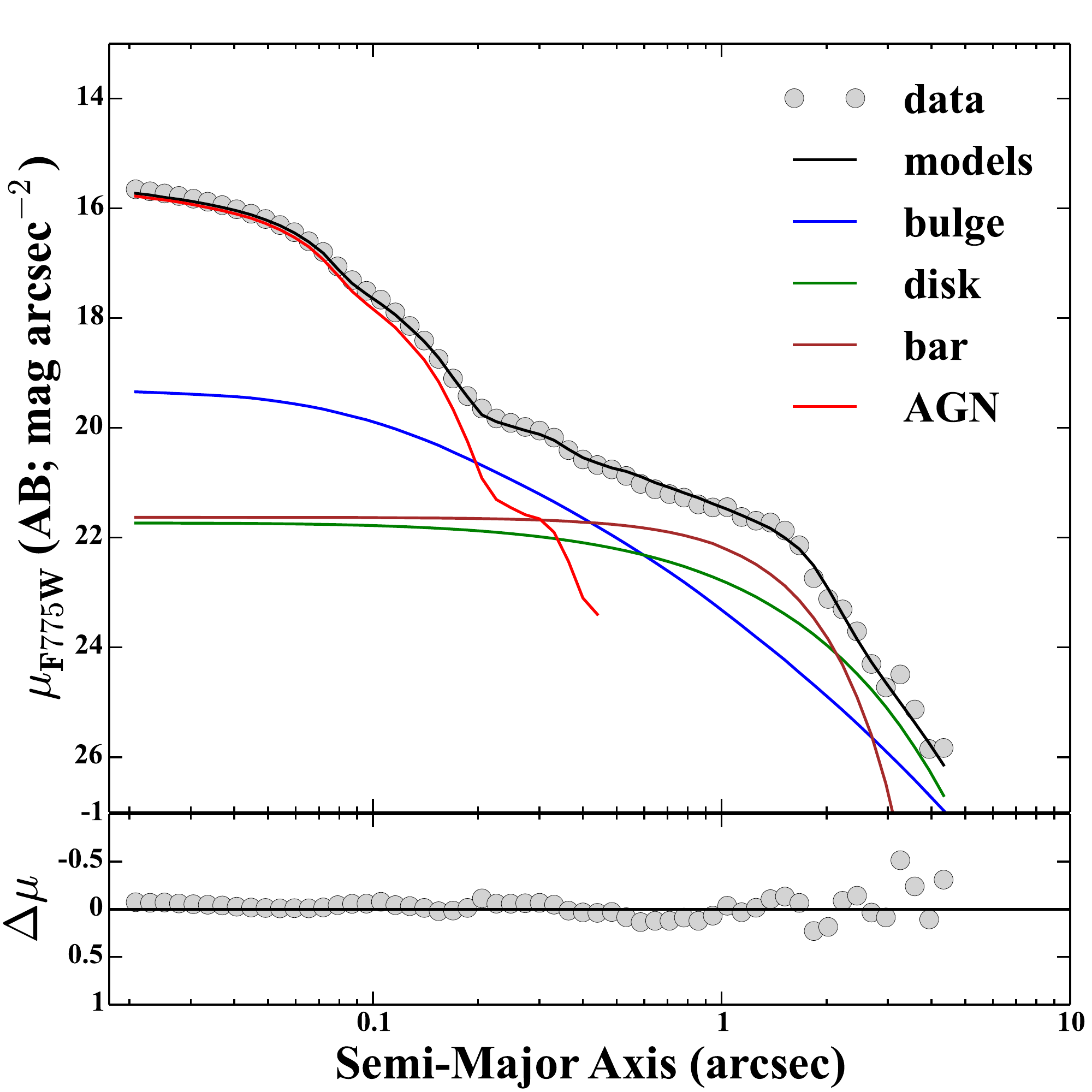}    
    \figurenum{A\-2}
        \caption{
        \it Continued.
        \label{fig:imgfit1dSBP_pre40_2}}
\end{figure*}

\begin{figure*}
\centering
    \includegraphics[width=0.75\textwidth]{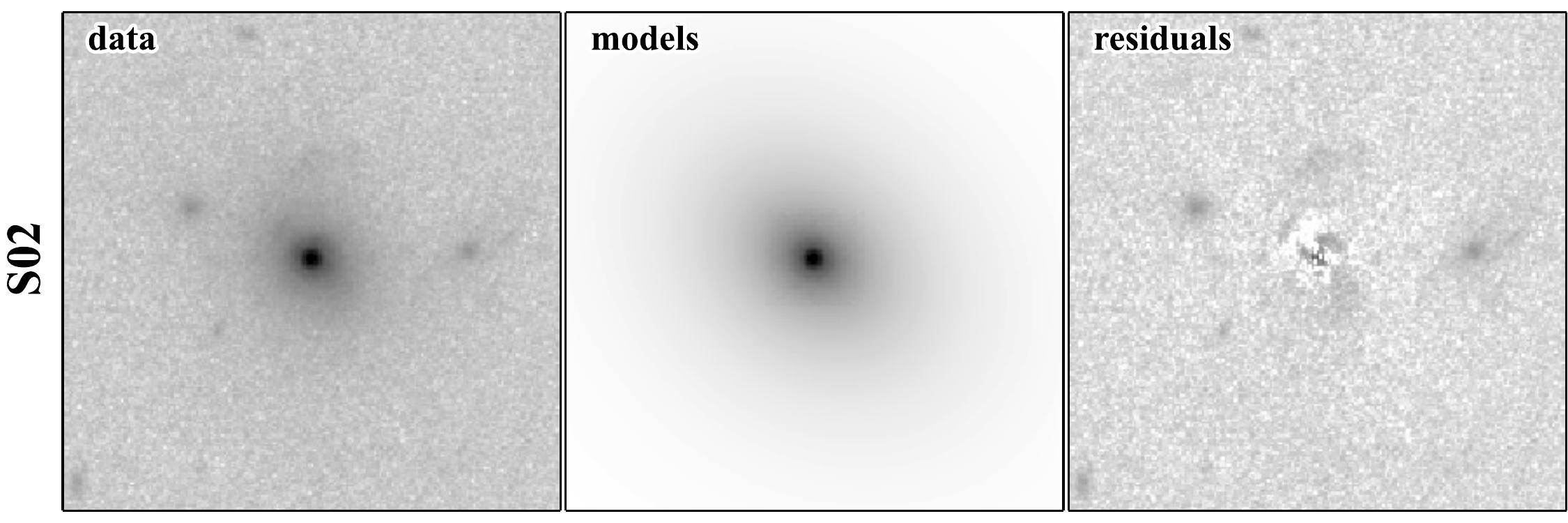}\includegraphics[width=0.25\textwidth]{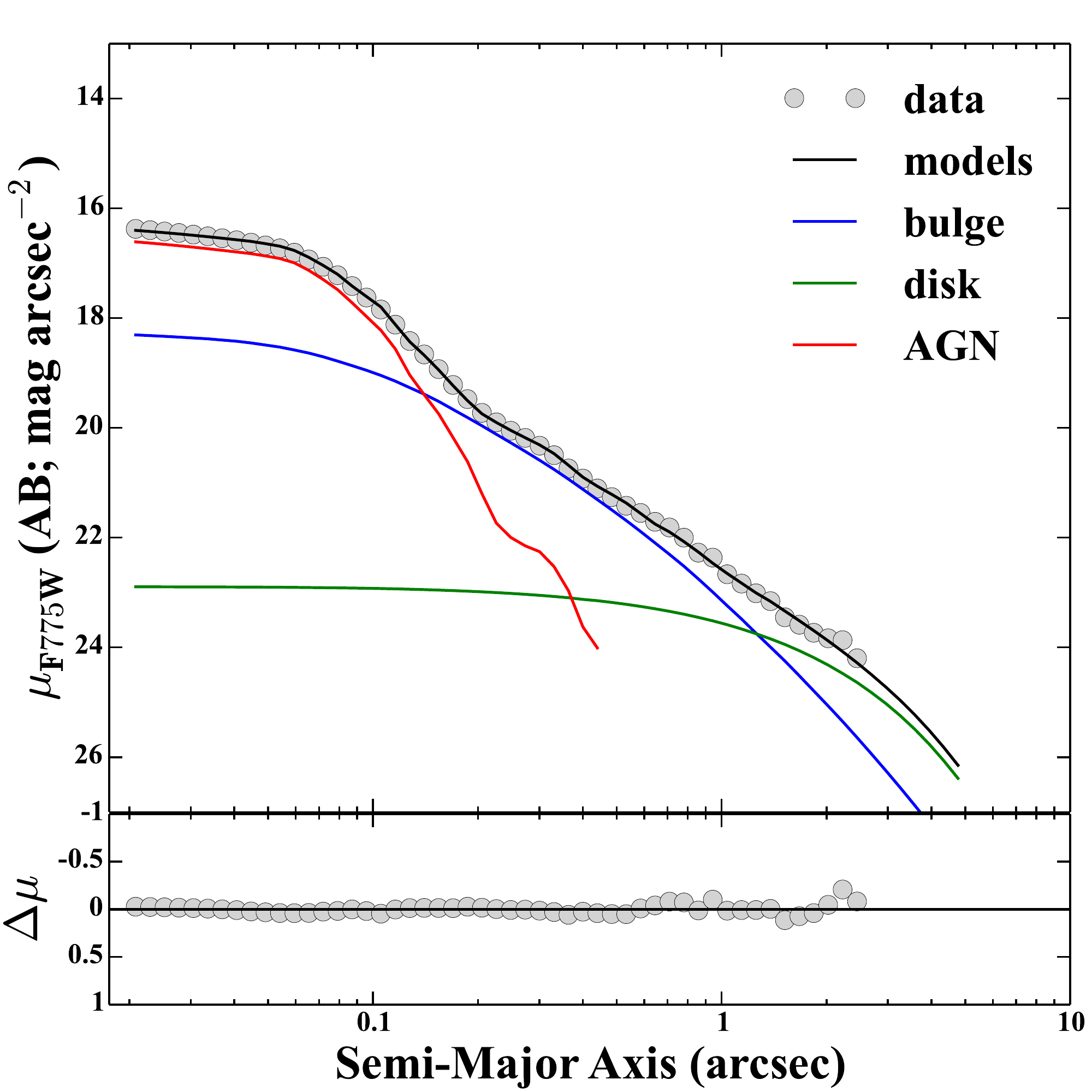}\\
    \includegraphics[width=0.75\textwidth]{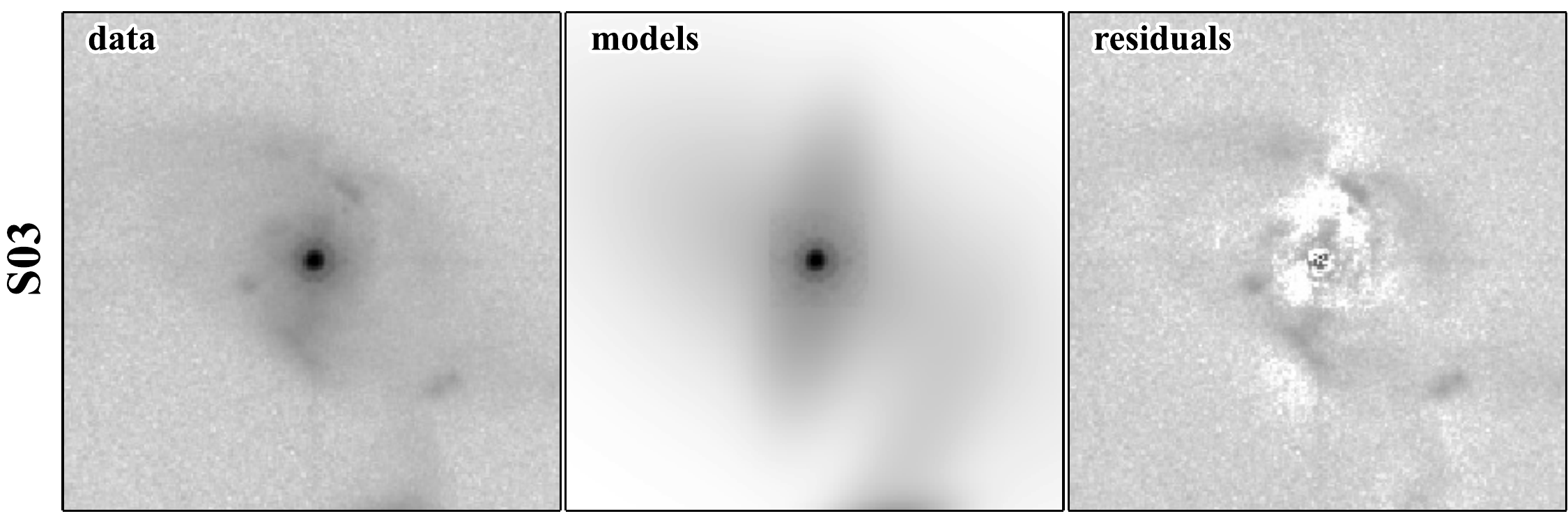}\includegraphics[width=0.25\textwidth]{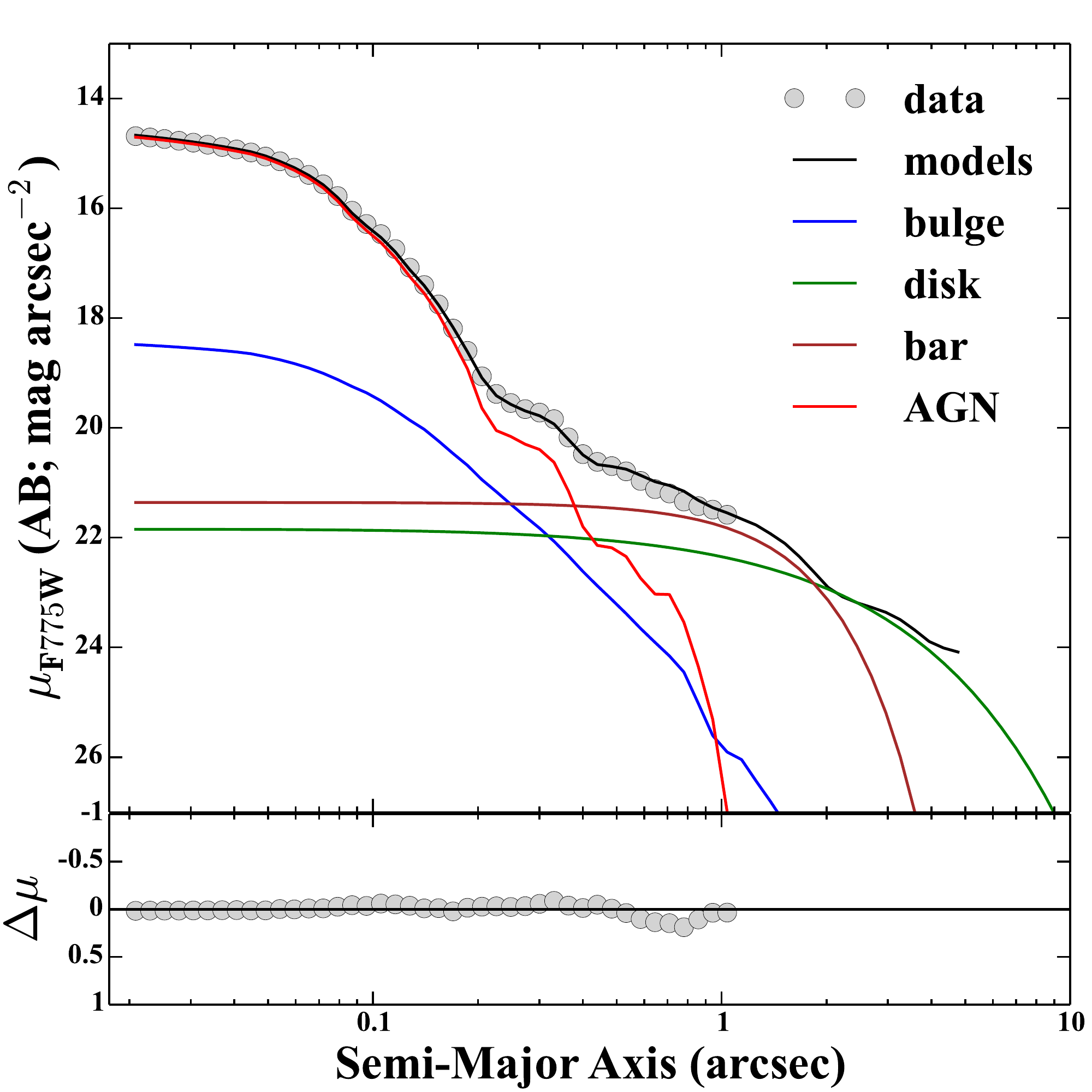}\\    \includegraphics[width=0.75\textwidth]{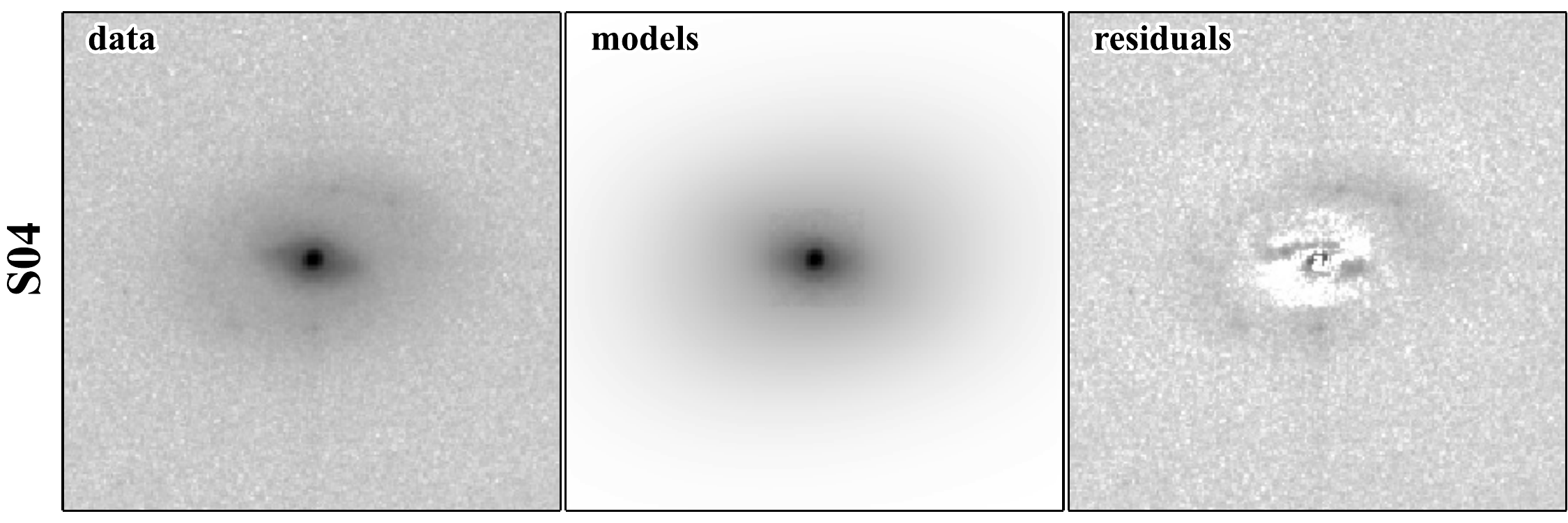}\includegraphics[width=0.25\textwidth]{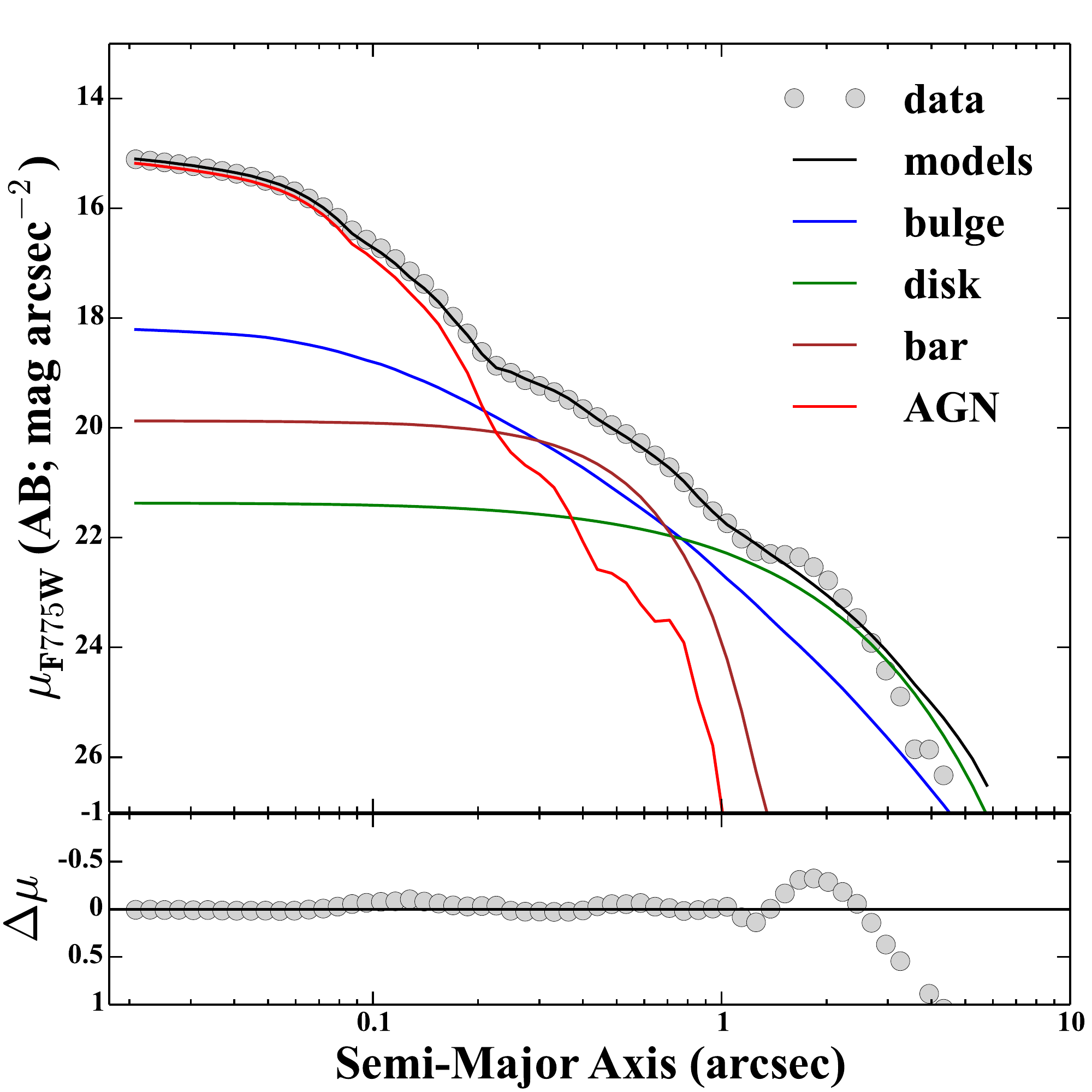}\\
	\includegraphics[width=0.75\textwidth]{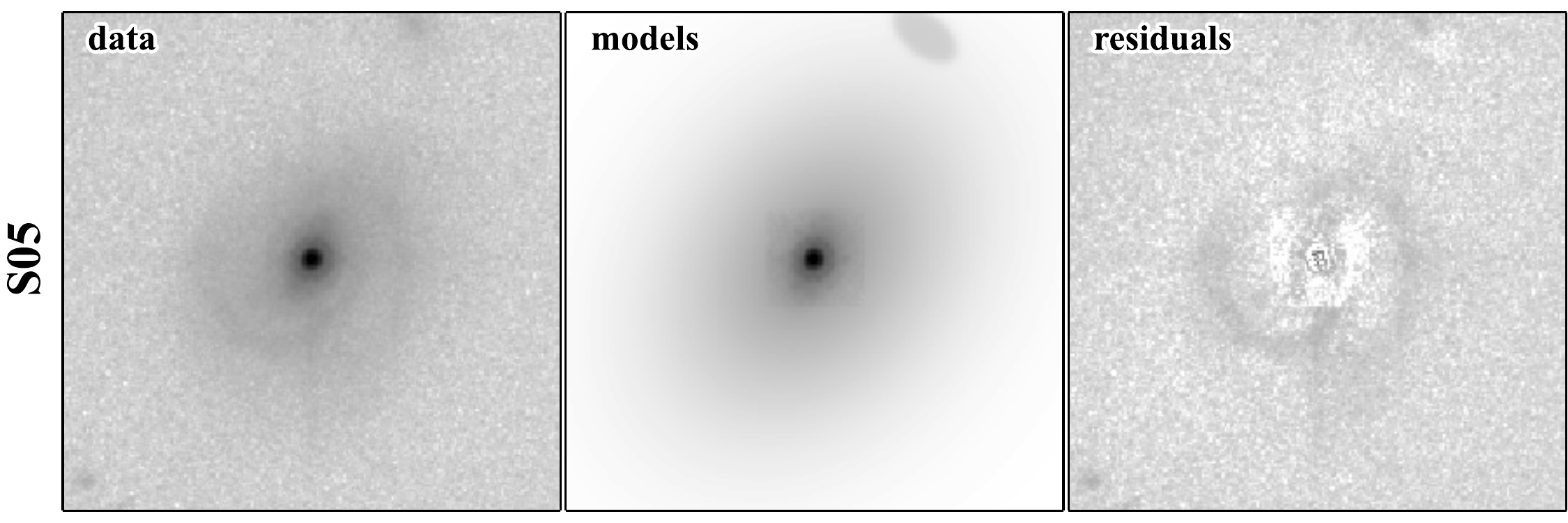}\includegraphics[width=0.25\textwidth]{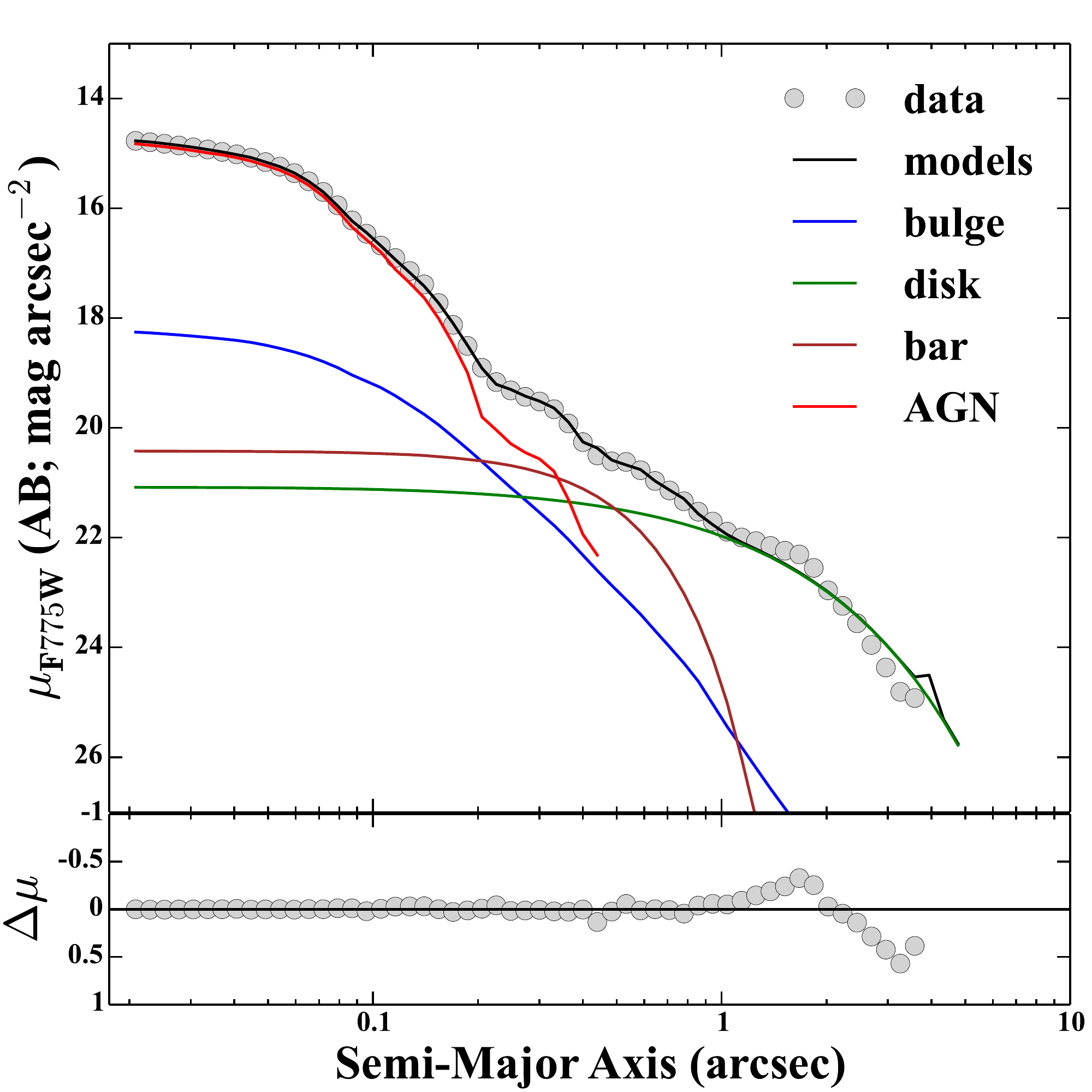}\\
	\includegraphics[width=0.75\textwidth]{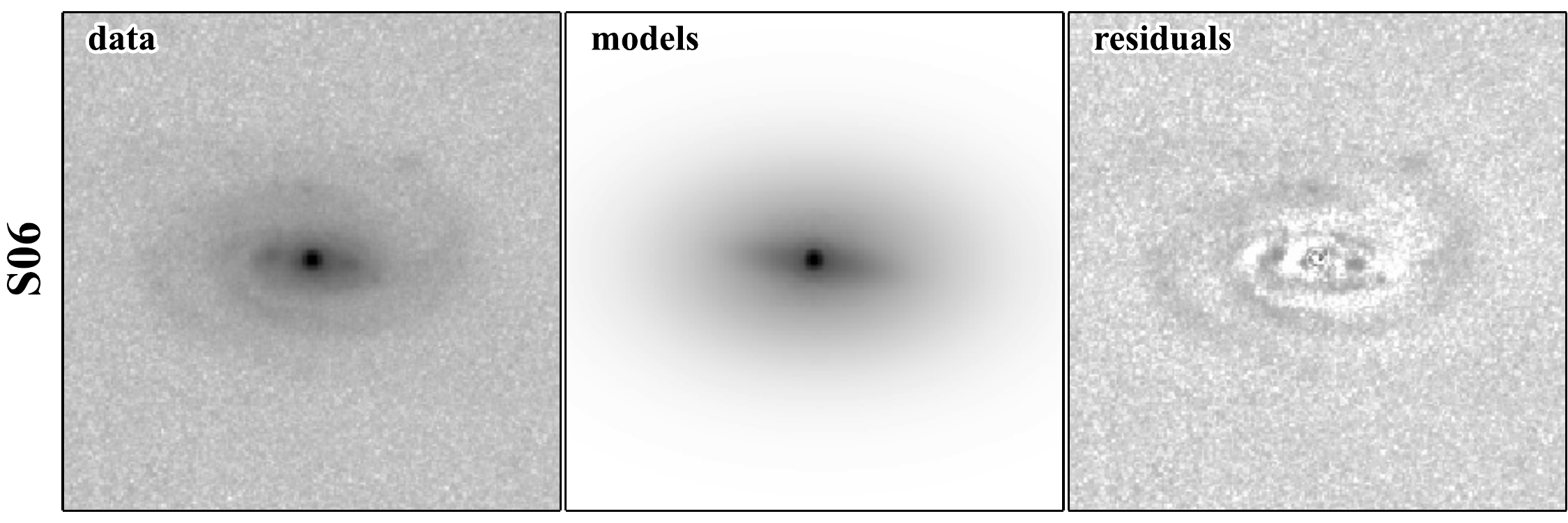}\includegraphics[width=0.25\textwidth]{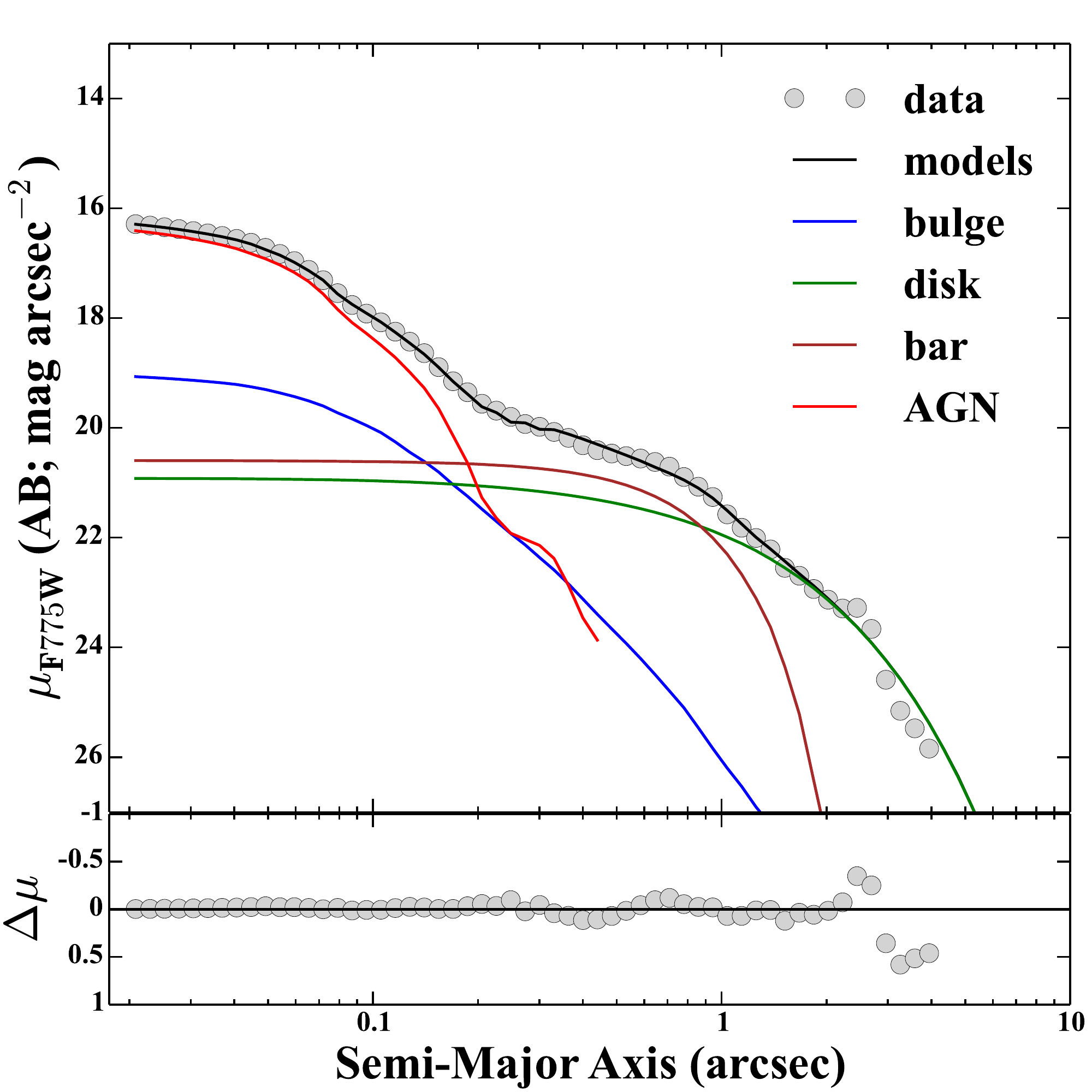}
    \figurenum{A\-2}
        \caption{
        \it Continued.
        \label{fig:imgfit1dSBP_pre40_3}}
\end{figure*}

\begin{figure*}
\centering
    \includegraphics[width=0.75\textwidth]{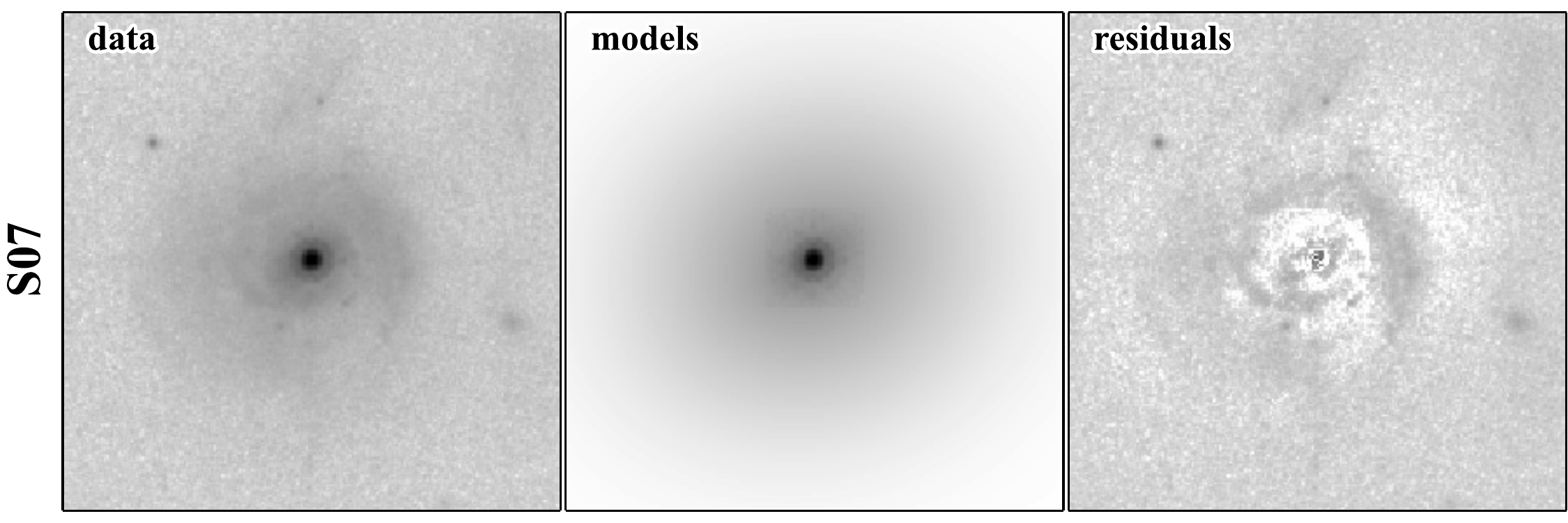}\includegraphics[width=0.25\textwidth]{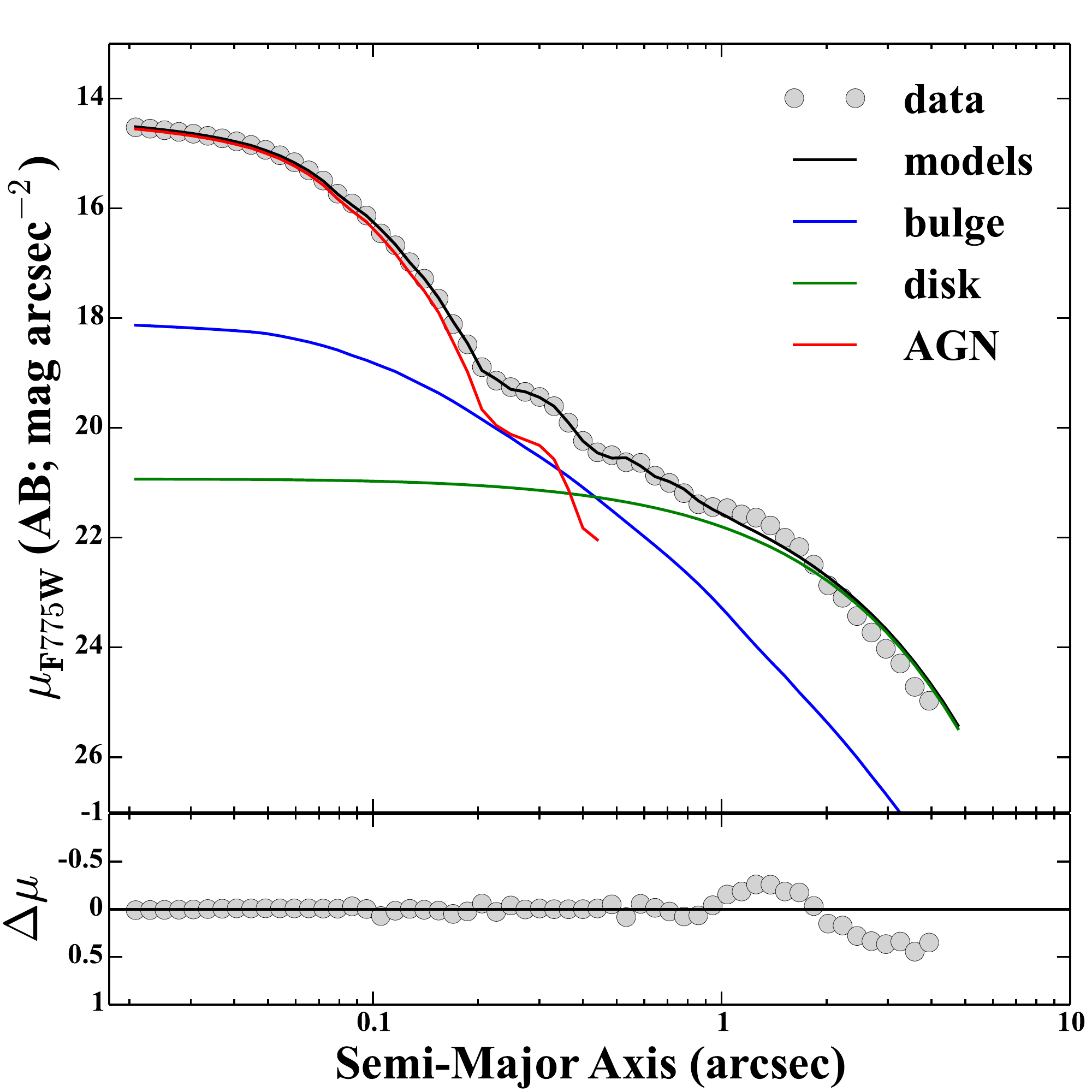}\\
    \includegraphics[width=0.75\textwidth]{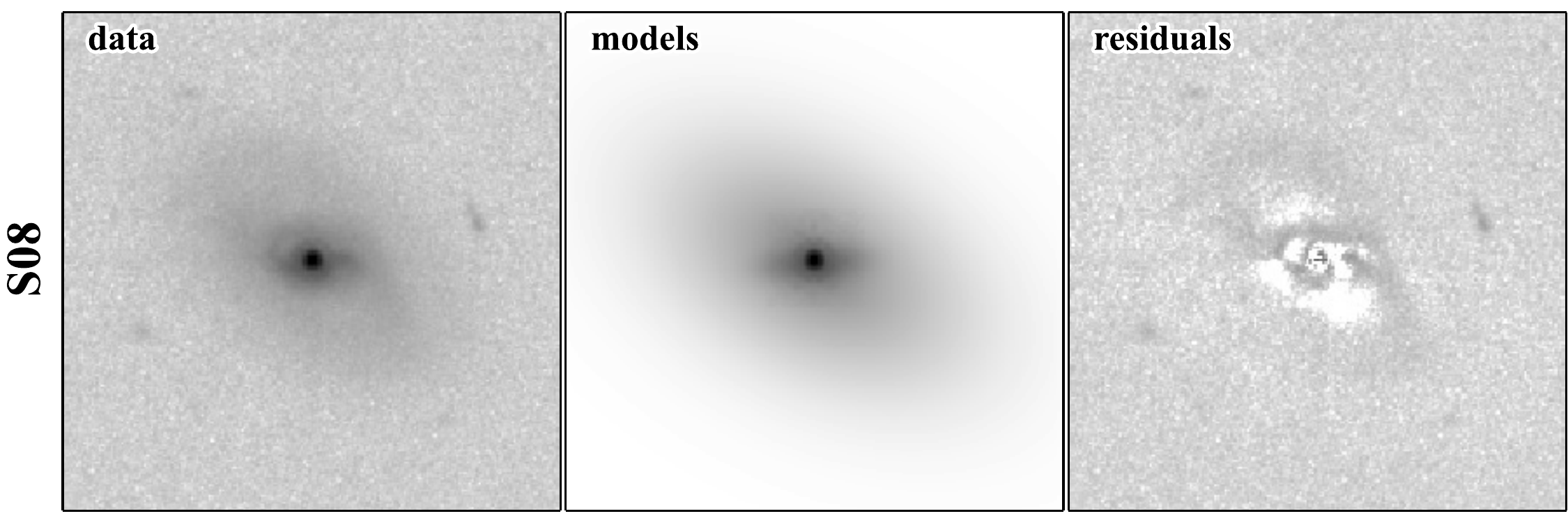}\includegraphics[width=0.25\textwidth]{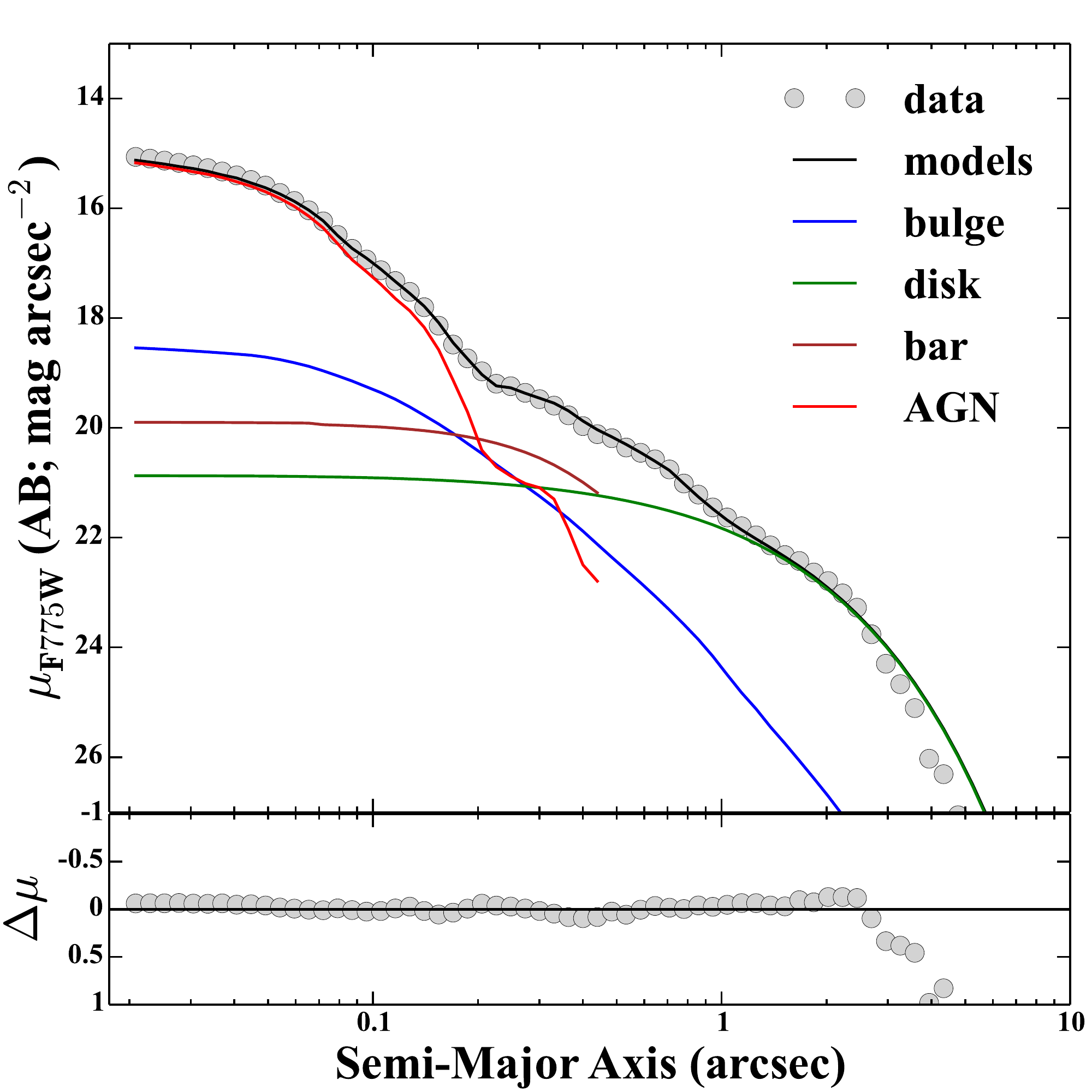}\\    \includegraphics[width=0.75\textwidth]{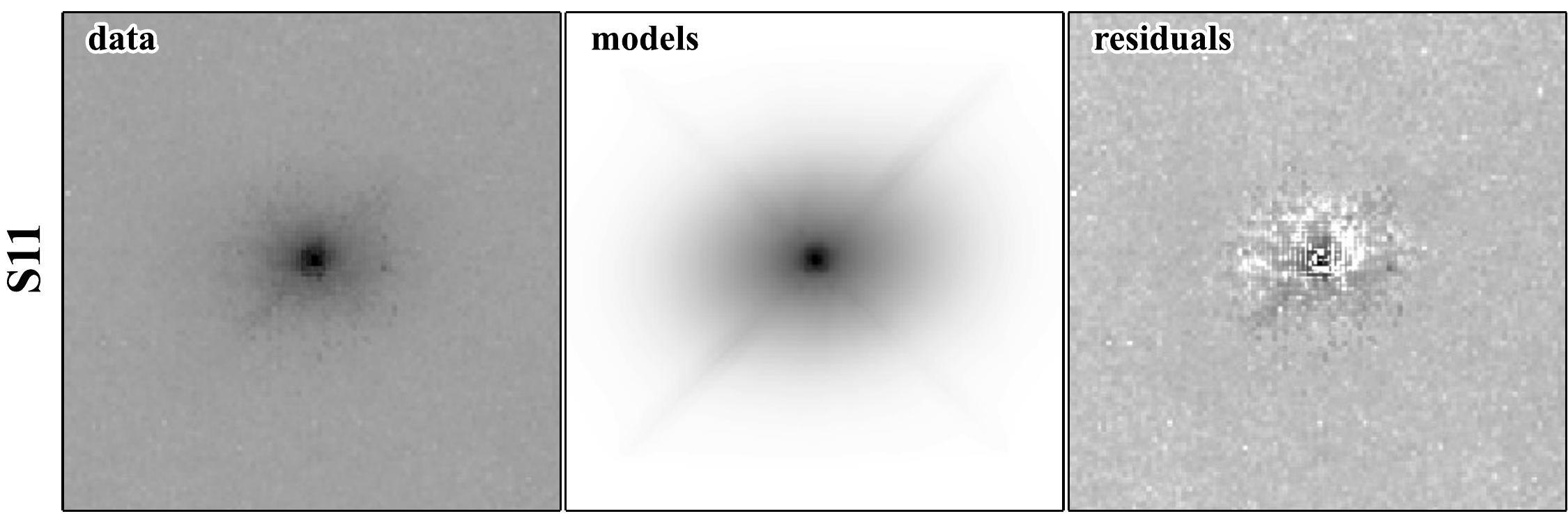}\includegraphics[width=0.25\textwidth]{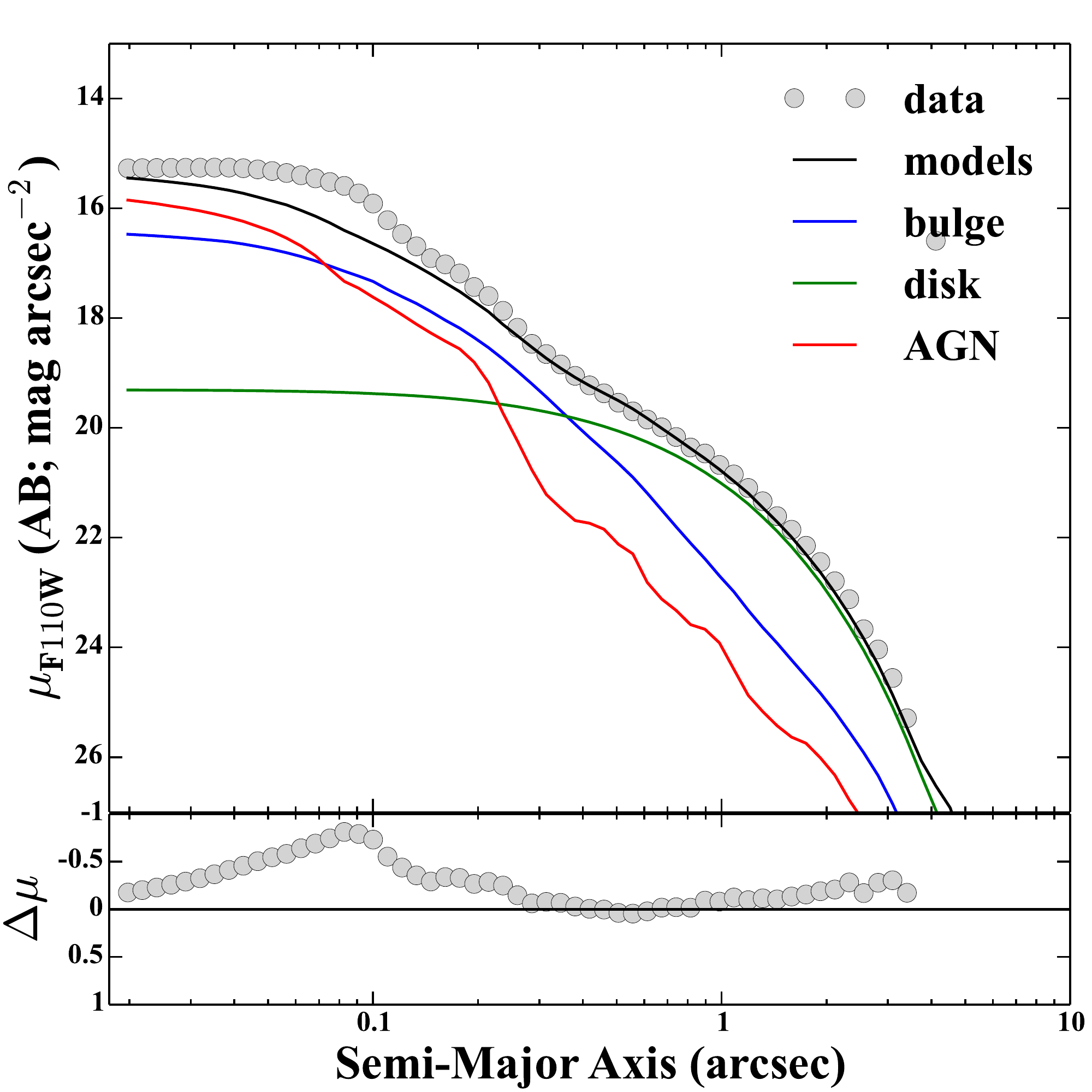}\\
	\includegraphics[width=0.75\textwidth]{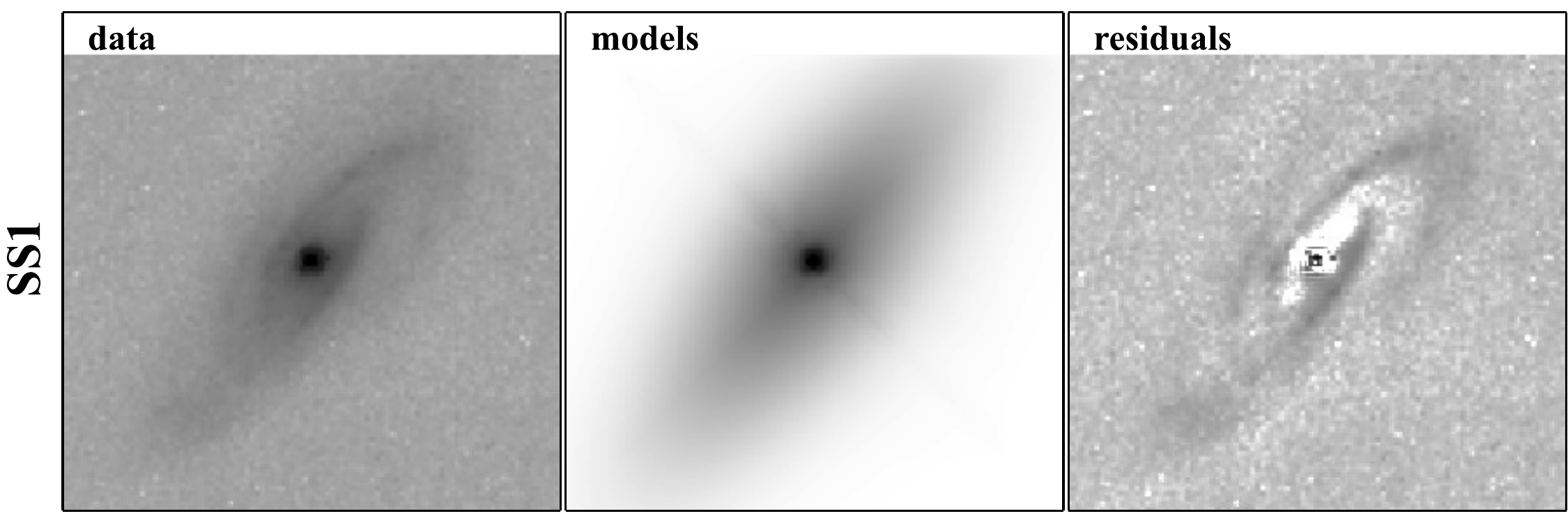}\includegraphics[width=0.25\textwidth]{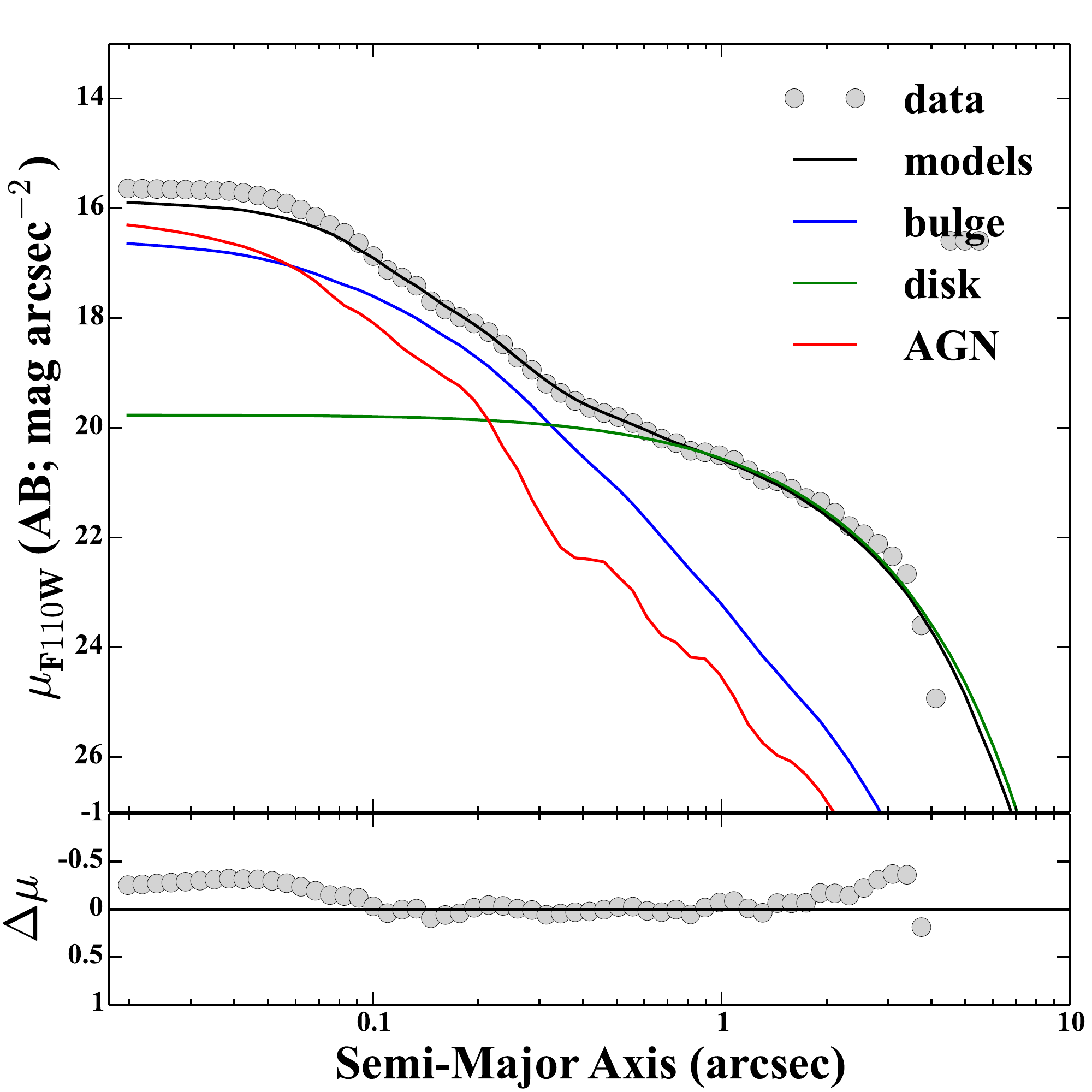}\\
	\includegraphics[width=0.75\textwidth]{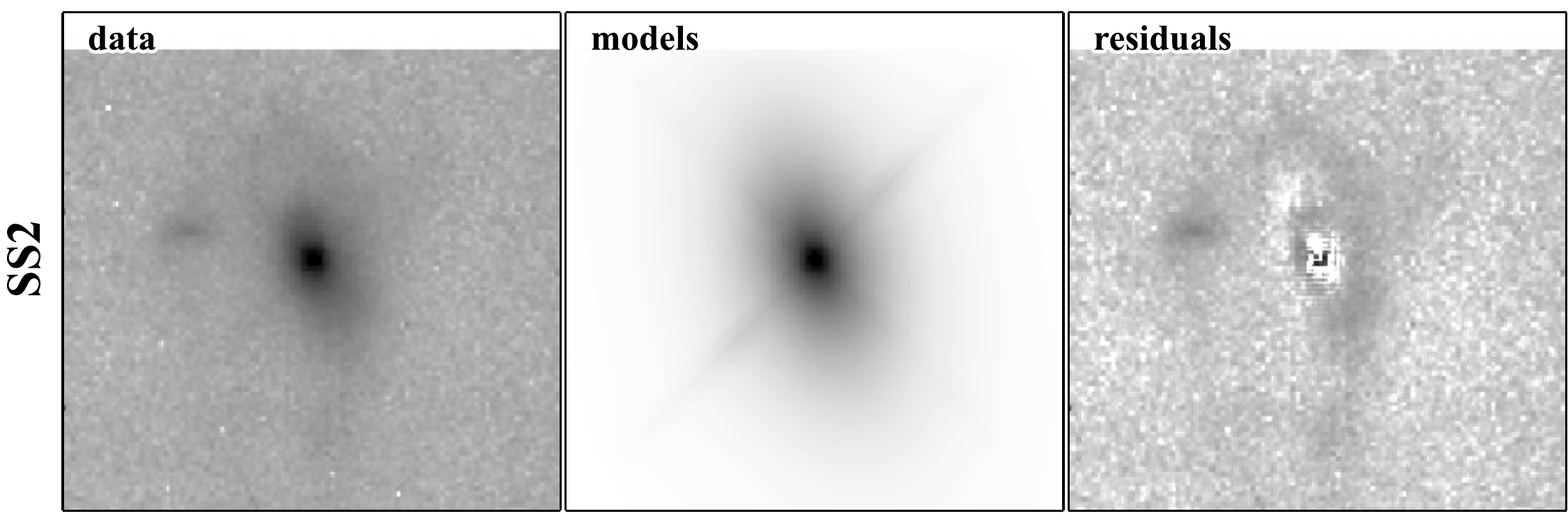}\includegraphics[width=0.25\textwidth]{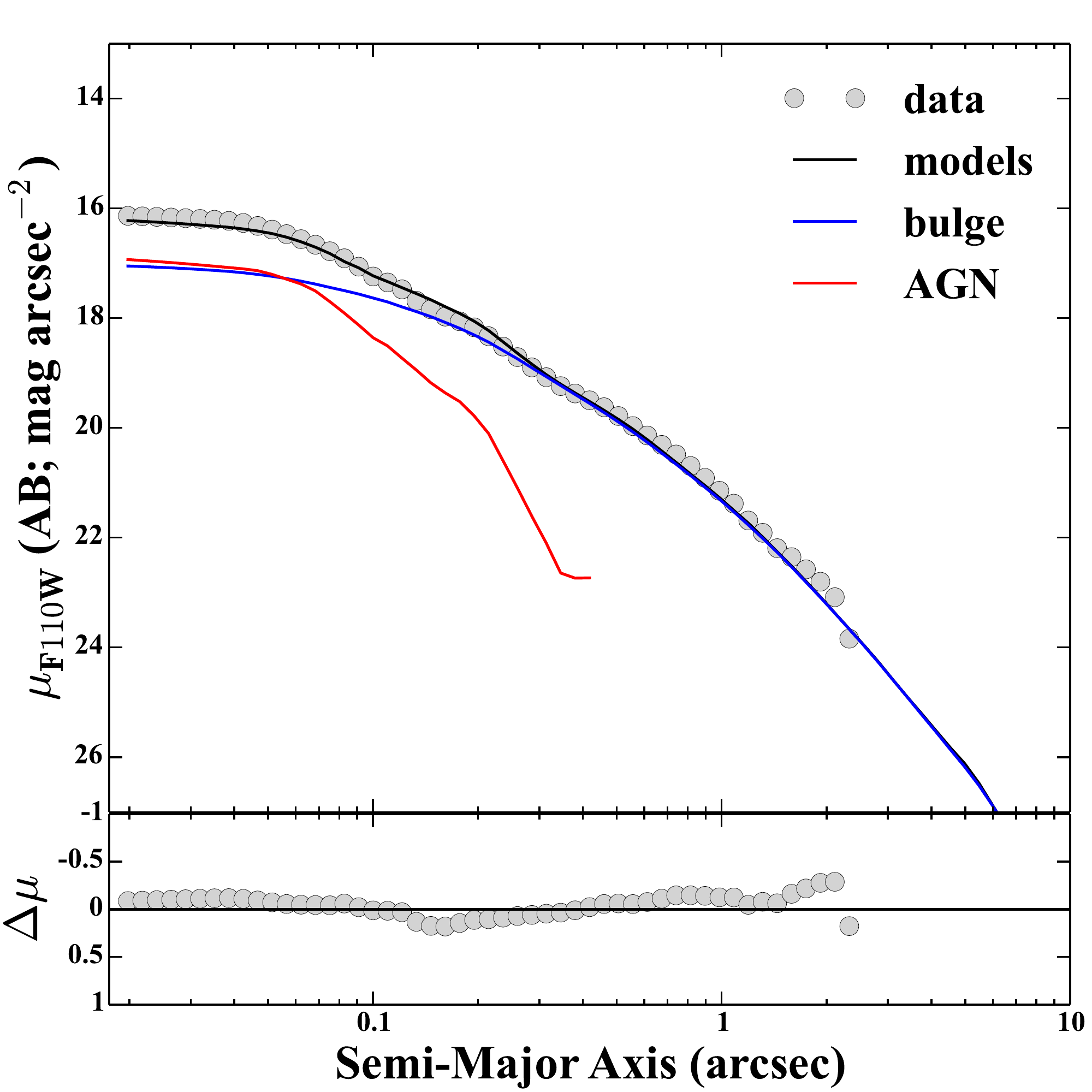}
    \figurenum{A\-2}
        \caption{
        \it Continued.
        \label{fig:imgfit1dSBP_pre40_4}}
\end{figure*}

\begin{figure*}
\centering
	\includegraphics[width=0.75\textwidth]{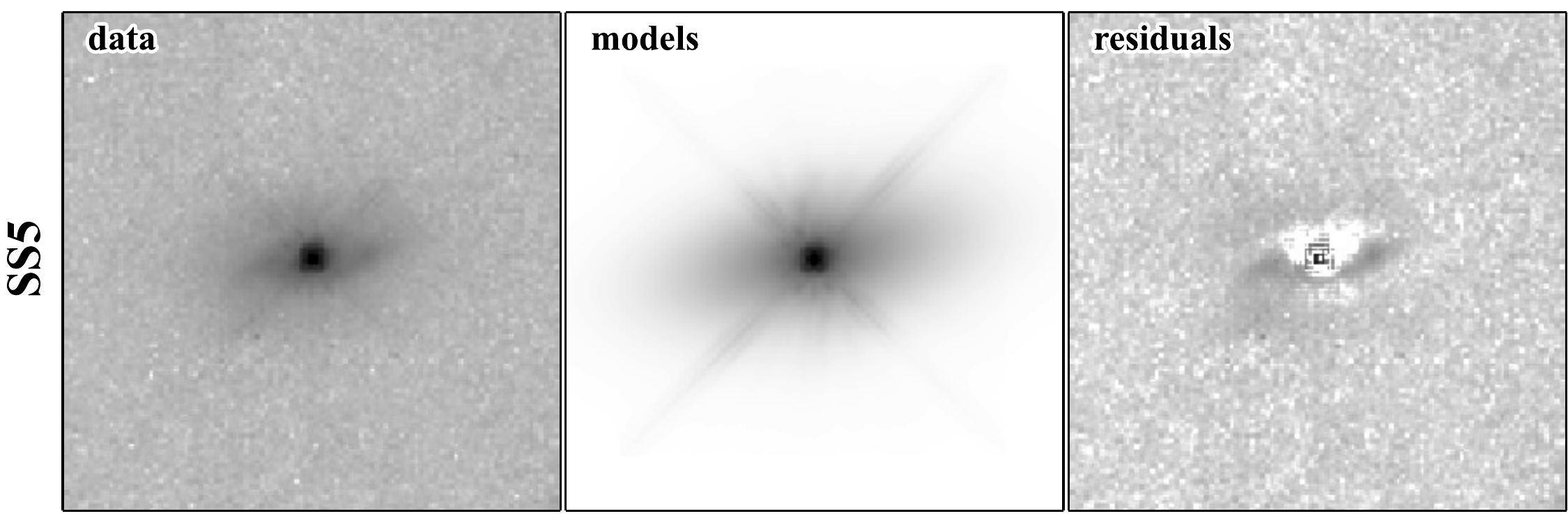}\includegraphics[width=0.25\textwidth]{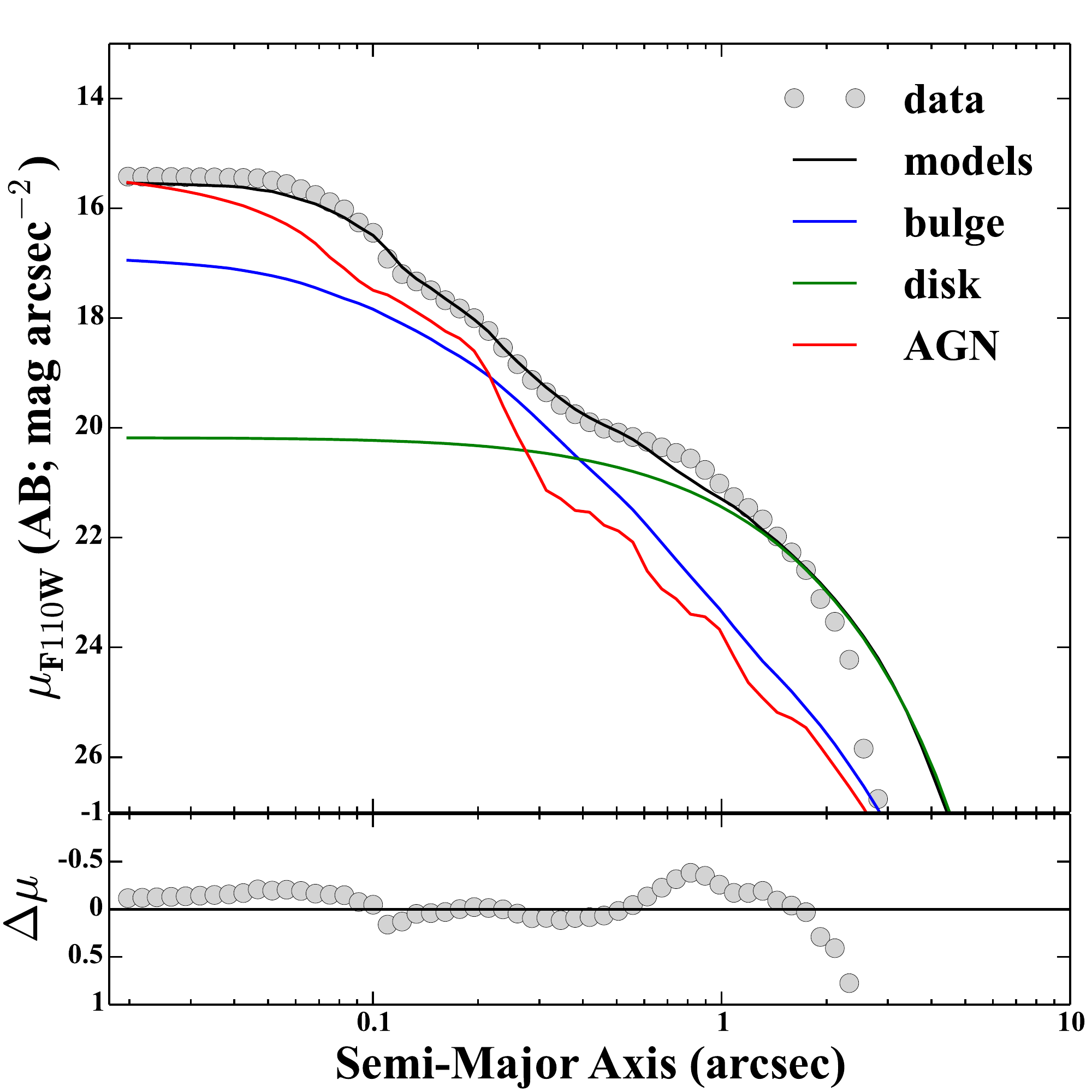}\\
	\includegraphics[width=0.75\textwidth]{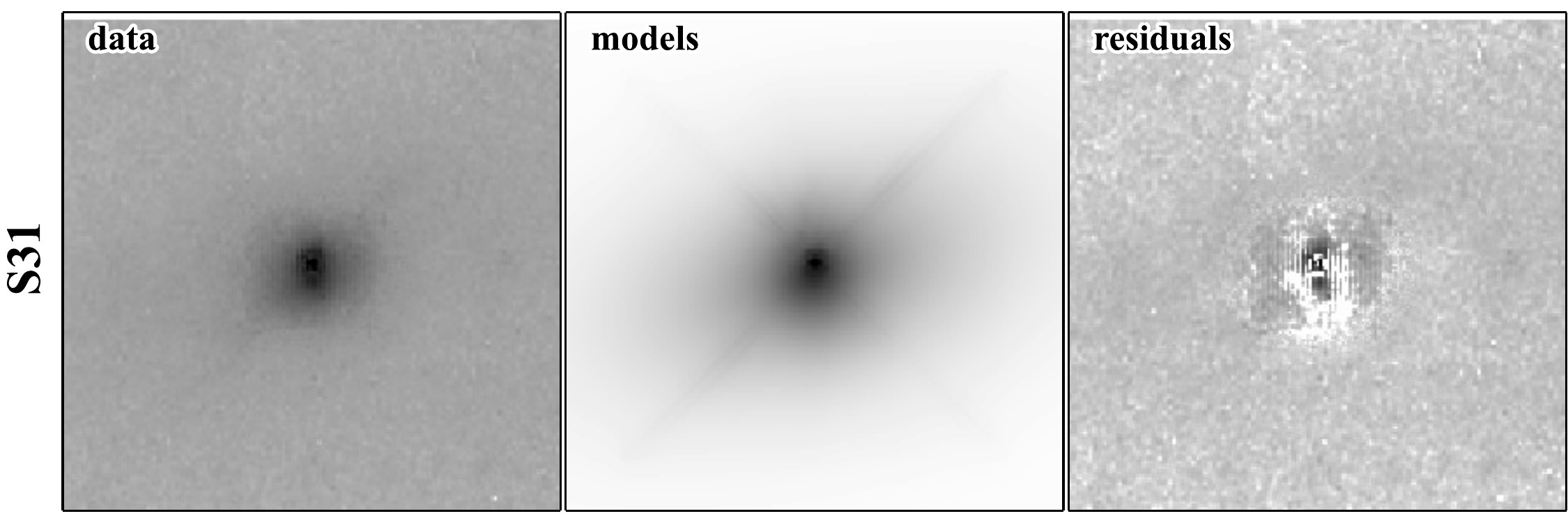}\includegraphics[width=0.25\textwidth]{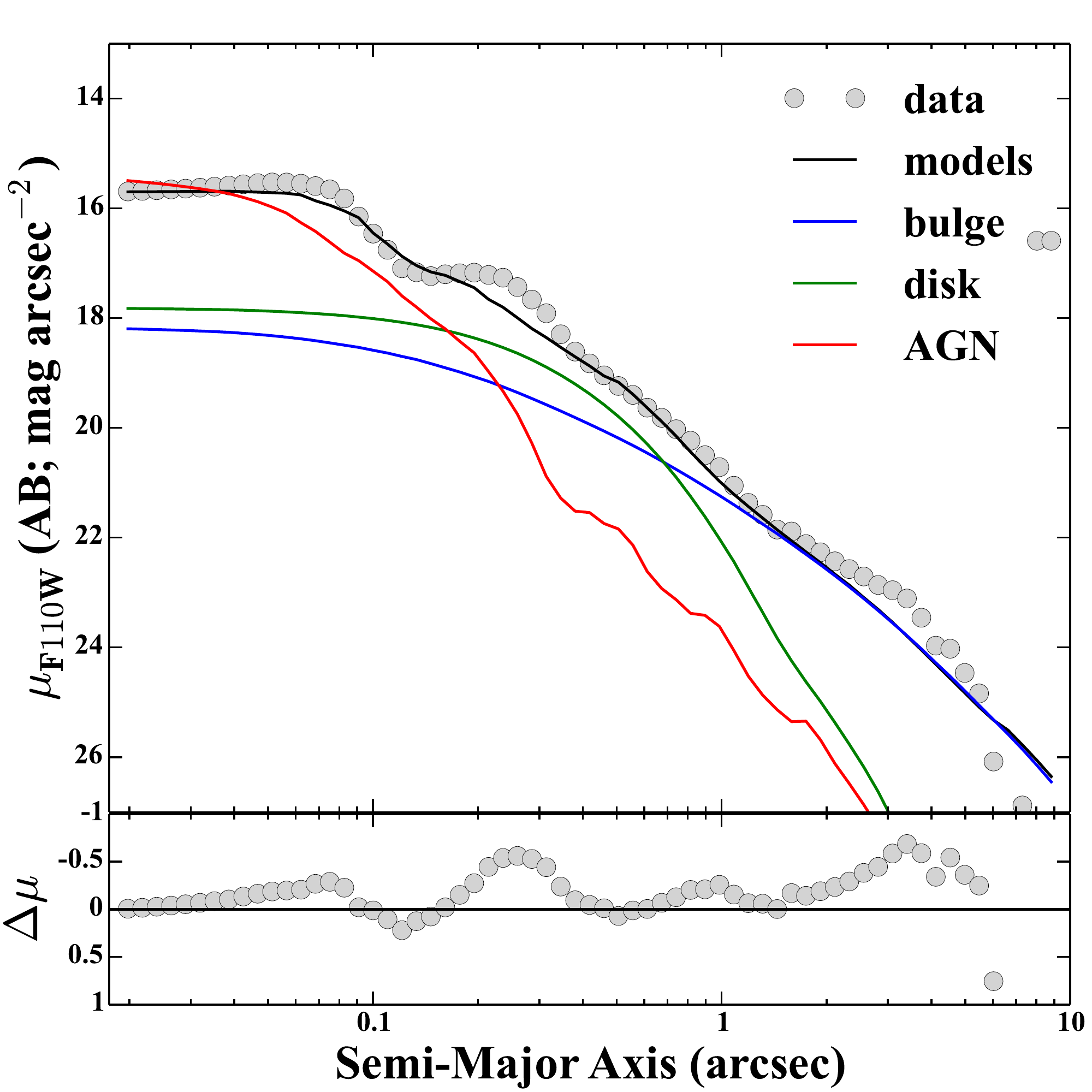}\\
	\includegraphics[width=0.75\textwidth]{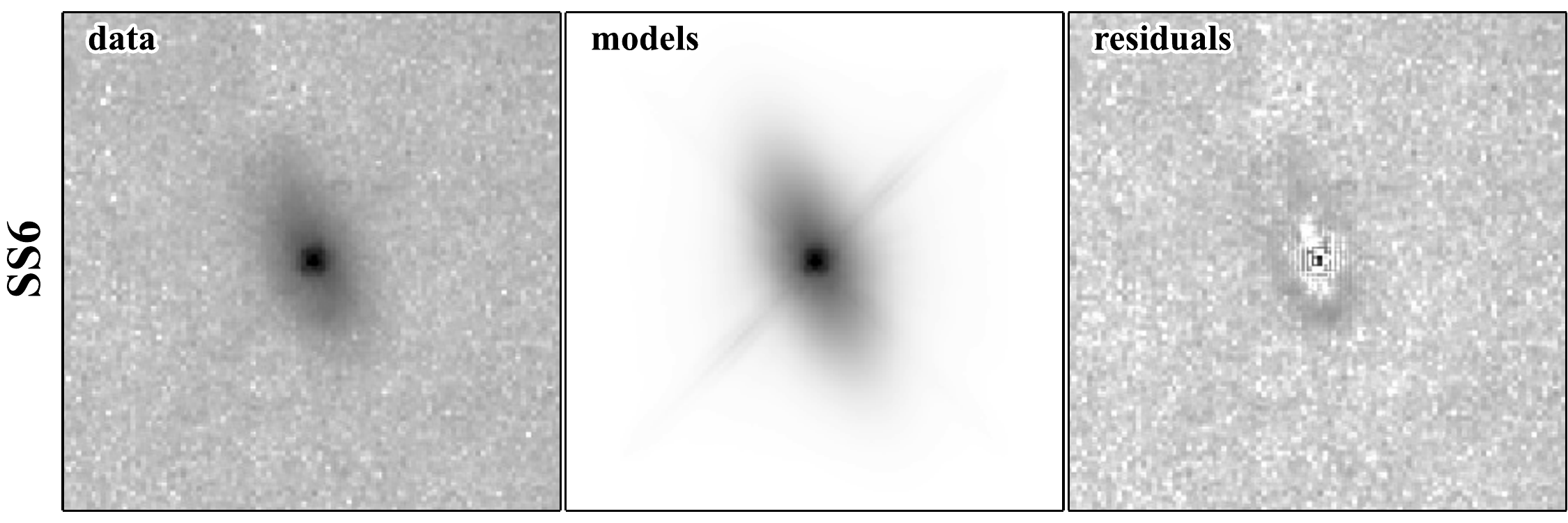}\includegraphics[width=0.25\textwidth]{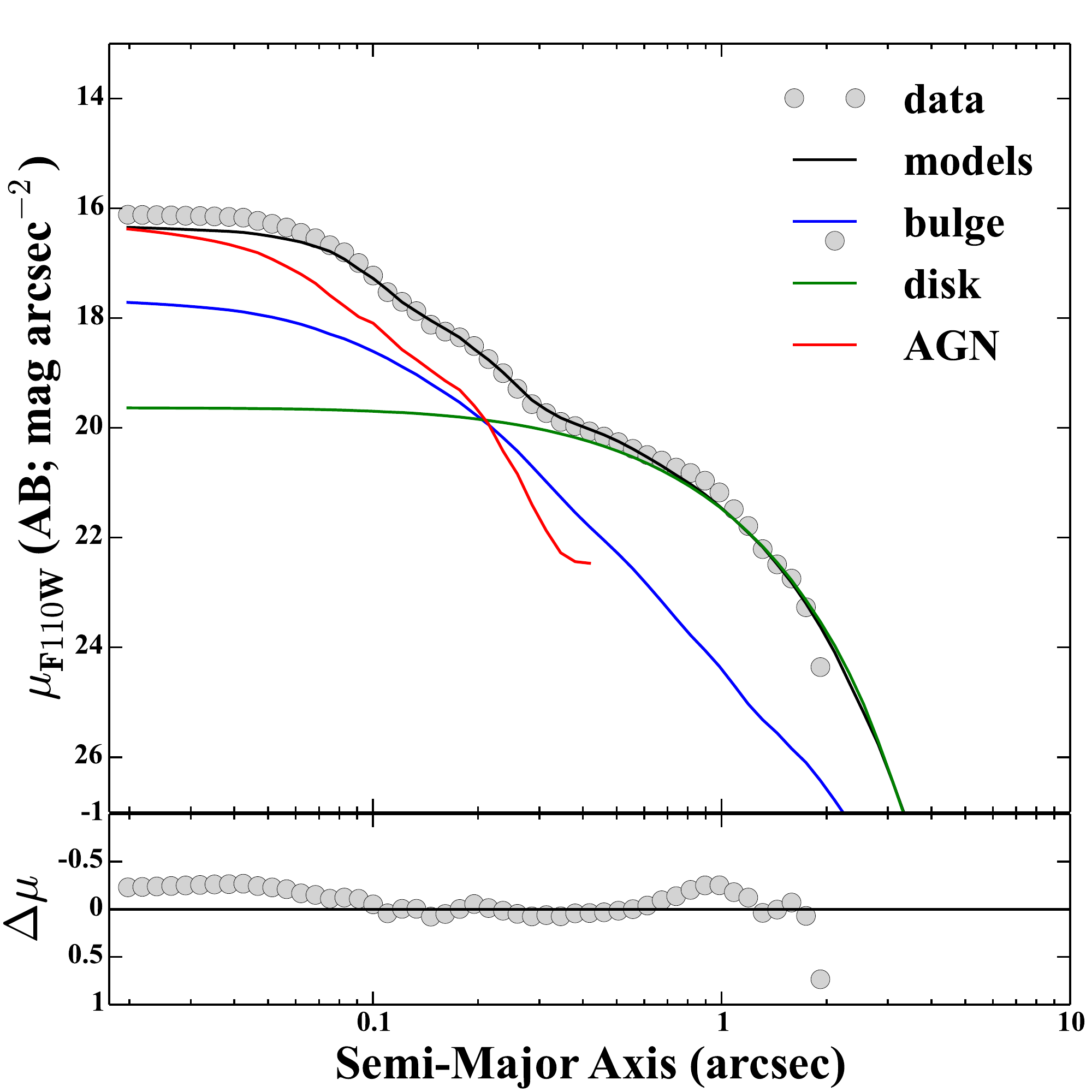}\\
	\includegraphics[width=0.75\textwidth]{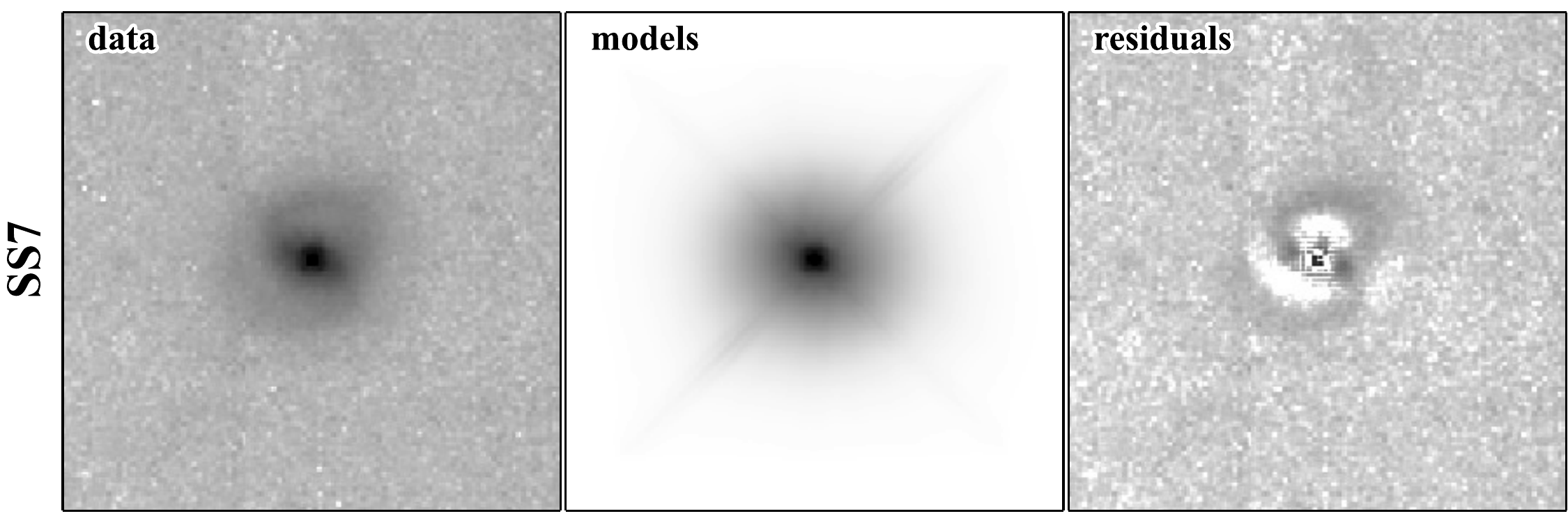}\includegraphics[width=0.25\textwidth]{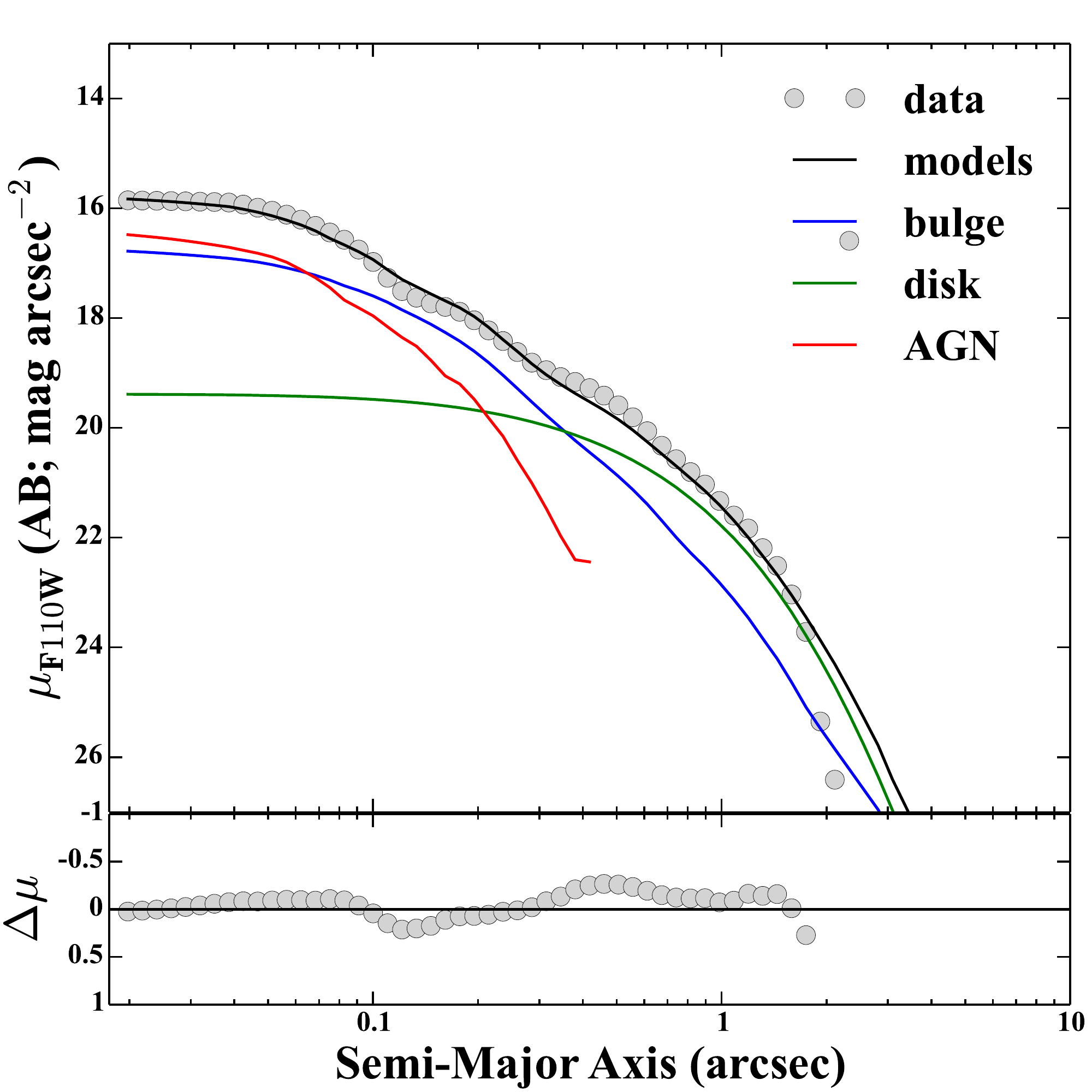}\\
	\includegraphics[width=0.75\textwidth]{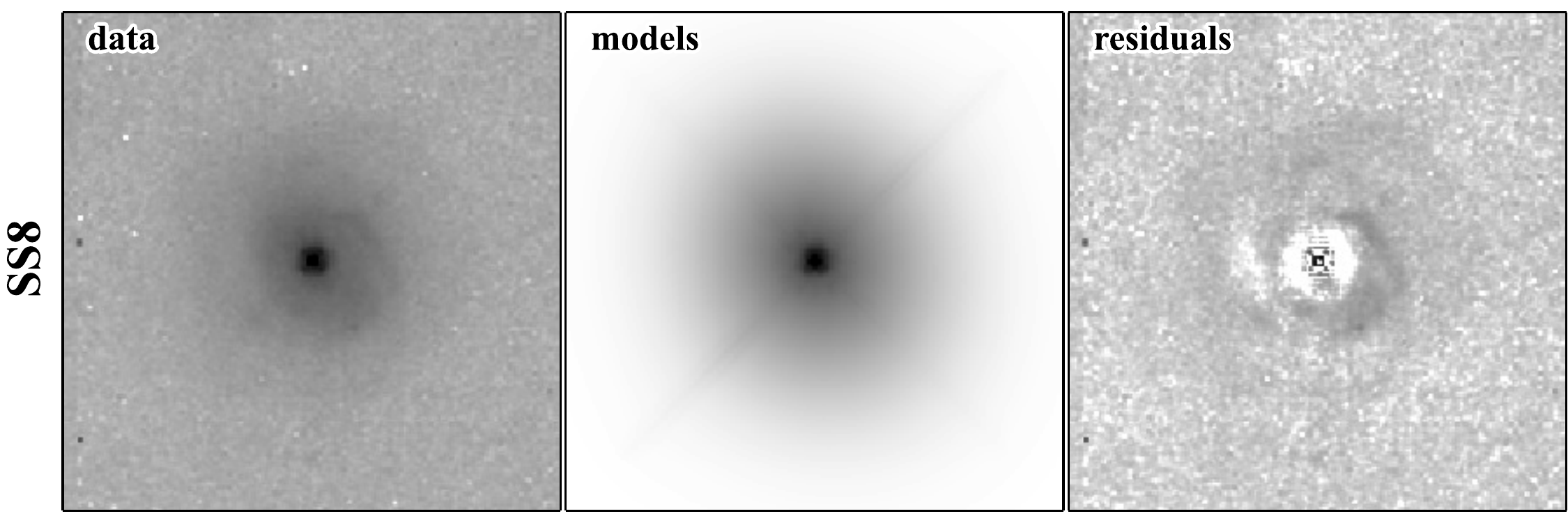}\includegraphics[width=0.25\textwidth]{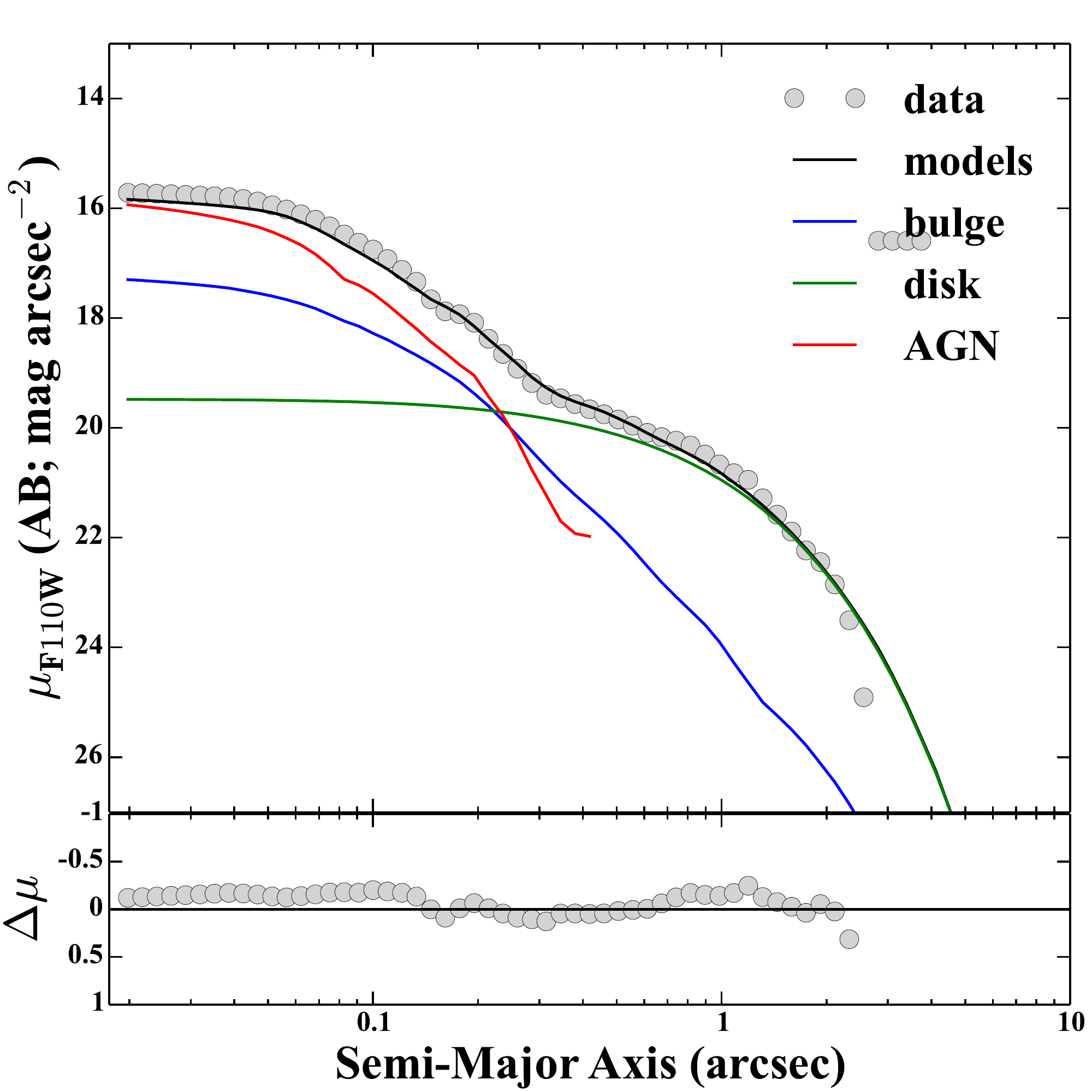}
    \figurenum{A\-2}
        \caption{
        \it Continued.
        \label{fig:imgfit1dSBP_pre40_5}}
\end{figure*}

\begin{figure*}
\centering
	\includegraphics[width=0.75\textwidth]{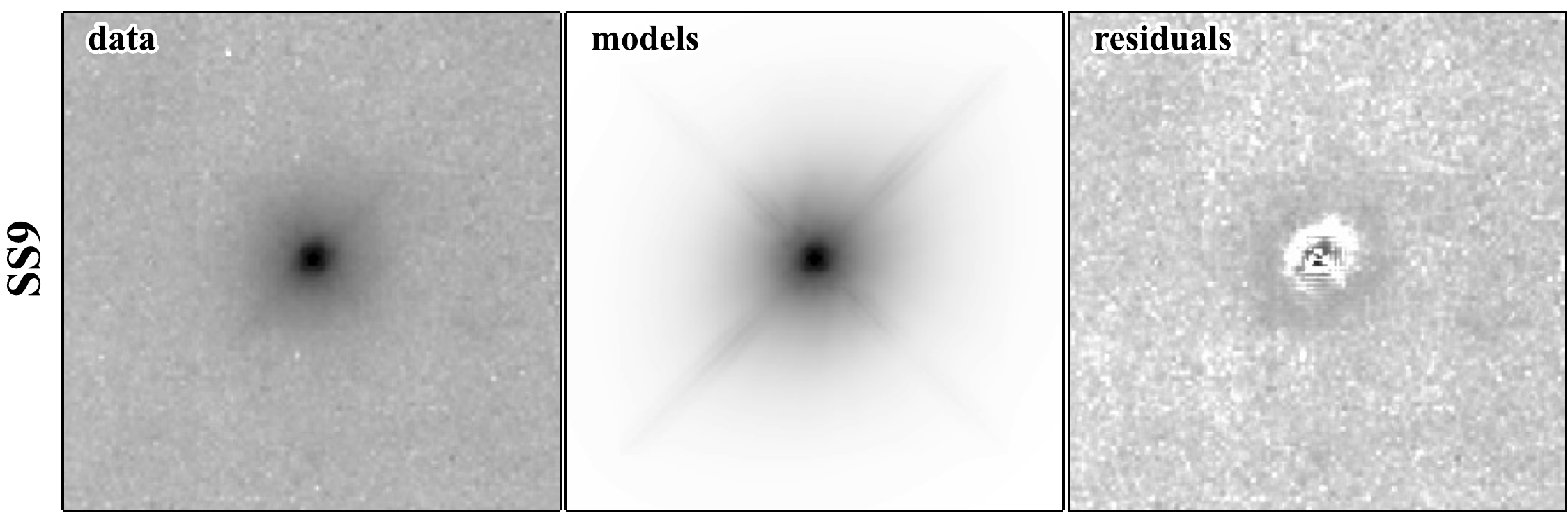}\includegraphics[width=0.25\textwidth]{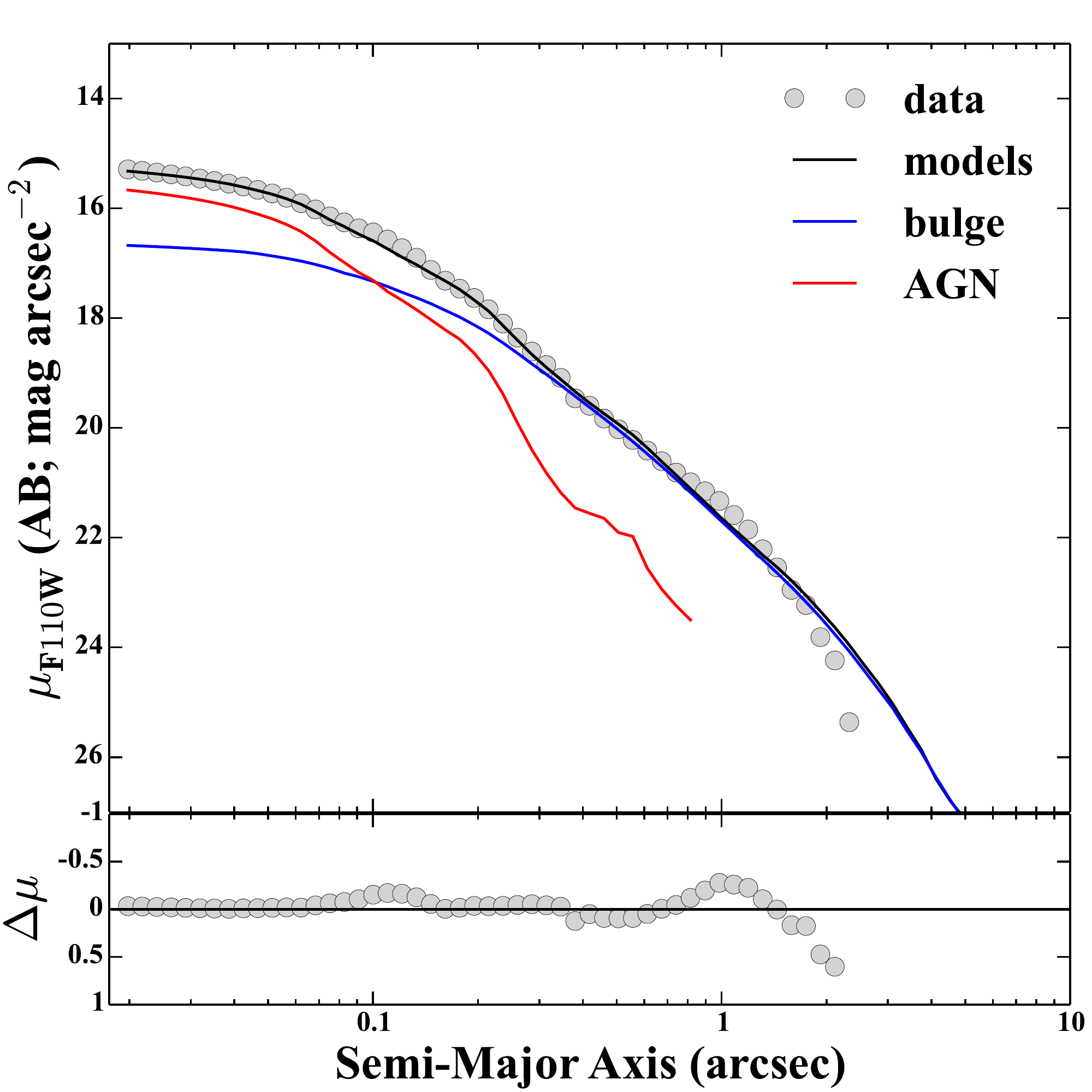}\\
	\includegraphics[width=0.75\textwidth]{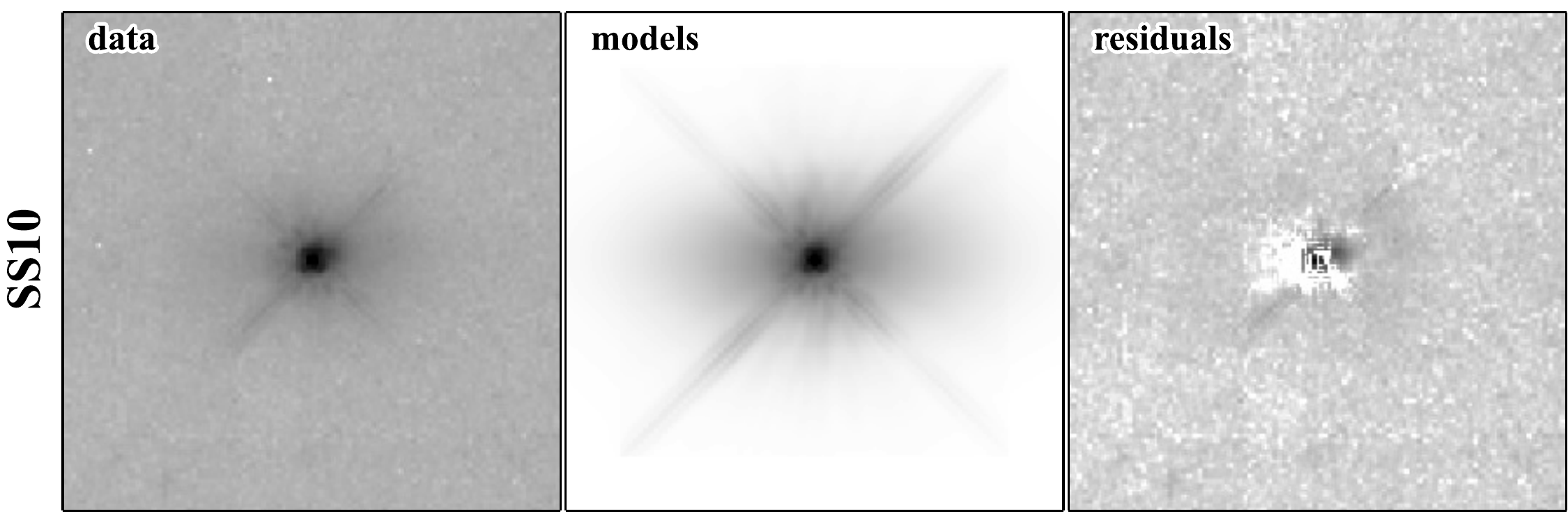}\includegraphics[width=0.25\textwidth]{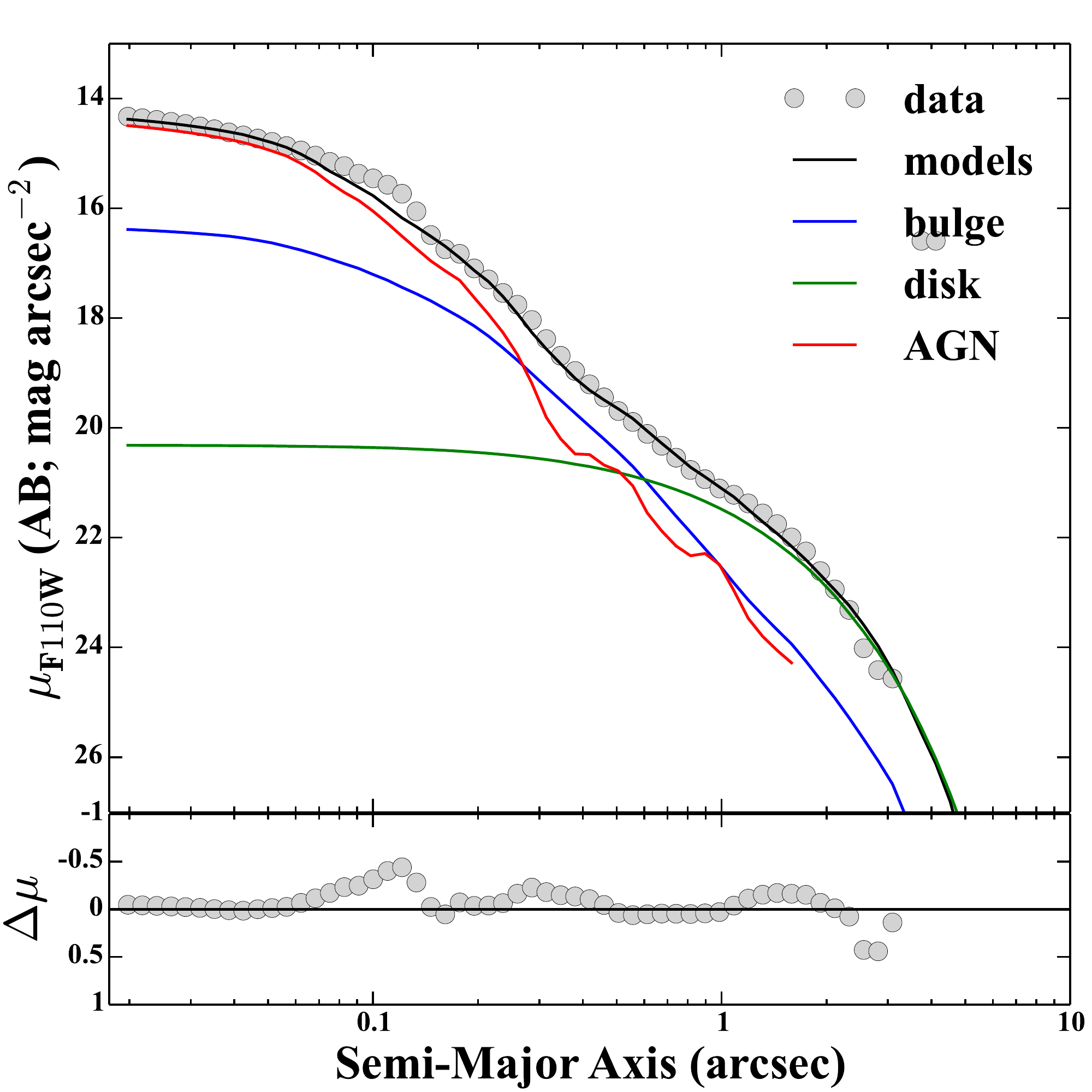}\\
	\includegraphics[width=0.75\textwidth]{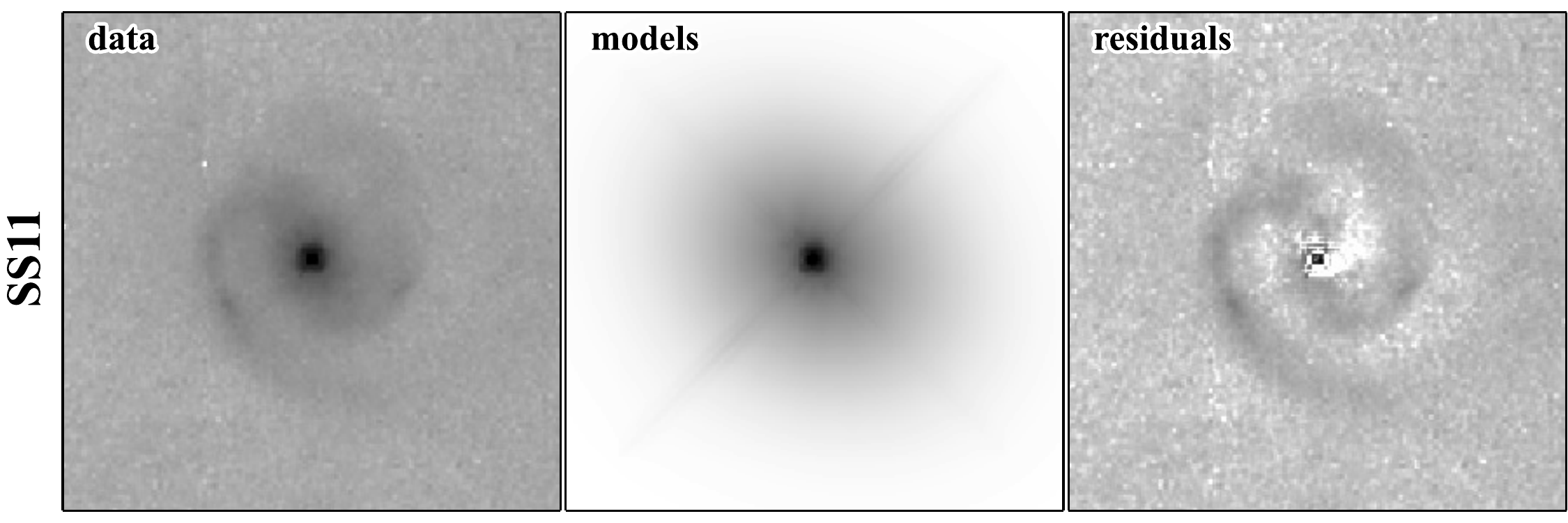}\includegraphics[width=0.25\textwidth]{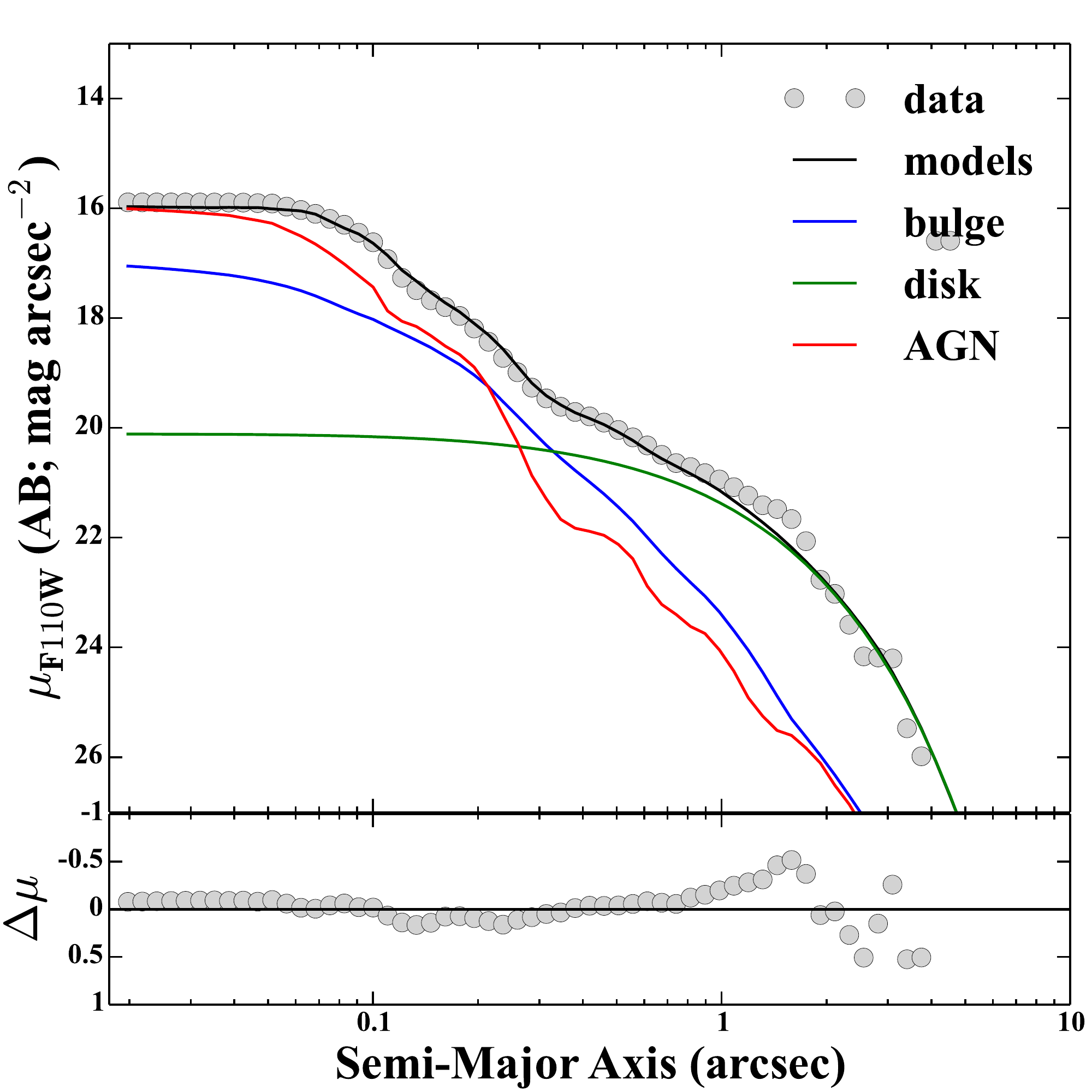}\\
	\includegraphics[width=0.75\textwidth]{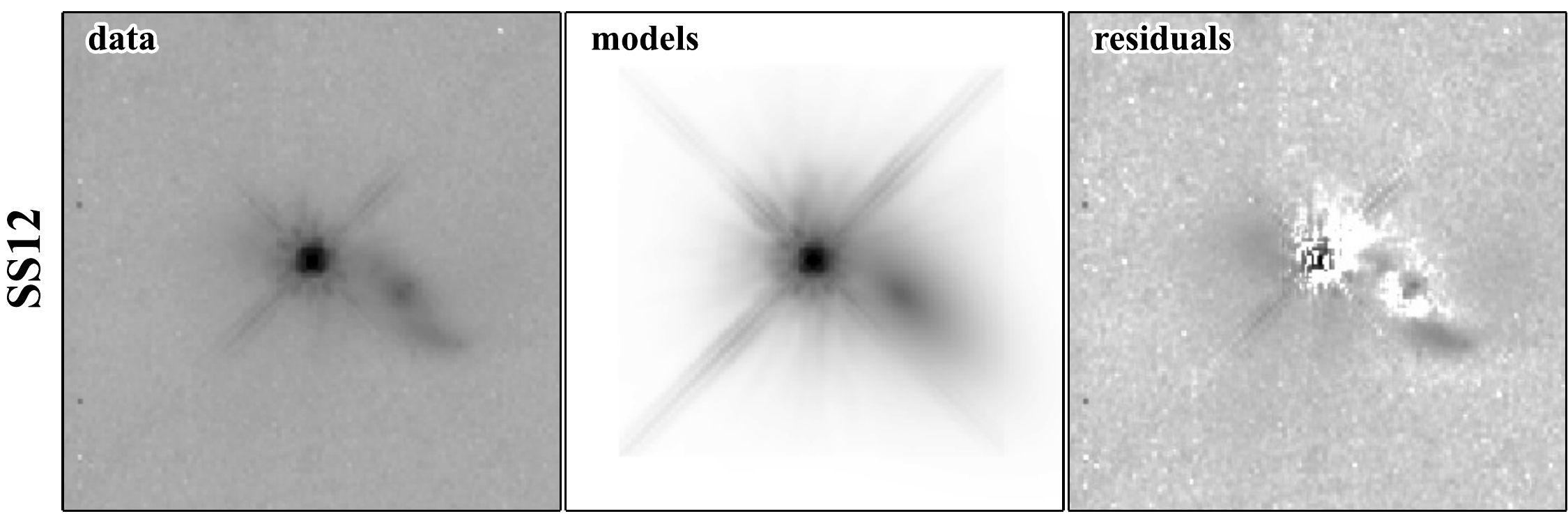}\includegraphics[width=0.25\textwidth]{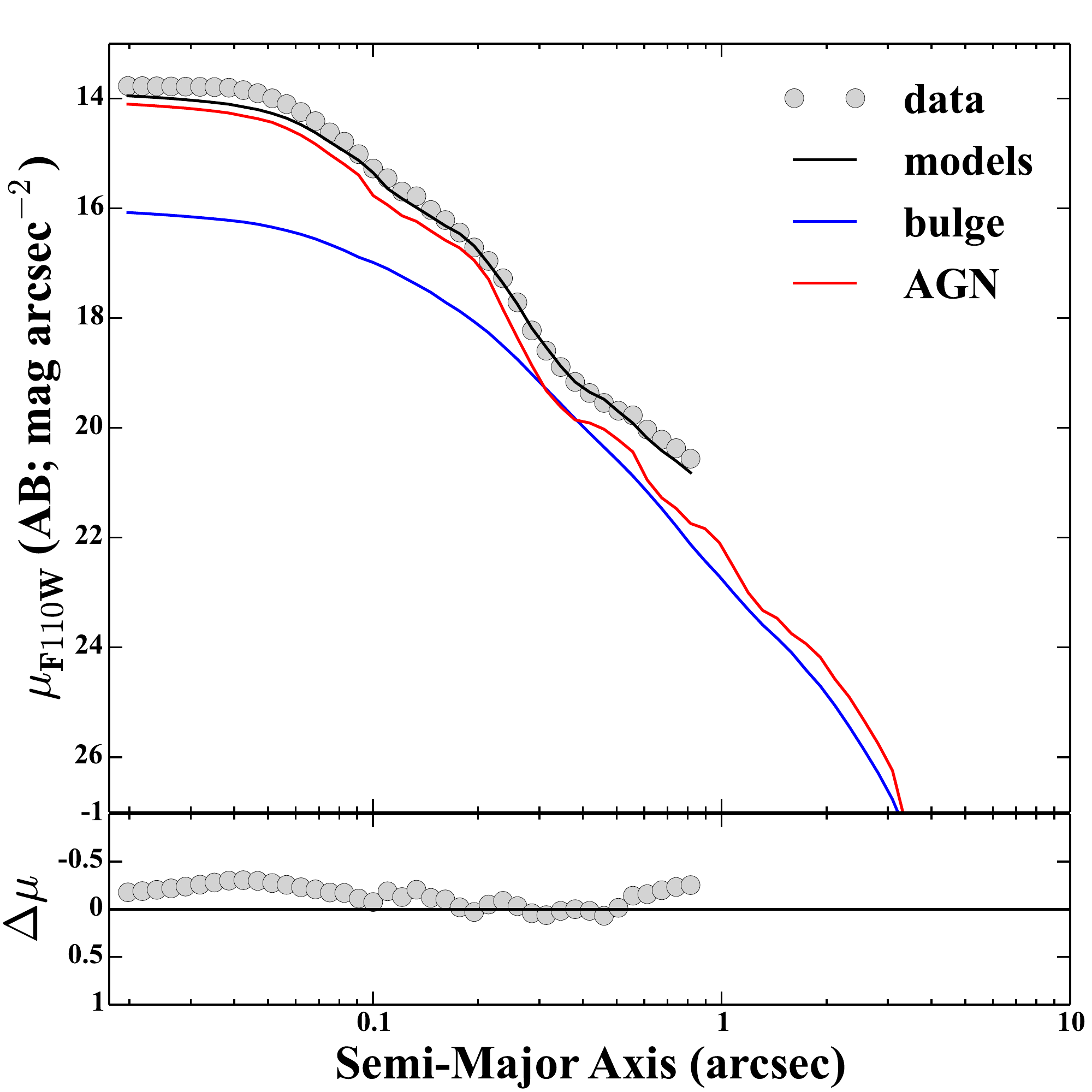}\\
	\includegraphics[width=0.75\textwidth]{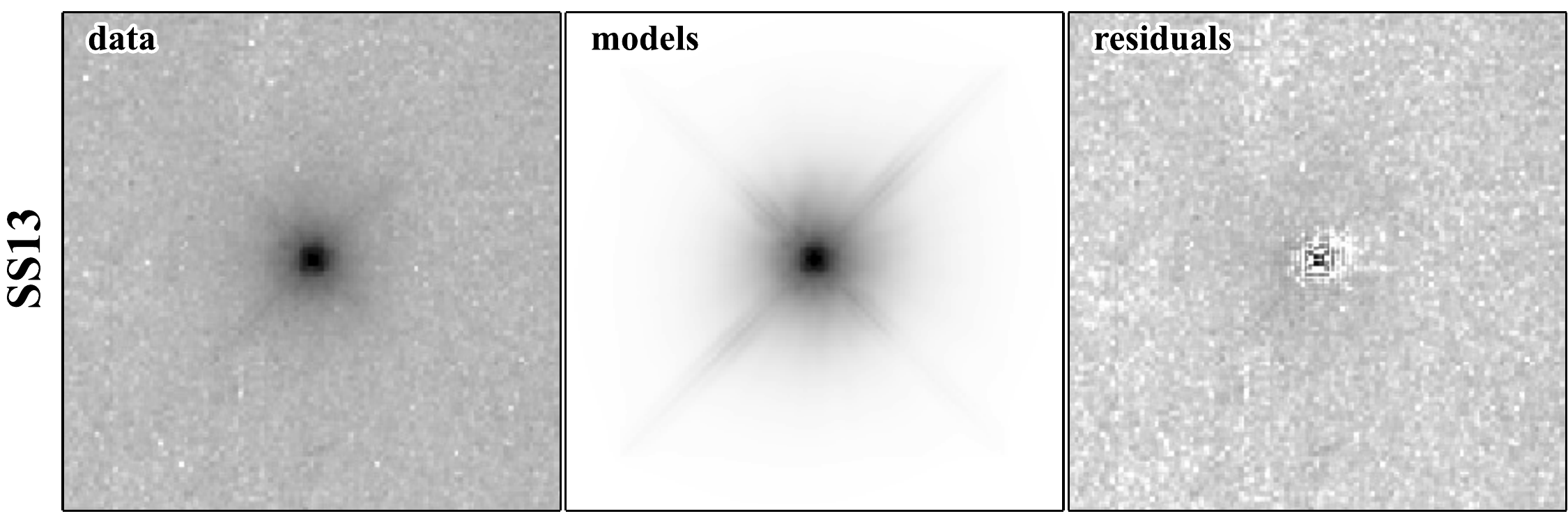}\includegraphics[width=0.25\textwidth]{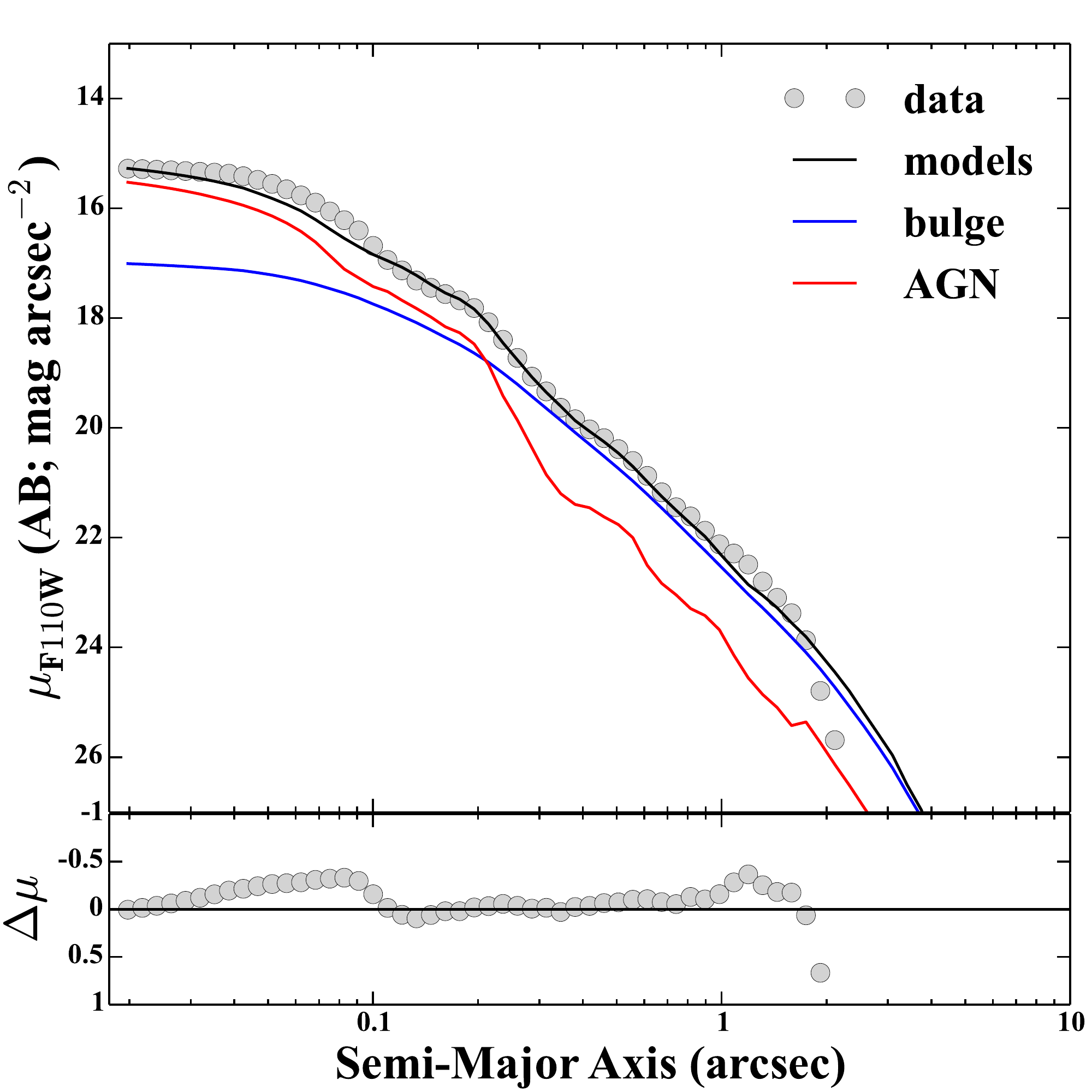}
    \figurenum{A\-2}
        \caption{
        \it Continued.
        \label{fig:imgfit1dSBP_pre40_6}}
\end{figure*}

\begin{figure*}
\centering
	\includegraphics[width=0.75\textwidth]{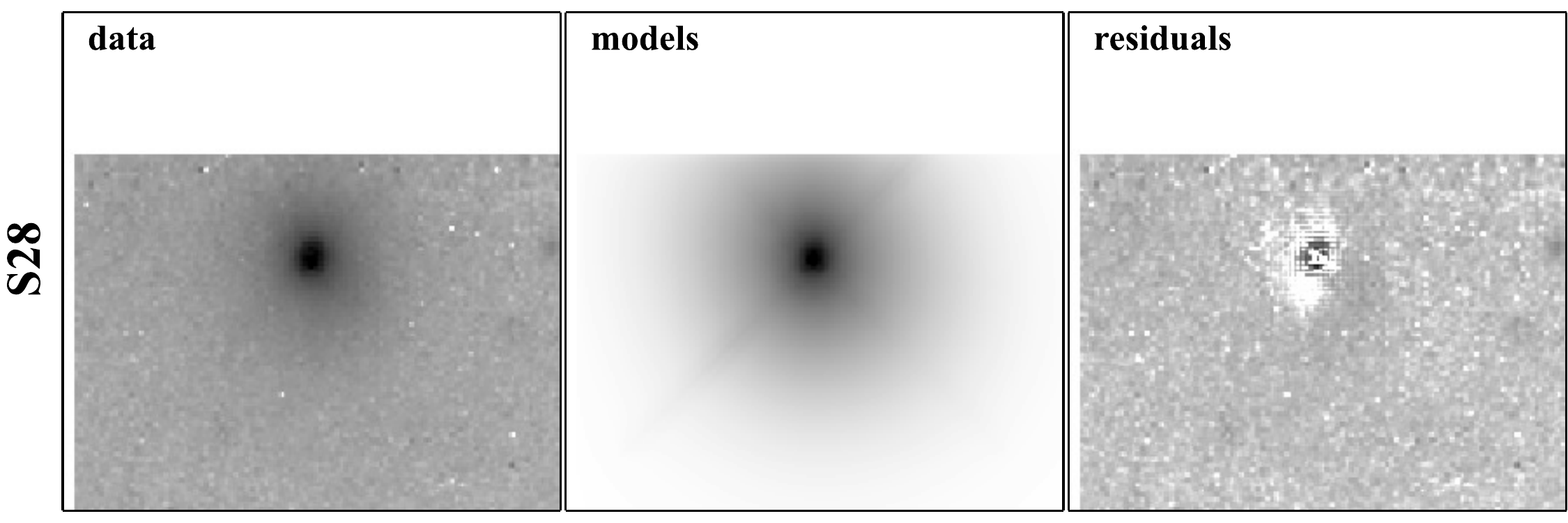}\includegraphics[width=0.25\textwidth]{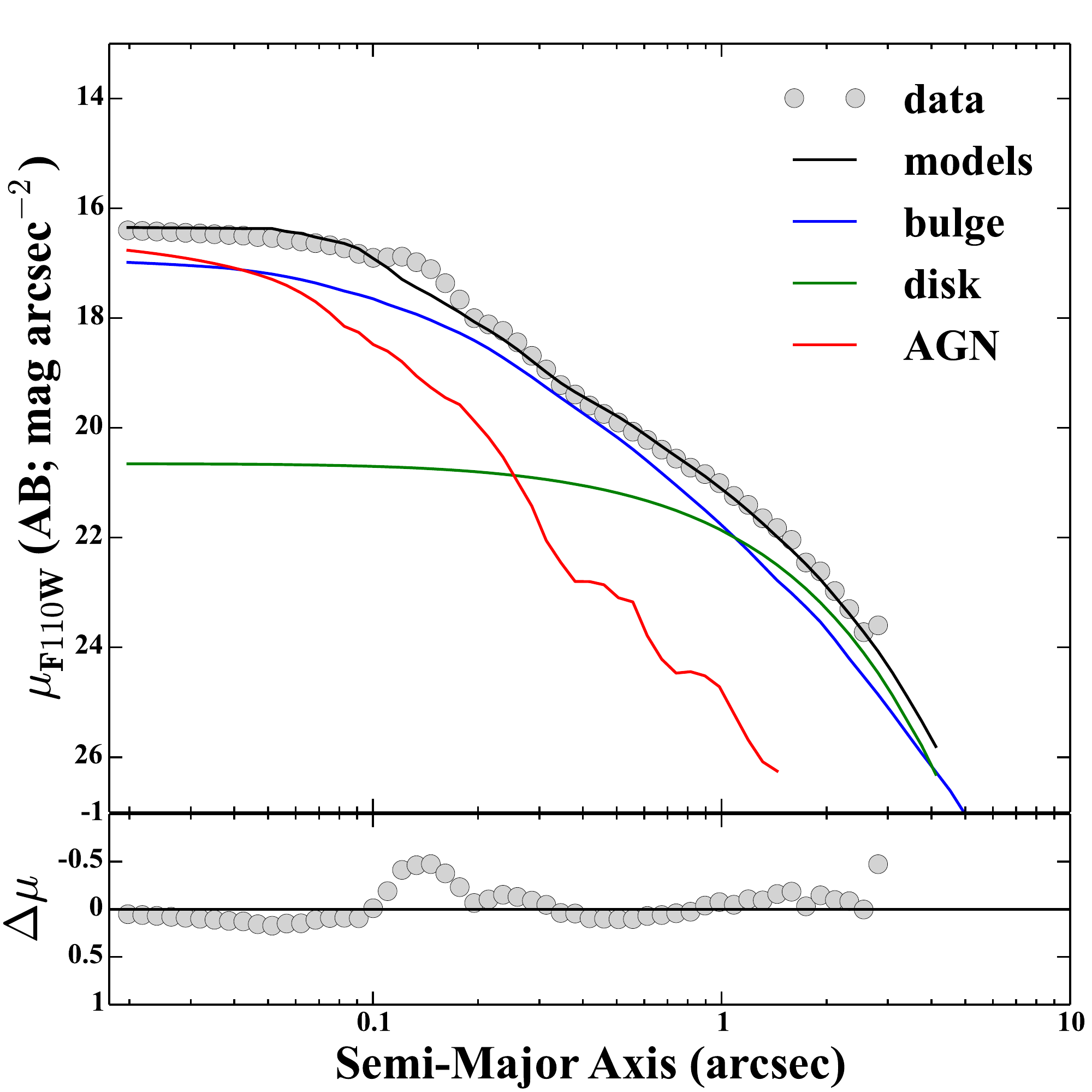}\\
	\includegraphics[width=0.75\textwidth]{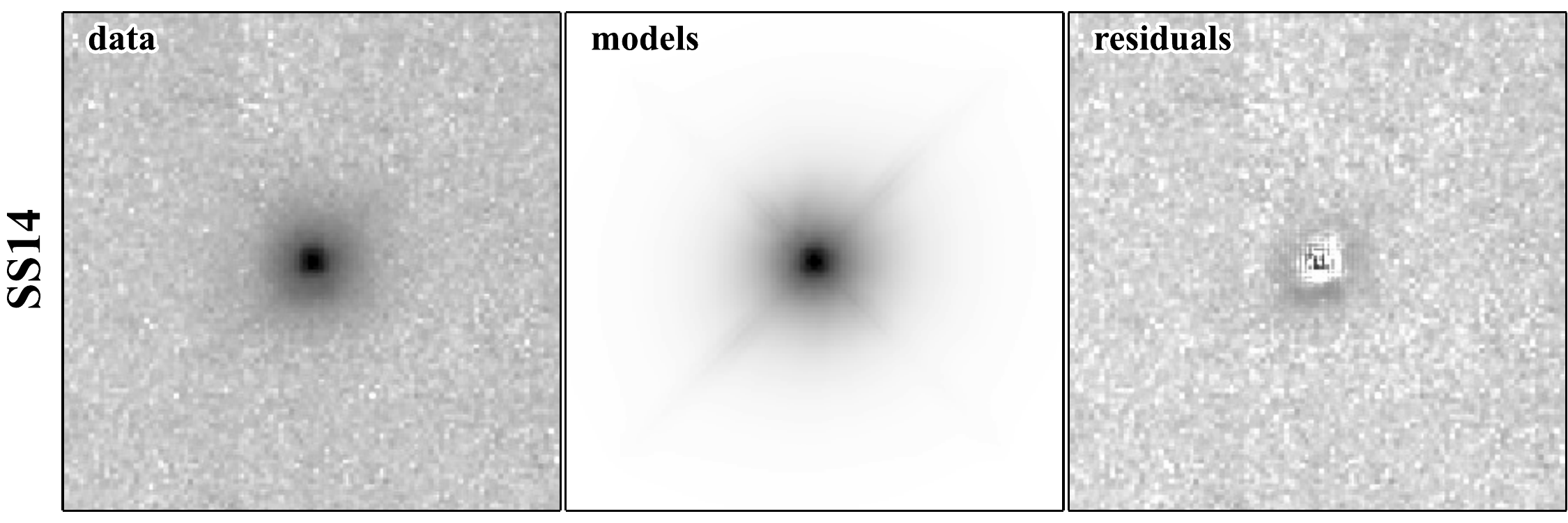}\includegraphics[width=0.25\textwidth]{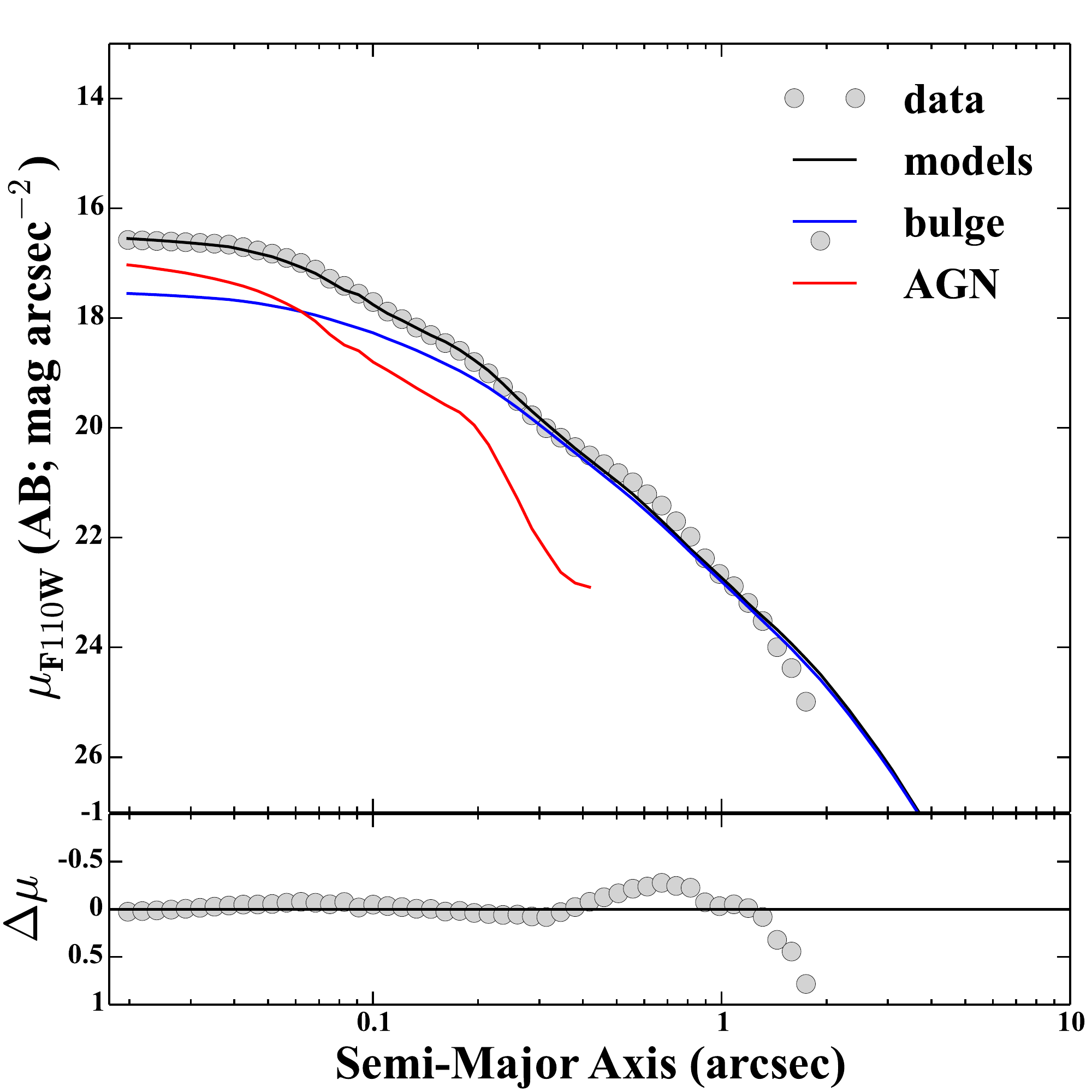}\\
	\includegraphics[width=0.75\textwidth]{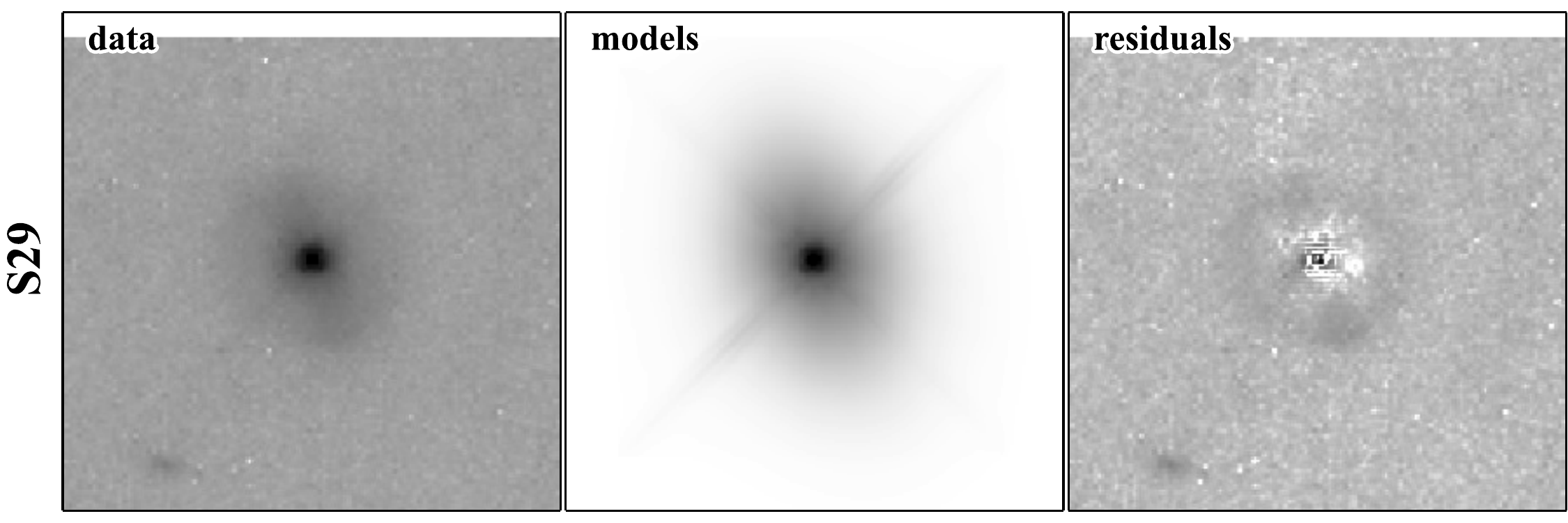}\includegraphics[width=0.25\textwidth]{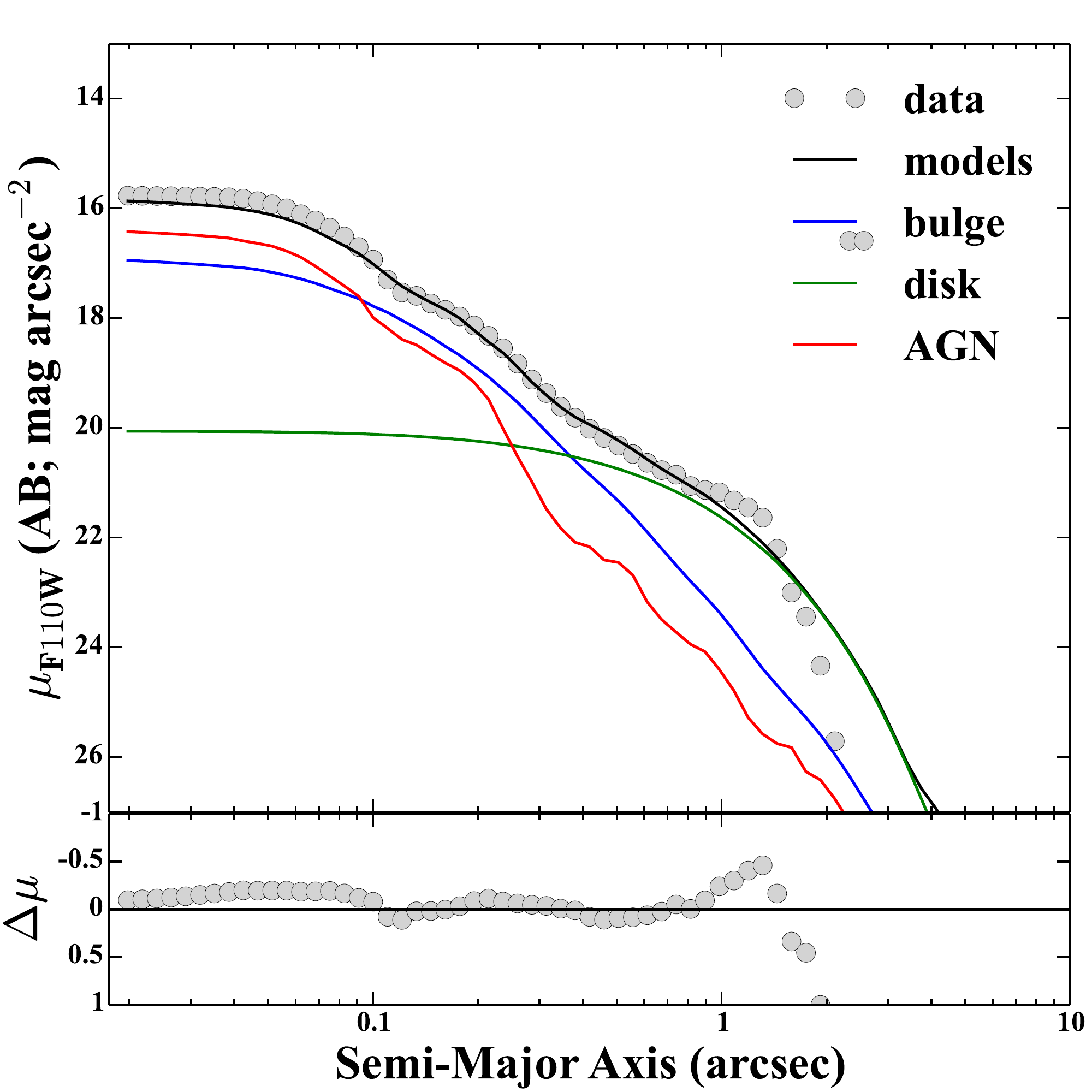}\\
	\includegraphics[width=0.75\textwidth]{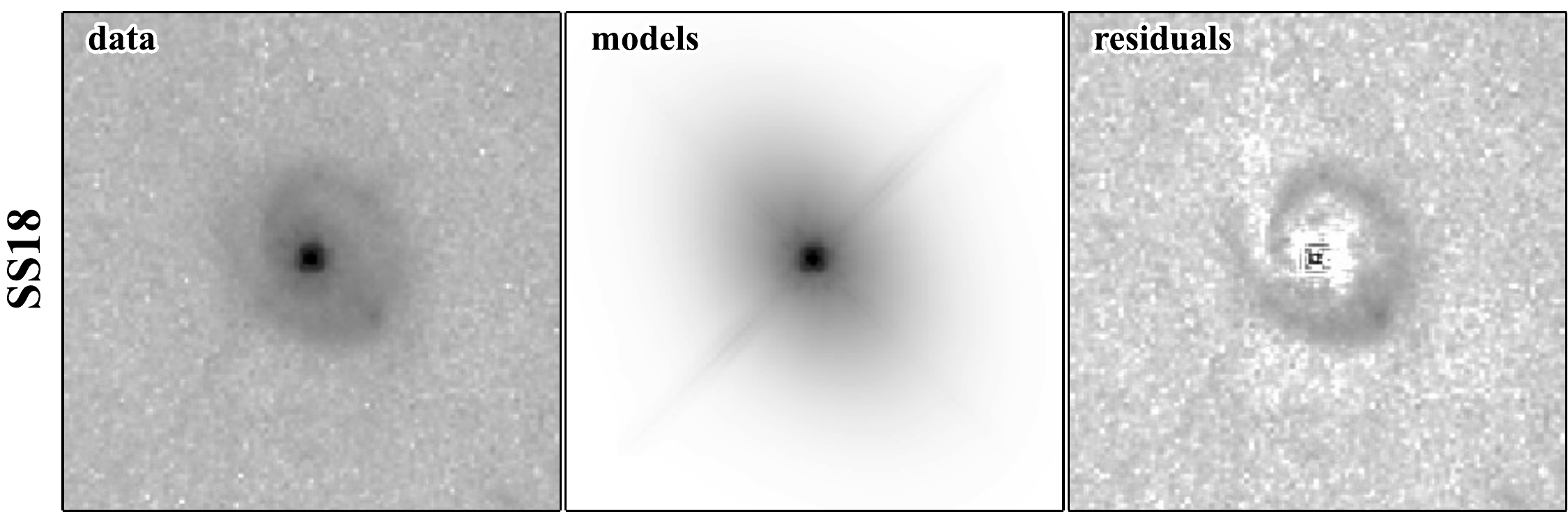}\includegraphics[width=0.25\textwidth]{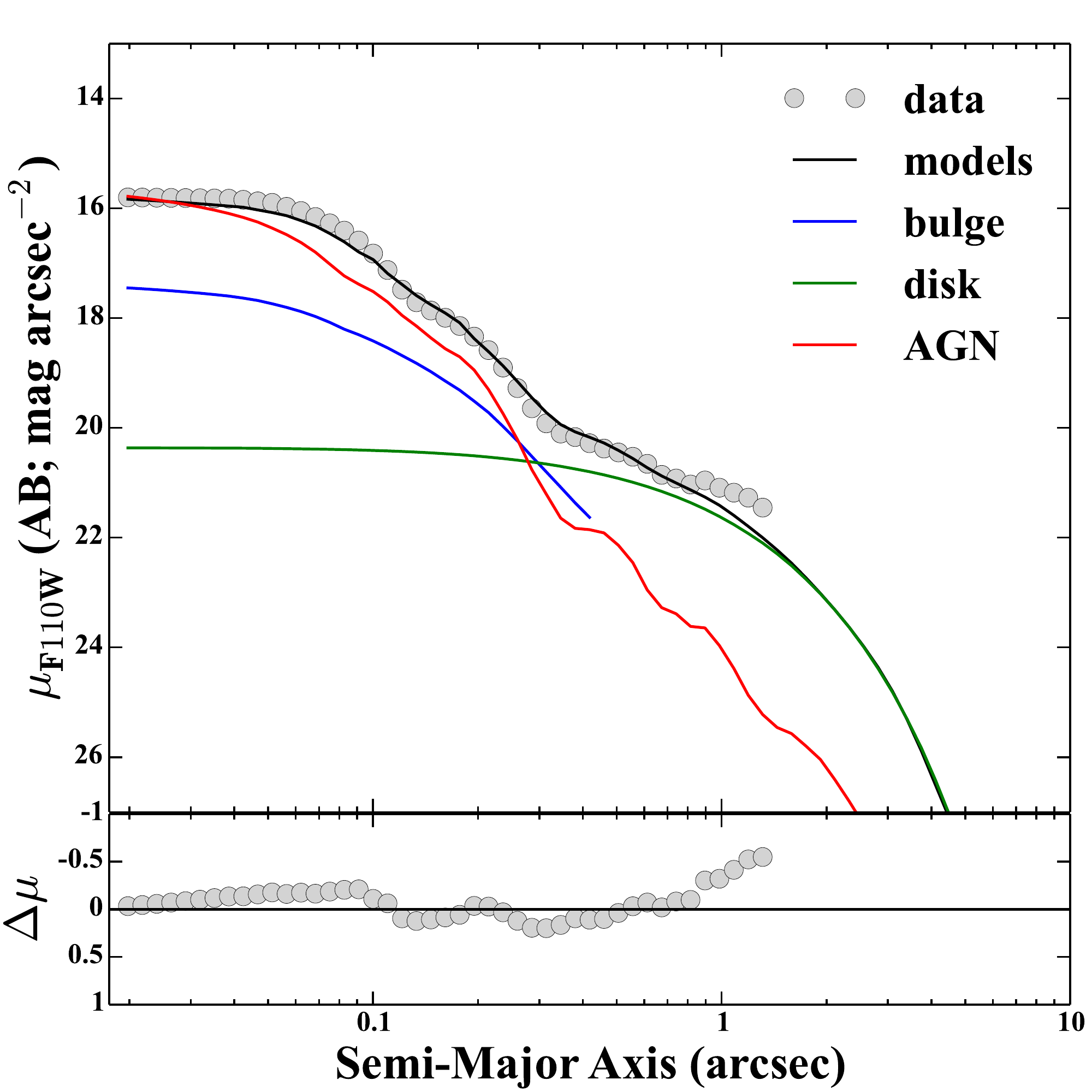}\\
	\includegraphics[width=0.75\textwidth]{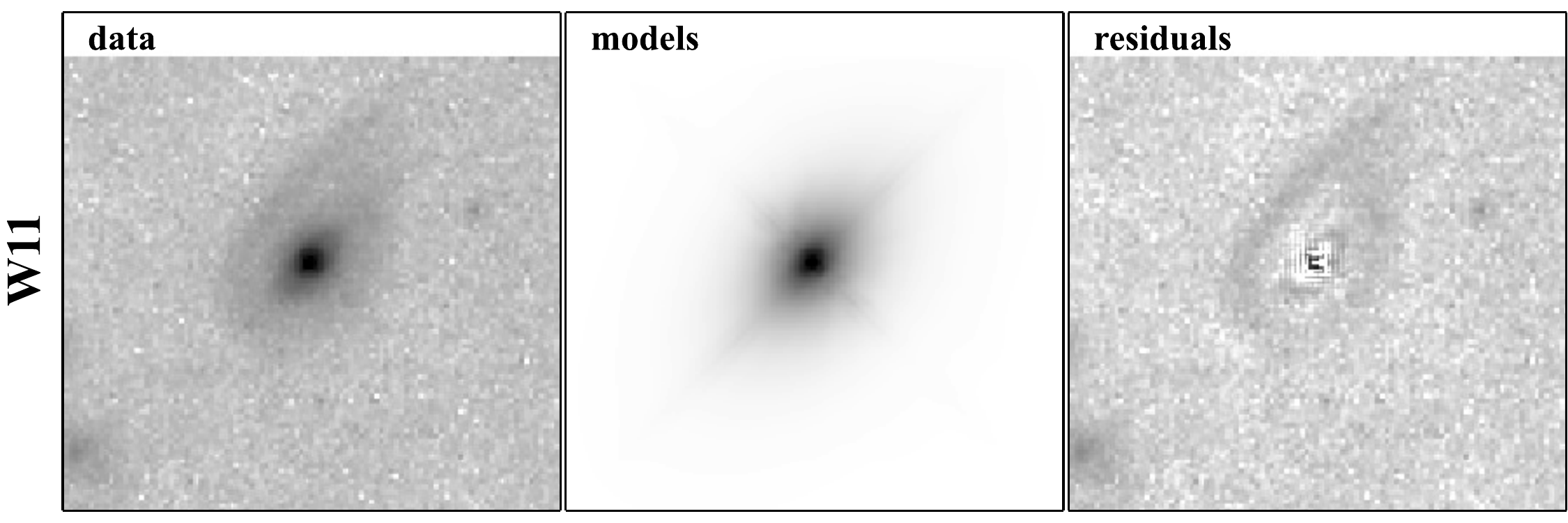}\includegraphics[width=0.25\textwidth]{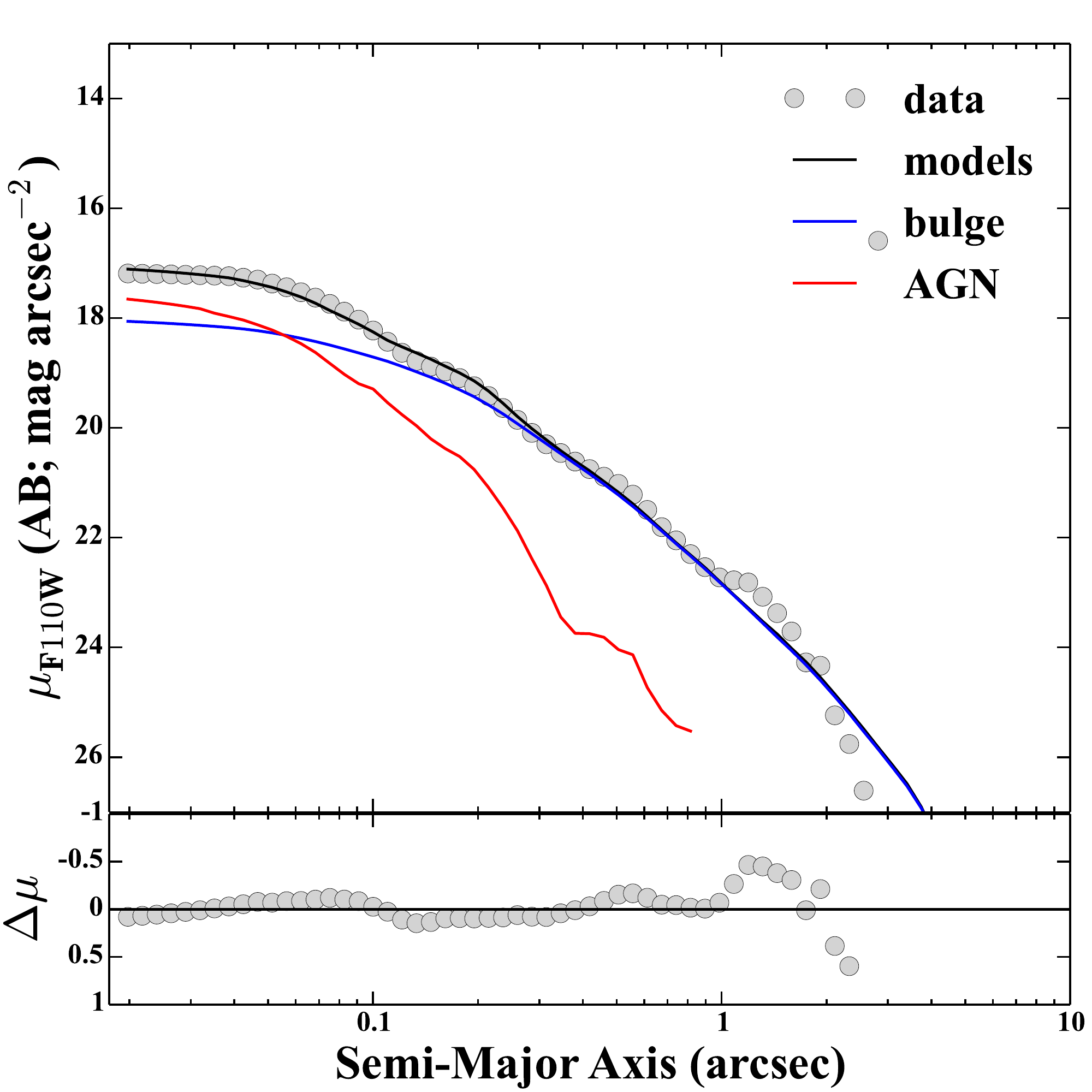}
    \figurenum{A\-2}
        \caption{
        \it Continued.
        \label{fig:imgfit1dSBP_pre40_7}}
\end{figure*}

\begin{figure*}
\centering
	\includegraphics[width=0.75\textwidth]{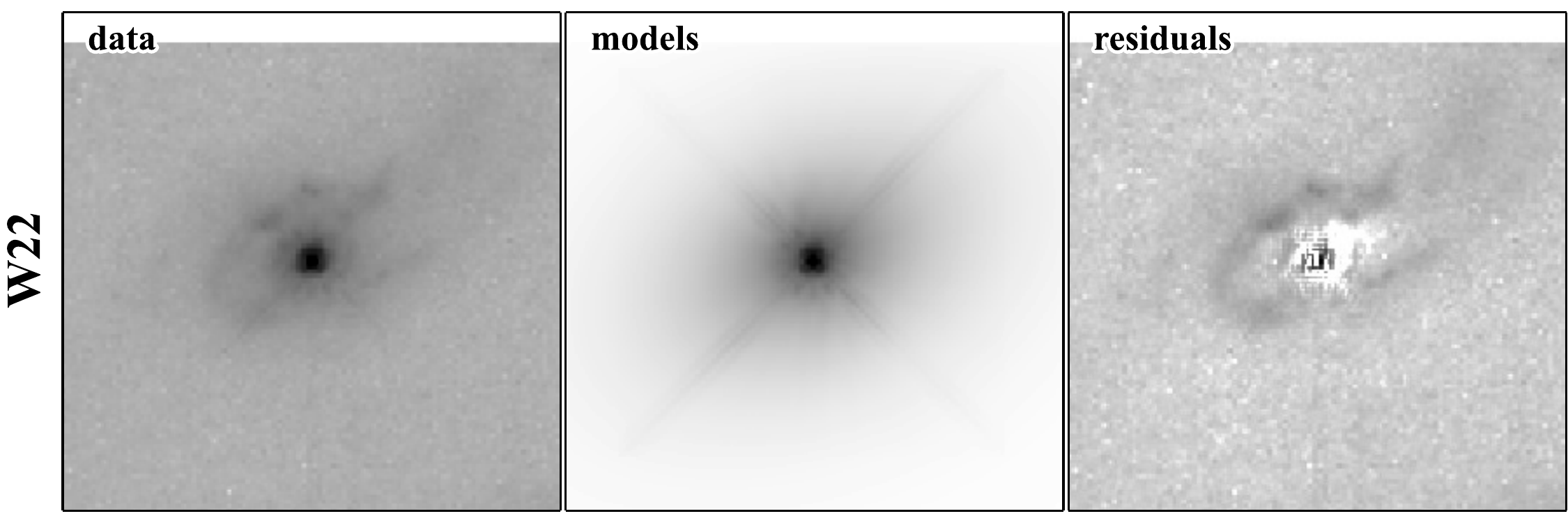}\includegraphics[width=0.25\textwidth]{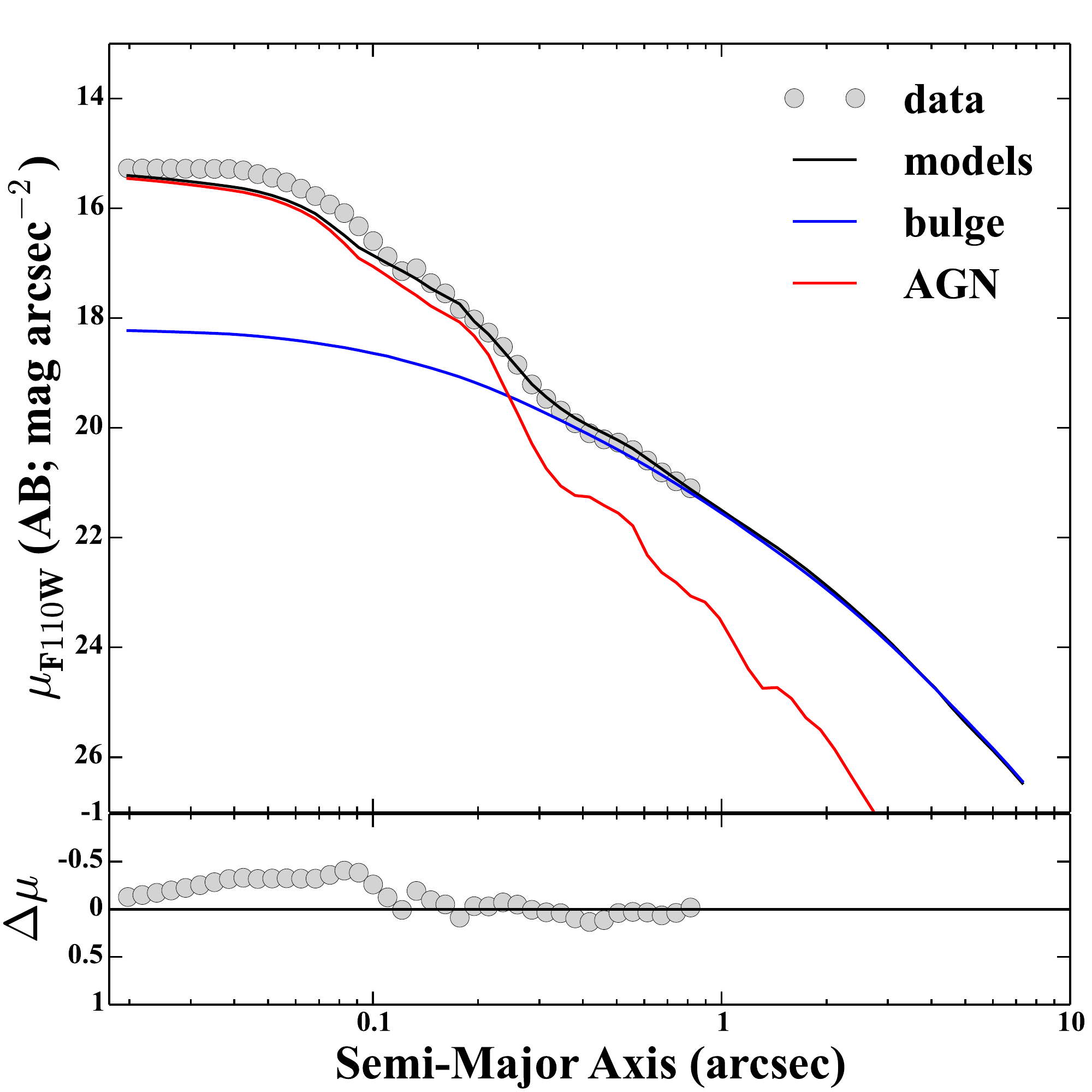}\\
	\includegraphics[width=0.75\textwidth]{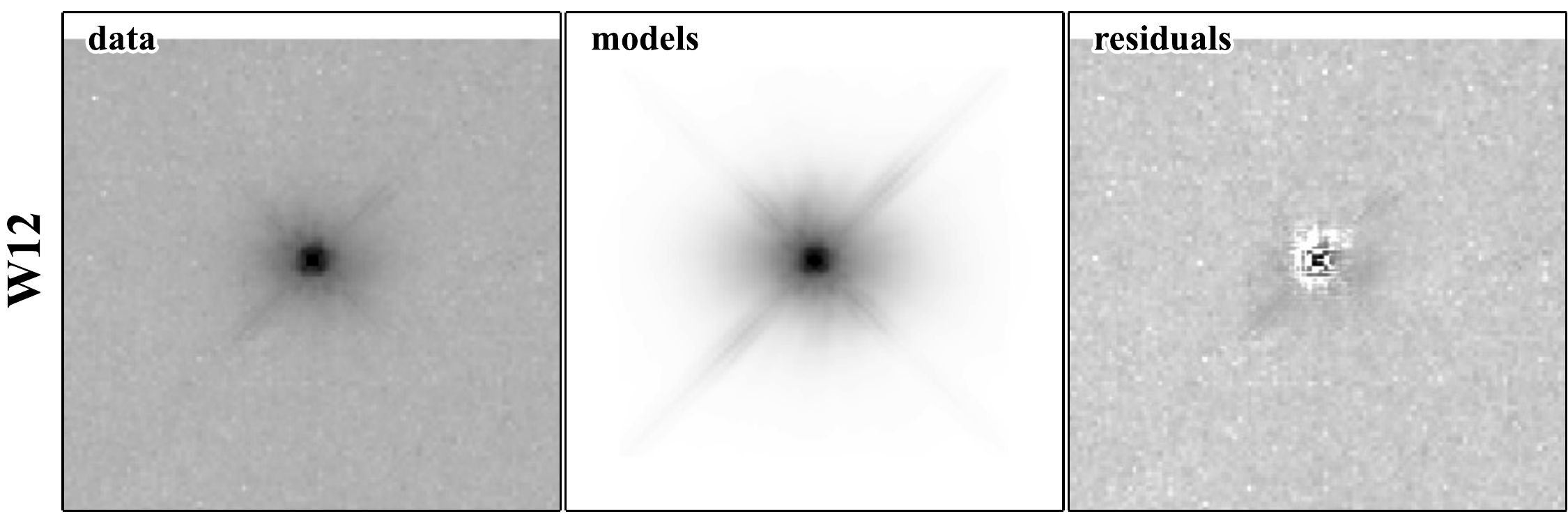}\includegraphics[width=0.25\textwidth]{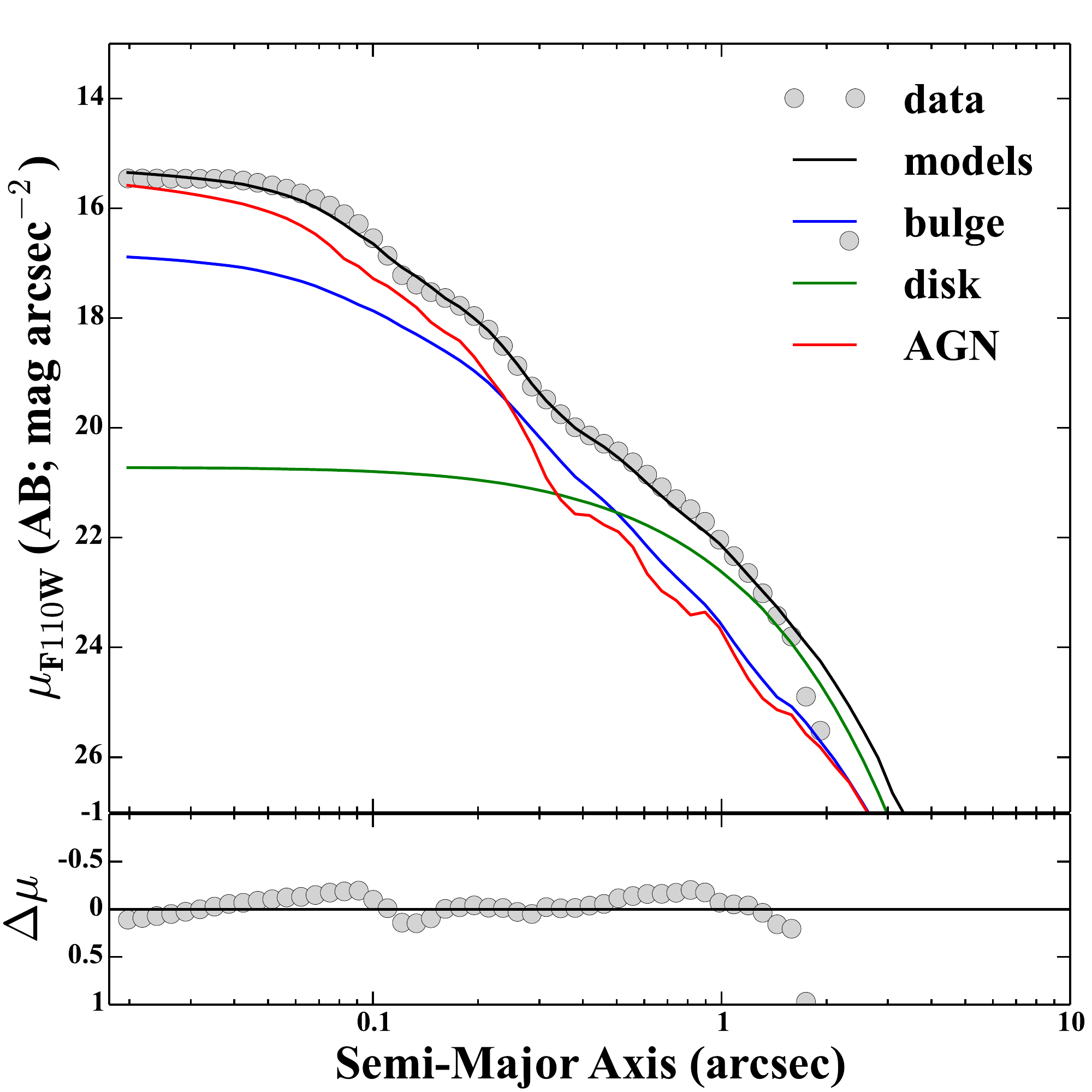}\\
	\includegraphics[width=0.75\textwidth]{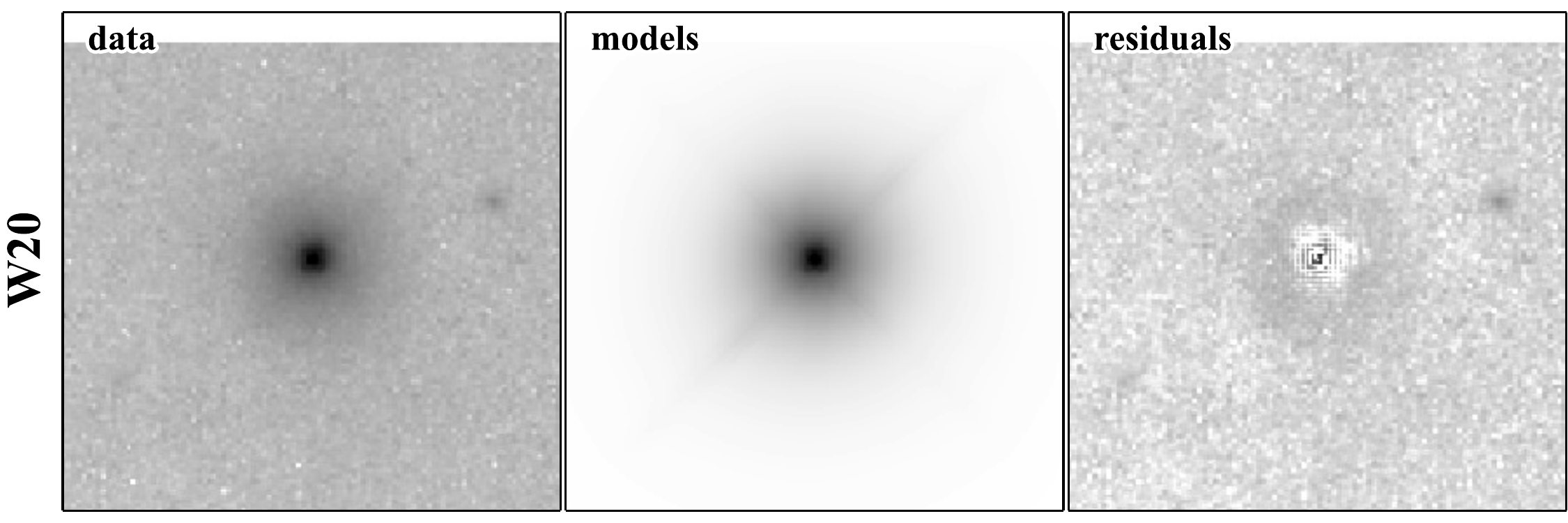}\includegraphics[width=0.25\textwidth]{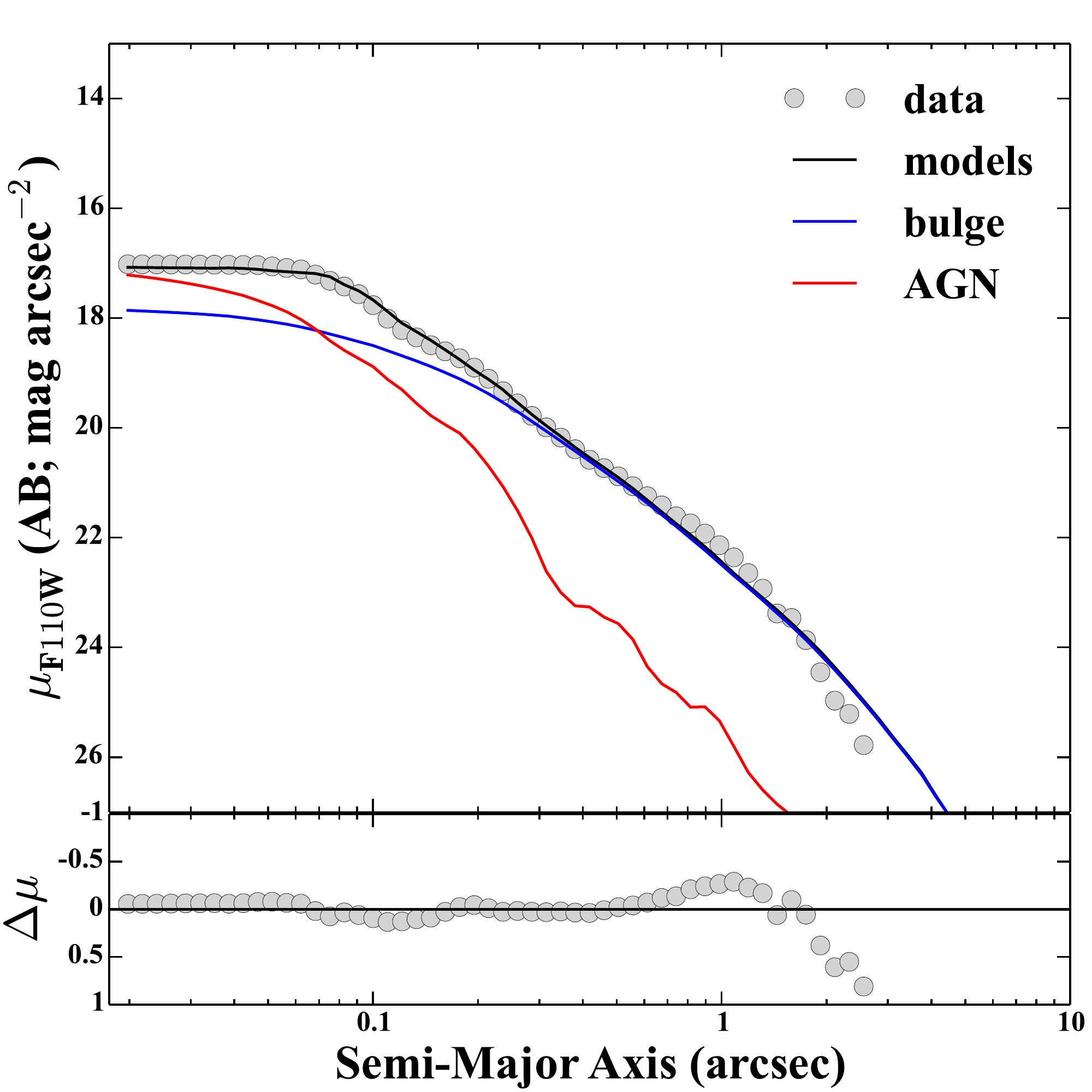}\\
	\includegraphics[width=0.75\textwidth]{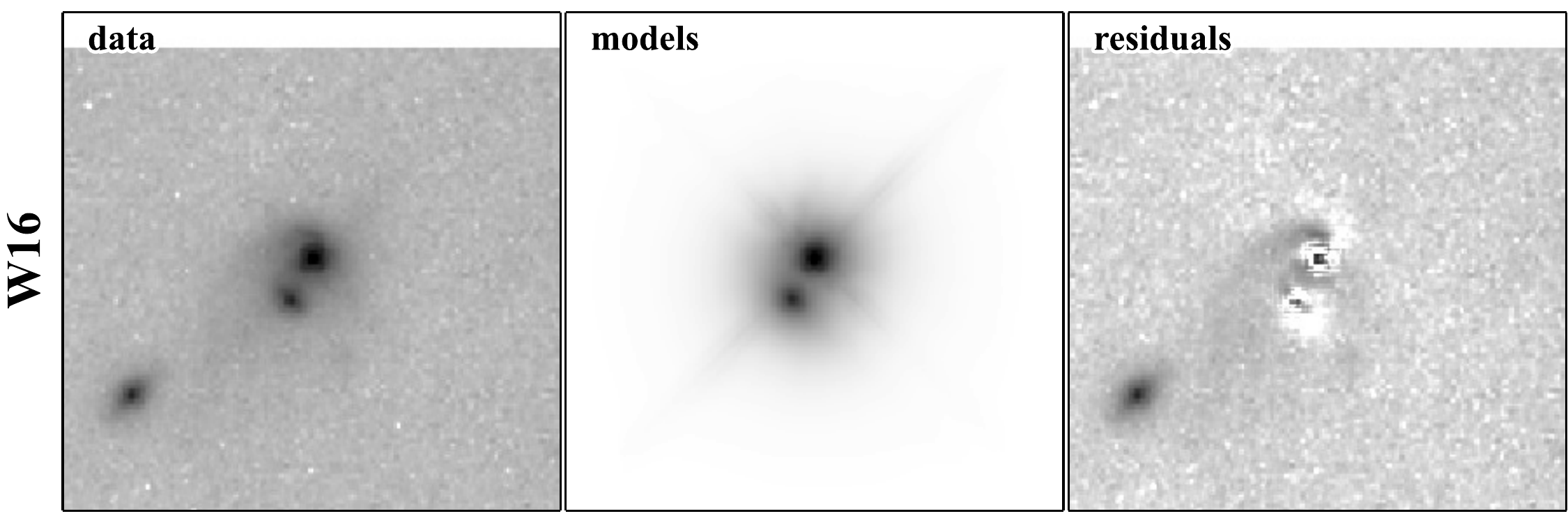}\includegraphics[width=0.25\textwidth]{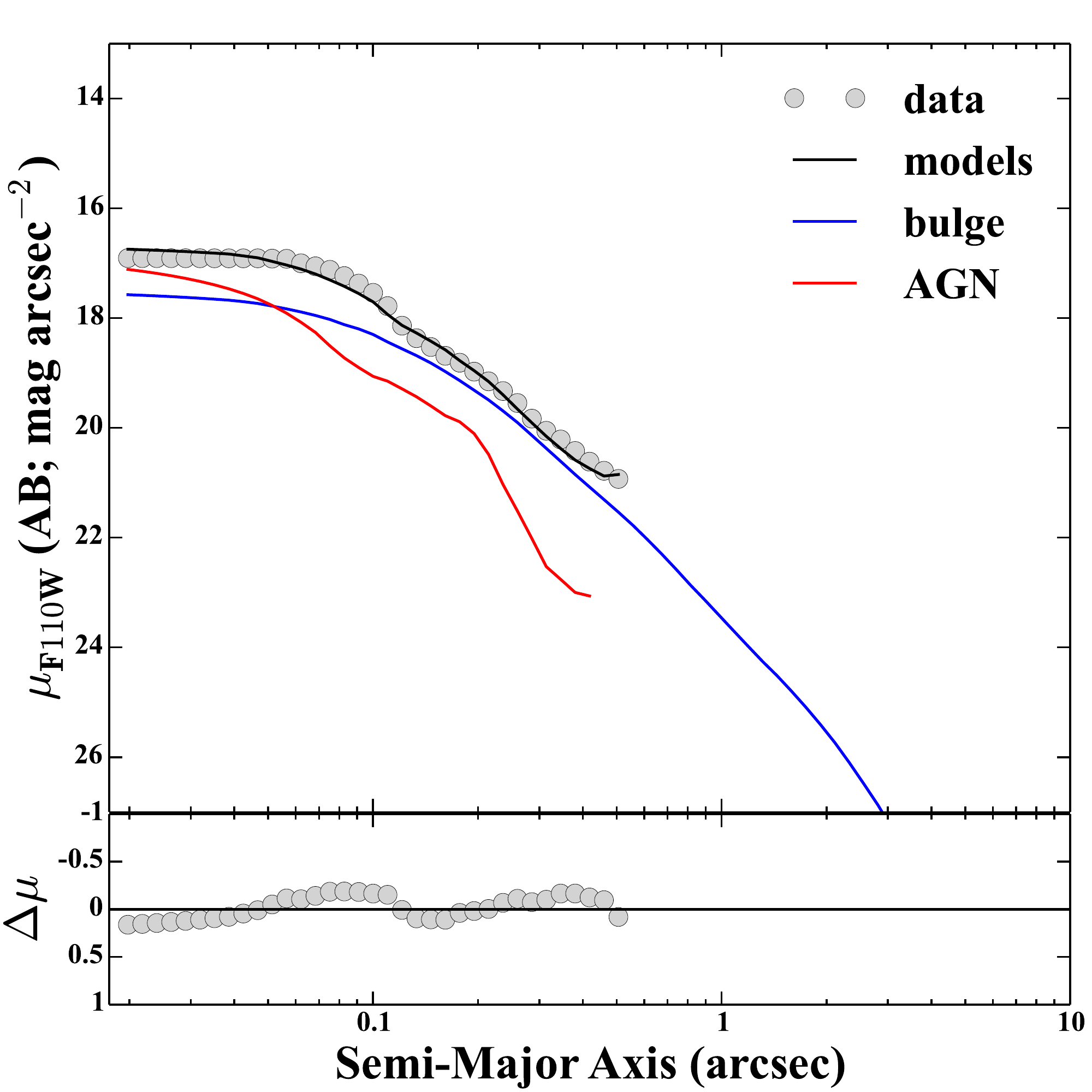}\\
	\includegraphics[width=0.75\textwidth]{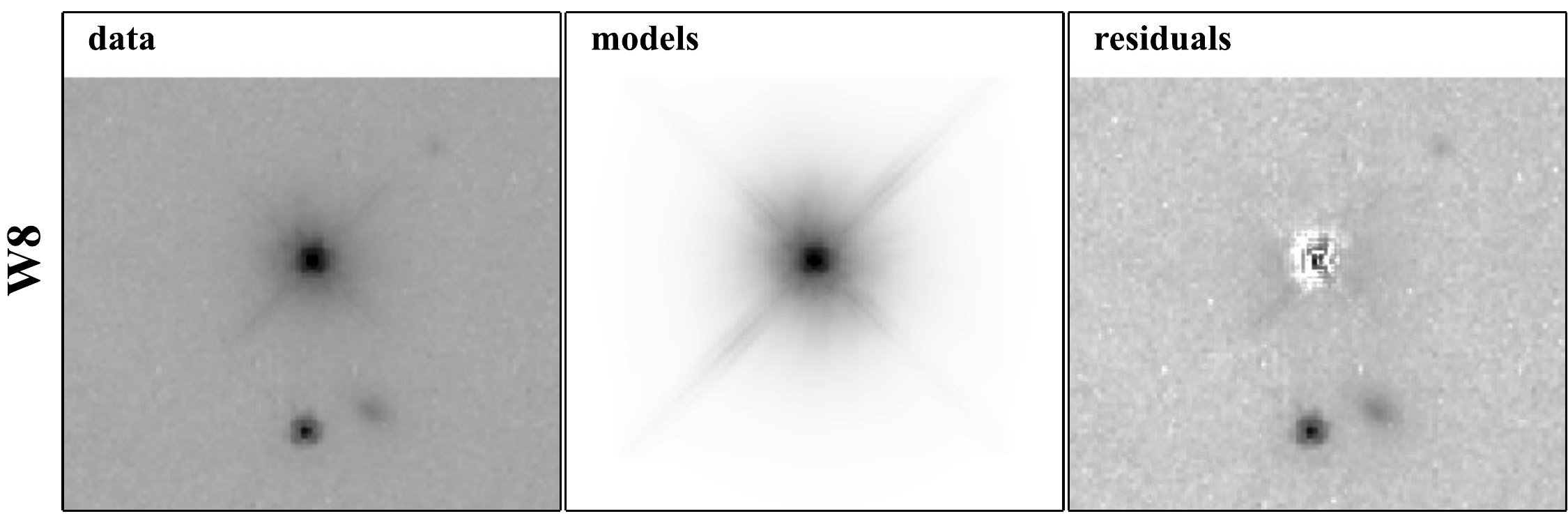}\includegraphics[width=0.25\textwidth]{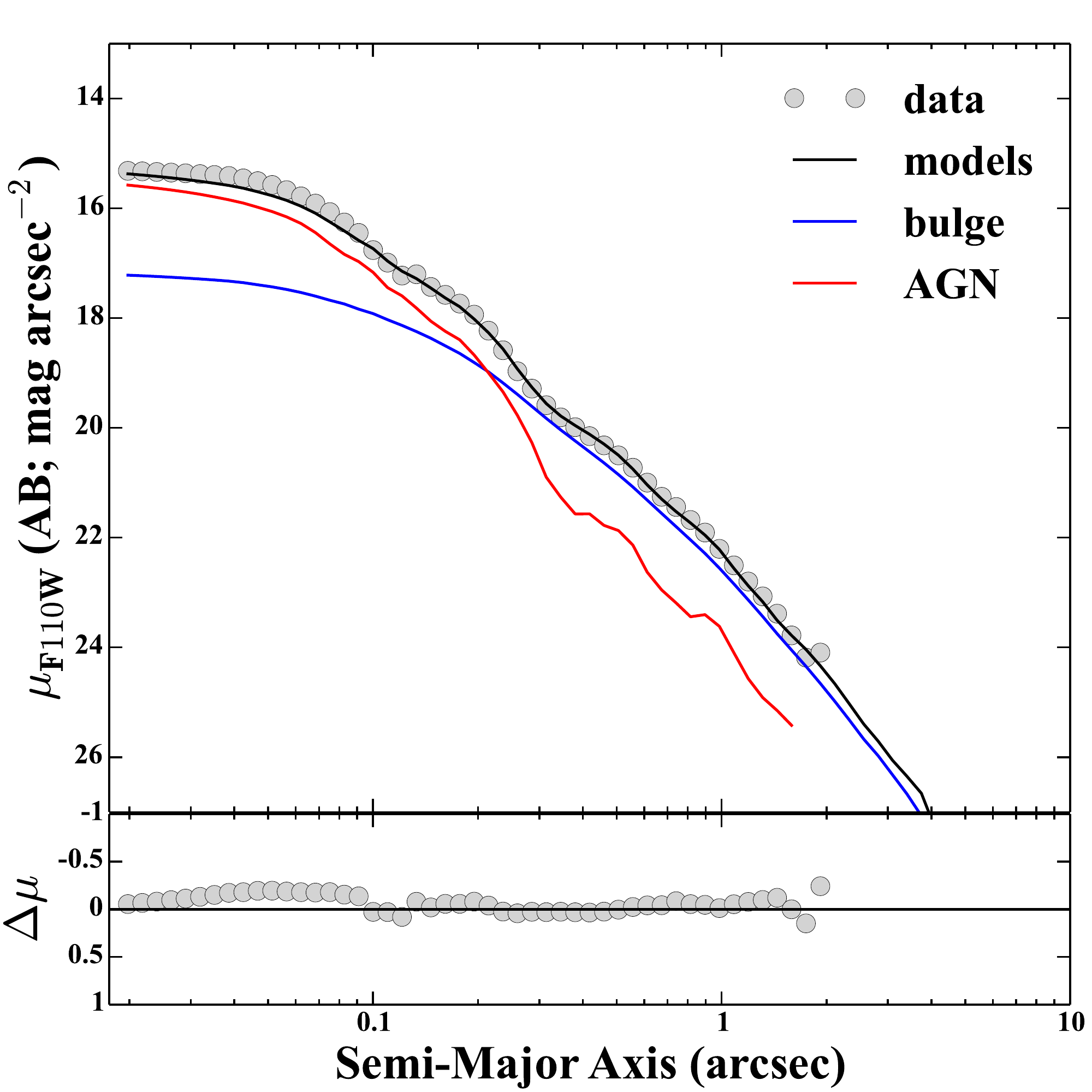}
    \figurenum{A\-2}
        \caption{
        \it Continued.
        \label{fig:imgfit1dSBP_pre40_8}}
\end{figure*}

\begin{figure*}
\centering
    \includegraphics[width=0.45\textwidth]{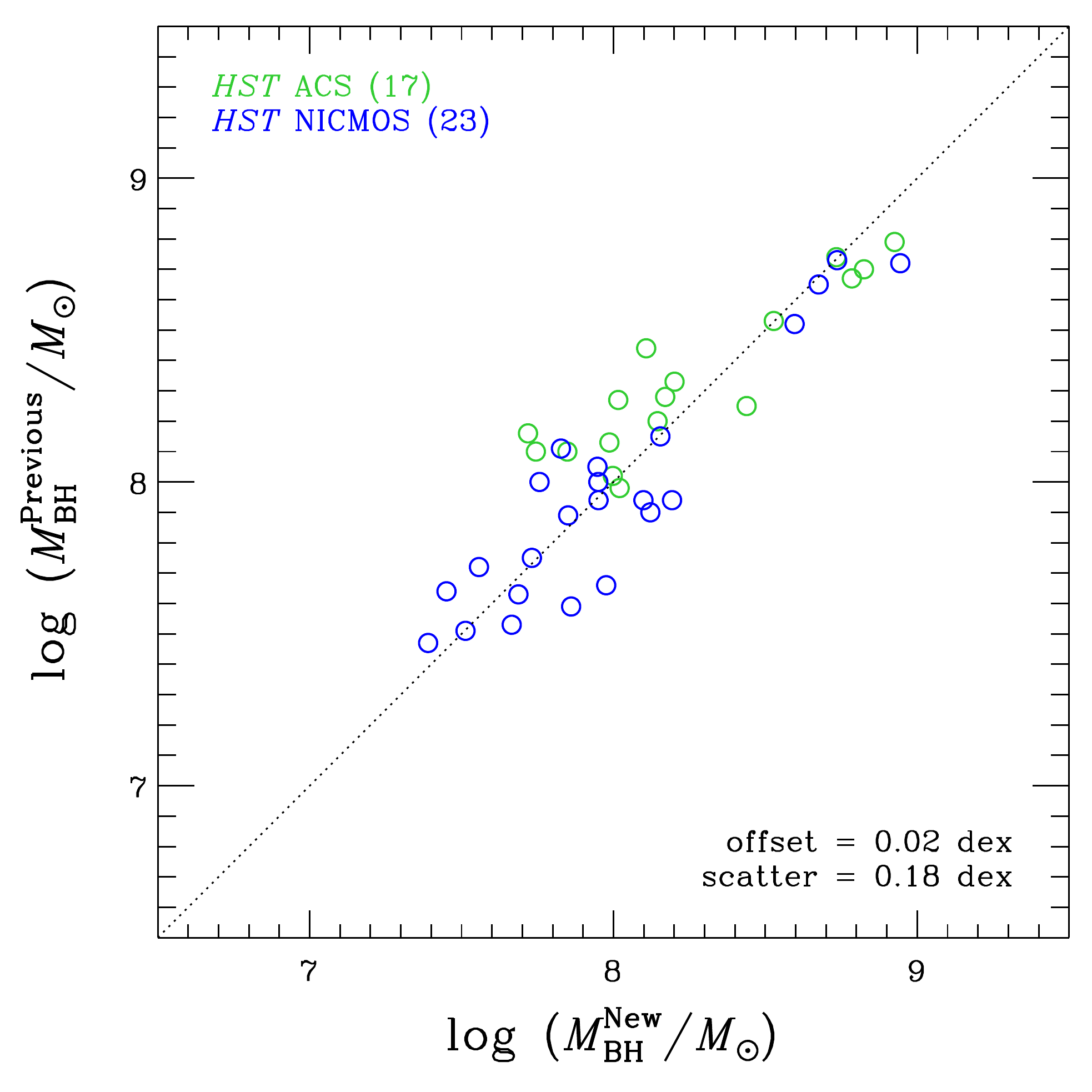}
    \includegraphics[width=0.45\textwidth]{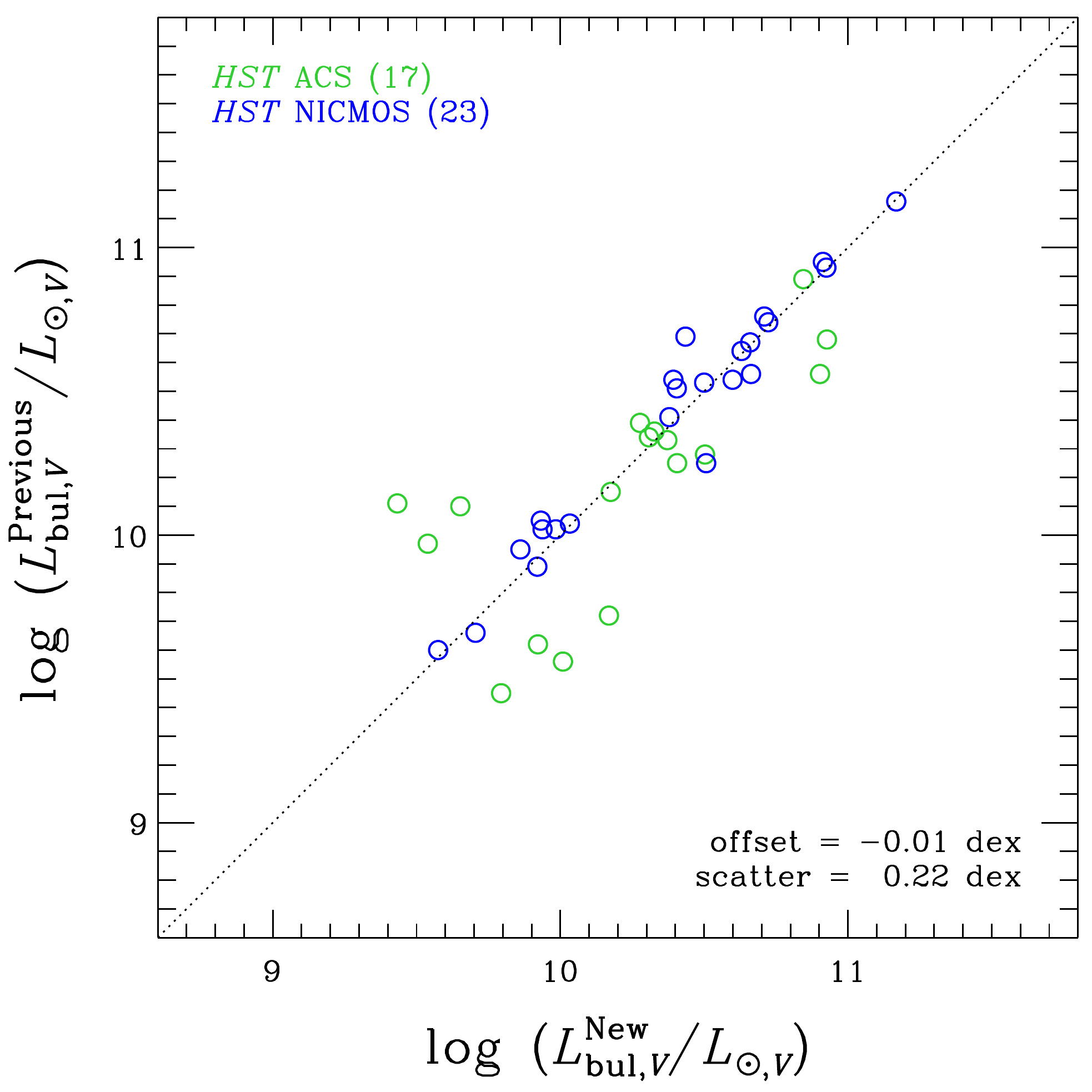}
    \caption{
    Difference of BH mass estimates (left) and bulge luminosity estimates (right) between previous results
\citep{Bennert+10} and new results presented here.
    \label{fig:compare_pre_new}}
\end{figure*}

\section{Comparison between $\lambda L_{5100}^{\lowercase{\rm spec}}$ and $\lambda L_{5100}^{\lowercase{\rm image}}$}\label{app:compare_L5100}

\begin{figure}
\centering
    \includegraphics[width=0.45\textwidth]{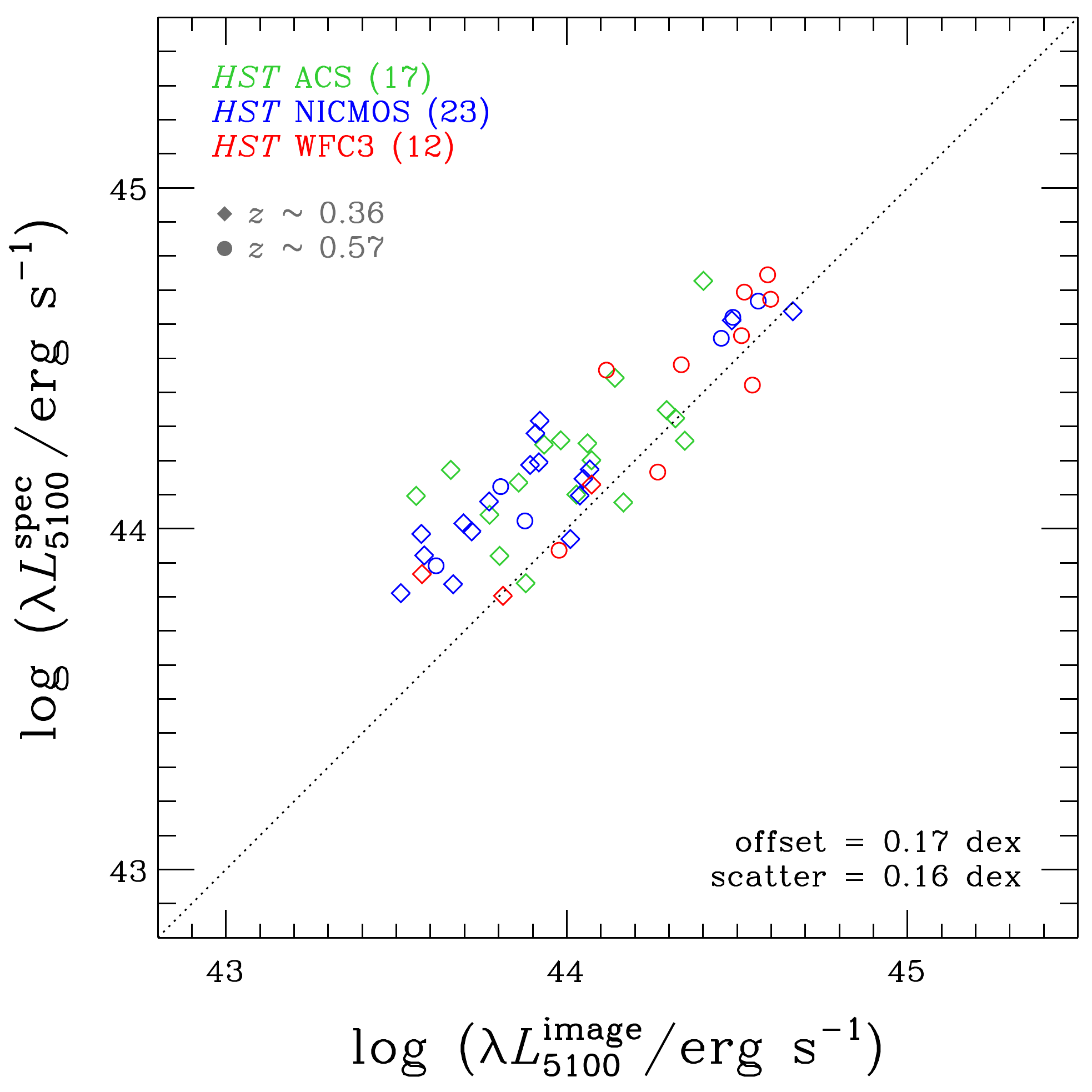}
    \caption{
    Difference of AGN continuum luminosity estimates from Keck spectra and \HST\ images for all 52 objects.
    \label{fig:compare_L5100}}
\end{figure}

Figure~\ref{fig:compare_L5100} compares AGN continuum luminosities, $\lambda L_{5100}$, measured from spectra and images. 
There are considerable offset and scatter between them due to several possible reasons.
In addition to AGN intrinsic variability and seeing effects,
the adopted single power-law SED when converting PSF magnitudes into luminosities at $5100$ \AA\ (Sec.~\ref{sec:Lbul}) will contribute some amount of the scatter. 
The AGN continuum luminosities measured from spectra are on average larger than those from images by $\sim0.17$ dex.
This is probably because the AGN luminosity measured from spectra could be overestimated from the different contribution of host galaxy starlights, which is stemming from aperture size difference between Keck slit and Sloan fiber spectra when performing flux (re-)calibration (Sec.~\ref{sec:obs_reduc}). Although the scatter between AGN luminosities estimated from spectra and images is reduced significantly (by $\sim0.2$ dex) after the renormalization, the overall flux scale could be increased against the genuine value due to the smaller contribution of host galaxy in Keck spectra than that of Sloan spectra if the amount of AGN variability is marginal.
There is another possibility of the overestimation when performing spectral decomposition in that the AGN power-law model could be contaminated with the contribution from young stellar population (if any) since it is not possible to decompose it unambiguously with this limited wavelength range of the spectra.

\end{appendix}
\end{CJK*}
\end{document}